\pdfoutput=1
\newcommand*{\ATLASLATEXPATH}{./}
\documentclass[PAPER, cernpreprint, atlasdraft=false, texlive=2016, UKenglish]{\ATLASLATEXPATH atlasdoc}

\usepackage{\ATLASLATEXPATH atlaspackage}
\usepackage{\ATLASLATEXPATH atlasbiblatex}

\usepackage{\ATLASLATEXPATH atlasphysics}

\usepackage{\ATLASLATEXPATH atlascontribute}

\graphicspath{{logos/}{./}}

\makeatletter
\DeclareOldFontCommand{\rm}{\normalfont\rmfamily}{\mathrm}
\DeclareOldFontCommand{\sf}{\normalfont\sffamily}{\mathsf}
\DeclareOldFontCommand{\tt}{\normalfont\ttfamily}{\mathtt}
\DeclareOldFontCommand{\bf}{\normalfont\bfseries}{\mathbf}
\DeclareOldFontCommand{\it}{\normalfont\itshape}{\mathit}
\DeclareOldFontCommand{\sl}{\normalfont\slshape}{\@nomath\sl}
\DeclareOldFontCommand{\sc}{\normalfont\scshape}{\@nomath\sc}
\makeatother

\usepackage{\ATLASLATEXPATH atlaspackage}
\usepackage{\ATLASLATEXPATH atlascover}
\usepackage{amssymb}
\usepackage{graphicx}
\usepackage{bm}
\usepackage{multirow}
\usepackage{bigdelim}
\usepackage{bigstrut}
\usepackage{multicol}
\usepackage{verbatim}
\usepackage{subfig}
\usepackage{amsmath}
\usepackage{rotating}
\usepackage{a4wide}

\AtlasTitle{Measurements of Higgs boson properties in the diphoton decay channel with 36 fb$^{-1}$ of $pp$ collision data at $\sqrt{s} = 13\,\TeV$ with the ATLAS detector}

\author{The ATLAS Collaboration}

\AtlasNote{HIGG-2016-21}

\PreprintIdNumber{CERN-EP-2017-288}

\AtlasJournalRef{Phys. Rev. D 98 (2018) 052005}
\AtlasDOI{10.1103/PhysRevD.98.052005}

\arXivId{1802.04146}

\AtlasAbstract{\footnotesize
Properties of the Higgs boson are measured in the two-photon final state using \lumi\,fb$^{-1}$ of proton--proton
collision data recorded at $\sqrt{s}=$ 13\,\TeV\ by the ATLAS experiment at the Large Hadron Collider. 
Cross-section measurements for the production of a Higgs boson through 
gluon--gluon fusion, vector-boson fusion, and in association with a vector boson
or a top-quark pair are reported. The signal strength, defined as the ratio of the observed to the expected
signal yield, is measured for each of these production processes as well as inclusively. The global signal strength
measurement of $0.99 \pm 0.14$ improves on the precision of the ATLAS measurement at $\sqrt{s}=7$ and 8~\TeV\ by a factor of two. 
Measurements of gluon--gluon fusion and vector-boson fusion productions yield signal strengths compatible with the Standard Model prediction.
Measurements of simplified template cross sections, designed to quantify the different Higgs boson production processes in specific
regions of phase space, are reported. 
The cross section for the production of the Higgs boson decaying to two isolated photons in a fiducial region closely matching
the experimental selection of the photons is measured to be $55 \pm 10 \fb$, which is in good agreement 
with the Standard Model prediction of $64 \pm 2 \fb$.
Furthermore, cross sections in fiducial regions enriched in Higgs boson production in vector-boson fusion or in
association with large missing transverse momentum, leptons or top-quark pairs are reported.
Differential and double-differential measurements are performed for several variables
related to the diphoton kinematics as well as the kinematics and multiplicity of the jets
produced in association with a Higgs boson. These differential cross sections are sensitive to
higher order QCD corrections and properties of the Higgs boson, such as its 
spin and CP quantum numbers. No significant deviations from a wide array of Standard Model predictions are observed.
Finally, the strength and tensor structure of the Higgs boson interactions are investigated
using an effective Lagrangian, which introduces additional CP-even and CP-odd interactions. 
No significant new physics contributions are observed.
}

\hypersetup{pdftitle={ATLAS document},pdfauthor={The ATLAS Collaboration}}

\usepackage{var-defs}

\begin{document}

\maketitle

\tableofcontents


\section{Introduction} \label{sec:intro}

In July 2012, the ATLAS~\cite{Aad:2008zzm} and CMS~\cite{Chatrchyan:2008aa} experiments announced the discovery of a
Higgs boson~\cite{Aad:2012tfa, Chatrchyan:2012xdj} using proton--proton collisions collected at center-of-mass
energies $\sqrt{s}=7~\TeV$ and 8~\TeV\ at the CERN Large Hadron Collider (LHC). Subsequent measurements of its
properties were found to be consistent with those expected for the Standard Model (SM) Higgs
boson~\cite{Khachatryan:2016vau} with a mass $m_H = 125.09 \pm 0.21{\rm (stat.)} \pm 0.11{\rm (syst.)}$~\GeV~\cite{Aad:2015zhl}.

Following the modifications of the LHC to provide proton--proton collisions at a center-of-mass energy of $\sqrt{s}= 13$~\TeV, the
Higgs sector can be probed more deeply:
the data set collected in 2015 and 2016 allows inclusive
Higgs boson measurements to be repeated with about two times better precision than
to those done at $\sqrt{s}=7$ and 8~\TeV\ with the Run~1 data set.
The increased center-of-mass energy results in much larger cross sections for events at high partonic
center-of-mass energy. This implies improved sensitivity to a variety of interesting physics processes, such as Higgs
bosons produced at high transverse momentum or Higgs bosons produced in association with a top--antitop quark pair. 
The Higgs boson decay into two photons ($H\rightarrow \gamma\gamma$) is a particularly attractive way to study the
properties of the Higgs boson and to search for deviations from the Standard Model predictions due to beyond-Standard
Model (BSM) processes. Despite the small branching ratio, $(2.27\pm 0.07)\times 10^{-3}$ for
$m_H=125.09$~\GeV~\cite{deFlorian:2016spz},
a reasonably large signal yield can be obtained thanks
to the high photon reconstruction and identification efficiency at the ATLAS experiment.
Furthermore, due to the excellent photon energy resolution of the ATLAS calorimeter,
the signal manifests itself as a narrow peak in the diphoton invariant mass (\mgg)
spectrum on top of a smoothly falling background, and the Higgs boson signal yield
can be measured using an appropriate fit to the \mgg\ distribution of the selected events.

In this paper, the results of measurements of the Higgs boson properties in the diphoton
decay channel are presented using \lumi~fb$^{-1}$ of $pp$ collision data collected at
$\sqrt{s}=13$~\TeV\ by the ATLAS detector in 2015 and 2016.
All the measurements are performed under the
assumption that the Higgs boson mass is $125.09$~\GeV,
and are compared to Standard Model predictions. 
Three types of measurements are presented in this paper and are summarized in the remainder of this section: 
(i) measurements of the total Higgs boson production-mode cross sections and ``signal strengths''; 
(ii) cross sections using the SM production modes as ``templates'' in simplified fiducial regions; and 
(iii) measurements of integrated or differential cross sections
in fiducial phase-space regions closely matched to the experimental selection.

The rest of this paper is organized as follows.
Section~\ref{sec:detector} provides a brief description of the ATLAS
detector, and Section~\ref{sec:dataset} describes the selected data set.
The generation of simulated event samples is described in Section~\ref{sec:mc}.
Section~\ref{sec:evsel} gives an overview of the event reconstruction and selection, and
Section~\ref{sec:sigext} explains the signal and background modeling used in the measurement.
The sources of systematic uncertainties are detailed in Section~\ref{sec:syst}.
Section~\ref{sec:methods_coup} describes the measurement of the
total Higgs boson production-mode cross sections, signal strengths, and simplified template cross sections (STXS).
Similarly, Section~\ref{sec:methods_fid} describes the measurement of the fiducial and differential cross sections.
Section~\ref{sec:conclusion} concludes with a brief summary of the main findings.

\subsection{Higgs boson production-mode cross sections and signal strengths}\label{subsec:xsec_intro}

In this paper, cross sections times branching ratio of the Higgs to two photons \bfhyy\ are
measured for inclusive Higgs boson production, as well as for several individual production processes:
gluon--gluon fusion (\ggH), vector-boson fusion (\VBF), Higgs
boson production in association with a vector boson (\VH), and
production of a Higgs boson in association with a top--antitop quark pair (\ttH) or a single top quark 
($t$-channel and W-associated, respectively denoted as \tHqb\ and \tHW, or in their sum as ``\tH'').
In the SM, gluon--gluon fusion is the dominant production mechanism at the LHC, contributing 
to about 87\% of the total cross section at $\sqrt{s} = 13$~\TeV~\cite{deFlorian:2016spz}. 
Vector-boson fusion and associated production with either a vector boson, with a top--antitop quark pair or a 
bottom--antibottom quark pair correspond to 6.8\%, 4.0\%, 0.9\%, and 0.9\%, respectively, of the total Higgs boson production 
cross section. 

The data are divided into 31 categories based on the reconstructed event properties to 
maximize the sensitivity to different production modes and the different regions of the 
simplified template cross sections, which are further described in Section~\ref{sec:intro_stxs}.
The categories are defined using the expected properties of the different production mechanisms:
10 categories aimed to measure gluon--gluon fusion properties, 4 categories to measure vector-boson fusion, 
8 categories that target associated production with vector bosons with different final states, 
and 9 categories that target associated production with a top--antitop quark pair or a single top-quark.
The definition of each category was optimized using simulated events
and a full summary of the categories can be found in Section~\ref{sec:methods_coup}.
In the sequence of the classification, priority is given to categories aimed at selecting
signal events from processes with smaller cross sections. 

In order to probe the production mechanisms independently of the $H \to \gamma\gamma$ branching ratio,
ratios of the different production-mode cross sections normalized to gluon--gluon fusion are also reported, with their full experimental
correlations. 
In addition, measurements of the signal strength $\mu$, which is the ratio of the measured cross section to the SM
prediction, are given for the different production processes as well as for the inclusive production. 
Finally, coupling-strength modifiers, which are scale factors of the tree-level Higgs boson couplings 
to the different particles or of the effective Higgs boson couplings to photons and gluons from 
loop-induced processes, are reported.

\subsection{Simplified template cross sections}\label{sec:intro_stxs}

The measurements of cross sections separated by the production mode
as presented in the previous section are extended to measurements in
specific regions of phase space using the framework of the ``simplified
template cross sections'' introduced in Refs.~\cite{LesHouches,deFlorian:2016spz}. These are
reported as cross section times $B(H \to \gamma\gamma)$ for a Higgs boson absolute rapidity 
\footnote{The ATLAS experiment uses a right-handed coordinate system with its origin at the nominal interaction point (IP) in the
center of the detector and the $z$-axis along the beam pipe. The $x$-axis points from the IP to the center of the LHC
ring, and the $y$-axis points upward. Cylindrical coordinates $(r,\phi)$ are used in the transverse plane, $\phi$ being
the azimuthal angle around the $z$-axis. The pseudorapidity is defined in terms of the polar angle $\theta$ as
$\eta=-\ln\tan(\theta/2)$. When dealing with massive particles, the rapidity $y=1/2 \ln[(E+p_z)/(E-p_z)]$ is used,
where $E$ is the energy and $p_z$ is the $z$-component of the momentum. Angular separation is expressed in terms of $\Delta R = \sqrt{ \left(\Delta \eta\right)^2 + \left(\Delta \phi\right)^2 }$.} 
$|y_H|$ less than 2.5 and with further particle-level requirements.
The different production modes are
separated in a theoretically motivated way using the SM modes \ggH,
\VBF, \VH\ and top-quark-associated production modes as ``templates''. The fiducial regions are defined in a
``simplified'' way using the measured kinematics and topology of the final
state, defined by the Higgs boson, the hadronic jets and the vector bosons or
top quarks in the event, to avoid large model-dependent extrapolations. The Higgs
boson is treated as a stable final-state particle, which allows an
easy combination with other decay channels. Similarly,
vector bosons or top quarks are treated as stable particles, but the cases of
leptonic and hadronic decays of the vector boson are distinguished.

In this paper a merged version of the so-called ``stage-1'' simplified template cross-section measurements are investigated.
These measurements provide more information for theoretical reinterpretation
compared to the signal strength measurements used in Run 1 and are defined 
to reduce the theoretical uncertainties typically folded into the signal strength results. 
In the full stage-1 proposal, template cross sections would be measured
in 31 regions of phase space for $|y_H|<2.5$, where the latter requirement reflects the experimental acceptance. 
The experimental categories used in this study (the same as those used for the
signal strength measurements) have been optimized to provide the
maximum sensitivity to such regions~\cite{LesHouches,deFlorian:2016spz}.

Since the current data set is not large enough to probe all of the stage-1 cross sections
with sufficiently small statistical uncertainties, regions with poor sensitivity
or with large anti-correlations are merged together into ten regions: Six 
regions probe gluon-fusion Higgs boson production with zero, one, and two jets associated with them.
Two regions probe \VBF\ Higgs boson production and Higgs boson production associated with vector bosons that
decay hadronically. A dedicated cross section is measured for Higgs boson production associated with vector bosons
that decay via leptonic modes.
The final cross section measures top-associated (\ttH\ and \tH) Higgs-boson production. To retain sensitivity to beyond the Standard Model Higgs boson production
, the $\ge 1$ jet, $\pT^H>\nobreak200 \,\GeV$ gluon--gluon fusion
and $\pT^j>\nobreak200 \, \GeV$ \VBF+\VH\ regions are not merged with other regions.
Here $\pT^H$ and $\pT^j$ denote the Higgs boson and leading jet transverse momenta, respectively, where the leading jet
is the highest transverse momentum jet in a given event. However, due to their large anti-correlation, only the cross section for the summed yield of these two regions is quoted here, and thus a total of nine kinematic regions are reported. The experimental sensitivity to the difference in the yields of these two regions is expected to be small, and the corresponding result is treated as a nuisance parameter rather than a measurement. 

Table~\ref{tab:STXS} summarizes the ten probes merged stage-1 cross sections and details which of the full 31  stage-1 cross sections were merged (middle and last column). A detailed description of the full 31 cross section proposal can be found in Appendix~\ref{sec:appendix_STXS}.

\begin{table}[!tp]
  \small
  \caption{The particle-level kinematic regions of the stage-1 simplified template cross sections,
    along with the intermediate set of regions used for the measurements presented in this paper.
    All regions require $|y_H|<2.5$. Jets are defined using the anti-$k_t$ algorithm with radius parameter $R=0.4$ and
    are required to have $\pT>30$~\GeV.
    The leading jet and Higgs boson transverse momenta are denoted by $\pT^j$ and $\pT^H$, respectively.
    The transverse momentum of the Higgs boson and the leading and subleading jet is denoted as $\pT^{Hjj}$ with the subleading
    jet being the second highest momentum jet in a given event. 
    Events are considered ``\VBF-like'' if they contain at least two jets with an invariant mass of $m_{jj} > 400$ \GeV\, and a rapidity 
    separation between the two jets of $|\Delta y_{jj}| > 2.8$. 
    Events are considered ``\VH-like'' if they contain at least two jets with an invariant mass of $60 \, \GeV < m_{jj} < 120 \, \GeV$.   
    All $qq' \to H qq'$ \VBF\ and \VH\ events (with the vector boson $V$ decaying hadronically) which are neither \VBF\ nor \VH-like 
    are part of the ``Rest'' selection.
    For the $\pT^H>200~\GeV$ gluon--gluon fusion and $\pT^j>200~\GeV$ \VBF\ + \VH\ regions, only the sum of
    the corresponding cross sections is reported while the difference of the two is profiled in the fit. 
    In total, the cross sections for nine kinematic regions are measured. The small contributions from \bbH\ are merged with \ggH.
    The process $gg\to ZH$ refers only to box and loop processes dominated by top and bottom quarks (see Section 4 for more details).
    }
\begin{center}
\begin{tabular}{lll}
\hline \hline
Process & Measurement region & Particle-level stage-1 region \\
\hline
\ggH\ + $gg\to Z(\to qq)H$ & 0-jet & 0-jet \\
                              & 1-jet, $p_{\textrm T}^H < 60$ \GeV & 1-jet, $p_{\textrm T}^H < 60$ \GeV \\
                              & 1-jet, $60 \leq p_{\textrm T}^H <120$ \GeV & 1-jet, $60 \leq p_{\textrm T}^H <120$ \GeV \\
                        & 1-jet, $120 \leq p_{\textrm T}^H <200$ \GeV& 1-jet, $120 \leq p_{\textrm T}^H <200$ \GeV \\
                        & $\geq 1$-jet, $p_{\textrm T}^H >200$ \GeV& 1-jet, $p_{\textrm T}^H >200$ \GeV \\
                        & & $\geq 2$-jet, $p_{\textrm T}^H >200$ \GeV \\
                        & $\geq 2$-jet, $p_{\textrm T}^H < 200$ \GeV or \VBF-like & $\geq 2$-jet, $p_{\textrm T}^H < 60$ \GeV \\
                        & & $\geq 2$-jet, $60 \leq p_{\textrm T}^H <120$ \GeV \\
                        & & $\geq 2$-jet, $120 \leq p_{\textrm T}^H <200$ \GeV \\
                        & & \VBF-like, $p_{\textrm T}^{Hjj} <25$ \GeV \\
                        & & \VBF-like, $p_{\textrm T}^{Hjj} \geq 25$ \GeV \\
\hline
$qq' \to H qq'$ (\VBF\ + \VH) & $p_{\textrm T}^j < 200$ \GeV & $p_{\textrm T}^j < 200$ \GeV, \VBF-like, $p_{\textrm T}^{Hjj} <25$ \GeV  \\
                             & & $p_{\textrm T}^j < 200$ \GeV, \VBF-like, $p_{\textrm T}^{Hjj} \geq 25$ \GeV \\
                             & & $p_{\textrm T}^j < 200$ \GeV, \VH-like                      \\
                             & & $p_{\textrm T}^j < 200$ \GeV, Rest                         \\
                             & $p_{\textrm T}^j > 200$ \GeV & $p_{\textrm T}^j > 200$ \GeV \\ 
\hline
$VH$ (leptonic decays) & $VH$ leptonic & $q\bar{q}\to ZH$, $\pT^Z<150$ \GeV \\
                       &               & $q\bar{q}\to ZH$, $150 <\pT^Z<250$ \GeV, 0-jet \\
                       &               & $q\bar{q}\to ZH$, $150 <\pT^Z<250$ \GeV, $\ge 1$-jet \\
                       &               & $q\bar{q}\to ZH$, $\pT^Z>250$ \GeV\\
                       &               & $q\bar{q}\to WH$, $\pT^W<150$ \GeV \\
                       &               & $q\bar{q}\to WH$, $150 <\pT^W<250$ \GeV, 0-jet \\
                       &               & $q\bar{q}\to WH$, $150 <\pT^W<250$ \GeV, $\ge 1$-jet \\
                       &               & $q\bar{q}\to WH$, $\pT^W>250$ \GeV\\
                       &               & $gg\to ZH$, $\pT^Z<150$ \GeV \\
                       &               & $gg\to ZH$, $\pT^Z>150$ \GeV, 0-jet \\
                       &               & $gg\to ZH$, $\pT^Z>150$ \GeV, $\ge 1$-jet \\
\hline
Top-associated production & top          & \ttH \\
                          &              & $W$-associated \tH (\tHW) \\
                          &              & $t$-channel \tH (\tHqb) \\
\hline
\bbH                      & merged w/ \ggH & \bbH \\
\hline \hline
\label{tab:STXS}
\end{tabular}
\end{center}
\end{table}

\subsection{Fiducial integrated and differential cross sections}\label{subsec:fix_xsec_intro}

Fiducial integrated and differential cross sections have previously been measured at $\sqrt{s}=8$~\TeV\ in the
$H\rightarrow \gamma\gamma$ decay channel by both the ATLAS~\cite{Aad:2014lwa} and the CMS~\cite{Khachatryan:2015rxa}
Collaborations. In this paper, fiducial cross sections are determined in a variety of phase-space regions sensitive to
inclusive Higgs boson production and to explicit Higgs boson production mechanisms.
The measurement of these cross sections provides an alternative way to study the properties of the Higgs boson and to
search for physics beyond the Standard Model. For each fiducial region of an integrated cross-section measurement or bin
of a differential distribution, the $H \to \gamma\gamma$ signal is extracted using a fit to the corresponding diphoton
invariant mass spectrum. The cross sections are determined by correcting these yields for experimental inefficiencies and
resolution effects, and by taking into account the integrated luminosity of the data. No attempt is made to separate 
individual production modes in favor of presenting fiducial regions enriched with a given production mode.

The inclusive fiducial region is defined at the particle level by two photons, not originating from the decay of a
hadron, that have absolute pseudorapidity $\left|\eta \right| < 2.37$, excluding the region $1.37 < \left|\eta \right| <
1.52$,\footnote{This pseudorapidity interval corresponds to the transition region between the barrel and endcap
sections of the ATLAS electromagnetic calorimeter, see Section~\ref{sec:detector}.}
with the leading (subleading) photon transverse momentum greater than 35\% (25\%) of $m_{\gamma\gamma}$. The two
photons are required to be isolated from hadronic activity by imposing that the summed transverse momentum of charged
stable particles (with a mean lifetime that satisfies $c \tau>10$~mm) with $\pt>1~\GeV$, within a cone of $\Delta R= 0.2$ centered on the
photon direction, be less than 5\% of the photon transverse momentum. This selection is applied to all the presented
fiducial integrated and differential cross section results and the isolation criterion was tuned to mimic  
the detector level selection. One additional cross section and three cross-section limits
are reported in smaller fiducial regions sensitive to specific Higgs boson production mechanisms:
\begin{itemize} 
\item a VBF-enhanced region with two jets with large invariant mass and rapidity separation,
\item a region of events containing at least one charged lepton\footnote{In this paper reconstructed charged leptons denote electrons and muons.},
\item a region of events with large missing transverse momentum,
\item and a region of events with a topology matching the presence of a top--antitop quark pair.
\end{itemize}

The fiducial cross section for different jet multiplicities are reported and compared to several 
predictions. Eleven fiducial differential cross sections are reported, for events belonging to the inclusive fiducial
region as a function of the following observables: 
\begin{itemize} 
\item \ptgg\ and \ygg, the transverse momentum and rapidity of the diphoton system,
\item \ptjl\ and \yjl, the transverse momentum and rapidity of the leading jet,
\item \ptjsl\ and \yjsl, the transverse momentum and rapidity of the subleading jet,
\item \costhetastar, the cosine of the angle between the beam axis and the diphoton system in the Collins--Soper frame~\cite{Collins:1984kg},
\item \dphijj\ and \deltayjj, the difference in azimuthal angle and in rapidity between the leading and subleading jets,
\item \dphiggjj, the difference in azimuthal angle between the dijet system formed by the leading and subleading jets and the diphoton system,
\item and \mjj, the invariant mass of the leading and subleading jets.
\end{itemize} 
Seven additional variables are reported in Appendix~\ref{app:add_fid_meas}.
Inclusive Higgs boson production is dominated by gluon--gluon fusion, for which the transverse momentum of the Higgs boson is largely balanced by the
emission of soft gluons and quarks. Measuring \ptgg\ probes the perturbative QCD modeling of this production mechanism
which is mildly sensitive to the bottom and charm quark Yukawa couplings of the Higgs boson~\cite{Bishara:2016jga}. The
distribution at high transverse momentum is sensitive to new heavy particles coupling to the Higgs boson and to the top
quark Yukawa coupling. The rapidity distribution of the Higgs boson is also sensitive to the modeling of the gluon--gluon
fusion production mechanism, as well as to the parton distribution functions (PDFs) of the colliding protons. The
transverse momentum and absolute rapidity of the leading and subleading jets probe the perturbative QCD modeling and are
sensitive to the relative contributions of the different Higgs production mechanisms. The angular variables \costhetastar\ and
\dphijj\ are sensitive to the spin and CP quantum numbers of the Higgs boson. The dijet rapidity separation \deltayjj,
the dijet mass \mjj\, and the azimuthal difference between the dijet and diphoton system \dphiggjj\ are sensitive to the
\VBF\ production mechanism. All fiducial differential cross
sections are reported with their full statistical and experimental correlations and are compared to several 
predictions.

The strength and tensor structure of the Higgs boson interactions are investigated using an effective Lagrangian, which
introduces additional CP-even and CP-odd interactions that can lead to deviations in the kinematic properties and event rates of the
Higgs boson and of the associated jets from those in the Standard Model. This is done by a simultaneous fit to five
differential cross sections, which are sensitive to the Wilson coefficients of four dimension-six CP-even or CP-odd
operators of the Strongly Interacting Light Higgs formulation~\cite{Giudice:2007fh}. A similar analysis was carried out
at $\sqrt{s}=8~\TeV$ by the ATLAS Collaboration~\cite{Aad:2015tna}.


\section{ATLAS detector} \label{sec:detector}

The ATLAS detector~\cite{Aad:2008zzm} covers almost the entire solid angle about the proton--proton interaction point.
It consists of an inner tracking detector, electromagnetic and hadronic calorimeters, and a muon spectrometer. 

Charged-particle tracks and interaction vertices are reconstructed using information from the inner detector (ID).
The ID consists of a silicon pixel detector (including the insertable B-layer \cite{ATLASDET-IBL} installed before
the start of Run 2),
of a silicon microstrip detector, and of a transition radiation tracker (TRT).
The ID is immersed in a 2\,T axial magnetic field provided by a thin superconducting solenoid.
The silicon detectors provide precision tracking over the pseudorapidity interval $|\eta|<2.5$, while the
TRT offers additional tracking and substantial discrimination between electrons and charged hadrons
for $|\eta|<2.0$.

The solenoid is surrounded by electromagnetic (EM) and hadronic sampling calorimeters allowing energy measurements
of photons, electrons and hadronic jets and discrimination between the different particle types.
The EM calorimeter is a lead/liquid-argon (LAr) sampling calorimeter. It consists of a barrel section,
covering the pseudorapidity region $|\eta|<1.475$, and of two endcap sections, covering $1.375<|\eta|<3.2$.
The EM calorimeter is divided in three layers, longitudinally in depth, for $|\eta|<2.5$, and in two layers for
$2.5 < |\eta| < 3.2$.
In the regions $|\eta| < 1.4$ and $1.5 < |\eta| < 2.4$, the first layer has a fine
$\eta$ segmentation to discriminate isolated photons from neutral hadrons decaying to 
pairs of close-by photons. It also allows, together with 
the information from the cluster barycenter in the second layer, where most of the
energy is collected, a measurement of the shower direction 
without assumptions on the photon production point.
In the range of $|\eta| < 1.8$ a presampler layer allows corrections to be made for energy
losses upstream of the calorimeter.
The hadronic calorimeter reconstructs hadronic showers using steel absorbers and
scintillator tiles ($|\eta|<1.7$), or either copper ($1.5<|\eta|<3.2$) or copper--tungsten
($3.1<|\eta|<4.9$) absorbers immersed in a LAr active medium.

A muon spectrometer surrounds the calorimeter. It comprises separate trigger ($|\eta|<2.4$)
and precision tracking chambers ($|\eta|<2.7$) in the magnetic field provided by three large air-core toroids.

A two-level trigger system~\cite{Aaboud:2016leb} was used during the $\sqrt{s} = 13$~\TeV\ data-taking period.
Dedicated hardware implements the first-level (L1) trigger selection, using only a subset of
the detector information and reducing the event rate to at most 100~kHz.
Events satisfying the L1 requirements are processed by a high-level trigger executing,
on a computer farm, algorithms similar to the offline reconstruction software, in order to
reduce the event rate to approximately 1~kHz.


\section{Data set} \label{sec:dataset}

Events were selected using a diphoton trigger requiring the presence in the EM calorimeter of two clusters of energy
depositions with transverse energy above 35~\GeV\ and 25~\GeV\ for the leading (highest transverse energy) and subleading
 (second highest transverse energy) cluster. In the high-level trigger 
the shape of the energy deposition of both clusters was required to be loosely consistent with that expected from an
electromagnetic shower initiated by a photon. The diphoton trigger has an efficiency greater than 99\% for events that
satisfy the final event selection described in Section~\ref{sec:evsel}.

After the application of data quality requirements, the data set amounts to an integrated luminosity of
\lumi\,fb$^{-1}$, of which 3.2~fb$^{-1}$ were collected in 2015 and 32.9\,fb$^{-1}$ were collected in 2016. The mean
number of proton--proton interactions per bunch crossing is 14 in the 2015 data set and 25 in the 2016 data set.


\section {Event simulation} \label{sec:mc}

Signal samples were generated for the main Higgs boson production
modes using Monte Carlo event generators as described in the following.
The mass and width of the Higgs boson were set in the simulation to $\mH=125$\,\GeV~and $\Gamma_H =
4.07$\,\MeV~\cite{Heinemeyer:2013tqa}, respectively.
The samples are normalized with the latest available theoretical calculations of the corresponding SM production
cross sections, as summarized in Ref.~\cite{deFlorian:2016spz} and detailed below.
The normalization of all Higgs boson samples also accounts for the $H\rightarrow\gamma\gamma$
branching ratio of 0.227\% calculated with \hdecay~\cite{Djouadi:1997yw,Djouadi:2006bz} and
\prophecy~\cite{Bredenstein:2006ha,Bredenstein:2006rh,Bredenstein:2006nk}.

Higgs boson production via \ggH\ is simulated at next-to-next-to-leading-order (NNLO) accuracy in QCD using
the \nnlops\ program~\cite{Hamilton:2013fea}, with the \pdflhc\ PDF set~\cite{Butterworth:2015oua}. The 
simulation achieves NNLO accuracy for arbitrary inclusive $gg \to H$ observables by reweighting the Higgs boson 
rapidity spectrum in Hj-MiNLO~\cite{Hamilton:2012rf} to that of \hnnlo~\cite{Catani:2007vq}. 
The transverse momentum spectrum of the Higgs boson obtained with this sample was found to be
compatible with the fixed-order \hnnlo\ calculation~\cite{Catani:2007vq} and the \hres 2.3 calculation~\cite{Bozzi:2005wk,deFlorian:2011xf} performing
resummation at next-to-next-to-leading-logarithm accuracy matched to a NNLO fixed-order calculation (NNLL+NNLO).
The \hres\ prediction includes the effects of the top and bottom quark
masses up to NLO precision in QCD and uses dynamical renormalization ($\mu_\mathrm{R}$) and
factorization ($\mu_F$) scales, $\mu_F = \mu_\mathrm{R} = 0.5 \sqrt{m_H^2 + {\pT^H}^2}$.
The parton-level events produced
by the \nnlops\ program are passed to \pythia~\cite{Sjostrand:2007gs} to provide parton showering, hadronization and
underlying event, using the AZNLO set of parameters that are tuned to data~\cite{Aad:2014xaa}. The
sample is normalized such that it reproduces the total cross section predicted by a
next-to-next-to-next-to-leading-order (N$^3$LO) QCD calculation with NLO electroweak corrections
applied~\cite{Anastasiou:2015ema, Anastasiou:2016cez, Actis:2008ug, Anastasiou:2008tj}.

Higgs boson production via \VBF\ is generated at NLO accuracy in QCD using the \powhegbox\ program
\cite{Nason:2004rx,Frixione:2007vw,Alioli:2010xd,Nason:2009ai} with the \pdflhc\ PDF set. The parton-level
events are passed to \pythia\ to provide parton showering, hadronization and the underlying event, using the AZNLO parameter set. The VBF
sample is normalized with an approximate-NNLO QCD cross section with NLO electroweak corrections
applied~\cite{Ciccolini:2007jr,Ciccolini:2007ec,Bolzoni:2010xr}.

Higgs boson production via \VH\ is generated at NLO accuracy in QCD through $qq$/$qg$-initiated production, denoted as $q\bar{q}'\to VH$, and through $gg \to ZH$ production using \powhegbox ~\cite{Mimasu:2015nqa} with the \pdflhc\ PDF set. 
Higgs boson production through $gg \to ZH$ has two distinct sources: a contribution with two additional partons, $gg \to ZH q \bar q$, and a contribution without any additional partons in the final state, including box and loop processes dominated by top and bottom quarks. In the following, the $gg\to ZH$ notation refers only to this latter contribution. \pythia\ is used for parton showering, hadronization and the underlying event using the AZNLO parameter set. The samples are normalized with cross sections calculated at NNLO in QCD and NLO electroweak corrections for $q\bar{q}'\to VH$ and at NLO and next-to-leading-logarithm accuracy in QCD for $gg \to ZH$~\cite{Brein:2003wg,Altenkamp:2012sx,Denner:2011id}.

Higgs boson production via \ttH\ is generated at NLO accuracy in QCD using
\amc\ \cite{Alwall:2014hca} with the \nnpdfthree\ PDF set~\cite{Ball:2014uwa} and interfaced to \pythia\ to provide parton
showering, hadronization and the underlying event, using the A14 parameter set~\cite{ATL-PHYS-PUB-2014-021}. The \ttH\ sample is normalized with a cross section
calculation accurate to NLO in QCD with NLO electroweak corrections
applied~\cite{Beenakker:2002nc,Dawson:2003zu,Yu:2014cka,Frixione:2015zaa}.

Higgs boson production via \bbH\ is simulated using \amc~\cite{Wiesemann:2014ioa} interfaced to \pythia\ with the
\ctten\ PDF set~\cite{Lai:2010vv}, and is normalized with the cross-section calculation obtained by matching, using the
\emph{Santander} scheme, the five-flavor scheme cross section accurate to NNLO in QCD with the four-flavor scheme cross section accurate to NLO in
QCD~\cite{Dawson:2003kb,Dittmaier:2003ej,Harlander:2011aa}. The sample includes the effect of interference with
the gluon--gluon fusion production mechanism.

Associated production of a Higgs boson with a single top-quark and a $W$-boson (\tHW) is generated at NLO accuracy,
removing the overlap with the \ttH\ sample through a diagram regularization technique,
using \amc\ interfaced to \herwigpp~\cite{Gieseke:2003hm,Bellm:2013hwb,Bahr:2008pv}, with the
\herwigpp~UEEE5 parameter set for the underlying event and the \ctten\ PDF set using the five-flavor scheme. 
Simulated Higgs boson events in association with a single top-quark, a $b$-quark and a light quark (\tHqb) 
are produced at LO accuracy in QCD using \amc\ interfaced to \pythia\ with 
the \ctten\ PDF set within the four-flavor scheme and using the A14
parameter set. The \tHW\ and \tHqb\ samples are normalized with calculations accurate to NLO in
QCD~\cite{Demartin:2015uha}.

\begin{table*}[!tp] 
 \caption{ Summary of the event generators and PDF sets used to model the signal and the main background processes. The SM
           cross sections $\sigma$ for the Higgs production processes with ${\mH=125.09}$~\GeV\ are also given separately for
           ${\sqrt{s} = 13}$~\TeV, together with the orders of the calculations corresponding to the quoted cross sections, 
           which are used to normalize the samples, after multiplication by the Higgs boson branching ratio to diphotons, 
           0.227\%. The following versions were used: \pythia\ version 8.212 (processes) and 8.186 (pile-up overlay); \herwigpp\ version 2.7.1; \powhegbox\ version 2; \amc\ version 2.4.3; \sherpa\ version 2.2.1
         } 
\label{mc_table} 
\begin{center}
\resizebox{0.95\textwidth}{!}{ \begin{tabular}{lccccc} \hline\hline \multirow{2}{*}{Process} &
\multirow{2}{*}{Generator} & \multirow{2}{*}{Showering} & \multirow{2}{*}{PDF set} & $\sigma\ [\mathrm{pb}] $ & \multirow{2}{*}{Order of calculation of $\sigma$}  \\ 
& &                   &                  &  $\sqrt{s}=13$~\TeV & \\ \hline 
\ggH & \nnlops & \pythia  & \pdflhc       &  48.52    & N$^3$LO(QCD)+NLO(EW)  \\ 
\VBF  & \powhegbox      & \pythia  & \pdflhc    & 3.78  & NNLO(QCD)+NLO(EW)   \\ 
\WH   & \powhegbox      & \pythia   & \pdflhc        &  1.37  & NNLO(QCD)+NLO(EW)  \\ 
\qqZH & \powhegbox      & \pythia   & \pdflhc         &  0.76  & NNLO(QCD)+NLO(EW)  \\ 
\ggZH & \powhegbox & \pythia & \pdflhc         &  0.12 & NLO+NLL(QCD)   \\ 
\ttH  & \amc            &  \pythia  & \nnpdfthree     & 0.51 & NLO(QCD)+NLO(EW)    \\ 
\bbH  & \amc            & \pythia   & \ctten         &  0.49  & 5FS(NNLO)+4FS(NLO)  \\
$t$-channel \tH  & \amc            & \pythia   & \ctten                  &  0.07  & 4FS(LO)    \\ 
$W$-associated \tH  & \amc & \herwigpp & \ctten             &  0.02& 5FS(NLO)     \\ 
\hline 
$\gamma\gamma$                   & \textsc{Sherpa}  & \textsc{Sherpa} & \ctten \\ 
$V\gamma\gamma$                  & \textsc{Sherpa}  & \textsc{Sherpa} & \ctten \\ 
\hline \hline \end{tabular}
}
\end{center} 
\end{table*}

The generated Higgs boson events are passed through a \geant~\cite{geant-1} simulation of the ATLAS
detector~\cite{SOFT-2010-01} and reconstructed with the same analysis software used for the data. 

Background events from continuum $\gamma\gamma$ production and $V\gamma\gamma$ production are simulated using the \sherpa\ event generator~\cite{Gleisberg:2008ta}, with the \ctten\ PDF set and the \sherpa\ default 
parameter set for the underlying-event activity. The corresponding matrix elements for $\gamma\gamma$ and $V\gamma\gamma$ are calculated at leading order (LO) in the strong coupling constant $\alpha_\mathrm{S}$ with the real emission of up to three or two additional partons, respectively, and are merged with the \sherpa\ parton shower~\cite{Schumann:2007mg} using the \mepslo\ prescription~\cite{Hoeche:2009rj}.

The very large sample size required for the modeling of the $\gamma\gamma$ background processes is obtained through a fast parametric simulation of the ATLAS detector response~\cite{Aad:2010ah}. For $V\gamma\gamma$ events the same full detector simulation as for the signal samples is used.

Additional proton--proton interactions (pileup) are included in the simulation for all generated events such that the
average number of interactions per bunch crossing reproduces that observed in the data. The inelastic proton--proton
collisions were produced using \pythia\ with the A2 parameter set~\cite{ATL-PHYS-PUB-2012-003} that are tuned to data
and the \mstwlo\ PDF set~\cite{Martin:2009iq}. A summary of the used signal and background samples is shown in
Table~\ref{mc_table}.


\section{Event reconstruction and selection}
\label{sec:evsel}

\subsection{Photon reconstruction and identification} \label{sec:photons}

The reconstruction of photon candidates is seeded by energy clusters in the electromagnetic calorimeter with a size of $\Delta\eta\times\Delta\phi =$ 0.075$\times$0.125, with transverse energy $E_{\rm T}$ greater than 2.5\,\GeV\ \cite{Aaboud:2016yuq}.
The reconstruction is designed to separate electron from photon candidates, and to classify the latter as \emph{unconverted} or \emph{converted} photon candidates. Converted photon candidates are associated with the conversion of photons into electron--positron pairs in the material upstream the electromagnetic calorimeter. Conversion vertex candidates are reconstructed from either two tracks consistent with originating from a photon conversion, or one track that does not have any hits in the innermost pixel layer. These tracks are required to induce transition radiation signals in the TRT consistent with the electron hypothesis, in order to suppress backgrounds from non-electron tracks. 
Clusters without any matching track or conversion vertex are classified as unconverted photon candidates, while clusters with a matching conversion vertex are classified as converted photon candidates.
In the simulation, the average reconstruction efficiency for photons with generated $\ET$ above 20~\GeV\ and generated pseudorapidity $|\eta|<2.37$ is 98\%.

The energy from unconverted and converted photon candidates is measured from an electromagnetic cluster of size $\Delta\eta\times\Delta\phi =$ 0.075$\times$0.175 in the barrel region of the calorimeter, and $\Delta\eta\times\Delta\phi =$ 0.125$\times$0.125 in the calorimeter endcaps. The cluster size is chosen sufficiently large to optimize the collection of energy of the particles produced in the photon conversion. The cluster electromagnetic energy is corrected in four steps to obtain the calibrated energy of the photon candidate, using a combination of simulation-based and data-driven correction factors~\cite{Aad:2014nim}. 
The simulation-based calibration procedure was re-optimized for the 13\,\TeV\ data. 
Its performance is found to be similar with that of Run~1~\cite{Aad:2014nim} in the full pseudorapidity range, and is improved in the barrel--endcap transition region, due to the use of information from additional scintillation detectors in this region~\cite{phot2015}.
The uniformity corrections and the intercalibration of the longitudinal calorimeter layers are unchanged compared to Run~1~\cite{Aad:2014nim}, and the data-driven calibration factors used to set the absolute energy scale are determined 
from $Z\to e^+e^-$ events in the full 2015 and 2016 data set. 
The energy response resolution is corrected in the simulation to match the resolution observed in data. This correction is derived simultaneously with the energy calibration factors using $Z\to e^+e^-$ events by adjusting the electron energy resolution such that the width of the reconstructed $Z$ boson peak in the simulation matches the width observed in data~\cite{phot2015}.

Photon candidates are required to satisfy a set of identification criteria to reduce the contamination from the background, 
primarily associated with neutral hadrons in jets decaying into photon pairs, based on the lateral and longitudinal shape 
of the electromagnetic shower in the calorimeter \cite{photon-pub-2015}. Photon candidates are required to deposit only a 
small fraction of their energy in the hadronic calorimeter, and to have a lateral shower shape consistent with that expected 
from a single electromagnetic shower. Two working points are used: a \emph{loose} criterion, primarily used for triggering and 
preselection purposes, and a \emph{tight} criterion. The tight selection requirements are tuned separately for unconverted 
and converted photon candidates. 
Corrections are applied to the electromagnetic shower shape variables of simulated photons, to account for small differences observed between data and simulation.
The variation of the photon identification efficiency associated with the different reconstruction of converted photons in the 2015 and 2016 data sets, due to the different TRT gas composition, has been studied with simulated samples and shown to be small. The efficiency of the tight identification criteria ranges from 84\% to 94\% (87\% to 98\%) for unconverted (converted) photons with transverse energy between 25~\GeV\ and 200~\GeV.

To reject the hadronic jet background, photon candidates are required to be isolated from any other activity in the calorimeter and the tracking detectors.
The calorimeter isolation is computed as the sum of the transverse energies of positive-energy topological clusters~\cite{Aad:2016upy} in the calorimeter within a cone of $\Delta R =$ 0.2 centered around the photon candidate. The transverse energy of the photon candidate is removed. 
The contributions of the underlying event and pileup are subtracted according to the method suggested in Ref.~\cite{Cacciari:2009dp}.
Candidates with a calorimeter isolation larger than 6.5\% of the photon transverse energy are rejected.
The track isolation is computed as the scalar sum of the transverse momenta of all tracks in a cone of $\Delta R=0.2$ with $\pt> 1$~\GeV\ which satisfy some loose track-quality criteria and originate from the diphoton primary vertex, i.e. the most likely production vertex of the diphoton pair (see Section~\ref{sec:evselPV}). For converted photon candidates, the tracks associated with the conversion are removed. Candidates with a track isolation larger than 5\% of the photon transverse energy are rejected.

\subsection{Event selection and selection of the diphoton primary vertex}
\label{sec:evselPV}

Events are preselected by requiring at least two photon candidates with $\ET> 25$~\GeV\ and $|\eta|<2.37$ (excluding the transition
region between the barrel and endcap calorimeters of $1.37 <|\eta|<1.52$) that fulfill the loose photon identification criteria~\cite{Aaboud:2016yuq}. The two
photon candidates with the highest $\ET$ are chosen as the diphoton candidate, and used to identify the \emph{diphoton
primary vertex} among all reconstructed vertices, using a neural-network algorithm based on track and primary vertex
information, as well as the directions of the two photons measured in the calorimeter and inner
detector~\cite{Aad:2014eha}. The neural-network algorithm selects a diphoton vertex within 0.3~mm of the true
$H\to\gamma\gamma$ production vertex in 79\% of simulated gluon--gluon fusion events. For the other Higgs production modes this
fraction ranges from 84\% to 97\%, increasing with jet activity or the presence of charged leptons. The performance of
the diphoton primary vertex neural-network algorithm is validated using $Z\to e^+e^-$ events in data and simulation, by
ignoring the tracks associated with the electron candidates and treating them as photon candidates. Sufficient agreement between
the data and the simulation is found. The diphoton primary vertex is used to redefine the direction of the photon
candidates, resulting in an improved diphoton invariant mass resolution. The invariant mass of the two photons is given
by $m_{\gamma \gamma} = \sqrt{ 2 E_1 E_2 \left( 1 - \cos \alpha \right)
}$,  where $E_1$ and $E_2$ are the energies of the leading and
subleading photons and $\alpha$ is the opening angle of the two photons with respect to the selected production vertex.

\begin{figure}[!tbp] 
  \begin{center} 
    \includegraphics[width=.6\textwidth]{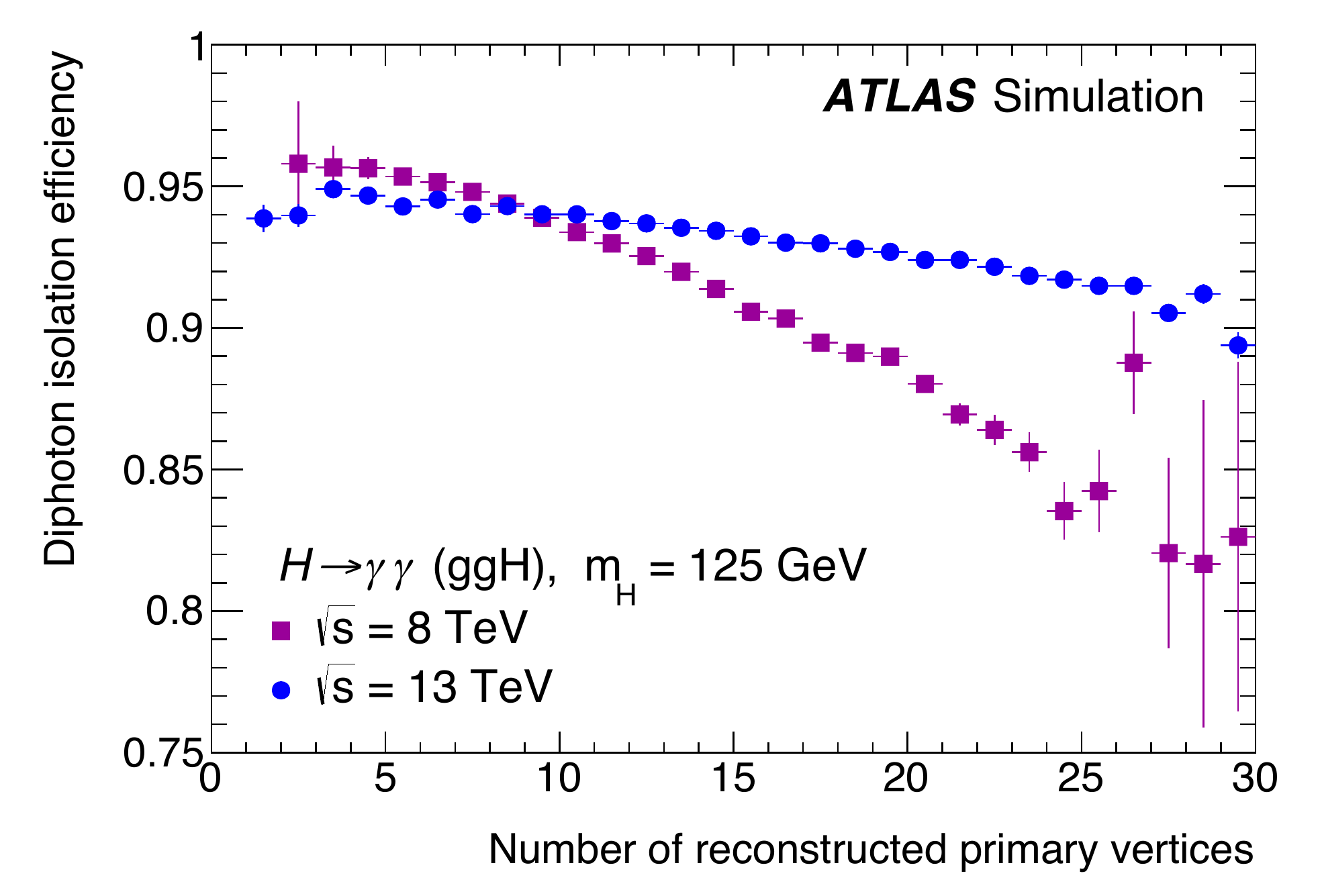}
    \caption{
    Efficiency for both photons to fulfill the isolation requirement as a function of the number of primary vertex
candidates in each event, determined with a sample of simulated Higgs bosons with $m_H$~=~125~\GeV, produced in gluon--gluon
fusion and decaying into two photons. Events are required to satisfy the kinematic selection described in
Section~\ref{sec:evselPV} for the 8~\TeV\ (violet squares) and 13~\TeV\ (blue circles) simulated sample. The error bars
show the statistical uncertainty of the generated samples.
The Run~2 (Run~1) isolation requirement is based on the transverse energy deposited 
in the calorimeter in a $\Delta R=0.2$ ($\Delta R=0.4$) cone around the photon candidates.
Both the Run~1 and Run~2 algorithms also use tracking information in a $\Delta R=0.2$ cone around the photon 
candidates. 
}
\label{fig:photonIso}
\end{center} 
\end{figure}

Following the identification of the diphoton primary vertex, the leading and subleading photon candidates in the
diphoton candidate are respectively required to have $\ET / m_{\gamma\gamma}>$ 0.35 and 0.25, and to both satisfy the
tight identification criteria as well as the calorimeter and track isolation requirements. Figure~\ref{fig:photonIso}
compares the simulated per-event efficiency of the track- and calorimeter-based isolation requirement
as a function of the number of primary vertex candidates with the per-event efficiency of the Run~1 algorithm described in Ref.~\cite{Aad:2014eha}, by using a MC sample of Higgs bosons produced by gluon-gluon fusion and decaying into two photons.
The re-optimization of the thresholds applied to the transverse energy sum of the calorimeter energy deposits 
and to the transverse momentum scalar sum of the tracks in the isolation cone, as well as the reduction of the size of the 
isolation cone for the calorimeter-based isolation, greatly reduces the degradation of the efficiency 
as the number of reconstructed primary vertices increases in comparison to the Run~1 algorithm.

In total 332030 events are selected with diphoton candidates with invariant mass $m_{\gamma\gamma}$ between 105\,\GeV\
and 160\,\GeV. The predicted signal efficiency, assuming the SM and including the acceptance of the kinematic selection, is 42\% (with the acceptance of the kinematic selection being 52\%).

\subsection{Reconstruction and selection of hadronic jets, $b$-jets, leptons and missing transverse momentum}

Jets are reconstructed using the \antikt\ algorithm~\cite{Cacciari:2008gp} with a radius parameter of 0.4 via the 
\fastjet\ package~\cite{Cacciari:2005hq,Fastjet}. The inputs to
the algorithm are three-dimensional topological clusters of energy deposits in the calorimeter cells~\cite{Aad:2016upy}. Jets are corrected
on an event-by-event basis for energy deposits originating from pileup~\cite{Aad:2015ina}, then calibrated using a
combination of simulation-based and data-driven correction factors, which correct for different responses to electromagnetic and hadronic showers of the calorimeter and inactive regions of the calorimeter~\cite{jes2015,Aad:2014bia}. 
Jets are required to have $\pt>25$~\GeV\ for $|\eta|<2.4$.
The jet selection is tightened to $\pt>30$~\GeV\ within $|y|<$ 4.4 for most event reconstruction categories 
 and  the measurement of fiducial integrated and differential cross sections (with exceptions noted in Sections~\ref{sec:categories} and ~\ref{sec:details_fidreg}).  Jets that do not originate from the
diphoton primary vertex are rejected, for $|\eta|<2.4$, using the jet vertex tagging algorithm (JVT)~\cite{jvt},
which combines tracking information into a multivariate likelihood. For jets with $\pt<60$~\GeV\ and $|\eta|<2.4$ a
\emph{medium} working point is used, with an efficiency greater than 92\% for non-pileup jets with $\pt >$ 30\,\GeV.
The efficiency of the JVT algorithm is corrected in the simulation to match that observed in the data. Jets are discarded if
they are within $\Delta R = 0.4$ of an isolated photon candidate, or within $\Delta R = 0.2$ of an isolated electron
candidate.

Jets consistent with the decay of a $b$-hadron are identified using a multivariate discriminant, having as input
information from track impact parameters and secondary vertices~\cite{btag2015, btag2016}. The chosen identification
criterion has an efficiency of 70\% for identifying jets originating from a $b$-quark. The efficiency is determined using
a \ttbar\ control region, with rejection factors of about 12 and 380 for jets originating from $c$-quarks and light quarks, respectively. Data-driven correction factors
are applied to the simulation such that the $b$-tagging efficiencies of jets originating from $b$-quarks, $c$-quarks and
light quarks are consistent with the ones observed in the data.

The reconstruction and calibration of electron candidates proceeds similarly as for photon candidates. Electromagnetic
calorimeter clusters with a matching track in the inner detector are reconstructed as electron candidates and calibrated
using dedicated corrections from the simulation and from data control samples. Electron candidates are required to have
$\pt > 10$~\GeV\ and $|\eta|<2.47$, excluding the region $1.37 < |\eta|<1.52$. Electrons must satisfy {\em medium}
identification criteria~\cite{electrons} using a likelihood-based discriminant. 

Muon candidates are primarily built from tracks reconstructed in the inner detector and the muon spectrometer, but are
complemented by candidates reconstructed only in the muon spectrometer that are compatible with originating from the
interaction point~\cite{Aad:2016jkr}. Muon candidates are required to have $\pt>10$~\GeV\ and $|\eta|<2.7$, and satisfy {\em
medium} identification criteria based on the number of hits in the silicon detectors, in the TRT and in the muon
spectrometer. For the measurements of fiducial cross sections the electron and muon selections are tightened to $\pT>15$~\GeV.

Lepton candidates are discarded if they are within $\Delta R = 0.4$ of an isolated photon candidate or a jet. Isolation
requirements are applied to all lepton candidates. Electron candidates are required to satisfy loose criteria for the
calorimeter and track isolation, aimed at a combined efficiency of 99\% independently of the candidate transverse
momentum. Muon candidates are similarly required to satisfy loose criteria for the calorimeter and track isolation, in this
case depending on the candidate transverse momentum, and aimed at a combined efficiency ranging from 95--97\% at $\pT =$ 10--60~\GeV\ to 99\% for $\pT > 60$~\GeV.

Tracks associated with both the electron and muon candidates are required to be consistent with originating from the diphoton
primary vertex by requiring their longitudinal impact parameter $z_0$ to satisfy $| z_0 \sin\theta | <0.5$~mm and
their unsigned transverse impact parameter $|d_0|$ relative to the beam axis to be respectively smaller than five or
three times its uncertainty.

The lepton efficiency as well as energy/momentum scale and resolution are determined using the decays of $Z$ bosons and
$J/\psi$ mesons in the full 2015 and 2016 data set using the methods described in Refs.~\cite{Aad:2016jkr,electrons}.
Lepton efficiency correction factors are applied to the simulation to improve the agreement with the data. 

The magnitude of the missing transverse momentum $\ET^\mathrm{miss}$ is measured from the negative vectorial sum of the
transverse momenta of all photon, electron and muon candidates and of all hadronic jets after accounting for overlaps
between jets, photons, electrons, and muons, as well as an estimate of soft contributions based on tracks originating from the
diphoton vertex which satisfy a set of quality criteria. A full description of this algorithm can be found
in Refs.~\cite{met2015a,met2015b}. The $\ET^\mathrm{miss}$ significance is defined as $\ET^\mathrm{miss} / \sqrt{\sum
\ET}$, where $\sum \ET$ is the sum of the transverse energies of all particles and jets used in the estimation of the missing
transverse momentum in units of \GeV.


\section{Signal and background modeling of diphoton mass spectrum}
\label{sec:sigext}

The Higgs boson signal yield is measured through an unbinned maximum-likelihood fit to the diphoton invariant mass spectrum in the
range 105\,\GeV $<\mgg< $160\,\GeV\ for each event reconstruction category, fiducial region, or each bin of a fiducial differential
cross section, as further discussed in Sections~\ref{sec:methods_coup} and ~\ref{sec:methods_fid}.
The mass range is chosen to be large enough to allow a reliable determination of the background from the data,
and at the same time small enough to avoid large uncertainties from the choice of the background
parameterization. The signal and background shapes are modeled as described below, and the background model parameters
are freely floated in the fit to the $m_{\gamma\gamma}$ spectra.

\subsection{Signal model}
\label{sec:sigmod}

The Higgs boson signal manifests itself as a narrow peak in the \mgg\ spectrum. 
The signal distribution is empirically modeled as a double-sided Crystal Ball function, consisting of a Gaussian 
central part and power-law tails on both sides. The Gaussian core of the Crystal Ball function is parameterized by the peak position $(m_H + \Delta \mu_\mathrm{CB})$ and the width $(\sigma_\mathrm{CB})$. The non-Gaussian contributions to the mass resolution arise mostly from converted photons $\gamma\to e^+e^-$ with at least one electron losing a significant fraction of its energy through bremsstrahlung in the inner detector material. The parametric form for a given reconstructed category or bin $i$ of a fiducial cross section
measurement, for a Higgs boson mass $m_H$, can be written as: 

\begin{small}
\begin{multline}
f_{i}^{\mathrm{sig}}(\mgg;\Delta\mu_{\mathrm{CB},i},\sigma_{\mathrm{CB},i},\alpha_{\mathrm{CB},i}^\pm, n_{\mathrm{CB},i}^\pm)
= {\cal N}_c 
\begin{cases} 
  e^{-t^{2}/2} & - \alpha_{\mathrm{CB},i}^{-}  \le t \le \alpha_{\mathrm{CB},i}^{+}  \\ 
 \left(\frac{n_{\mathrm{CB},i}^-}{|\alpha_{\mathrm{CB},i}^-|} \right)^{n_{\mathrm{CB},i}^-} e^{-|\alpha_{\mathrm{CB},i}^-|^{2}/2}  \left( \frac{n_{\mathrm{CB},i}^-}{\alpha_{\mathrm{CB},i}^-} - \alpha_{\mathrm{CB},i}^- - t \right)^{-n_{\mathrm{CB},i}^-} & t <-\alpha_{\mathrm{CB},i}^- \\ 
 \left( \frac{n_{\mathrm{CB},i}^+}{|\alpha_{\mathrm{CB},i}^+|} \right)^{n_{\mathrm{CB},i}^+} e^{-|\alpha_{\mathrm{CB},i}^+|^{2}/2}  \left( \frac{n_{\mathrm{CB},i}^+}{\alpha_{\mathrm{CB},i}^+} - \alpha_{\mathrm{CB},i}^+ - t \right)^{-n_{\mathrm{CB},i}^+} & t > \alpha_{\mathrm{CB},i}^+ 
\end{cases}, \nonumber
\end{multline} 
\end{small}

where $t=(\mgg- m_H - \Delta\mu_{\mathrm{CB},i})/\sigma_{\mathrm{CB},i}$, and ${\cal N}_c$ is a normalization factor.
The non-Gaussian parts are parameterized by $\alpha_{\mathrm{CB},i}^\pm$ and $n_{\mathrm{CB},i}^\pm$ separately for the low- ($-$) and high-mass ($+$) tails.

\begin{figure*}[!tbp]
\begin{center} 
  \includegraphics[width=.65\textwidth]{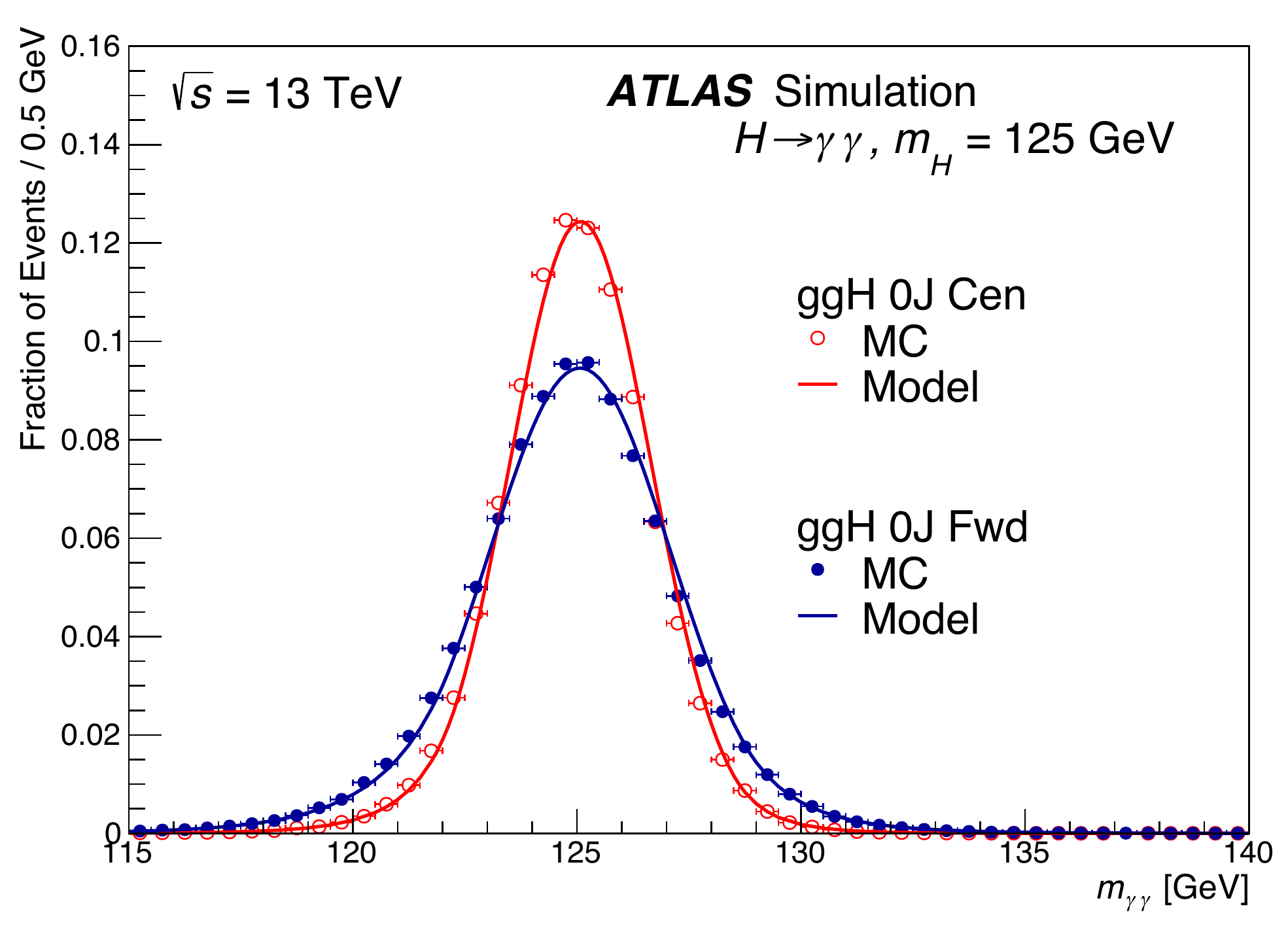}
\caption{ 
 The diphoton mass signal shapes of two gluon--gluon fusion categories that are later introduced in Section~\ref{sec:categories} are shown: ggH 0J Fwd aims to select gluon--gluon fusion events with no additional jet and at least one photon in the pseudorapidity
region $|\eta| > 0.95$; ggH 0J Cen applies a similar selection, but requires both photons to have 
$|\eta| \le 0.95$ in order to have a better energy resolution. The simulated sample (labeled as MC) is compared to the fit model and contains simulated events from all Higgs boson production processes
 described in Section~\ref{sec:mc} with $\mH = 125$\,\GeV.}
\label{fig:dscb} \end{center} 
\end{figure*}

The parameters of the model that define the shape of the signal distribution are determined through fits to the
simulated signal samples. The parameterization is derived separately for each reconstructed category or fiducial region
of the integrated or differential cross-section measurement. Figure~\ref{fig:dscb} shows an example for two categories
with different mass resolution: the improved mass resolution in the central region of the detector
(defined by requiring $|\eta| \le 0.95$ for both selected photons) with respect to the forward region
(defined by requiring one photon with $|\eta| \le 0.95$ and one photon with $0.95 < |\eta| < 2.37$) results in better discriminating power against the non-resonant background and in turn in a smaller statistical error of the extracted Higgs boson signal yield. The effective signal mass resolution of the two categories, defined as half the width containing 68\% (90\%) of the signal events, is 1.6 (2.7)~\GeV\ and 2.1 (3.8)~\GeV, respectively, and the mass resolution for all used categories can be found in Appendix~\ref{app:suppmaterial}.

\subsection{Background composition and model}
\label{sec:bkdmod}

The diphoton invariant mass model for the background used to fit the data is determined from studies of the bias in the
signal yield in signal+background fits to large control samples of data or simulated background events.

Continuum $\gamma\gamma$ production is simulated with the \sherpa\ event generator as explained in Section~\ref{sec:mc}, neglecting any interference effects with the $H \to \gamma\gamma$ signal. 
The $\gamma j$ and $jj$ backgrounds are obtained by reweighting this sample using an
$m_{\gamma\gamma}$ dependent linear correction function obtained from the fraction of $\gamma\gamma$ to $\gamma j$
and $\gamma\gamma$ to $jj$ background events in data, respectively. 

For very low rate categories targeting $\ttH$ or $tH$ events, in which the background
simulation suffers from very large statistical uncertainties, various background-enriched control samples are directly
obtained from the data by either reversing photon identification or isolation criteria, or by loosening or removing
completely $b$-tagging identification requirements on the jets, and normalizing to the data in the $\mgg$ sidebands of
the events satisfying the nominal selection.
For low rate categories targeting associated vector boson production, background control samples are obtained by summing
the distributions from the main background processes: Those of $\gamma \gamma$ and $V\gamma\gamma$ events are
obtained directly from the simulation, while the $m_{\gamma\gamma}$ distributions of $\gamma j$ and $jj$ events are
obtained from data control samples in which the nominal selection is applied except that at least one (for $\gamma j$)
or both (for $jj$) of the two photon candidates fail to meet either the identification or isolation criteria. Except for the
$V\gamma\gamma$ component, which is normalized with its theoretical cross section, the other contributions are normalized
according to their relative fractions determined in data as described in the following.

\begin{figure}[!tbp]
\subfloat[] {\includegraphics[width=.48\textwidth]{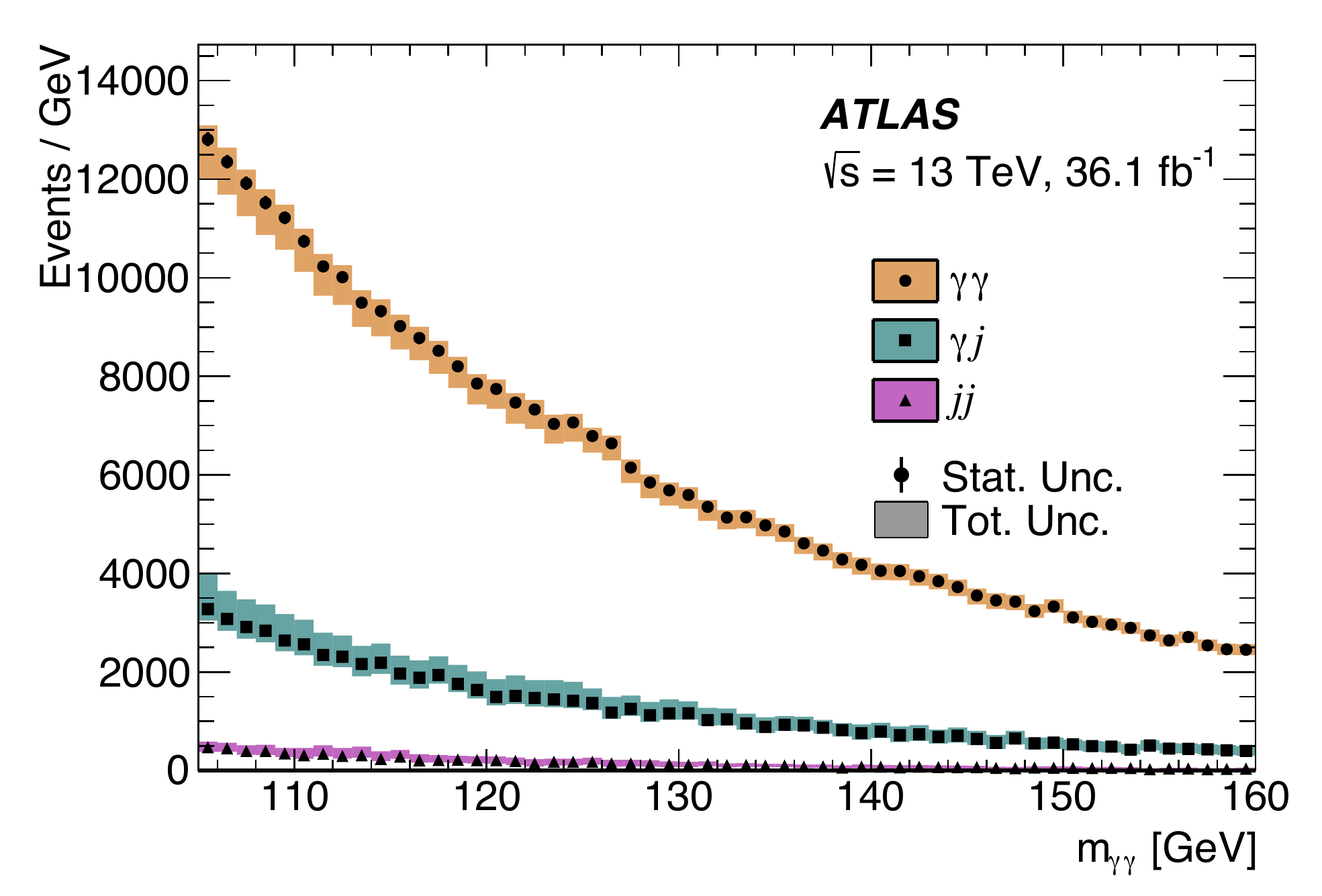}}
\subfloat[] {\includegraphics[width=.48\textwidth]{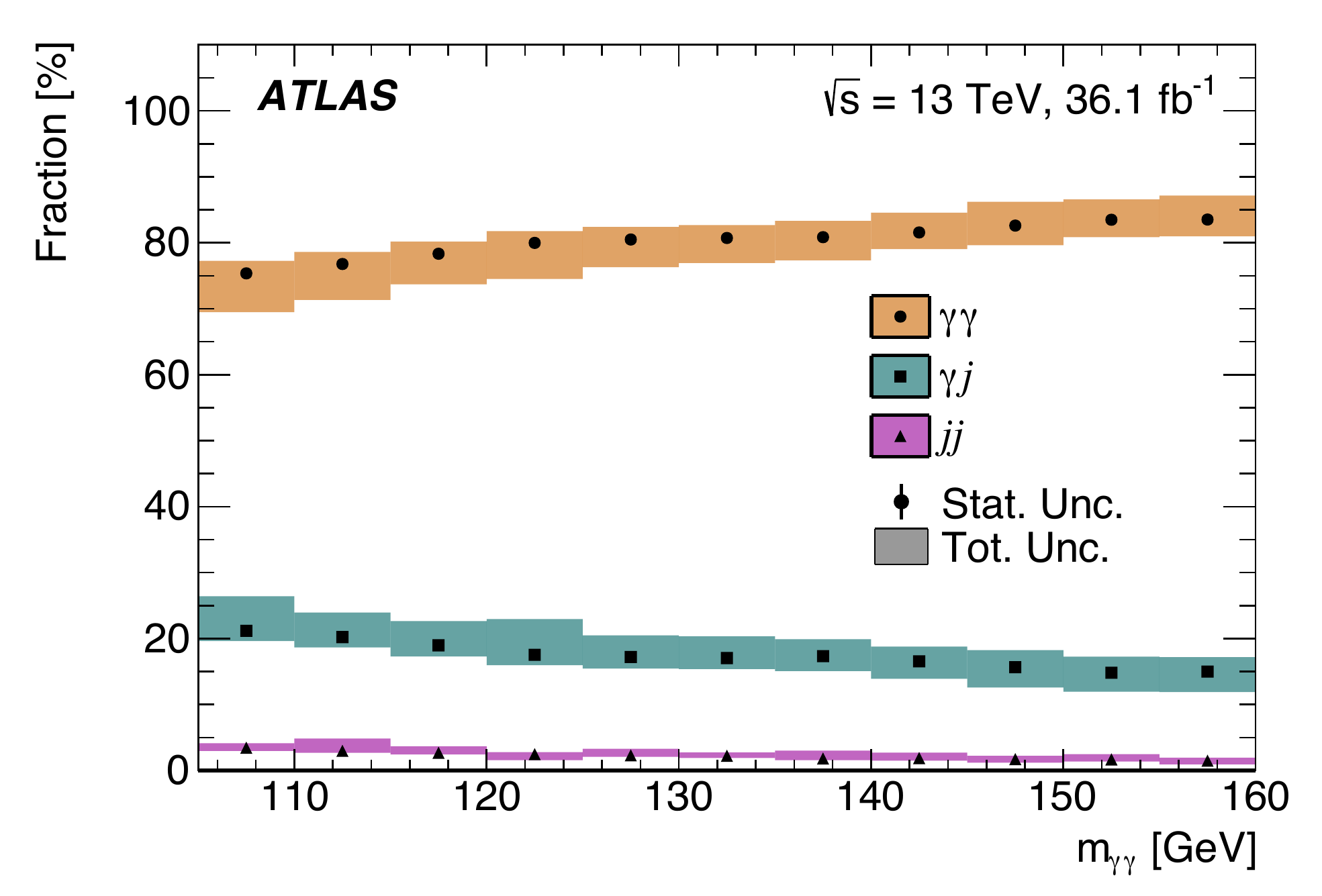}} \\
\subfloat[] {\includegraphics[width=.48\textwidth]{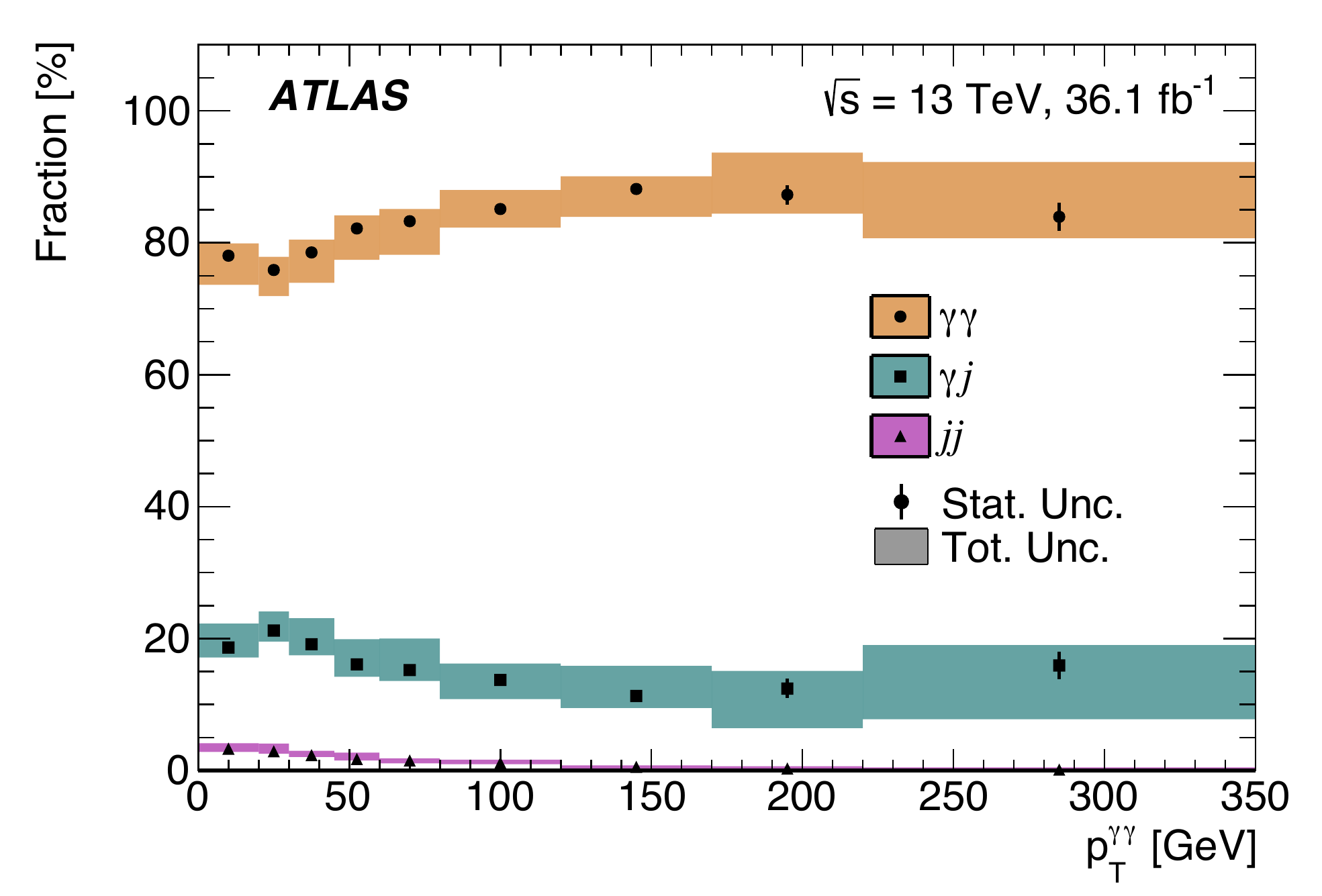}}
\subfloat[] {\includegraphics[width=.48\textwidth]{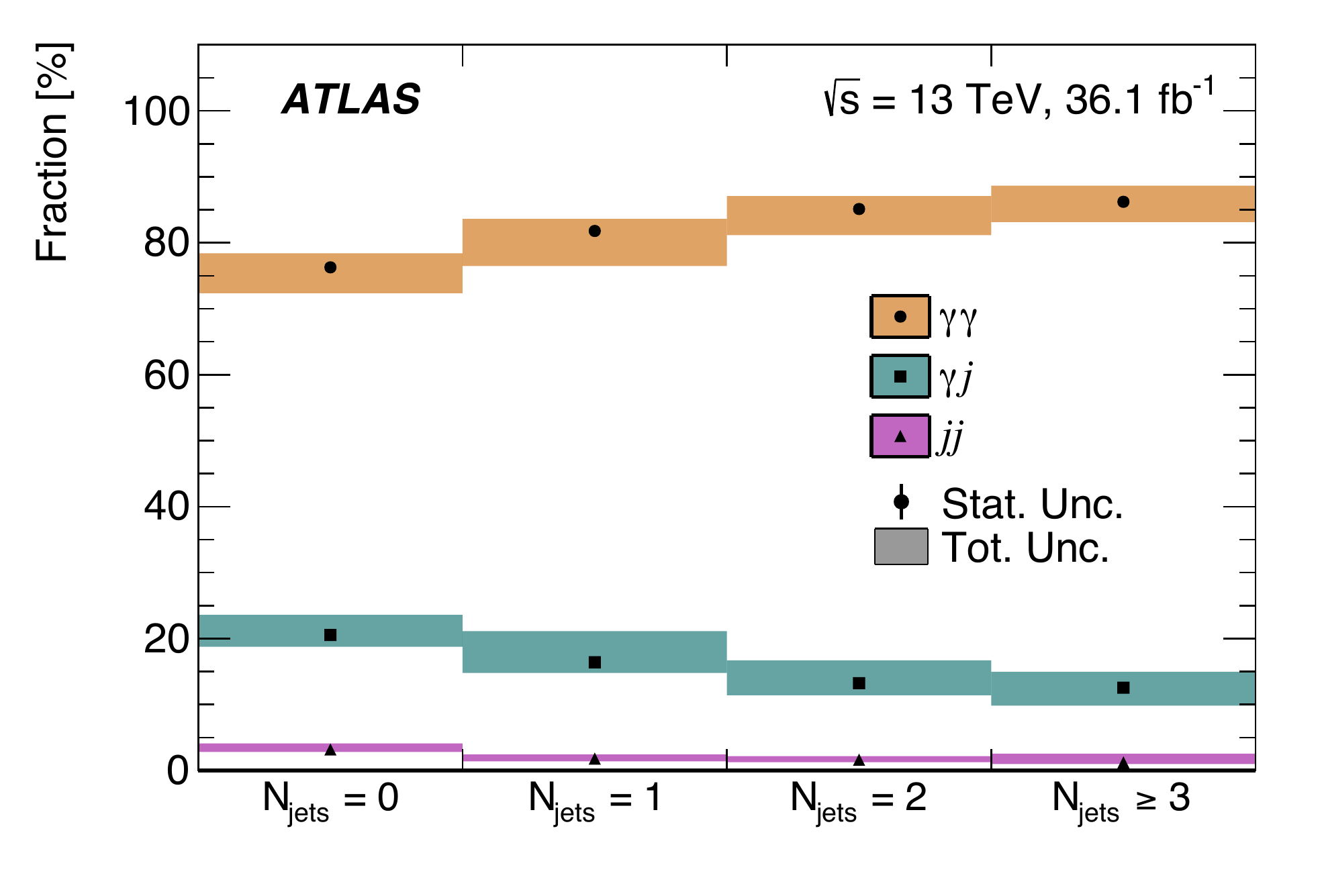}}
\caption{The data-driven determination of (a) event yields and (b) event
  fractions for $\gamma\gamma$, $\gamma j$ and $jj$ events as a
  function of \mgg\ after the final selection outlined in
  Section~\ref{sec:evsel}. The event fractions for two differential
  observables, (c) \ptgg\ and (d) \njet\ defined for jets with a $\pt > 30$~\GeV\ are
  shown as well. The shaded regions show the total uncertainty of the
  measured yield and fraction, and the error bars show the statistical
  uncertainties.}
\label{fig:bkdfrac}
\end{figure}

The measurement of the background fractions in data is performed for each category or fiducial region. The relative
fractions of $\gamma \gamma$, $\gamma j$ and $jj$ background events are determined using a double two-dimensional
sideband method~\cite{Aad:2011mh,Aad:2010sp}. The nominal identification and isolation requirement are loosened for both photon
candidates, and the data are split into 16 orthogonal regions defined by diphoton pairs in which one or both photons satisfy
or fail to meet identification and/or isolation requirements. The region in which both photons satisfy the nominal identification
and isolation requirements corresponds to the nominal selection of Section~\ref{sec:evsel}, while the other 15 regions
provide control regions, whose $\gamma\gamma$, $\gamma j$ and $jj$ yields are related to those in the signal region via
the efficiencies for photons and for hadronic jets to satisfy the photon identification and isolation requirements. The
$\gamma\gamma$, $\gamma j$ and $jj$ yields in the signal region are thus obtained, together with the efficiencies for
hadronic jets, by solving a linear system of equations using as inputs the observed yields in the 16 regions and the
photon efficiencies predicted by the simulation.
In the $VH$ categories, a small extra contribution from $V\gamma\gamma$ events with an electron originating from the decay of the
vector boson $V$ which is incorrectly reconstructed as a photon, is also estimated from the simulation and subtracted before applying the
two-dimensional sideband method.
The dominant systematic uncertainties in the measured background fractions are due to the definition of the background
control regions. The yields and relative fractions of the $\gamma \gamma$, $\gamma j$ and $jj$ backgrounds are shown in
Figure~\ref{fig:bkdfrac} as a function of \mgg\ for the selected events. The fractions of these background
sources in the inclusive diphoton sample are ($78.7\, ^{+1.8}_{-5.2}$)\%, ($18.6\, ^{+4.2}_{-1.6}$)\% and
($2.6\, ^{+0.5}_{-0.4}$)\%, respectively. The uncertainties in the measured background fractions are systematically dominated.
These results are comparable to previous results at $\sqrt{s}$ = 7 and 8 \TeV~\cite{Aad:2014lwa,Aad:2014eha}. In addition
the purity is shown as a function of the \pt\ of the diphoton system, and the number of reconstructed jets with $\pt > 30$~\GeV.

The functional form used to model the background $\mgg$ distribution in the fit to the data is chosen, in each region,
to ensure a small bias in the extracted signal yield relative to its experimental precision, following the procedure described in
Ref.~\cite{Aad:2012tfa}. The potential bias (\emph{spurious signal}) is estimated as the maximum of the absolute value
of the fitted signal yield, using a signal model with mass between 121 and 129~\GeV, in fits to the background control regions described before. 

The spurious signal is required, at 95\% confidence level (CL), to be less than 10\% of the expected SM signal yield
or less than 20\% of the expected statistical uncertainty in the SM signal yield. In the case when two or more
functions satisfy those requirements, the background model with the least number of parameters is chosen.

Prior to the final fit to the data, the selected model is tested against a model from the same family of functions but
with one more degree of freedom (for instance, the exponential of a second-order polynomial is tested against an exponential of a third-order polynomial) to check,
using only events in the diphoton invariant mass sidebands ({\em i.e.} excluding the range $121~\GeV < \mgg <
129~\GeV$), if the data favors a more complex model. A test statistic is built from the $\chi^2$ values and number of
degrees of freedom of two binned fits to the data with the two background models. The expected distribution of the test
statistic is built from pseudo-data assuming that the function with fewer degrees of freedom is the true underlying
model. The value of the test statistic obtained in the data is compared to such distribution, and the simpler model is
rejected in favor of the more complex one if the $p$-value of such comparison is lower than 5\%. The background
distribution of all regions is found to be well modeled by at least one of the following functions: an exponential of a first- or second-order polynomial, a power law, or a third-order Bernstein polynomial. 

\subsection{Statistical model} \label{sec:statmod}

The data are interpreted following the statistical procedure summarized in Ref.~\cite{Cowan:2010js} and described in
detail in Ref.~\cite{Aad:2012an}. An extended likelihood function is built from the number of observed events and
invariant diphoton mass values of the observed events using the analytic functions describing the distributions of
\mgg\ in the range 105--160~\GeV\ for the signal and the background.

The likelihood for a given reconstructed category, fiducial region, or differential bin $i$ of the integrated or differential
cross-section measurement is a marked Poisson probability distribution, 
\begin{equation*} 
\LL_i = \mathrm{Pois}(n_i |N_i(\vecth))\cdot\prod_{j=1}^{n_i}f_i(\mgg^j,\vecth) \cdot G(\vecth), 
\end{equation*} 
where $n_i$ ($N_i$) is the observed (expected) number of selected candidates, $f_i(\mgg^j)$ is the value of the probability
density function (pdf) of the invariant mass distribution evaluated for each candidate $j$, $\vecth$ are nuisance
parameters and $G(\vecth)$ is a set of unit Gaussian constraints on a subset of the nuisance parameters, as described in
the following.
The likelihood for the measurements of the total Higgs boson production-mode cross sections and signal strengths is given by the product of the likelihood functions of each event reconstruction category.
For the fiducial integrated and differential cross-section measurements the likelihood of all bins $i$ in a spectrum is taken.

The number of expected candidates is the sum of the signal and background yields, denoted by
$N_{i}^{\mathrm{sig}}$ and $N_{i}^{\mathrm{bkg}}$, and the fitted spurious signal yield,
$N_{i}^{\mathrm{spur}}\cdot\thetaspurc$, 
\begin{equation*} 
N_i= N_{i}^{\mathrm{sig}} + N_{i}^{\mathrm{bkg}} + N_{i}^{\mathrm{spur}}\cdot\thetaspurc. 
\end{equation*}

In more detail, the invariant mass distribution for each category has signal and background components,
\begin{eqnarray*} 
f_i(\mgg^j)= \left[( N_{i}^{\mathrm{sig}} + N_{i}^{\mathrm{spur}}\cdot\thetaspurc)\cdot f_{i}^{\mathrm{sig}}(\mgg^j,\vecth_{i}^\mathrm{sig}) + N_{i}^{\mathrm{bkg}} \cdot f_{i}^{\mathrm{bkg}}(\mgg^j,\vecth_{i}^\mathrm{bkg})\right]/N_i, 
\end{eqnarray*} 
where $\vecth_{i}^\mathrm{sig}$ and $\vecth_{i}^\mathrm{bkg}$ are nuisance parameters associated with systematic
uncertainties affecting the resolutions and positions (Section~\ref{sec:systshape}) of the invariant mass distributions of
the signal $f_{i}^{\mathrm{sig}}$ (further detailed in Section~\ref{sec:sigmod}) or the shape of the background
$f_{i}^{\mathrm{bkg}}$ (as explained in Section~\ref{sec:bkdmod}), respectively.

Systematic uncertainties are incorporated into the likelihood function by multiplying the
relevant parameter of the statistical model by a factor
\begin{equation*}
  F_\mathrm{G}(\sigma,\theta)=(1+\sigma\cdot\theta) 
\end{equation*}
in the case of a Gaussian pdf
for the effect of an uncertainty of size $\sigma$ or, for cases where a negative model parameter does not make physical
sense (e.g.\ the uncertainty in the integrated luminosity), by a factor
\begin{equation*}
  F_\mathrm{LN}(\sigma,\theta)=e^{\sqrt{\ln(1+\sigma^2)} \theta}
\end{equation*}
for a log-normal pdf. In both cases the corresponding component of the constraint product $G(\theta)$ is a unit Gaussian
centered at zero for the nuisance parameter $\theta$. The systematic uncertainties affecting the yield and mass resolution use the log-normal
form while a Gaussian form is used for all others. When two uncertainties are considered fully correlated they share the
same nuisance parameter. Systematic uncertainties with partial correlations
are decomposed into their uncorrelated and fully correlated components before being assigned to nuisance parameters.

All measured Higgs boson signal yields are determined with the profile likelihood ratio test statistic 
\begin{equation} \label{eq:likelihoodratio}
\Lambda(\nu) = -2\ln{\LL(\nu,\hat{\vecth}_\nu)\over \LL(\hat{\nu},\hat{\vecth})},  
\end{equation} 
where $\hat{\nu}$ and $\hat{\vecth}$ are the values of the parameter of interest (e.g. a signal strength or a simplified
template cross section) and nuisance parameters that unconditionally maximize the likelihood while $\hat{\vecth}_\nu$
are the values of the nuisance parameters that maximize the likelihood on the condition that the parameter of interest
is held fixed to a given value $\nu$.
In the asymptotic approximation, which is valid for all the results presented here, $\Lambda(\nu)$ may be
interpreted as an increase in $\chi^2$ from its minimum value~\cite{Cowan:2010js} such that approximate confidence
intervals are easily constructed.
The total uncertainty in $\nu$ is thus obtained from the $\nu$ values such that $\Lambda(\nu) = 1$ with all other parameters
``profiled'' ({\em i.e.} set to the values that maximize the likelihood for those values of $\nu$). 
Theory uncertainties in the parameters of interest are found by fixing the nuisance parameters
associated with experimental uncertainties and subtracting in quadrature the 
statistical uncertainty. The statistical uncertainty is similarly determined, by fixing all nuisance 
parameters to their best-fit values, except for those describing the background shape and normalization.
The experimental uncertainty is found by subtracting in quadrature the
theory and the statistical uncertainties from the total uncertainty. 

\subsection{Limit setting in the absence of a signal}

In the absence of a significant signal yield in the measured production process categories or fiducial regions, upper limits
on the corresponding signal strength or cross section are placed.
For production-mode measurements, the limit is set by treating all other parameters of the fit as nuisance parameters. 
For the fiducial regions, each measurement is split into two orthogonal categories, one of which contains the events in
the specified fiducial region and one that contains the events that are outside of it. The diphoton spectrum in both
sets of events are simultaneously analyzed to extract the desired limit. 

For category-based measurements the 95\% CL upper limit on the parameter of interest $\nu$ is determined using the $\mathrm{CL}_\mathrm{s}$ prescription~\cite{Read:2002hq}. For this, the agreement between data and the expected yield for the hypothesized value of the parameter of interest $\nu$ is quantified by the test statistic, $q_\nu$, defined as 
\begin{align} \label{eq:qmu}
 q_\nu = \bigg\{ \begin{matrix} \Lambda(\nu) & 0 < \hat \nu \le \nu \\ 
                     0 & \nu  < \hat \nu \end{matrix} \, , 
\end{align}
where $\hat \nu \ge 0$ is the fitted parameter of interest. The observed value of the test statistic, $q_{\rm obs}$, is
determined from the ratio of the likelihood obtained by fixing the number of signal events to that predicted for a given
value of the parameter of interest, to the likelihood normalized by allowing the number of signal events to float in the
fit. The asymptotic behavior of Eq.~(\ref{eq:qmu}) is well known~\cite{Cowan:2010js}. For fiducial measurements the 95\% CL upper limit are determined using a one-sided Gaussian interpretation of the observed cross section.


\section {Systematic uncertainties} \label{sec:syst}

Several sources of systematic uncertainty are considered in this measurement. They can be grouped into three
categories: (i) uncertainties associated with the parameterization of the signal and background when fitting the
\mgg\ spectrum, (ii) experimental uncertainties arising either from the extraction of the signal in a given category or from migrations between categories, and (iii) theoretical and modeling uncertainties in each category, causing migrations between categories, or affect the fiducial acceptance. 

The origin of the uncertainties and their treatment are discussed in detail below and summarized in Table~\ref{tab:tableNPs}.\footnote{The breakdown of uncertainties differs from those used in the Run 1 measurement Ref.~\cite{Aad:2014eha} as more updated recommendations for experimental and theory uncertainties are used.} 

The analysis based on event reconstruction categories
and those of fiducial cross sections treat yield and migration
uncertainties differently: whereas the former incorporate them directly into the likelihood function (cf. Section~\ref{sec:statmod}), the latter incorporate them at a later stage as part of the correction factor (introduced in Section~\ref{sec:fid_def}) or the luminosity. Modeling uncertainties were also estimated with different approaches as discussed further in Sections~\ref{sec:theo} and \ref{sec:theo_fid}.
A summary of the impact of the uncertainties on the measurement is given in Sections~\ref{sec:sig_strength} and~\ref{sec:fid_syst}.

\begin{table}[!tp]
\caption{
Summary of the sources of systematic uncertainties for results based on event reconstruction categories or fiducial integrated and differential cross sections. 
The columns labels ``Category Likelihood'' and ``Fiducial Likelihood'' provide an overview about which terms are part of the Likelihood
($\checkmark$) or incorporated at a later stage~\mbox{(-)}. Both sets of results incorporate uncertainties associated with the Higgs boson mass, photon energy scale and resolution, and uncertainties associated with the choice of
the background function into the likelihood function, either using log normal (${F_\mathrm{LN}(\sigma_i,\theta_i)}$) or Gaussian constraints (${F_\mathrm{G}(\sigma_i,\theta_i)}$) with $\sigma_i$ denoting the systematic uncertainty ($i$ is the index to each of the unique nuisance parameters~$\theta$).
When acting on $N_\mathrm{S}^\mathrm{tot}$ the uncertainty value is  the same for  all processes, whereas the uncertainty has a different value for each signal process for the case denoted~$N_\mathrm{S}^\mathrm{p}$. The number of nuisance parameters, $N_\mathrm{NP}$, for the spurious signal uncertainty varies, e.g. for the category-based results 31 independent error sources are present and for the differential measurements one source per measured bin is included. 
}
\begin{center} \footnotesize
{\def\arraystretch{1.2}
\begin{tabular}{cccrccc}
\hline\hline 
     & & Systematic uncertainty source &  $N_\mathrm{NP}$ & Constraint & Category & Fiducial  \\ 
     & &  &  &  & Likelihood & Likelihood  \\ 
\hline\hline 
\parbox[t]{1mm}{\multirow{6}{*}{\rotatebox[origin=c]{90}{Theory}}} &
\parbox[t]{1mm}{\multirow{3}{*}{\rotatebox[origin=c]{90}{}}}
       & ggH QCD                 & 9 &  $N_\mathrm{S}^\mathrm{ggH} \, F_\mathrm{LN}(\sigma_i,\theta_i)$  & \checkmark & -   \\ 
     & & Missing higher orders (non-\ggH)   &  6 & $N_\mathrm{S}^\mathrm{p} \, F_\mathrm{LN}(\sigma_i,\theta_i)$ & \checkmark & - \\
     & & $\bfhyy$                 &  1 & $N_\mathrm{S}^\mathrm{tot} \, F_\mathrm{LN}(\sigma_i,\theta_i)$  & \checkmark & - \\
   & \parbox[t]{1mm}{\multirow{3}{*}{\rotatebox[origin=c]{90}{}}}  
       & PDF                    & 30 & $N_\mathrm{S}^\mathrm{p} \, F_\mathrm{LN}(\sigma_i,\theta_i)$ & \checkmark & - \\
     & & $\alpha_\mathrm{S}$    &  1 & $N_\mathrm{S}^\mathrm{p} \, F_\mathrm{LN}(\sigma_i,\theta_i)$ & \checkmark & - \\
     & & UE/PS                  &  5 & $N_\mathrm{S}^\mathrm{p} \, F_\mathrm{LN}(\sigma_i,\theta_i)$ & \checkmark & - \\
\hline\hline 
\parbox[t]{1mm}{\multirow{13}{*}{\rotatebox[origin=c]{90}{Experimental}}} &
\parbox[t]{1mm}{\multirow{5}{*}{\rotatebox[origin=c]{90}{Yield}}}
       & Heavy flavor content &   1 & $N_\mathrm{S}^\mathrm{p} \, F_\mathrm{LN}(\sigma_i,\theta_i)$    & \checkmark & - \\
     & & Luminosity           &   1 & $N_\mathrm{S}^\mathrm{tot}  \, F_\mathrm{LN}(\sigma_i,\theta_i)$ & \checkmark & - \\
     & & Trigger              &   1 & $N_\mathrm{S}^\mathrm{tot}  \, F_\mathrm{LN}(\sigma_i,\theta_i)$ & \checkmark & - \\
     & & Photon identification&   1 & $N_\mathrm{S}^\mathrm{p}  \, F_\mathrm{LN}(\sigma_i,\theta_i)$   & \checkmark & - \\
     & & Photon isolation     &   2 & $N_\mathrm{S}^\mathrm{p}  \, F_\mathrm{LN}(\sigma_i,\theta_i)$   & \checkmark & - \\
   \cline{2-5}
 & \parbox[t]{1mm}{\multirow{8}{*}{\rotatebox[origin=c]{90}{Migration}}} 
       & Flavor tagging       & 14 & $N_\mathrm{S}^\mathrm{p}  \, F_\mathrm{LN}(\sigma_i,\theta_i)$   & \checkmark & - \\
     & & Jet                  & 20 & $N_\mathrm{S}^\mathrm{p}  \, F_\mathrm{LN}(\sigma_i,\theta_i)$   & \checkmark & - \\
     & & Jet flavor composition  & 7 & $N_\mathrm{S}^\mathrm{p}  \, F_\mathrm{LN}(\sigma_i,\theta_i)$ & \checkmark & - \\
     & & Jet flavor response  & 7 & $N_\mathrm{S}^\mathrm{p}  \, F_\mathrm{LN}(\sigma_i,\theta_i)$    & \checkmark & - \\
     & & Electron             &  3 & $N_\mathrm{S}^\mathrm{p}  \, F_\mathrm{LN}(\sigma_i,\theta_i)$   & \checkmark & - \\
     & & Muon                 & 11 & $N_\mathrm{S}^\mathrm{p}  \, F_\mathrm{LN}(\sigma_i,\theta_i)$   & \checkmark & - \\
     & & Missing transverse momentum  &  3 & $N_\mathrm{S}^\mathrm{p}  \, F_\mathrm{LN}(\sigma_i,\theta_i)$   & \checkmark & - \\
     & & Pileup               &  1 & $N_\mathrm{S}^\mathrm{p}  \, F_\mathrm{LN}(\sigma_i,\theta_i)$   & \checkmark & - \\
     & & Photon energy scale  &  40 & $N_\mathrm{S}^\mathrm{p}  \, F_\mathrm{LN}(\sigma_i,\theta_i)$  & \checkmark & -   \\     
 \hline\hline 
  \parbox[t]{1mm}{\multirow{3}{*}{\rotatebox[origin=c]{90}{Mass}}}
     & &   ATLAS-CMS $m_H$           &  1  & $\mu_\mathrm{CB} \, F_\mathrm{G}(\sigma_i,\theta_i)$     & \checkmark & \checkmark    \\
     & &   Photon energy scale       &  40 & $\mu_\mathrm{CB} \, F_\mathrm{G}(\sigma_i,\theta_i)$     & \checkmark & \checkmark  \\
     & &   Photon energy resolution  &  9 & $\sigma_\mathrm{CB}  \, F_\mathrm{LN}(\sigma_i,\theta_i)$ & \checkmark & \checkmark  \\
   \cline{2-5}
 \hline\hline 
  Background   & & Spurious signal       & Varies & $N_{\mathrm{spur},c}\, \theta_{\mathrm{spur},c}$      & \checkmark & \checkmark        \\ 
\hline\hline 
\end{tabular}}
\end{center}
\label{tab:tableNPs} 
\end{table}

\subsection{Systematic uncertainties in the signal and background modeling from fitting the
\mgg\ spectrum} \label{sec:systshape}

Systematic uncertainties associated with the signal and background parameterizations are treated in a similar way for all
the measurements. These include systematic uncertainties in the photon energy scale and resolution, and the uncertainties
due to the specific choice of background model. 

The fit to the \mgg\ spectra is performed for a Higgs boson mass of $\mH = 125.09 \pm 0.24$\,\GeV~\cite{Aad:2015zhl}.
The uncertainties in the photon energy scale and resolution impact the signal model, as the photon energy scale shifts
the position of the peak and the assumed energy resolution broadens or narrows the signal peak relative to its nominal width.
Uncertainties in the photon energy scale are included as nuisance parameters associated with Gaussian
constraint terms in the likelihood functions. Uncertainties in the photon energy resolution are included as nuisance
parameters, and are typically among the dominant sources of systematic
uncertainty in the measurement. The systematic uncertainties in the photon energy resolution and scale follow those in
Refs.~\cite{Aad:2014nim,phot2015}. The overall energy scale factors and their uncertainties have been determined using
$Z\rightarrow e^+ e^-$ events collected during 2015 and 2016. Compared to Ref.~\cite{phot2015}, several systematic
uncertainties were re-evaluated with the 13~\TeV\ data, including uncertainties related to the observed LAr cell
non-linearity, the material simulation, the intercalibration of the first and second layer of the calorimeter, and the
pedestal corrections. The typical impact of the photon energy scale uncertainties is to shift the peak position by
between $\pm~0.21\%$ and $\pm~0.36\%$ of the nominal peak position, whereas the typical impact of the photon energy
resolution uncertainty is to change the width of the signal distribution by between $\pm~6\%$ and $\pm~13\%$ of the
nominal width. The size of both uncertainties is dependent on the energy, rapidity and jet activity of the selected photon pair.

An additional uncertainty in the signal peak position is added as a nuisance parameter in the fit, reflecting the
uncertainty in the measurement of the Higgs boson mass of 0.24\,\GeV~\cite{Aad:2015zhl}. The uncertainty in the Higgs
boson mass is dominated by the statistical component, and the systematic component has contributions from both the 
ATLAS and the CMS muon momentum and electromagnetic energy scale uncertainties. Therefore, the correlation between this
uncertainty and the photon energy scale uncertainty in the measurements presented here is considered negligible. A variation of the signal mass by $\pm~0.24$\,\GeV\ (without including this uncertainty in the fit) is found to impact the measured global
signal strength or the diphoton fiducial cross section by less than $\pm~0.1\%$.

The uncertainty due to the choice of background function is taken to be the spurious signal yield obtained when 
fitting the \mgg\ spectrum reconstructed from background-only simulated samples (or signal-suppressed control regions
in data), as discussed in Section~\ref{sec:sigext}.

\subsection{Experimental systematic uncertainties affecting the expected event yields} \label{sec:systyield}

There are two categories of uncertainties: 1) those in the expected overall signal yield and 2) those that cause migrations of events between categories and bins, as well as into and out of the photon fiducial selection.

The sources of uncertainties in the expected signal yield consist of:
\begin{itemize}
\item The luminosity delivered to the ATLAS experiment. 
The uncertainty in the combined 2015+2016 integrated luminosity is 3.2\%.
It is derived, following a methodology similar to that detailed in Ref.~\cite{Aaboud:2016hhf},
from a calibration of the luminosity scale using $x$--$y$ beam-separation scans performed in August 2015 and May 2016.
\item The efficiency of the diphoton trigger. Its uncertainty is estimated 
to be 0.4\%  by comparing the trigger efficiencies determined using
a bootstrap method~\cite{Aaboud:2016leb} in data and simulation.
\item The photon identification efficiency. Its uncertainty is estimated to be 1.6\% and is obtained by varying the efficiency scale factors within their
uncertainties, derived from control samples of photons from radiative $Z$ boson decays and from inclusive $\gamma$ events, 
and of electrons from $Z \to e^+ \, e^-$ decays. 
\item The photon track isolation efficiency. Its uncertainty is estimated to be 0.8\% and is derived from measurements of the
efficiency correction factors using inclusive photon control samples.
\item The photon calorimeter isolation efficiency. Its uncertainty is estimated to be 0.1\% and is obtained from the difference
between applying and not applying corrections derived from inclusive photon events to the calorimeter isolation variable 
in the simulation.
\end{itemize}
Uncertainties which affect the calibration of photons, jets, and leptons cause migrations between categories and bins, as well as migrations into and out of the fiducial acceptance.
These include:
\begin{itemize}
\item The modeling of pileup in the simulation. The corresponding uncertainty is derived by varying the average number of pileup events in the simulation by an amount consistent with data. The typical size ranges from 1.4\% up to 5.6\% depending on the category or fiducial cross section bin.
\item Uncertainties in the photon energy scale and resolution. These uncertainties cause
migrations into and out of the fiducial volume or between the event reconstruction categories and impact the expected number of
events. The calibration of the absolute energy scale is derived using $Z\to e^+\,e^-$ decays.
The impact of the corresponding uncertainties is small, however, for all measurements, and ranges for instance between 0.2\% for events with a low diphoton \pt\ up to 1.9\% for events with a high diphoton \pt .
\item Uncertainties in the jet energy calibration and the jet energy resolution.
Uncertainties in the jet energy scale 
and resolution are estimated by varying the jet energies by an amount commensurate with the differences
observed between 13~\TeV\ data and simulation in the transverse momentum balance in dijet, $\gamma+{\rm jet}$ and
$Z+{\rm jet}$ events~\cite{Aaboud:2017jcu,jes2015,Aad:2014bia}. The typical size of this uncertainty ranges from 2.8\% to 15\%.
\item Uncertainties due to the efficiency of the jet vertex
tagger. Such uncertainties are estimated by shifting the associated corrections applied to the simulation by an amount allowed by the data.
For the measurement of the fiducial integrated and differential cross sections, uncertainties associated with the modeling of
pileup jets in the simulation are estimated by recalculating the correction factor after removing 20\% of pileup jets
at random, which is commensurate with the observed differences in data and simulation for jets tagged as low-JVT
(pileup) and high-JVT (hard scatter). Its typical size ranges from nil to 0.3\%.
\item Uncertainties associated with the efficiency of the $b$-tagging algorithm. They have been
estimated to be typically of the order of 3\% and are determined using $t\bar{t}$ events in 13~\TeV\ data for jets containing the decay of a $b$-quark, using the method outlined in
Ref.~\cite{Aad:2015ydr}. The corresponding uncertainties in the identification of jets originating from $c$-quarks,
light quarks and gluons are taken directly from Run~1 studies~\cite{Aad:2015ydr}, with additional uncertainties to cover the
extrapolation to Run~2 conditions. 
\item Uncertainties in the electron~\cite{electrons} and muon~\cite{Aad:2016jkr}
reconstruction, identification and isolation efficiencies. They have been obtained from
dilepton decays of $Z$ bosons and $J/\psi$ mesons collected in Run~2, using
a tag-and-probe technique. The typical size of these uncertainties is about 0.6\% for electrons and about 0.5\% for muons in the relevant categories
or fiducial regions. 
\item Uncertainties in the electron~\cite{ATL-PHYS-PUB-2016-015} and muon~\cite{Aad:2016jkr} energy and momentum scale and resolution. 
They are determined from comparisons between the reconstructed invariant mass in data and simulation
of dileptons from decays of $Z$ bosons or $J/\psi$ mesons. The impact is negligible for all measurements.
\item Uncertainties associated with energy scales and resolutions of photons, jets and leptons are propagated to the
$\ET^\mathrm{miss}$ uncertainty, together with the uncertainty in the contribution to $\ET^\mathrm{miss}$ from 
charged-particle tracks not associated with high-$\pT$ leptons, jets, or photon conversions~\cite{met2015b}. This results
in a typical migration uncertainties ranging from 4.0\% to 4.8\% for relevant categories or fiducial regions.

\end{itemize}

\subsection{Theoretical and modeling uncertainties for results based on event reconstruction categories} \label{sec:theo}

The overall theoretical cross-section uncertainties affect the signal strength measurements, which are ratios of the observed to
predicted event yields, but not the cross-section measurements which do not rely on absolute predictions.
Modeling uncertainties that alter the kinematic properties of the events, such as the Higgs boson transverse momentum or
the jet multiplicity, have an impact on both types of measurements.

The theoretical modeling uncertainties in the per-category acceptance of each production process affect the measurement
of production-mode cross sections and signal strengths. 
Uncertainties due to the choice of parton
distribution functions and the value of $\alpha_\mathrm{S}$ are estimated using the PDF4LHC15 recommendations~\cite{Butterworth:2015oua} with the
nominal \texttt{PDF4LHC\_nlo\_30\_as} PDF set. Samples using the \ctten\ PDF set are reweighted to PDF4LHC15 to estimate these uncertainties.
For the gluon--gluon fusion process, the
total production-mode cross section has been calculated at N${}^3$LO precision in QCD and has an uncertainty of 3.9\%, as determined by QCD-scale variations and including top, bottom, and charm quark mass effect uncertainties. However, the
perturbative uncertainty becomes significantly larger in different kinematic regions, e.g. when requiring additional
jets or high Higgs boson $\pT$. To take this effect into account nine uncertainty sources are included:
\begin{itemize}
\item Four sources~\cite{deFlorian:2016spz} account for uncertainties in the jet multiplicities due to 
  missing higher-order corrections: two accounting for yield uncertainties (with uncertainties up to 8.9\% in each STXS region) and two accounting for migrations between jet multiplicity bins (with uncertainties up to 18\% in each STXS region),
using the STWZ~\cite{Stewart:2013faa} and BLPTW~\cite{Liu:2013hba, Stewart:2013faa, Boughezal:2013oha} predictions as an
input. For more details see Table 20 of Ref.~\cite{deFlorian:2016spz}.
\item Three uncertainty sources parameterize modeling uncertainties in the Higgs boson $\pT$.
The first two encapsulate the migration uncertainty between the intermediate and high $\pT$ region with events with at least one jet. The third 
uncertainty parameterizes top-quark mass effects in the gluon--gluon fusion loop, where the difference between the LO and NLO predictions is taken
as an uncertainty due to missing higher-order corrections. This introduces a negligible uncertainty at low Higgs boson $\pT$ and a
sizable uncertainty of the order of 30\% at $\pT > 500$~\GeV.
\item Two sources account for the uncertainty in the acceptance of gluon--gluon fusion events in the VBF categories,
due to missing QCD higher-orders in the calculation. Such uncertainties are estimated by variations of the renormalization and factorization scales in \mcfm~\cite{Campbell:2010ff} by a factor of two around the nominal scale of $\mu_r = \mu_f = m_H$. The two sources account for the uncertainty in the overall normalization of $H+2~{\rm jet}$ and $H\ +\ge 3~{\rm jet}$ events 
as well as for the uncertainty due to the multivariate requirement on $\dphiggjj$ (cf. Section~\ref{sec:vbf_category}), which suppresses additional jet activity.
The uncertainty estimation uses an extension of the Stewart--Tackmann method~\cite{Stewart:2011cf,Gangal:2013nxa} and typically ranges between 20\% and 32\%. 
\end{itemize}
The applicability of these uncertainties to \nnlops\ was tested by comparing the STWZ+BLPTW and the \mcfm\ cross section predictions 
in variables relevant for the definition of the simplified cross-section bins, and reasonable agreement was found. In addition, 
the \ggH\ acceptance of \nnlops\ of all categories based on BDT classifiers is compared to the acceptance derived from the \amc\ prediction or Refs.~\cite{Alwall:2014hca, Frederix:2016cnl} which includes up to two jets at NLO accuracy using the \fxfx\ merging scheme~\cite{Frederix:2012ps}. 
Sufficient agreement was found for all categories and no additional modeling uncertainties are assigned based on these comparisons.  
\footnote{Recent measurements of QCD and electroweak (VBF) $Z$-boson production in association with two jets in Ref.~\cite{Aaboud:2017emo} show large deviations between the data and the predictions for the QCD $Zjj$ background at large $m_{jj}$. These differences are significantly larger than the 30–40\% uncertainties assigned here to the \ggH\ background in the experimental categories targeting \VBF\ Higgs boson production. Increasing this uncertainty to 100\% results in an increase in the theory uncertainty in the \VBF\ signal strengths or simplified template cross sections by a factor of about two, while the increase in the total uncertainty is about 10\%, as it is dominated by the statistical component.}

Finally, in the categories targeting production in association with top quarks, the normalization of
each of the \ggH, \VBF, and \VH\ production mechanisms is assigned an uncertainty of 100\%, motivated by comparisons of data with 
simulation in $t\bar{t}b\bar{b}$~\cite{Aad:2015yja} and $Vb$~\cite{Aad:2014dvb, Aad:2013vka} productions, but this has little impact on the final results.

The uncertainty in the modeling of the parton shower, underlying event and hadronization affects all measurements (labeled as ``UE/PS'' in the following).
It is estimated by taking the relative difference in acceptance at particle level after
switching the parton showering algorithm from \pythia\ to \herwigsev\
in the \ggH, \VBF, and \VH\ samples and from \herwigpp\ to \pythia\ in the \ttH\ sample, respectively. These differences are treated
as four independent uncertainty sources. Additionally, for \ggH\ the effect of the eigenvector tunes from the AZNLO set are merged to provide one additional uncertainty component. 

The theoretical modeling uncertainties in the measurement of signal strengths include all of the sources that affect the
measurement of the production-mode cross sections, plus additional uncertainties in the overall normalization of each
production mechanism. Uncertainties in the overall normalization of each production process from missing higher-order
QCD effects and the choice of parton distribution function are specified as part of the theoretical calculations used to
normalize the simulated samples. 
The normalization uncertainty from the $H\rightarrow\gamma\gamma$ branching ratio is taken from \hdecay\ and \prophecy.

\subsection{Theoretical and modeling uncertainties for fiducial integrated and differential results} \label{sec:theo_fid}

The theoretical modeling uncertainty in the detector correction factor (introduced in Section~\ref{sec:fid_def})
used to measure the fiducial integrated and differential cross sections is taken to be the envelope of the following three sources:
\begin{enumerate}
 \item The uncertainty in the relative contributions of the different Higgs boson production mechanisms. This uncertainty is estimated by varying the fraction of the \ggH, \VBF, \VH\ and \ttH\ processes by an amount commensurate with the 68\% confidence levels of the measured production mode cross-section ratios~\cite{Khachatryan:2016vau}. The variations of each production mechanism are carried out simultaneously and include the known correlations between the measured production mode cross-section ratios. These uncertainties range from 0.1\% to 31\%, depending on the fiducial region or differential variable, increasing typically in bins and regions sensitive to \ttH-production.
 \item  The uncertainty in the detector correction factor due to a possible mismodeling of the Higgs boson transverse momentum and rapidity
distributions is estimated by reweighting the Higgs boson distributions in simulation to match those observed in the
data. The resulting uncertainties range from 0.1\% to 4.5\%, increasing in fiducial regions and bins with high jet multiplicities.
 \item The uncertainty in the modeling of the parton shower, underlying event, and hadronization. This uncertainty is derived as described in Section~\ref{sec:theo} and the size of this uncertainty ranges from from 0.1\% up to 30\%, with the highest uncertainties in fiducial regions with large missing transverse energy.
\end{enumerate}
Typically differential measurements involving only the photon kinematics are less affected by these model uncertainties than
measurements using selections on jets or missing transverse momentum.
  
\subsection{Illustration of model errors for simplified template cross section and fiducial cross section measurements} \label{sec:theo_fid_stxs}

To illustrate the difference between the two approaches of assigning theory and model errors used for category based results and the fiducial cross section results, the theoretical modeling uncertainties in the corresponding zero-jet \ggF-dominated and \VBF-dominated regions are compared.

The simplified template cross section defined as $gg \to H$ events with $\left| y_{H} \right| < 2.5$ and no jets derives its sensitivity from the two categories requiring no jet and either one or both photons reconstructed in the barrel region of the electromagnetic calorimeter (defined by $\left| \eta \right| \leq 0.95 $). The total theory uncertainty is dominated by the uncertainty in the choice of parton distribution functions (1.5\%), in the value of $\alpha_\mathrm{S}$ (1.4\%), and in the modeling of the parton shower, underlying event, and hadronization (1.7\%), and amounts to a relative uncertainty of 2.7\%. The fiducial zero-jet cross section, in contrast, has only a modeling uncertainty of 0.1\%, dominated by the possible mismodeling of the Higgs boson transverse momentum and rapidity distributions.

The simplified template cross section defined as $qq \to Hqq$ events with $\left| y_{H} \right| < 2.5$ and no jets with $\pt^j < 200$~\GeV\ derives its sensitivity from four VBF categories employing multivariate methods to distinguish \VBF\ events from background processes. The total theory uncertainty is dominated by the modeling uncertainties in the parton shower, underlying event, and hadronization (9\%) and in the remaining \ggH\ contamination (5\%), and amounts to a relative uncertainty of 13\%. In contrast, the fiducial VBF cross section, defined by a dijet mass $m_{jj}$ of at least 400~\GeV, a large rapidity separation $\left| \Delta y_{jj}\right| > 2.8$, and an azimuthal difference between the Higgs boson and the dijet pair of $\dphiggjj > 2.6$, has only a modeling uncertainty of 4.5\%, dominated by the composition variation of the fractions of the \ggH, \VBF, \VH\ and \ttH\ processes (4.5\%).


\section{Measurement of total production-mode cross sections, signal strengths, and simplified template cross sections}
\label{sec:methods_coup}

\subsection{Event categorization}
\label{sec:categories}

The events satisfying the diphoton selection discussed in Section~\ref{sec:evselPV} are classified, in accord with the
reconstructed event kinematics and topology, into 31 exclusive categories that are optimized for the best separation of
the Higgs boson production processes and for the maximum sensitivity to the phase space regions defined by the stage 1
of the simplified template cross-section framework. A combined fit to the event reconstruction categories is then performed to
determine nine simplified template cross sections (with $|y_H|<$ 2.5), as well as production-mode cross sections and
signal strength interpretations of the data. The categorization proceeds from the production modes with the smallest
expected cross sections to the production modes with largest expected cross sections, in the order described below. In
categories with definitions based on jet properties, jets with transverse momenta greater than 30~\GeV\ are used, unless
explicitly stated otherwise. 

\subsubsection{\ttH\ and \tH\ enriched categories}

Nine categories enriched in events produced in association with a top quark are defined to target the \ttH, \tHqb, and
\tHW\ production modes. These categories are separated into a hadronic channel, where top quarks in the event decay to
hadrons via $t \to Wb \to qq'b$; and a leptonic channel, where at least one top quark decays to a charged lepton via $t
\to Wb \to \ell\nu b$. The single top quark categories are optimized for sensitivity to SM \tH\ production,
and are expected to provide additional sensitivity to anomalous values of the top quark Yukawa coupling.

Three categories target the leptonic channel by requiring the presence of at least one prompt lepton and at least
one $b$-tagged jet with transverse momentum greater than 25\,\GeV. Two of these categories target \tH\ production while
the third one is optimized for \ttH\ events. Both \tH\ categories veto events with more than one prompt lepton. The
first of these categories (``tH lep 0fwd'') contains events with at most three central jets ($|\eta| < 2.5$) and a veto on
forward jets ($|\eta| > 2.5$). The second \tH\ category (``tH lep 1fwd'') is defined by events with at most four central
jets and at least 1 forward jet. The ``ttH lep'' category includes events with at least two central jets, while no
requirement is applied to the forward jets. To suppress $ZH$ events with $Z\to \ell\ell$, same-flavor dilepton
candidates with an invariant mass within 10~\GeV\ of the $Z$ boson mass are vetoed.

\begin{figure*}[!tbp]
 \centering
\subfloat[] {\includegraphics[width=.5\textwidth]{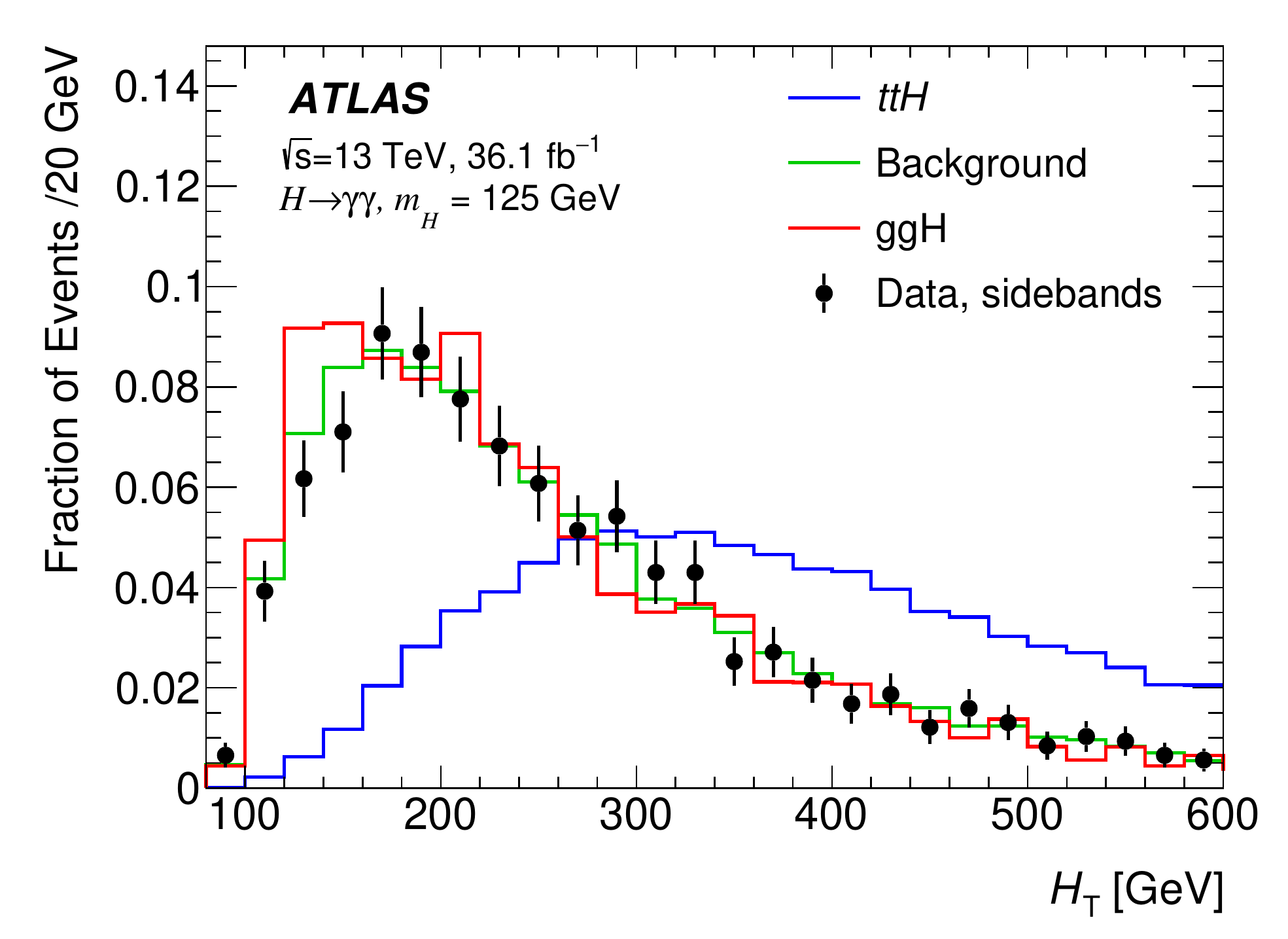}}
\subfloat[] {\includegraphics[width=.5\textwidth]{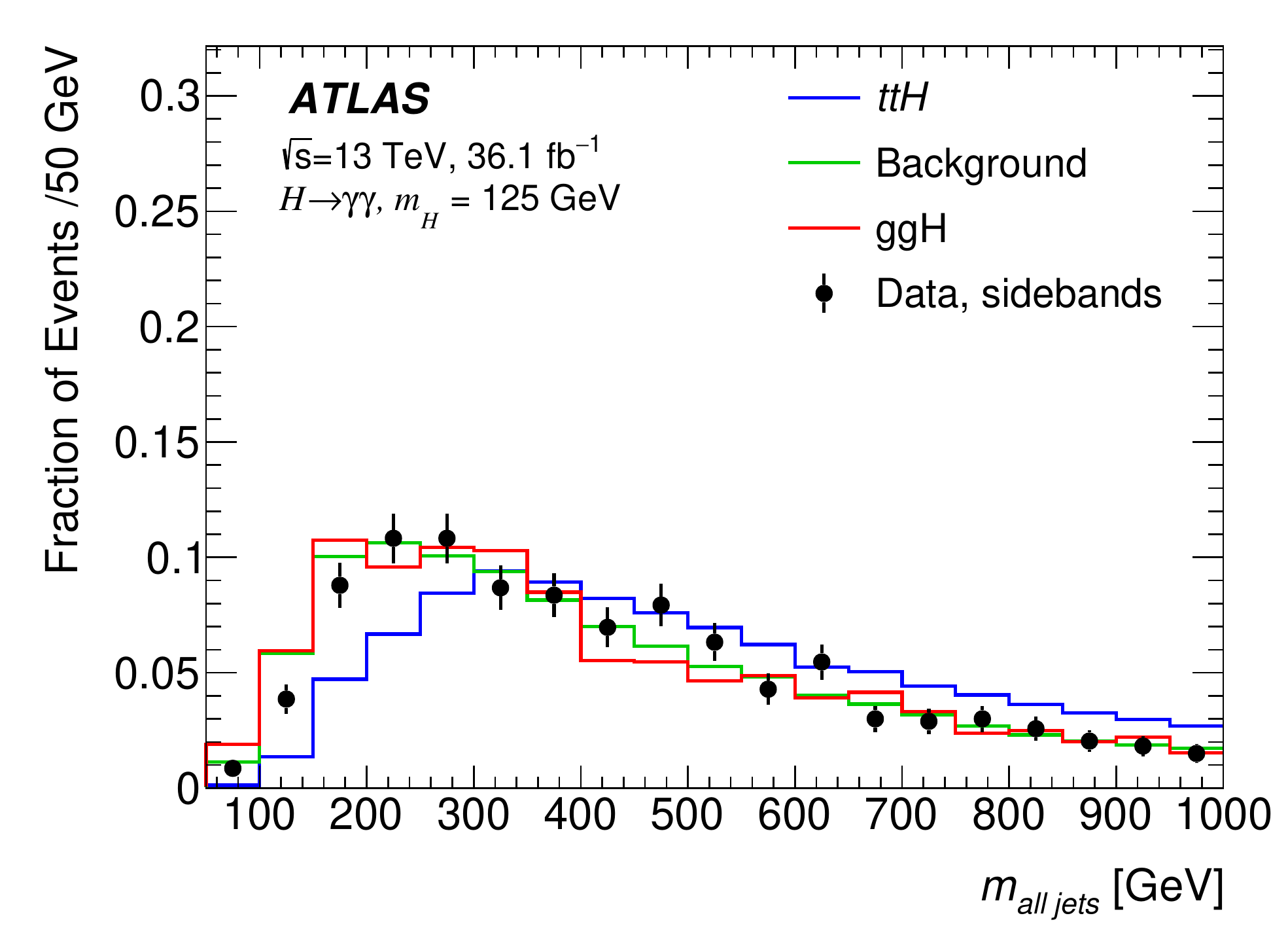}}
\caption{The normalized distributions for the expected background of two kinematic variables used for 
  the selection of the hadronic \ttH\ categories: 
  (a) $H_\mathrm{T}$ and (b) $m_\mathrm{all \, jets}$ for events after the first step of the selection (see
  text) for simulated $H\to\gamma\gamma$ events produced via \ttH\ (blue) and gluon--gluon fusion (red), for the
  expected background derived from the data control region (green) described in the text 
  and events from data with 105\,\GeV $<m_{\gamma\gamma}<$ 120\,\GeV\ or 130\,\GeV $<m_{\gamma\gamma}<$ 160\,\GeV\ 
  (black dots with error bars showing the statistical uncertainty).
}
\label{fig:ttHvars}
\end{figure*}

Six categories target the hadronic decay channel by selecting events with no prompt leptons and at least three jets, of
which at least one is $b$-tagged. Four of these categories (``ttH had BDT1'' to ``ttH had BDT4'') are defined by means
of a boosted decision tree (BDT) trained to identify \ttH\ signal against \ggH\ and multijet background. The BDT
exploits five kinematic variables: $H_\mathrm{T}$, the scalar sum of jet transverse momenta, $m_\mathrm{all\ jets}$,
the mass of all jets, as well as the number
of all jets, central jets ($|\eta| < 2.5$), and $b$-tagged jets. 
The training uses \ttH\ and \ggH\ simulated events and
a data-driven multijet background sample defined by diphoton events with at least three jets and in which at least one
photon fails to meet either identification or isolation requirements. Using the BDT response as a discriminating variable,
events are separated into four categories with an expected fraction of \ttH\ events (among all Higgs boson events in
this category) of 95\%, 89\%, 86\%, and 79\%, respectively. Two additional hadronic categories enhanced in \tH\
production  (``tH had 4j1b'' and ``tH had 4j2b'') are included, defined by events with exactly four jets with transverse
momentum greater than 25~\GeV\ and split by events with exactly one or two $b$-tagged jets, respectively.
The distributions of two of the discriminating variables are shown in Figure~\ref{fig:ttHvars}.

\subsubsection{$VH$ leptonic enriched categories}
\label{sec:VHcats}

Five categories are enriched in Higgs boson production in association with a vector boson, based on different decays of
the vector bosons.

The \VH\ dilepton category (``VH dilep'') targets \ZH\ production with \Zll\ by requiring the presence of two
same-flavor opposite-sign leptons with an invariant mass between 70~\GeV\ and 110~\GeV. Two additional 
\VH\ one-lepton categories target $WH$ production with $W\to\ell\nu$. Events are requested to contain exactly one
selected electron or muon. To suppress $ZH$ events with $Z$ bosons decaying to $ee$, in which an electron is
misidentified as a photon, a veto is applied to events in which the invariant mass of the selected electron and any of
the two signal photons is between 84~\GeV\ and 94~\GeV. Events are then split into two regions, where the $\pT$ of the
lepton+\met\ system is higher (``VH lep High'') or lower (``VH lep Low'') than 150~\GeV. An additional requirement on
the \met\ significance, defined as $\ET^\mathrm{miss} / \sqrt{\sum\ET}$, of at least 1.0 is applied to events in the low lepton+\met\ $\pT$ category.

Two \VH\ missing transverse momentum categories target \ZH\ production with $Z\to\nu\nu$ and $W\to\ell\nu$ where the
lepton was not reconstructed or failed to meet the selection criteria. One category (``VH MET Low'') requires
$80~\GeV<\met<150~\GeV$ and \met\ significance greater than 8. The other category (``VH MET High'') requires
$\met>150~\GeV$ and \met\ significance greater than 9, or $\met>250~\GeV$.

\subsubsection{BSM enriched and $VH$ hadronic categories}
\label{sec:BSM_VHcats}

To provide sensitivity to potential beyond SM contributions, a category (``jet BSM'') defined by events with
a leading jet with transverse momentum greater than 200~\GeV\ is included in the event selection. This category
includes SM events in the typical VBF topology, boosted $V(\to jj)H$ production where the vector boson is reconstructed
as a single jet, as well as events produced in gluon--gluon fusion with an energetic jet.

Two $VH$ hadronic categories target $VH$ production with a hadronically decaying vector boson. Events are required to
have at least two jets with $60 < \mjj < 120$\,\GeV. A BDT classifies the events using the following information: the
dijet invariant mass, the component \pttgg\ of the diphoton $\vec{p}_\mathrm{T}$ transverse to its thrust 
axis in the transverse plane, the rapidity difference between the dijet and the diphoton system, and the 
cosine $\cos\theta^*_{\gamma\gamma,jj}$ where $\theta^*_{\gamma\gamma,jj}$ is the angle between the diphoton systems momentum and the direction of
motion of the diphoton--dijet system in the Collins--Soper frame. The training uses \VH\ events as signal, and a mixture
of simulated signals (everything except \VH\ events), simulated $\gamma\gamma$ events,
and $\gamma j$ and $jj$ data control samples as background. Using the BDT response as a discriminating variable, events
are classified into two categories (``$VH$ had tight'' and ``$VH$ had loose'') with an expected fraction of signal events
due to $VH$ production of 42\% and 25\%, respectively.

\begin{figure*}[!tbp]
 \centering
\subfloat[] {\includegraphics[width=.5\textwidth]{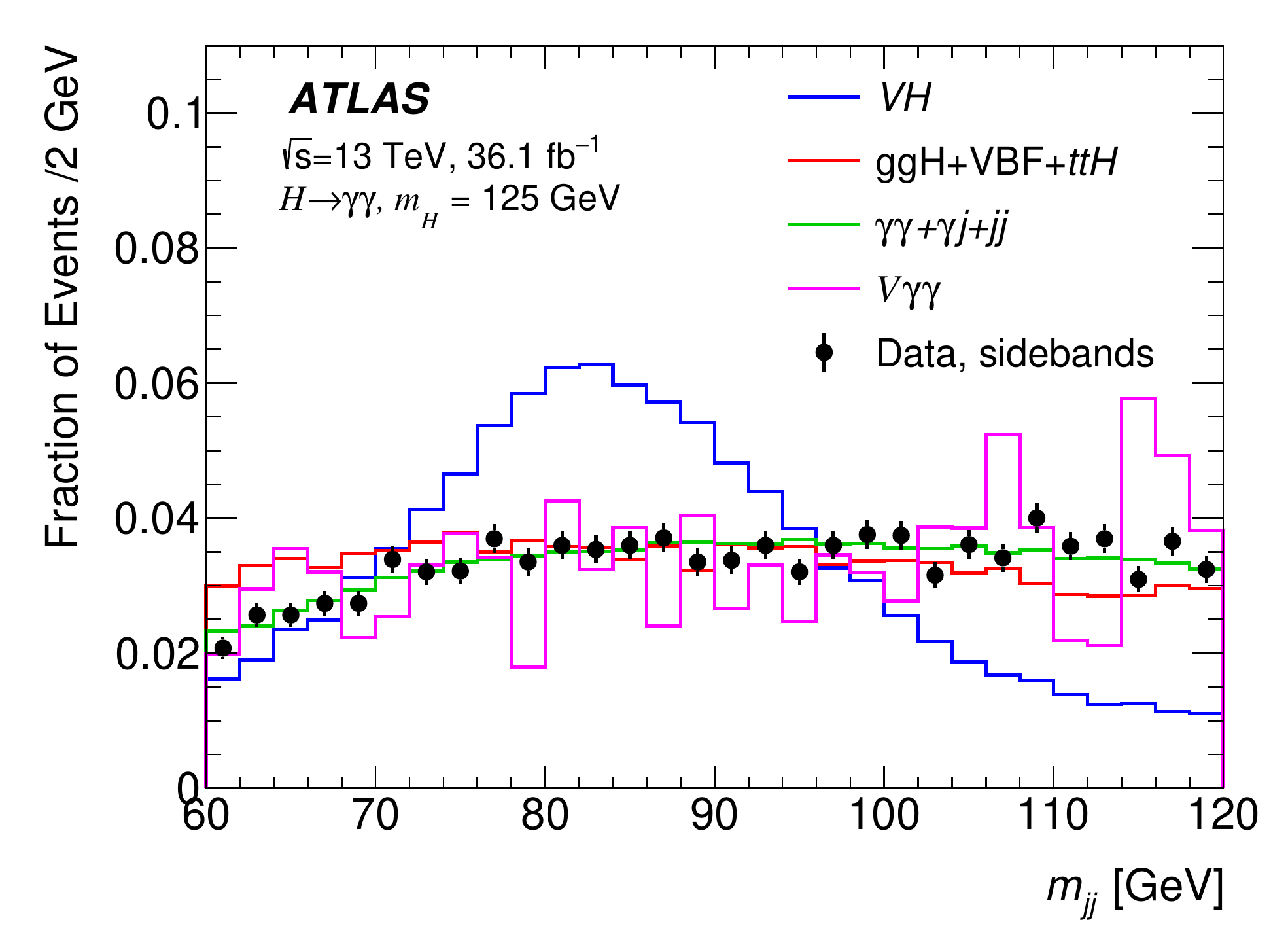}}
\subfloat[] {\includegraphics[width=.5\textwidth]{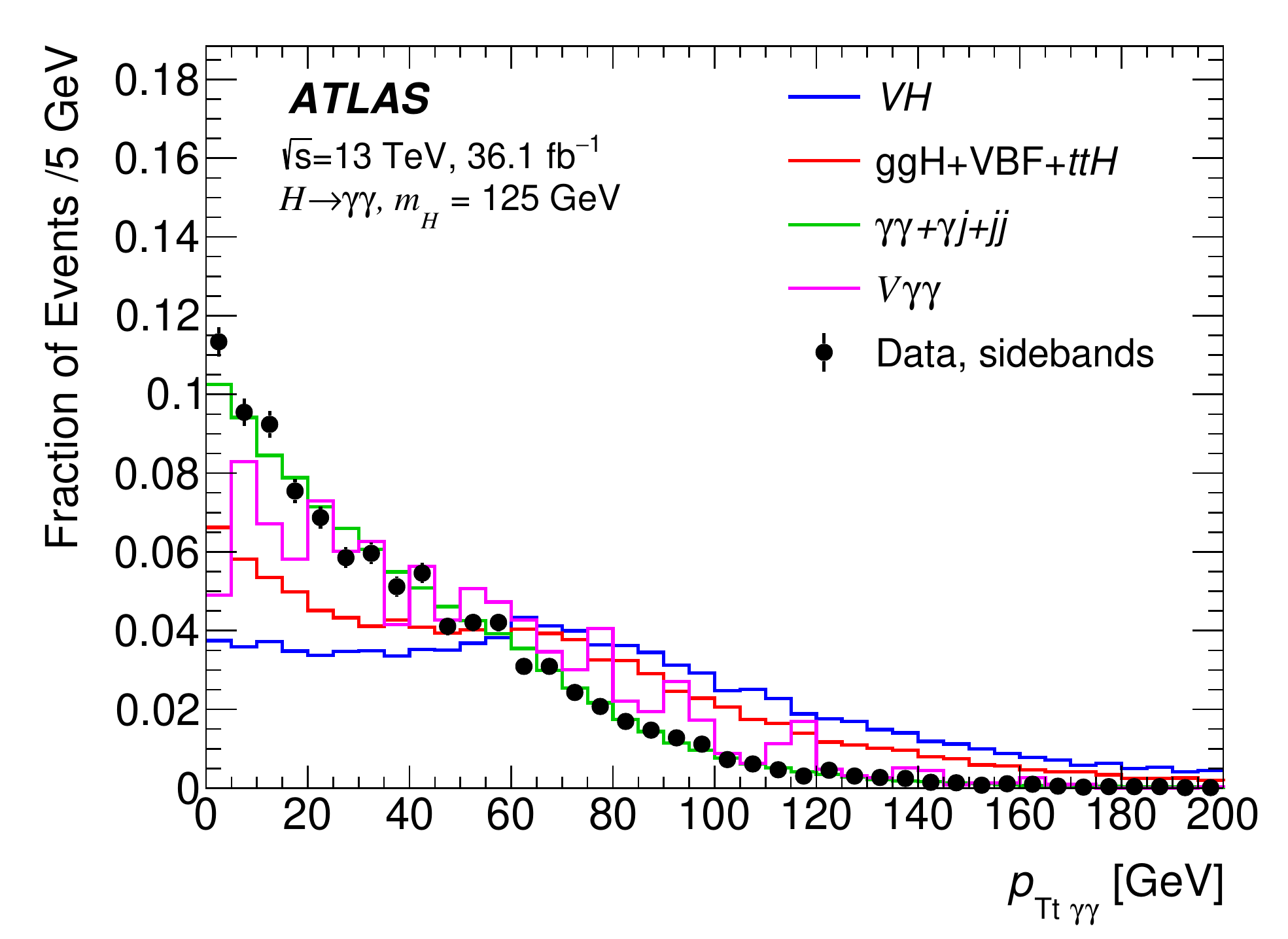}}
\caption{The normalized distributions of two kinematic variables used for the selection of the \VH\ hadronic categories: 
  (a) $m_{jj}$ and (b) $\pttgg$ for events after the first step of
  the $\VH$ hadronic category preselection (see text)
  for simulated $H\to\gamma\gamma$ events produced in association
  with hadronically decaying vector bosons (blue) or through
  \ggH, \VBF\ or \ttH\ processes (red), for the expected background
  from data ($\gamma j$, $jj$) and simulation ($\gamma\gamma$, $V\gamma\gamma$)
  control samples (green, purple), and for events from data with 105\,\GeV $<m_{\gamma\gamma}<$ 120\,\GeV\ or 
  130\,\GeV $<m_{\gamma\gamma}<$ 160\,\GeV\ (black dots with error bars showing the statistical uncertainty).
}
\label{fig:VHvars}
\end{figure*}

Figure~\ref{fig:VHvars} shows the distributions of
$m_{jj}$ and \pttgg\ in signal and background events and in events selected in data from the
$m_{\gamma\gamma}$ sidebands. The variables show good separation between $VH$ events and both the other signal events and
background events.

\subsubsection{VBF enriched categories}\label{sec:vbf_category}

Four categories are defined to enhance the sensitivity to vector boson
fusion production.
Events are required to contain at least two hadronic jets, and
the selections applied are based on the two leading jets ($j_1$, $j_2$) in
the event.
The pseudorapidity separation \deltarapjj\ between the two
leading jets is required to be greater than 2. In addition
$|\eta_{\gamma\gamma}-0.5 (\eta_{j1}+\eta_{j2})|$
is required to be less than 5, with $\eta_{\gamma\gamma}$ denoting the pseudorapidity
of the diphoton system. The events are first split into two regions based on the value of
the transverse momentum $\pTHjj$ of the vector sum of the momenta of the reconstructed Higgs boson and of the two leading jets. This variable is highly correlated with the $\pt$ of the third jet due to momentum 
balance. The signal in the $\ptggjj < 25~\GeV$ ``low \pTHjj'' region is dominated by 
exclusive two-jet-like events, while the signal in the $\ptggjj > 25~\GeV$ ``high
\pTHjj'' region is dominated by inclusive $\ge 3$-jet like events.
This choice minimizes the otherwise large \ggH\ jet-migration uncertainties in this phase space and 
is similar to a central-jet veto that separates contributions from \ggH\ and \VBF.

A BDT is then used to classify events in each region, using six kinematic variables: \mjj, \deltarapjj, \pttgg, the
absolute azimuthal difference of the diphoton and the dijet system \dphiggjj, the minimum angular separation between either of the two signal photons and either of the two leading jets $\Delta R^\mathrm{min}_{\gamma j}$, and \mbox{$|\eta_{\gamma\gamma}-0.5 (\eta_{j1}+\eta_{j2})|$}.
A requirement of \dphiggjj\ to be near $\pi$ effectively vetoes 
additional jets in the event by restricting the phase space for
additional emissions and, to avoid large theoretical uncertainties, the BDT does not use shape information for
events with $\dphiggjj > 2.94$ by merging these events into one bin.
The training of the BDT uses VBF events as signal, and a mixture of simulated
gluon--gluon fusion and $\gamma\gamma$ events and of $\gamma j$ and $jj$
data control samples as background.
Four exclusive categories are defined with "loose" and "tight" requirements on the BDT classifier in the two \pTHjj\ regions.
The "tight" category in the  $\pTHjj > 25~\GeV$ region has an
expected fraction of VBF events among all Higgs boson events in this
category of 49\%, while the "loose" category has an expected
fraction of VBF events of 20\%. 
In the $\pTHjj < 25~\GeV$ region the "tight" category has an expected fraction
of VBF events of 85\%, whereas the "loose" category has an expected fraction of 61\%. 

\begin{figure*}[!tbp]
 \centering
\subfloat[] {\includegraphics[width=.5\textwidth]{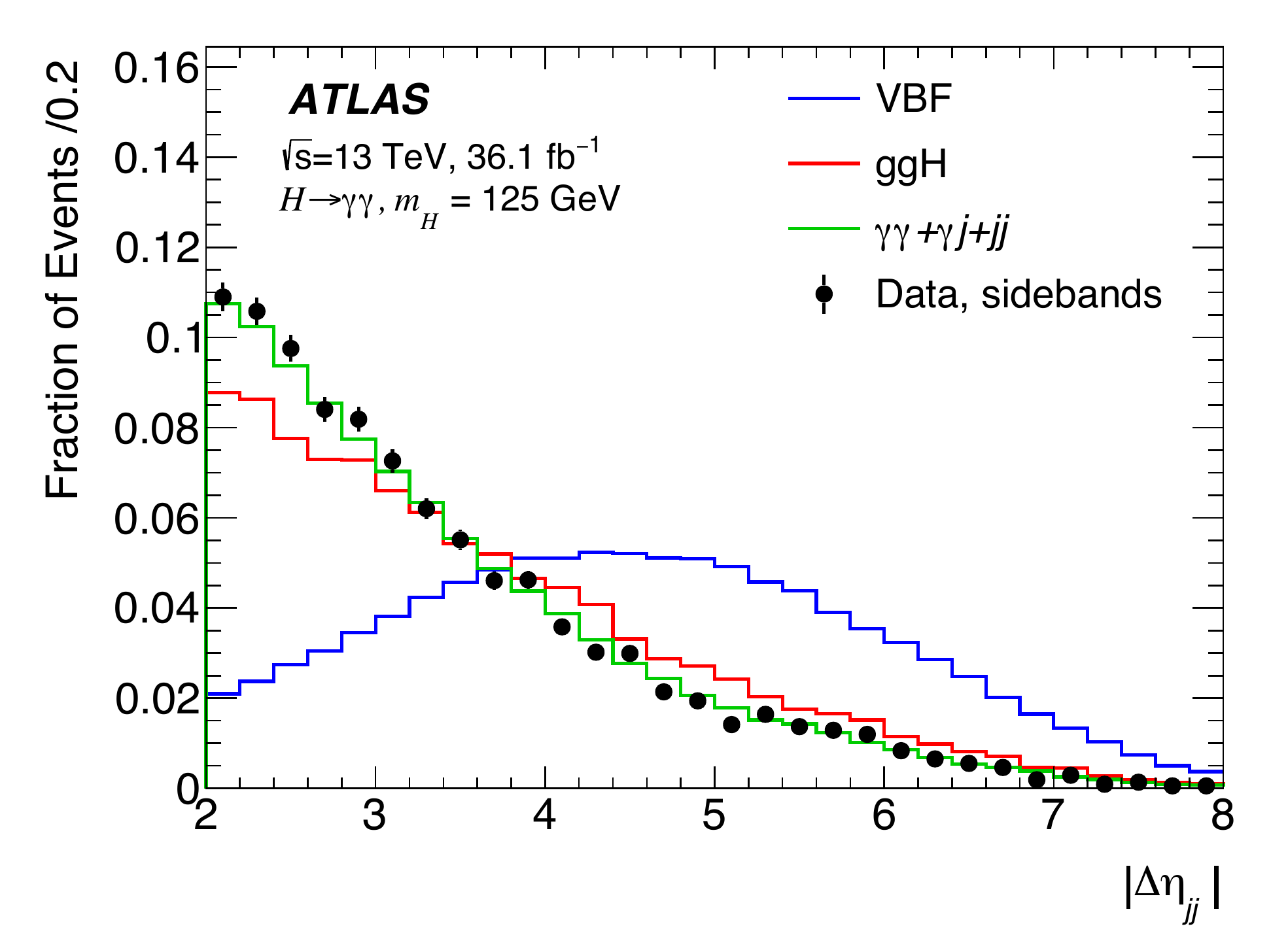}}
\subfloat[] {\includegraphics[width=.5\textwidth]{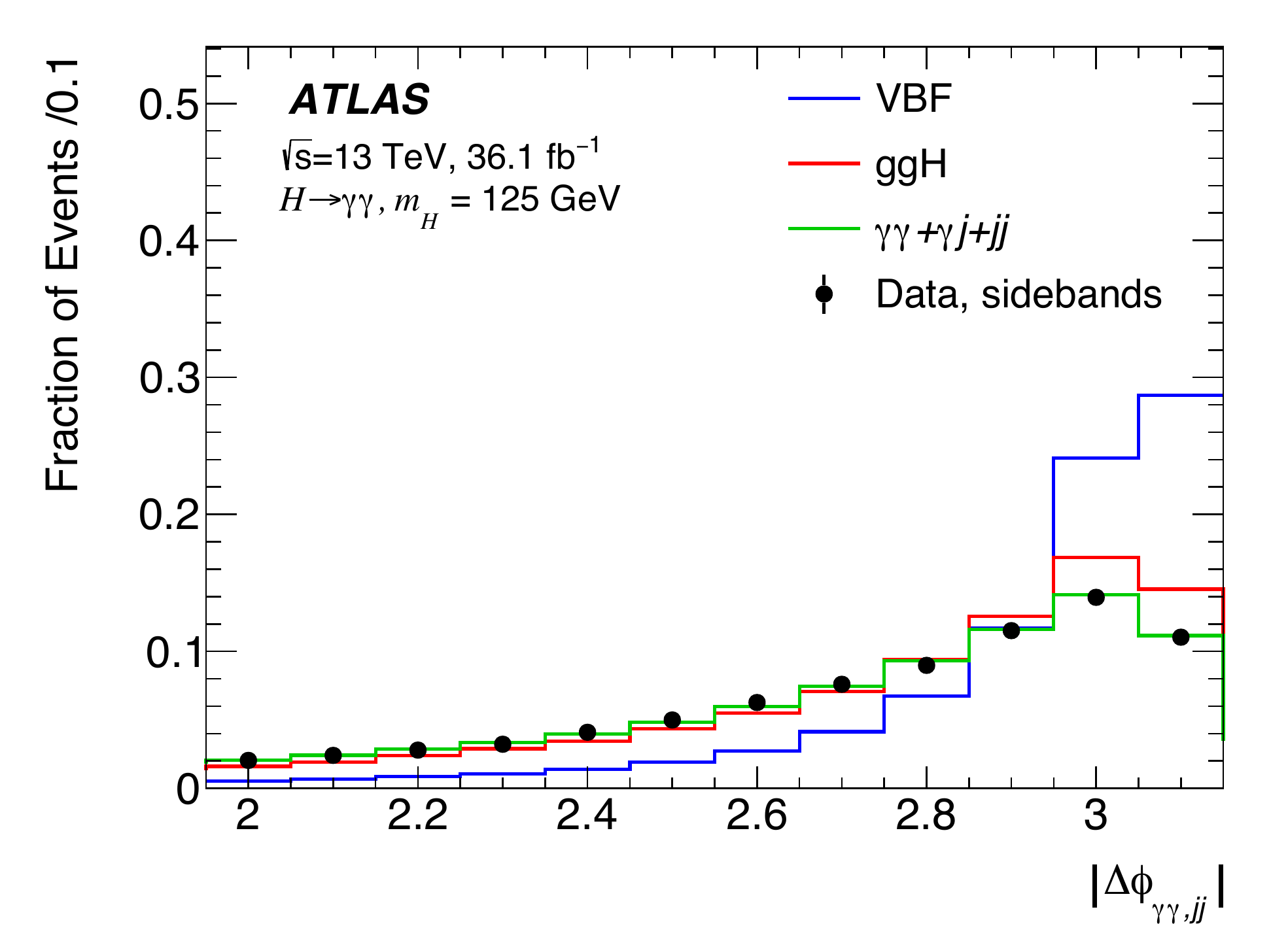}}
\caption{The normalized distributions for the expected background of two kinematic variables  used for the selection of the VBF categories: 
  (a) \deltarapjj\ and (b) \dphiggjj\ for events after the first step of the selection (see
  text) for simulated $H\to\gamma\gamma$ events produced via vector-boson fusion (blue) and gluon--gluon fusion (red), for the
  expected background from data ($\gamma j$, $jj$) and simulation ($\gamma\gamma$) control samples (green), and for
  events from data with 105\,\GeV $<m_{\gamma\gamma}<$ 120\,\GeV\ or 130\,\GeV $<m_{\gamma\gamma}<$ 160\,\GeV\ (black dots with error bars showing the statistical uncertainty).
}
\label{fig:VBFvars}
\end{figure*}

Figure~\ref{fig:VBFvars} shows the distributions of \deltarapjj\ and \dphiggjj\ in simulated
$H\to\gamma\gamma$ events, background events from simulated diphotons and data control samples of $\gamma j$ and $jj$ events, and events selected from the
$m_{\gamma\gamma}$ sidebands in data. The variables show good separation between VBF events and both gluon--gluon fusion
events and background events.

\subsubsection{Untagged categories}
\label{sec:ggH}

The remaining ``untagged'' events are dominated by events produced through
gluon--gluon fusion and they are further split into ten categories. 
The untagged events are first separated by jet multiplicity into events 
with zero jets, exactly one jet, or at least two jets. The zero-jet events 
are split into two categories with either two photons in the ``central''
pseudorapidity region $|\eta| < 0.95$, in which the energy resolution is better (``ggH 0J Cen''),
or with at least one photon in the ``forward'' region $|\eta| > 0.95$ which has worse energy resolution
(``ggH 0J FWD''). The exclusive one-jet (``ggH 1J'') and inclusive two-jet (``ggH 2J'') categories are
further split into regions of diphoton transverse momentum with $\ptgg \in [0,60)$ (``Low''),
$[60,120)$ (``Med''), $[120,200)$ (``High'') or $> 200$~\GeV (``BSM''),
the latter of which is particularly sensitive to the presence of BSM physics in the
loop diagrams associated with the gluon--gluon fusion production mode.

\subsubsection{Categorization summary}\label{sec:cat_summary}

\begin{figure}[!tbp]
  \begin{center}
    \includegraphics[width=.85\textwidth]{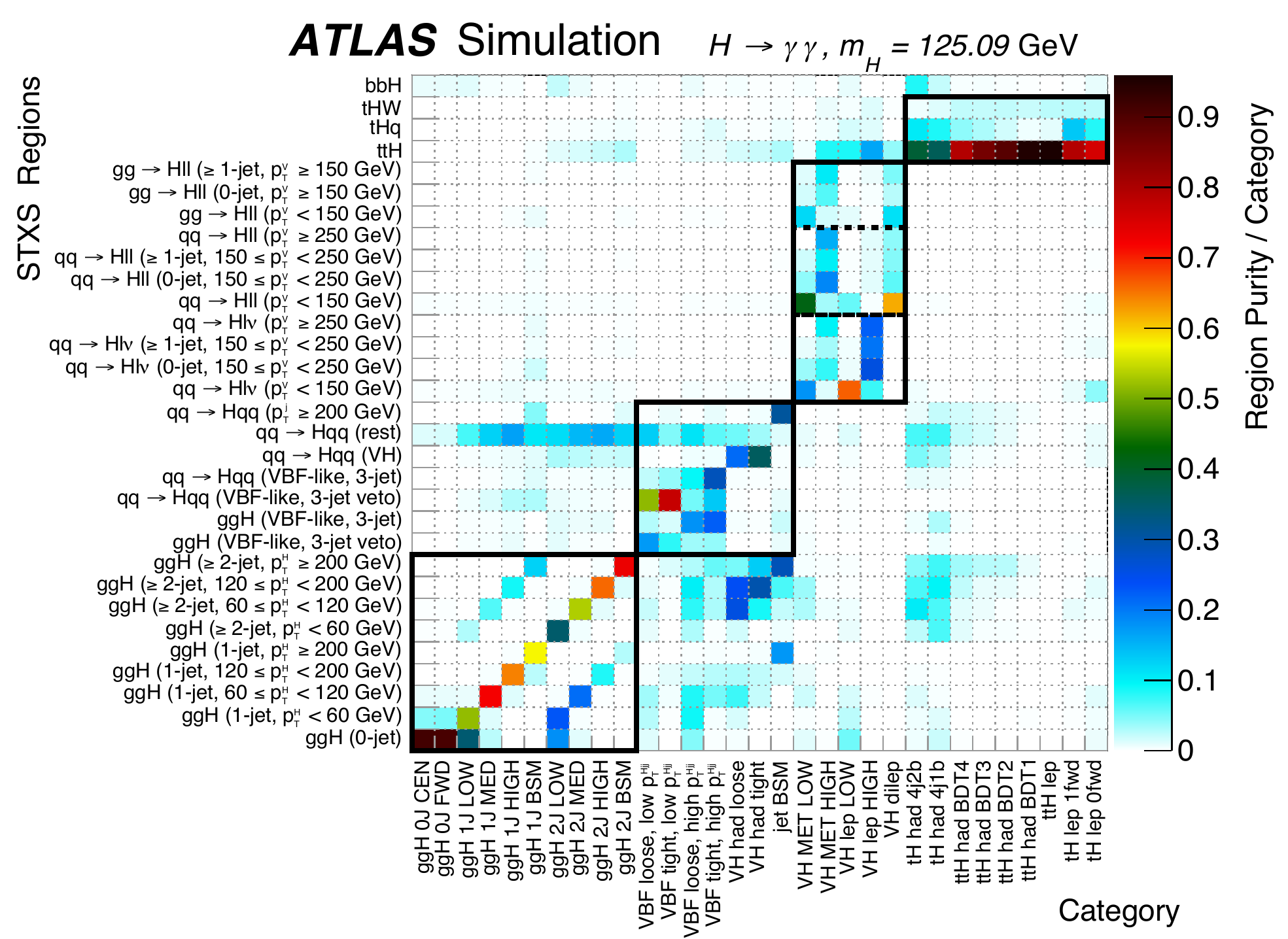}
  \end{center}
  \caption{
    The fraction of signal events assigned to each reconstructed category ($x$ axis and listed in Table~\ref{tab:cat-summary}) 
    and originating from a given region (listed in Table~\ref{tab:STXS}) of the stage-1 simplified template cross section 
    framework ($y$ axis). The black lines separate the \ttH\ and \tH, $VH$ leptonic, $VH$ hadronic and VBF enriched, 
    and untagged categories, along with the simplified template cross-section regions they are most sensitive to.
    The color shows the purity of the region per category. 
  }
  \label{fig:purity2D}
\end{figure}

\begin{figure}[!tbp]
  \begin{center}
    \includegraphics[width=.65\textwidth]{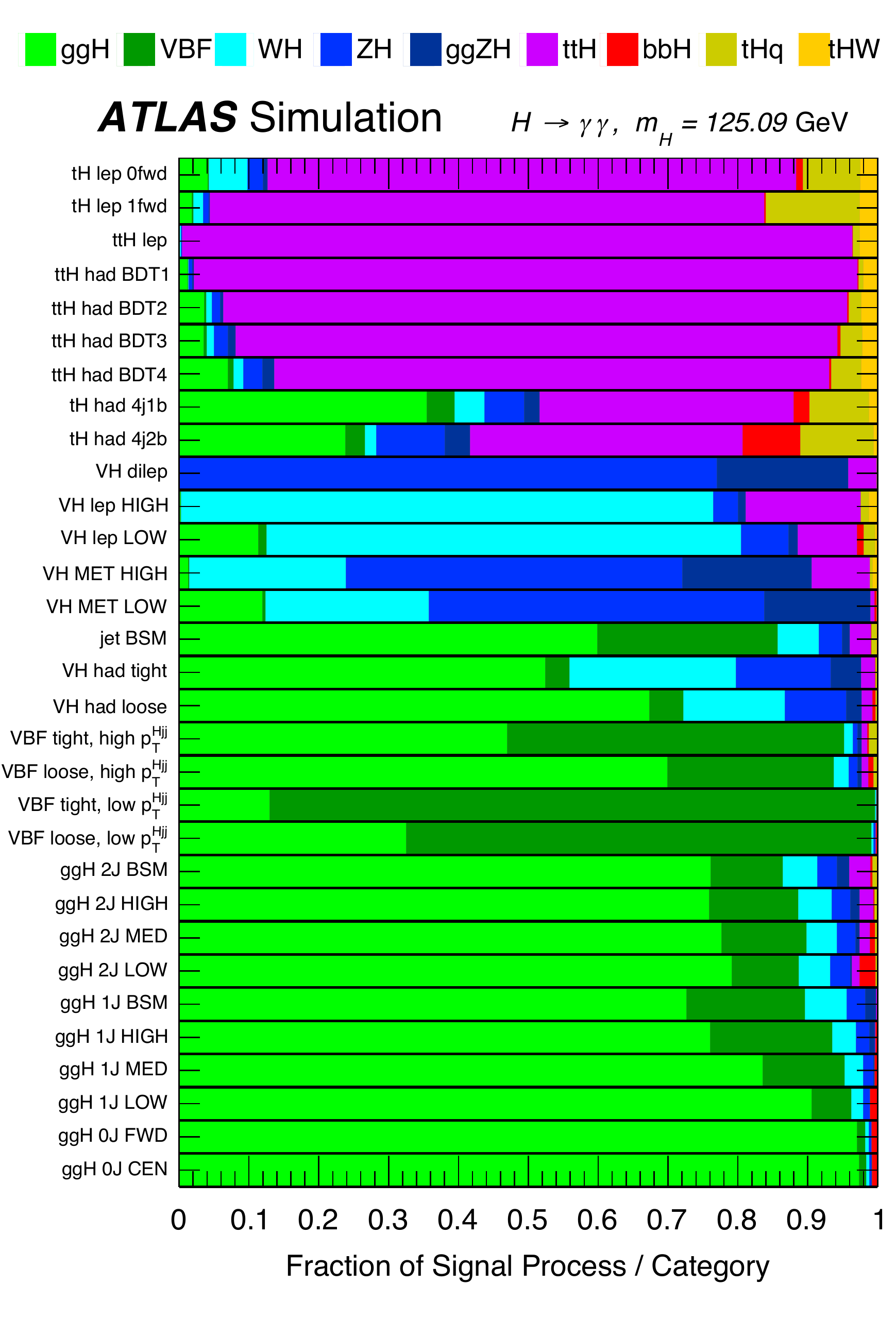}
  \end{center}
  \caption{The expected composition of the selected Higgs boson events, in terms of the different production modes, for each reconstructed category.}
  \label{fig:purity1D}
\end{figure}

A summary of the selection requirements defining each category is provided in Table~\ref{tab:cat-summary}.
The predicted signal efficiencies times acceptance and the event fractions per production mode for each category
are given in Table~\ref{tab:eff_signal_prod_modes}. The fractions of signal events in each reconstructed category
originating from a given simplified template cross-section region are shown in Figure~\ref{fig:purity2D}.
The defined \ggH\ categories exhibit high purities as they are defined with a near one-to-one correspondence
with the STXS regions despite small contaminations arising primarily from pileup and selection inefficiencies.
The \qqH\ bins are more ambiguous however still retain much of the diagonal structure. On the other hand, the \VH\ leptonic
and top categories are not sensitive to all of the STXS regions of interest, necessitating a merging.
Finally, the fractions of signal events in each category from a given production mode are 
shown in Figure~\ref{fig:purity1D}.

More information about the number of background events, the purity and the SM signal composition can be found in Tables~\ref{tab:SB90_tableExp} and~\ref{tab:purity_table} in Appendix~\ref{app:suppmaterial}.

\begin{table}[!htp]
  \caption{
    Shorthand label and event selection defining each of the 31 event reconstruction categories for the measurement of the signal 
    strengths and simplified template cross sections. 
    The labels denote the predominant production process or kinematic properties the category targets.
    Jets are required to have $\pT>30$~\GeV\ unless otherwise noted.
    The categories are mutually exclusive and the criteria are applied in descending order of the shown categories.
  }
\begin{center}
\label{tab:cat-summary}
\footnotesize
\begin{tabular}{cp{11cm}}
\hline \hline
 Category & Selection \\
 \hline \hline
tH lep 0fwd  & $N_\mathrm{lep}$ = 1, $N_\mathrm{jets}^\mathrm{cen}$ $\le$ 3, $N_{b\mathrm{-tag}}$ $\ge$ 1, $N_\mathrm{jets}^\mathrm{fwd}$ = 0 ($\pt^\mathrm{jet} >$ \SI{25}{\GeV}) \\
tH lep 1fwd  & $N_\mathrm{lep}$ = 1, $N_\mathrm{jets}^\mathrm{cen}$ $\le$ 4, $N_{b\mathrm{-tag}}$ $\ge$ 1, $N_\mathrm{jets}^\mathrm{fwd}$ $\ge$ 1 ($\pt^\mathrm{jet} >$ \SI{25}{\GeV}) \\
ttH lep  & $N_\mathrm{lep}$ $\ge$ 1, $N_\mathrm{jets}^\mathrm{cen}$ $\ge$ 2, $N_{b\mathrm{-tag}}$ $\ge$ 1, $Z_{\ell \ell}$ veto  ($\pt^\mathrm{jet} >$ \SI{25}{\GeV}) \\ 
ttH had BDT1 & $N_\mathrm{lep}$ = 0, $N_\mathrm{jets} \ge$ 3, $N_{b\mathrm{-tag}}$ $\ge$ 1, $\mathrm{BDT}_\mathrm{ttH} > 0.92$ \\
ttH had BDT2 & $N_\mathrm{lep}$ = 0, $N_\mathrm{jets} \ge$ 3, $N_{b\mathrm{-tag}}$ $\ge$ 1, $0.83 < \mathrm{BDT}_\mathrm{ttH}< 0.92$ \\
ttH had BDT3 & $N_\mathrm{lep}$ = 0, $N_\mathrm{jets} \ge$ 3, $N_{b\mathrm{-tag}}$ $\ge$ 1, $0.79 < \mathrm{BDT}_\mathrm{ttH} < 0.83$ \\
ttH had BDT4 & $N_\mathrm{lep}$ = 0, $N_\mathrm{jets} \ge$ 3, $N_{b\mathrm{-tag}}$ $\ge$ 1, $0.52 < \mathrm{BDT}_\mathrm{ttH} < 0.79$ \\
tH had 4j1b & $N_\mathrm{lep}$ = 0, $N_\mathrm{jets}^\mathrm{cen}$ = 4, $N_{b\mathrm{-tag}}$ = 1 ($\pt^\mathrm{jet} > \SI{25}{\GeV}$) \\
tH had 4j2b & $N_\mathrm{lep}$ = 0, $N_\mathrm{jets}^\mathrm{cen}$ = 4, $N_{b\mathrm{-tag}}$ $\ge$ 2 ($\pt^\mathrm{jet} > \SI{25}{\GeV}$) \\
\hline
 VH dilep & $N_\mathrm{lep} \geq 2$, $\SI{70}{\GeV} \leq m_{\ell\ell} \leq \SI{110}{\GeV}$ \\
 VH lep High &  $N_\mathrm{lep} = 1$, $|m_{e\gamma} - \SI{89}{\GeV}| > \SI{5}{\GeV}$, $\pT^{\ell+\MET{}} > \SI{150}{\GeV}$\\        
 VH lep Low & 
  $N_\mathrm{lep} = 1$, $|m_{e\gamma} - \SI{89}{\GeV}| > \SI{5}{\GeV}$,\ \ 
  $\pt^{\ell+\MET{}} < \SI{150}{\GeV},  \met{}\ \text{significance} > 1$\\        
 VH MET High & 
  $\SI{150}{\GeV} < \MET{} < \SI{250}{\GeV},  \met{}\ \text{significance} > 9$ or $\MET{} > \SI{250}{\GeV}$\\
 VH MET Low & 
  $\SI{80}{\GeV} < \MET{} < \SI{150}{\GeV},  \met{}\ \text{significance} > 8$\\
\hline
 jet BSM &
  $p_\mathrm{T,j1} > \SI{200}{\GeV} $\\
 VH had tight & $\SI{60}{\GeV} < m_\mathrm{jj} < \SI{120}{\GeV}$, $\mathrm{BDT}_\mathrm{VH} > 0.78$ \\
 VH had loose & $60~\GeV < m_\mathrm{jj} < \SI{120}{\GeV}$, $0.35 < \mathrm{BDT}_\mathrm{VH} < 0.78$ \\
 VBF tight, high $\pt^{Hjj}$ &
  $\deltarapjj > 2$, $|\eta_{\gamma\gamma} - 0.5(\eta_\mathrm{j1} + \eta_\mathrm{j2})| < 5$, $\pt^{Hjj} >$ \SI{25}{\GeV}, $\mathrm{BDT}_\mathrm{VBF} > 0.47$ \\
 VBF loose, high $\pt^{Hjj}$ &
  $\deltarapjj > 2$, $|\eta_{\gamma\gamma} - 0.5(\eta_\mathrm{j1} + \eta_\mathrm{j2})| < 5$, $\pt^{Hjj} >$ \SI{25}{\GeV}, $-0.32 < \mathrm{BDT}_\mathrm{VBF} < 0.47$ \\
 VBF tight, low $\pt^{Hjj}$ & 
  $\deltarapjj > 2$, $|\eta_{\gamma\gamma} - 0.5(\eta_\mathrm{j1} + \eta_\mathrm{j2})| < 5$, $\pt^{Hjj} <$ \SI{25}{\GeV}, $\mathrm{BDT}_\mathrm{VBF} > 0.87$ \\
 VBF loose, low $\pt^{Hjj}$ & 
  $\deltarapjj > 2$, $|\eta_{\gamma\gamma} - 0.5(\eta_\mathrm{j1} + \eta_\mathrm{j2})| < 5$, $\pt^{Hjj} <$ \SI{25}{\GeV}, $0.26 < \mathrm{BDT}_\mathrm{VBF} < 0.87$ \\
\hline
 ggH 2J BSM  & $\ge 2$ jets, $p_\mathrm{T}^{\gamma\gamma}\ge \SI{200}{\GeV}$ \\
 ggH 2J High & $\ge 2$ jets, $p_\mathrm{T}^{\gamma\gamma}\in [120, 200]$ \GeV \\
 ggH 2J Med  & $\ge 2$ jets, $p_\mathrm{T}^{\gamma\gamma}\in [60, 120]$ \GeV \\
 ggH 2J Low  & $\ge 2$ jets, $p_\mathrm{T}^{\gamma\gamma}\in [0, 60]$ \GeV \\
 ggH 1J BSM  &   $= 1$ jet, $p_\mathrm{T}^{\gamma\gamma}\ge \SI{200}{\GeV}$ \\
 ggH 1J High &   $= 1$ jet, $p_\mathrm{T}^{\gamma\gamma}\in [120, 200]$ \GeV \\
 ggH 1J Med  &   $= 1$ jet, $p_\mathrm{T}^{\gamma\gamma}\in [60, 120]$ \GeV \\
 ggH 1J Low  &   $= 1$ jet, $p_\mathrm{T}^{\gamma\gamma}\in [0, 60]$ \GeV \\
 ggH 0J Fwd  &   $= 0$ jets, one photon with $|\eta|>0.95$\\
 ggH 0J Cen  &   $= 0$ jets, two photons with $|\eta|\le 0.95$\\
\hline \hline
\end{tabular}
\end{center}
\end{table}

\begin{table}[!htbp]
\tiny
\begin{center}
  \caption{Signal efficiencies times acceptance, $\epsilon$, and expected signal event
    fractions per production mode, $f$, in each category for ${\sqrt{s} = 13}$~\TeV\ and $m_{H} = 125.09$~\GeV. The
    second-to-last row shows the total efficiency per production process summed over the categories. Values labeled as 
    'nil' correspond to efficiencies or fractions that are smaller than 0.05\%.
    The total number of expected signal events, $N_\mathrm{S}$, in the last row corresponds to an integrated luminosity of \SI{36.1}{\per\fb}.
    }
\label{tab:eff_signal_prod_modes}
\begin{tabular}{lrrrrrrrrrrrrrrrrr}
\hline\hline
& \multicolumn{2}{c}{\ggH} & \multicolumn{2}{c}{\VBF} & \multicolumn{2}{c}{\WH} & \multicolumn{2}{c}{\ZH} & \multicolumn{2}{c}{ttH} & \multicolumn{2}{c}{\bbH} & \multicolumn{2}{c}{\tHqb} & \multicolumn{2}{c}{\tHW} & All\\ 
Category & $\epsilon$[\%] & $f$[\%]& $\epsilon$[\%] & $f$[\%]& $\epsilon$[\%] & $f$[\%]& $\epsilon$[\%] & $f$[\%]& $\epsilon$[\%] & $f$[\%]& $\epsilon$[\%] & $f$[\%]& $\epsilon$[\%] & $f$[\%]& $\epsilon$[\%] & $f$[\%]& $N_\mathrm{S}$ \\
\hline \hline
ggH 0J Cen & 8.9 & 97.3 & 1.2 & 1.1 & 1.4 & 0.4 & 1.9 & 0.4 & nil & nil & 8.2 & 0.9 & nil & nil & nil & nil& 333.5 \\
ggH 0J Fwd & 15.5 & 97.0 & 2.4 & 1.2 & 3.0 & 0.5 & 3.7 & 0.4 & nil & nil & 14.7 & 0.9 & 0.2 & nil & 0.1 & nil& 579.5 \\
ggH 1J Low & 7.2 & 90.5 & 5.7 & 5.7 & 5.0 & 1.7 & 4.4 & 1.0 & 0.1 & nil & 9.1 & 1.1 & 0.5 & nil & 0.2 & nil& 289.9 \\
ggH 1J Med & 3.6 & 83.5 & 6.4 & 11.7 & 4.2 & 2.6 & 4.1 & 1.6 & 0.1 & nil & 1.9 & 0.4 & 0.6 & nil & 0.3 & nil& 156.2 \\
ggH 1J High & 0.7 & 76.0 & 1.9 & 17.5 & 1.1 & 3.4 & 1.4 & 2.7 & 0.1 & 0.1 & 0.3 & 0.3 & 0.2 & nil & 0.1 & nil& 31.5 \\
ggH 1J BSM & nil & 72.4 & 0.1 & 16.9 & 0.1 & 6.0 & 0.2 & 4.2 & nil & 0.3 & nil & nil & nil & 0.1 & nil & nil& 2.2 \\
ggH 2J Low & 1.8 & 79.1 & 2.7 & 9.6 & 3.7 & 4.5 & 4.1 & 3.1 & 2.2 & 1.1 & 5.4 & 2.3 & 3.9 & 0.3 & 1.9 & nil& 81.1 \\
ggH 2J Med & 1.5 & 77.6 & 3.1 & 12.2 & 3.2 & 4.4 & 3.8 & 3.2 & 2.6 & 1.5 & 1.6 & 0.7 & 4.5 & 0.4 & 2.4 & nil& 72.4 \\
ggH 2J High & 0.6 & 75.8 & 1.3 & 12.8 & 1.4 & 4.9 & 1.9 & 4.0 & 1.4 & 2.0 & 0.1 & 0.1 & 2.2 & 0.4 & 1.6 & 0.1& 29.2 \\
ggH 2J BSM & 0.2 & 76.2 & 0.3 & 10.3 & 0.4 & 4.9 & 0.6 & 4.6 & 0.6 & 3.0 & 0.1 & 0.2 & 0.8 & 0.6 & 1.3 & 0.2& 7.6 \\
VBF Hjj Low loose & 0.2 & 32.3 & 4.5 & 66.7 & 0.1 & 0.3 & 0.1 & 0.3 & nil & nil & 0.1 & 0.2 & 0.3 & 0.1 & nil & nil& 19.4 \\
VBF Hjj Low tight & nil & 12.9 & 4.2 & 86.7 & nil & 0.1 & nil & 0.1 & nil & nil & nil & nil & 0.3 & 0.1 & nil & nil& 13.8 \\
VBF Hjj High loose & 0.3 & 69.9 & 1.4 & 23.8 & 0.4 & 2.2 & 0.5 & 1.8 & 0.4 & 0.9 & 0.4 & 0.7 & 1.8 & 0.6 & 0.5 & nil& 16.5 \\
VBF Hjj High tight & 0.3 & 47.0 & 3.4 & 48.2 & 0.2 & 1.2 & 0.4 & 1.3 & 0.4 & 0.8 & 0.2 & 0.3 & 4.4 & 1.2 & 0.6 & nil& 20.2 \\
VHhad loose & 0.3 & 67.2 & 0.3 & 4.9 & 2.4 & 14.6 & 2.9 & 11.0 & 0.6 & 1.6 & 0.2 & 0.4 & 0.8 & 0.3 & 0.8 & 0.1& 16.5 \\
VHhad tight & 0.2 & 52.4 & 0.1 & 3.4 & 3.0 & 23.8 & 3.5 & 18.0 & 0.6 & 1.9 & nil & 0.1 & 0.5 & 0.2 & 1.0 & 0.1& 12.3 \\
jet BSM & 0.4 & 59.9 & 2.4 & 25.8 & 1.6 & 5.9 & 1.9 & 4.4 & 2.0 & 3.0 & 0.1 & 0.1 & 3.1 & 0.6 & 5.1 & 0.2& 26.7 \\
VHMET Low & nil & 11.9 & nil & 0.4 & 0.1 & 23.4 & 0.6 & 63.2 & nil & 0.5 & nil & 0.3 & nil & 0.2 & nil & nil& 0.6 \\
VHMET High & nil & 1.3 & nil & 0.1 & 0.3 & 22.8 & 1.4 & 66.2 & 0.3 & 8.3 & nil & nil & 0.1 & 0.6 & 0.8 & 0.7& 1.3 \\
VHlep Low & nil & 11.4 & nil & 1.1 & 4.4 & 68.0 & 0.8 & 8.1 & 1.3 & 8.5 & 0.2 & 0.9 & 1.8 & 1.6 & 2.2 & 0.4& 6.4 \\
VHlep High & nil & 0.2 & nil & nil & 1.2 & 76.5 & 0.1 & 4.6 & 0.6 & 16.2 & nil & nil & 0.3 & 1.2 & 1.6 & 1.3& 1.5 \\
VHdilep & nil & nil & nil & nil & nil & nil & 1.4 & 95.8 & 0.1 & 4.0 & nil & nil & nil & nil & 0.1 & 0.2& 0.9 \\
tHhad 4j2b & nil & 23.8 & nil & 2.8 & nil & 1.6 & 0.1 & 13.5 & 0.6 & 39.0 & 0.1 & 8.2 & 1.2 & 10.5 & 0.3 & 0.6& 0.6 \\
tHhad 4j1b & nil & 35.4 & nil & 4.0 & 0.1 & 4.3 & 0.3 & 7.9 & 2.2 & 36.3 & 0.2 & 2.2 & 3.8 & 8.5 & 2.6 & 1.3& 2.5 \\
ttHhadBDT4 & nil & 7.0 & nil & 0.8 & nil & 1.4 & 0.2 & 4.5 & 4.8 & 79.4 & nil & 0.3 & 1.9 & 4.3 & 4.7 & 2.4& 2.5 \\
ttHhadBDT3 & nil & 3.5 & nil & 0.5 & nil & 1.0 & nil & 3.1 & 1.3 & 86.1 & nil & 0.5 & 0.3 & 3.1 & 1.1 & 2.2& 0.6 \\
ttHhadBDT2 & nil & 3.6 & nil & 0.3 & nil & 0.8 & nil & 1.6 & 3.8 & 89.3 & nil & 0.2 & 0.6 & 1.8 & 3.4 & 2.4& 1.8 \\
ttHhadBDT1 & nil & 1.2 & nil & 0.1 & nil & 0.1 & nil & 0.7 & 3.4 & 95.0 & nil & 0.1 & 0.2 & 0.7 & 2.5 & 2.1& 1.4 \\
ttHlep & nil & nil & nil & nil & nil & 0.2 & nil & 0.1 & 5.6 & 96.0 & nil & 0.1 & 0.4 & 1.0 & 5.0 & 2.6& 2.4 \\
tHlep 1fwd & nil & 1.8 & nil & 0.2 & nil & 1.4 & nil & 0.9 & 2.1 & 79.4 & nil & 0.2 & 2.6 & 13.5 & 2.3 & 2.6& 1.1 \\
tHlep 0fwd & nil & 4.1 & nil & 0.2 & 0.1 & 5.6 & nil & 2.8 & 1.9 & 75.7 & nil & 0.9 & 1.5 & 8.2 & 2.1 & 2.5& 1.0 \\
\hline
 Total $\epsilon$ [\%] &41.8 & - &41.3 & - &37.6 & - &40.5 & - &39.1 & - &42.8 & - &38.9 & - &44.5 & - &41.8 \\
\hline 
 Events & \multicolumn{2}{c}{1518.4}& \multicolumn{2}{c}{119.1}& \multicolumn{2}{c}{37.1}& \multicolumn{2}{c}{25.2}& \multicolumn{2}{c}{16.0}& \multicolumn{2}{c}{14.8}& \multicolumn{2}{c}{2.2}& \multicolumn{2}{c}{0.5}& 1733.2\\
\hline
\hline
\end{tabular}
\end{center} 
\end{table}

\clearpage


\subsection{Production mode measurements}
\label{sec:results_coup}

Using the 31 categories, total and production mode specific signal strength measurements are carried out. Measurements of total production cross sections and simplified template cross sections are reported. The simplified template cross sections are measured in a merged scheme introduced in Section~\ref{sec:intro_stxs} and summarized in Table~\ref{tab:STXS}. In addition, the result of coupling-strength fits are reported.

\subsubsection{Observed Data}

\begin{figure}[!tbp]
  \centering
  \includegraphics[width=0.65\columnwidth]{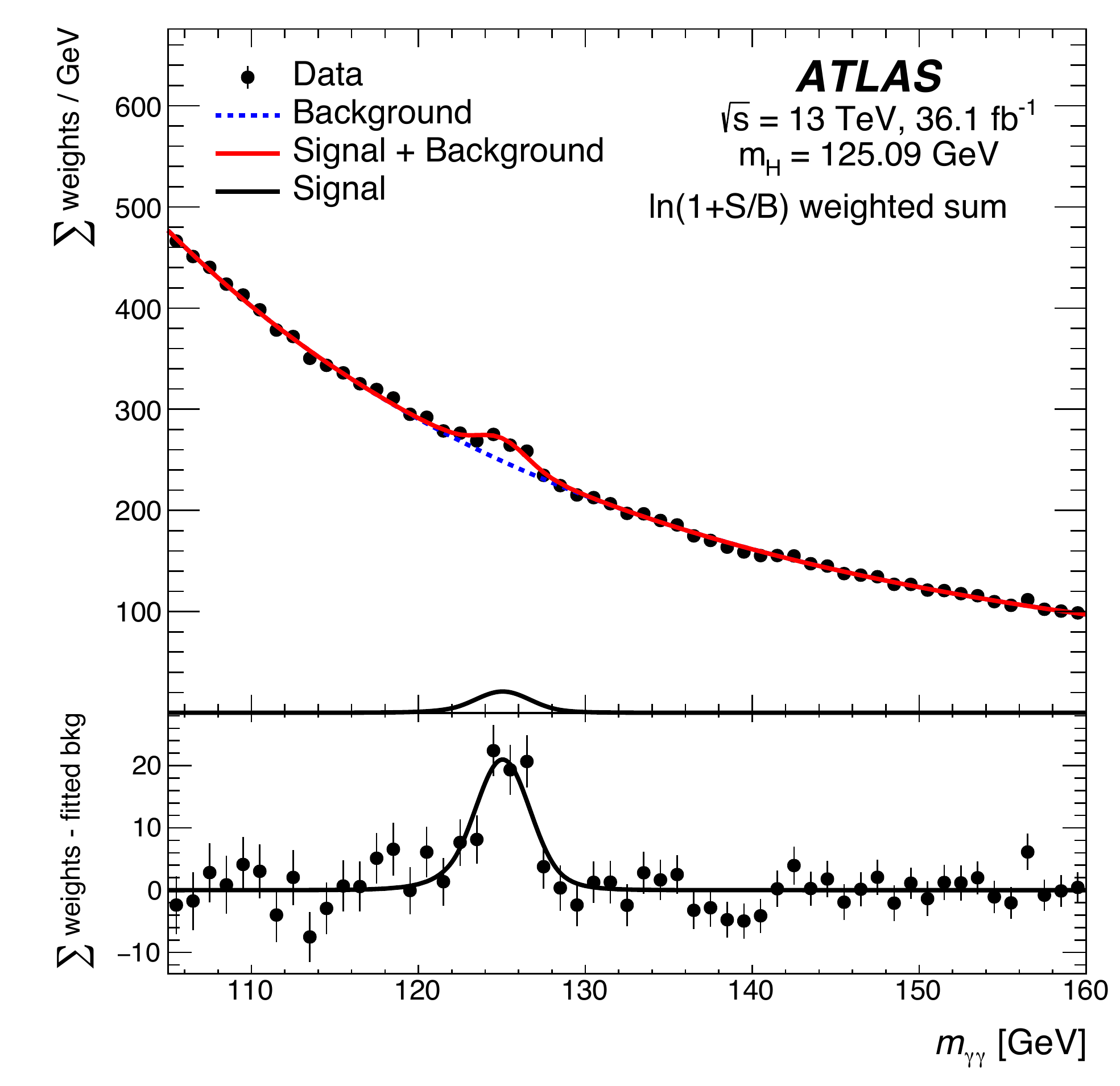}
  \caption{Weighted diphoton invariant mass spectrum observed in the 2015 and 2016 data at 13~\TeV. 
    Each event is weighted by the $\ln(1+ S_{\hspace{-0.4ex} 90}/B_{90})$ ratio of the expected signal ($S_{\hspace{-0.4ex} 90}$) and background ($B_{90}$)
    of the 90\% signal quantile in the category to which it belongs to. 
    The values of $S_{90}$ and $B_{90}$ used for each category are shown in Table~\ref{tab:SB90_tableExp}
    of Appendix~\ref{app:suppmaterial}.
    The error bars represent 68\% confidence intervals of the weighted sums.
    The solid red curve shows the fitted signal-plus-background model when the Higgs boson 
    mass is constrained to be $125.09\pm 0.24$~\GeV. The background component of the fit is shown with the 
    dotted blue curve. The signal component of the fit is shown with the solid black curve.
    Both the signal-plus-background and background-only curves reported here are 
    obtained from the sum of the individual curves in each category weighted by the logarithm of unity plus the signal-to-background ratio.
    The bottom plot shows the residuals between the data and the background component of the fitted model. 
    } 
  \label{fig:weightedmyy_all}
\end{figure}

\begin{figure*}[!tbp]
  \centering
	\subfloat[] {\includegraphics[width=0.50\columnwidth]{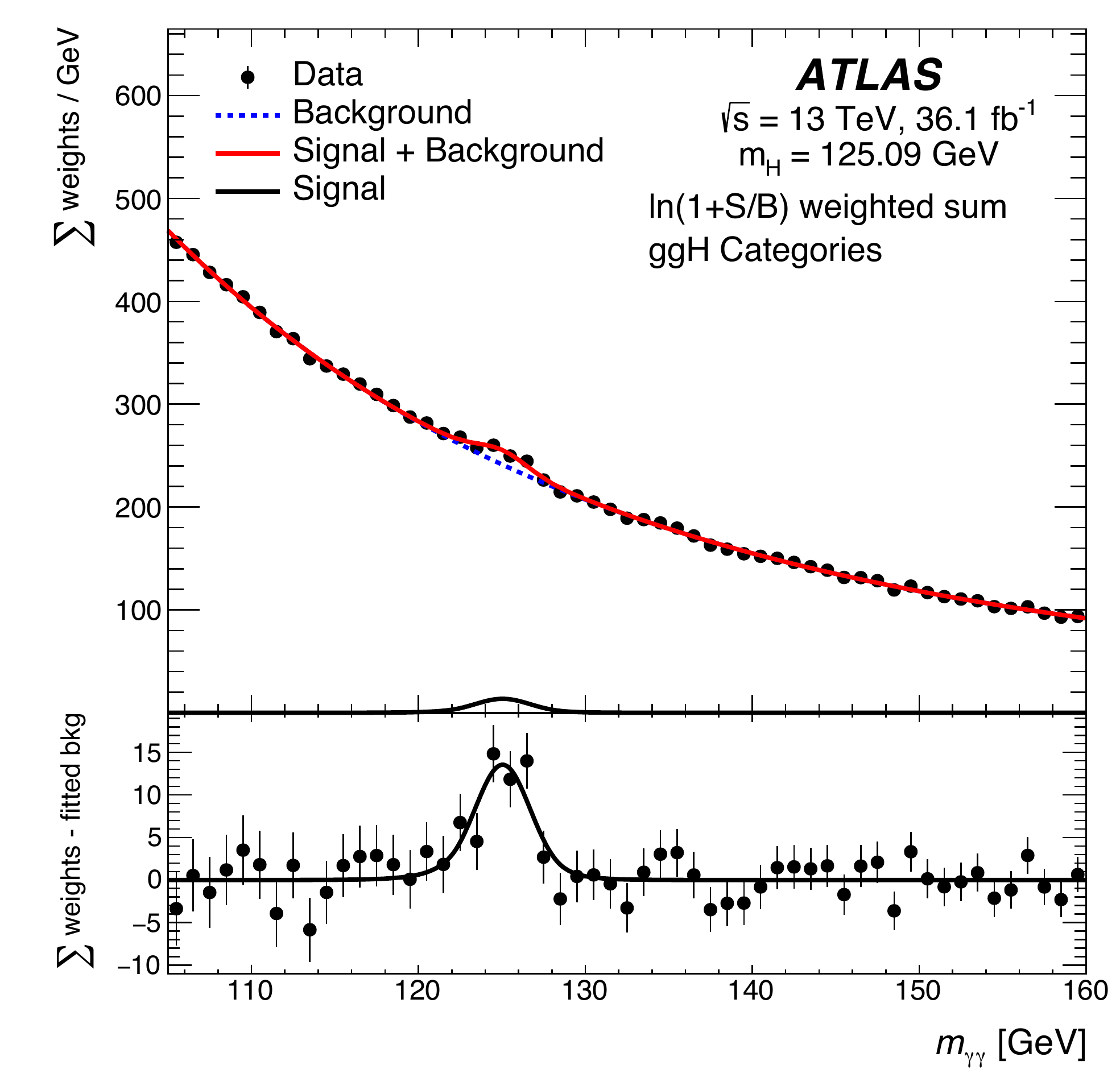} \label{invmass_untagged}}
	\subfloat[] {\includegraphics[width=0.50\columnwidth]{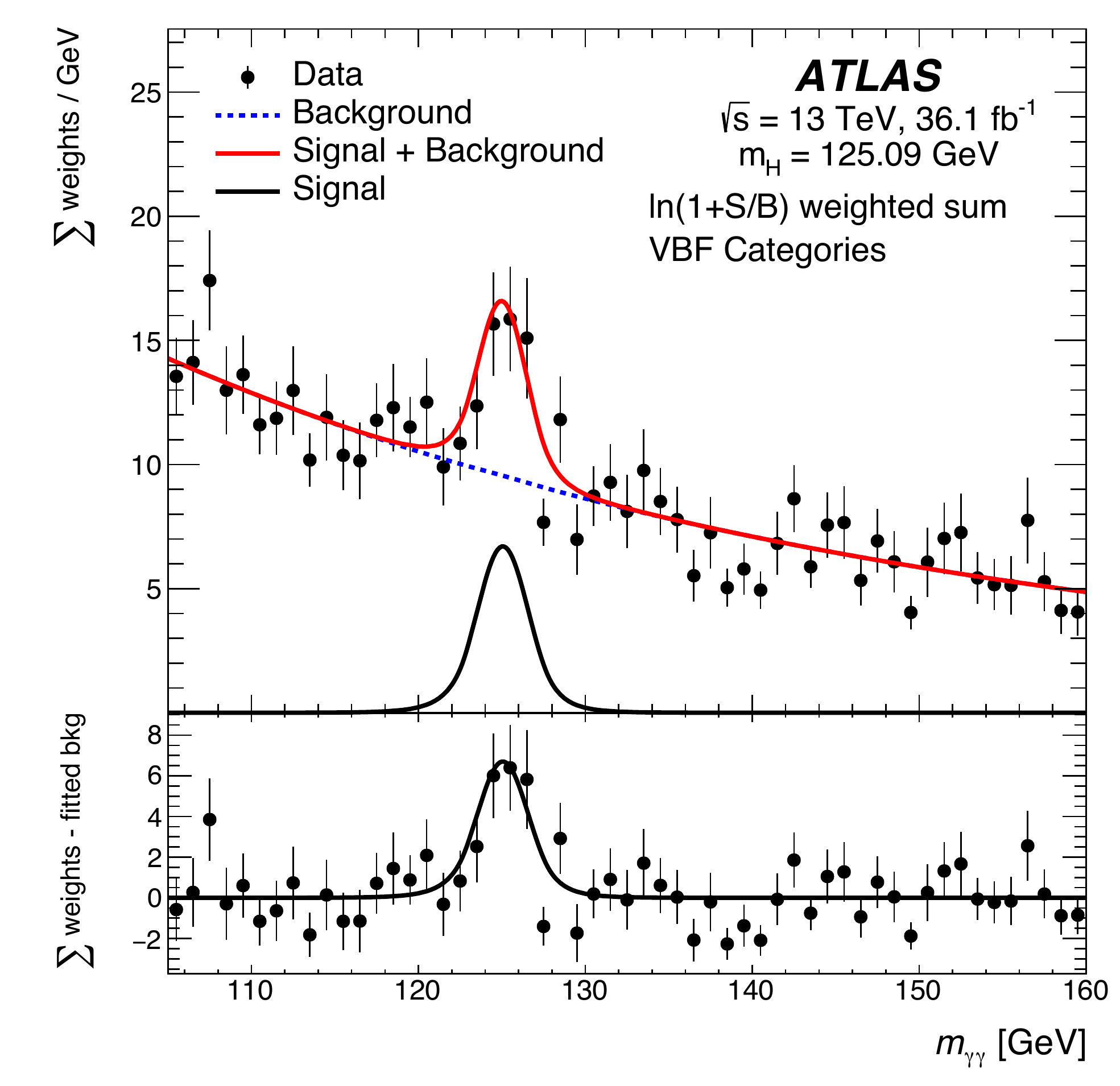} \label{invmass_VBF}} \\
	\subfloat[] {\includegraphics[width=0.50\columnwidth]{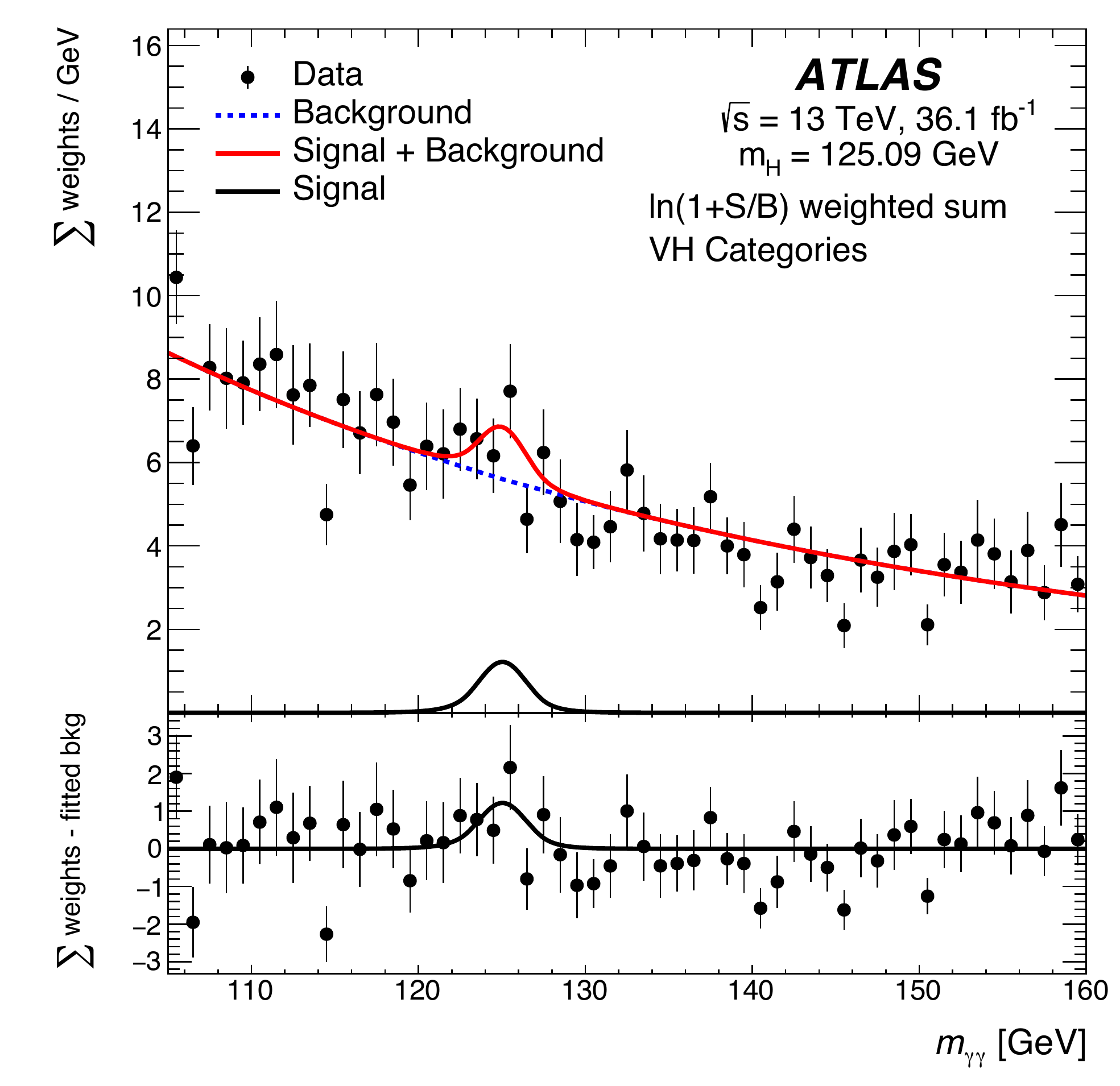}  \label{invmass_VH}}
	\subfloat[] {\includegraphics[width=0.50\columnwidth]{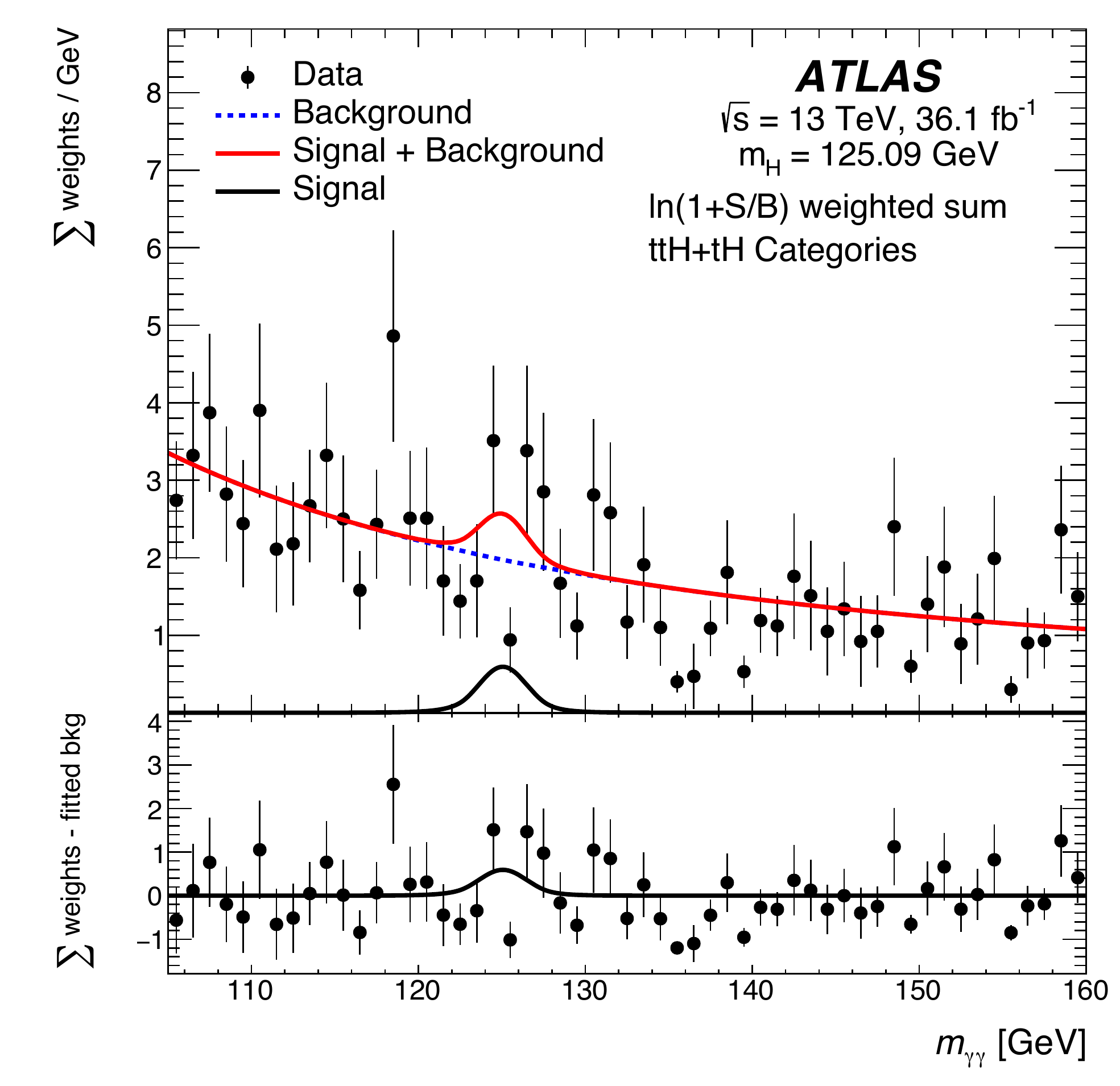} \label{invmass_ttH}} 
        \caption{Weighted diphoton invariant mass spectra observed in the 13~\TeV\ data for events belonging to: 
          (a) ``untagged'' categories and the ``jet BSM'' category, in which the expected signal is produced
          mainly through gluon--gluon fusion, (b) \VBF\ categories, (c) \VH\ categories and (d) \ttH\ categories.
          Each event is weighted by the $\ln(1 + S_{90}/B_{90})$ ratio of the expected signal ($S_{90}$) and background ($B_{90}$)
          of the 90\% signal quantile in the category it belongs to. 
          The values of $S_{90}$ and $B_{90}$ used for each category are shown in Table~\ref{tab:SB90_tableExp}
          of Appendix~\ref{app:suppmaterial}.
          The error bars represent 68\% confidence intervals of the weighted sums.
          The solid red curve shows the fitted signal-plus-background model when the Higgs boson 
          mass is constrained to be $125.09\pm 0.24$~\GeV. The background component of the fit is shown with the 
          dotted blue curve. The signal component of the fit is shown with the solid black curve.          
          Both the signal-plus-background and background-only curves reported here are 
          obtained from the sum of the individual curves in each category weighted by the logarithm of unity plus the signal-to-background ratio.
          The bottom plot shows the residuals between the data and the background component of the fitted model.}
        \label{fig:weightedmyy_cats}
\end{figure*}

The observed invariant mass distribution of the selected diphoton pairs of all categories as defined in
Table~\ref{tab:cat-summary}, is shown in Figure~\ref{fig:weightedmyy_all}.
Figure~\ref{fig:weightedmyy_cats} shows the invariant mass
distributions for the sums of the categories most sensitive to the different production modes. In all cases, for
illustration purposes, events in each category are weighted according to the expected signal ($S_{90}$) to background
($B_{90}$) ratio in a $\mgg$ region containing 90\% of the expected signal yield, using a weight of the form $\ln \left(
1 + S_{90}/B_{90}\right)$. The results of signal-plus-background fits to these spectra, displaying both the total sum
and the background-only components, are shown, as well as the residuals between the data and the background component.
Both the signal-plus-background and background-only distributions shown are obtained from the sum of the individual
distributions in each category weighted in the same way as the data points.
In the fit of Figure~\ref{fig:weightedmyy_all} a single signal strength $\mu$ affecting
simultaneously all production modes has been assumed, while in the fits of Figure~\ref{fig:weightedmyy_cats} 
the four signal strengths $\mu_\mathrm{ggH}$,
$\mu_\mathrm{VBF}$, $\mu_\mathrm{VH}$ and $\mu_\mathrm{ttH+tH}$ are allowed to vary separately, 
as described in the following section. The observed mass peak of the Higgs boson, constrained in the fit as $m_H =125.09 \pm 0.24~\GeV$, is well within 68\% CL of the Run~1 ATLAS+CMS combined measurement.

\subsubsection{Signal strengths}\label{sec:sig_strength}

The signal strengths, {\em i.e.} the ratios of the measured Higgs
boson production-mode cross sections times diphoton branching ratio to the
SM predictions for each production mode, are measured with the extended likelihood
analysis described in Section~\ref{sec:statmod}.
In the likelihood the signal yield $N_{\mathrm{sig,}m}^i$ in each category $i$ for a
particular production mode $m$ is expressed
as the product of the integrated luminosity $\int L\,\mathrm{d}t$, the signal strength
$\mu_m$ for that production mode, the expected SM Higgs boson production mode cross
section times branching ratio to diphotons, and the acceptance times
efficiency $\epsilon$ (Table~\ref{tab:eff_signal_prod_modes}) for
signal events from that production mode in the selected category 
($N_\mathrm{sig,m}^i = \mu_m\times \int L\,\mathrm{d}t \times \sigma_m^\mathrm{SM} \times
\bfhyy\times \epsilon_m^i$).

\begin{figure}[!tbp]
\centering
\includegraphics[width=.82\textwidth]{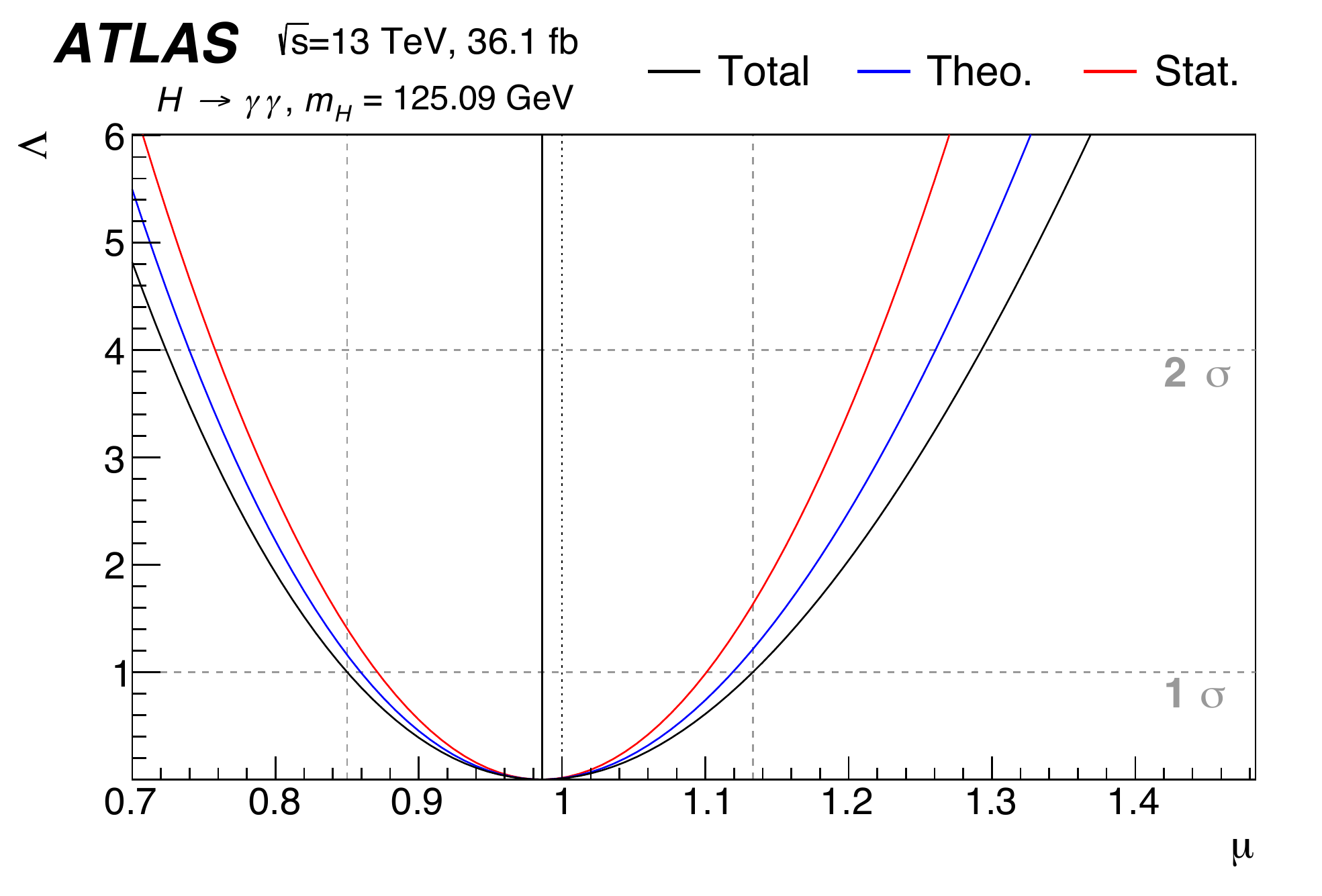}
\caption{Observed negative log-profile likelihood $\Lambda$ of the global signal strength $\mu$. The three likelihood contours shown 
correspond to all theory and experimental nuisance parameters fixed (Stat.), all experimental nuisance parameters fixed (Theo.), and with all nuisance parameters floating (Total). The intersections of the solid curves and horizontal lines at $\Lambda = 1$ and $\Lambda = 4$ indicate the 1 and 2 $\sigma$ confidence intervals of the corresponding result.
  }
\label{fig:nll_globMu}
\end{figure}

A global signal strength $\mu$ is measured assuming the ratios between different production processes 
to be as predicted by the SM. The profile of the negative log-likelihood ratio $\lambda(\mu)$
of the global signal strength of all Higgs
processes $\mu$ for \higgsMass\ is shown in Figure~\ref{fig:nll_globMu}.

The measured central value and 68\% CL interval for $\mu$ is found to be:
\begin{align*}
\mu &= 0.99\ ^{+0.15}_{-0.14} = 0.99\ \pm0.12\,\mathrm{(stat.)}\ ^{+0.06}_{-0.05}\,\mathrm{(exp.)}\ ^{+0.07}_{-0.05}\,\mathrm{(theo.)} \ ,
\end{align*}
well compatible with the SM prediction ($\mu = 1$).
This result confirms the ATLAS Run-1 diphoton signal strength measurement of 
$\mu=1.17 \pm 0.23\,\mathrm{(stat.)}\ ^{+0.10}_{-0.08}\,\mathrm{(exp.)}\ ^{+0.12}_{-0.08}\,\mathrm{(theo.)}$
with around a factor of two improvement in each component of the uncertainty.
The Run-1 result was obtained using the NNLO SM prediction for \ggH\ production~\cite{Heinemeyer:2013tqa,deFlorian:2012yg}, 
which is about 10\% lower than the N${}^{3}$LO calculation used here (see Section~\ref{sec:mc}).
The impact of the main sources of systematic uncertainty (presented in Table~\ref{tab:tableNPs} and Section~\ref{sec:syst}) in the measured global signal strength is summarized in Table~\ref{tab:impact_globMu}. The distinction between yield and migration uncertainties adopted in Table~\ref{tab:tableNPs} is used and the uncertainties are grouped into theory uncertainties, experimental uncertainties, mass resolution and scale, background shape, and luminosity.

\begin{table}[!tp]
\begin{center}
\caption{
  Main systematic uncertainties $\sigma_\mu^\mathrm{syst.}$ in the combined signal strength parameter $\mu$. 
  The values for each group of uncertainties are determined by subtracting in quadrature from the total uncertainty
  the change in the 68\% CL range of $\mu$ when the corresponding nuisance parameters are fixed to their best fit values.
  The experimental uncertainty in the yield does not include the luminosity contribution, which is accounted for
  separately. The uncertainties correspond to the sources detailed in Table~\ref{tab:tableNPs}.
}
\label{tab:impact_globMu}
\begin{tabular}{lc}
\hline
\hline
Uncertainty Group & $\sigma_{\mu}^\mathrm{syst.}$ \\
\hline
\hline
Theory (QCD) & 0.041 \\
Theory ($B(H \to \gamma\gamma)$) & 0.028 \\
Theory (PDF+$\alpha_S$) & 0.021 \\
Theory (UE/PS) & 0.026 \\
\hline
Luminosity & 0.031 \\
Experimental (yield) & 0.017 \\
Experimental (migrations) & 0.015 \\
\hline
Mass resolution & 0.029 \\
Mass scale & 0.006 \\
\hline
Background shape & 0.027 \\
\hline
\hline
\end{tabular}
\end{center}
\end{table}

\begin{figure}[!tbp]
\centering
\includegraphics[width=.82\textwidth]{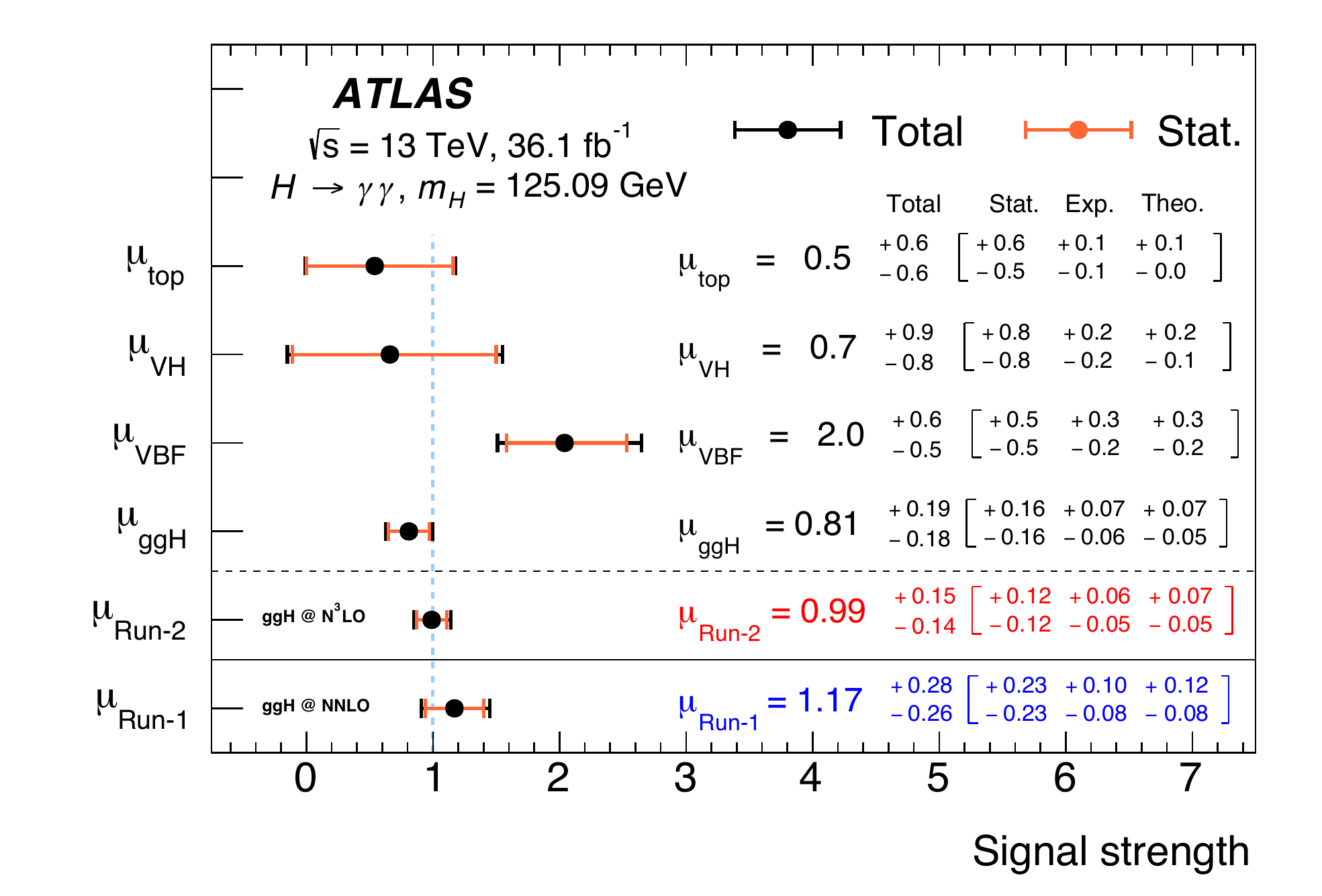}
\caption{
Summary of the signal strengths measured for the different production processes (\ggH, \VBF, \VH\ and top) and
globally ($\mu_\mathrm{Run-2}$), compared to the global signal strength measured at 7 and 8~\TeV\
($\mu_\mathrm{Run-1}$)~\cite{Aad:2014eha}. The black and orange error bars show the total and statistical uncertainties. The
signal strength $\mu_\mathrm{Run-1}$ was derived assuming the Higgs production-mode cross section based on
Refs.~\cite{Heinemeyer:2013tqa,deFlorian:2012yg}. Uncertainties smaller than 0.05 are displayed as 0.0.
In the more recent theoretical predictions used in this
analysis~\cite{Anastasiou:2016cez,deFlorian:2016spz}, the gluon--gluon fusion production-mode cross section is larger by
approximately 10\%. In this measurement, the \bbH\ contributions are scaled with \ggH\ ($\mu_\mathrm{bbH} =
\mu_\mathrm{ggH}$), and the \tH\ and \ttH\ productions are measured together ($\mu_\mathrm{top} =
\mu_\mathrm{ttH+tH}$). Associated production with $Z$ or $W$ bosons is assumed to be scaled by a single signal
strength parameter ($\mu_\mathrm{VH} = \mu_\mathrm{ZH} = \mu_\mathrm{WH}$).
}
\label{fig:summary_muLegacy}
\end{figure}

In addition to the global signal strength, the signal strengths of the
primary production processes are evaluated by exploiting the
sensitivities of the analysis categories of Table~\ref{tab:cat-summary} to specific production
processes. The measured signal strengths are shown together with the global signal strengths discussed above in Figure~\ref{fig:summary_muLegacy} and found to be:

\begin{align*}
\mu_\mathrm{ggH} &= 0.81\ ^{+0.19}_{-0.18} = 0.81\ \pm0.16\,\mathrm{(stat.)}\ ^{+0.07}_{-0.06}\,\mathrm{(exp.)}\ ^{+0.07}_{-0.05}\,\mathrm{(theo.)}\\
\mu_\mathrm{VBF} &= 2.0\ ^{+0.6}_{-0.5} = 2.0\ \pm0.5\,\mathrm{(stat.)}\ ^{+0.3}_{-0.2}\,\mathrm{(exp.)}\ ^{+0.3}_{-0.2}\,\mathrm{(theo.)}\\
\mu_\mathrm{VH}  &= 0.7\ ^{+0.9}_{-0.8} = 0.7\ \pm0.8\,\mathrm{(stat.)}\ ^{+0.2}_{-0.2}\,\mathrm{(exp.)}\ ^{+0.2}_{-0.1}\,\mathrm{(theo.)}\\
\mu_\mathrm{top} &= 0.5\ ^{+0.6}_{-0.6} = 0.5\ ^{+0.6}_{-0.5}\,\mathrm{(stat.)}\ ^{+0.1}_{-0.1}\,\mathrm{(exp.)}\ ^{+0.1}_{-0.0}\,\mathrm{(theo.)}\\
\end{align*}

For Higgs boson production via \VH\ the signal strength is assumed to be scaled by a single parameter ({\em i.e.} $\mu_\mathrm{VH} = \mu_\mathrm{ZH} = \mu_\mathrm{WH}$). The \bbH\ contributions are scaled with \ggH\ ({\em i.e.} $\mu_\mathrm{bbH} = \mu_\mathrm{ggH}$), and the \tH\ and \ttH\ productions are measured together rather than separately ({\em i.e.} $\mu_\mathrm{top} = \mu_\mathrm{ttH+tH}$). 

The \ggH\ signal strength is $1~\sigma$ below the Standard Model prediction, while the \VBF\ signal strength is $2.2~\sigma$ above the prediction.
The expected and observed significances $Z_0$ of \VBF\ production are reported in Table~\ref{tab:muLegacy_Z0}:
the significance of the observed \VBF\ signal is close to $5~\sigma$.

\begin{figure}[!tbp]
\centering
\includegraphics[width=.82\textwidth]{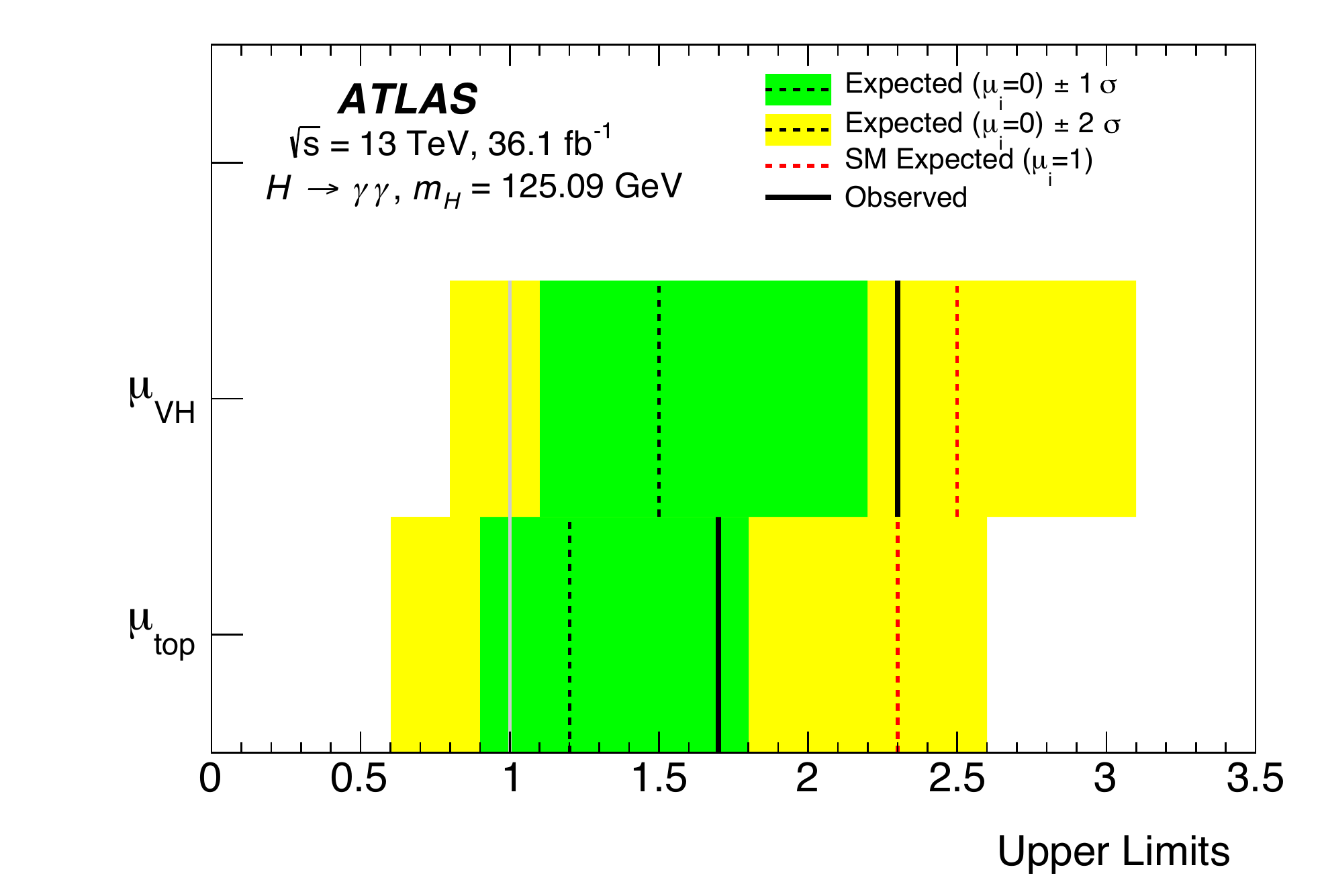}
\caption{ Summary of asymptotic limits for the signal strengths of the associated production processes (\VH\ and top). }
\label{fig:limits_muLegacy}
\end{figure}

Since no significant evidence is observed for \VH\ and top-associated Higgs boson production, upper limits at 95\% CL are reported for their signal strengths, as shown in Table~\ref{tab:muLegacy_limits} and Figure~\ref{fig:limits_muLegacy}. The accuracy of the asymptotic approximation was validated using ensembles of pseudo-experiments. Appendix~\ref{app:mu_WH_n_mu_ZH} provides separate limits on $\mu_\mathrm{ZH}$ and $\mu_\mathrm{WH}$, and Appendix~\ref{app:summary_coup:mu}
shows the expected uncertainties for the inclusive and production-mode specific signal strengths reported in Figure~\ref{fig:summary_muLegacy}. 

\begin{table}[!tp]
\begin{center}
\caption{Expected and observed significances of the VBF, \VH\ and
  top quark associated production mode signal strengths.}
\label{tab:muLegacy_Z0}
\begin{tabular}{c|c|c}
  \hline
  \hline
   Measurement          &   Expected $Z_{0}$  & Observed $Z_{0}$    \\
\hline \hline
   $\mu_{\mathrm{VBF}}$ &   $2.6~\sigma$  &   $4.9~\sigma$  \\
   $\mu_{\mathrm{VH}}$  &   $1.4~\sigma$  &   $0.8~\sigma$  \\
   $\mu_{\mathrm{top}}$ &   $1.8~\sigma$  &   $1.0~\sigma$  \\
\hline
\hline
\end{tabular}
\end{center}
\end{table}

\begin{table}[!tp]
\begin{center}
\caption{Observed and expected upper limits at 95\% CL on the signal strengths
  $\mu_\mathrm{VH}$ and $\mu_\mathrm{top}$.
  The median expected limits are given for either the case when the
  true value of the signal strength under study is the SM value ($\mu_i=1$)
  or zero. The $\pm 1~\sigma$ and $\pm 2~\sigma$ intervals for the
  expected upper limit in the case $\mu_i=0$ are also reported.
}
\label{tab:muLegacy_limits}
\begin{tabular}{c|c|c|c|c|c|c|c}
\hline
\hline  
  Measurement          &   Observed &   Expected Limit    &   Expected Limit  &   $+2\sigma$ &   $+1\sigma$ &   $-1\sigma$ &   $-2\sigma$ \\
                       &            &  ($\mu_i = 1$)  & ($\mu_i = 0$) &              &              &              &              \\
\hline \hline
  $\mu_{\mathrm{VH}}$  &        2.3 &             2.5 &           1.5 &          3.1 &          2.2 &          1.1 &          0.8 \\
  $\mu_{\mathrm{top}}$ &        1.7 &             2.3 &           1.2 &          2.6 &          1.8 &          0.9 &          0.6 \\
\hline
\hline
\end{tabular}
\end{center}
\end{table}

\clearpage

\subsubsection{Production-mode cross sections}

The production-mode cross sections for \higgsMass\ in a region with Higgs-boson rapidity $|y_H| < 2.5$, multiplied by the
branching ratio of the Higgs boson decay to diphotons, are evaluated in the following way.
The fitted value of $\sigma_\mathrm{top}$ corresponds to the sum of \ttH,
\tHqb, and \tHW\ production-mode cross sections under the assumption that
their relative ratios are as predicted by the SM.
The \VH\ production-mode cross section value is fitted under the
assumption that the ratio of the \WH\ and \ZH\ production mode cross
sections is as predicted by the SM and includes
both production from quark and gluon initial states.  
Such results are obtained through signal+background fits to
the diphoton invariant mass distribution in each category by expressing,
in the likelihood, the signal yield $N_{\mathrm{sig,}m}^i$
in each category $i$ for a particular production mode $m$
as $N_{\mathrm{sig,}m} = \int L\,\mathrm{d}t \times \sigma_m^\mathrm{SM} \times
\bfhyySM\times \epsilon_m^i$ using the same notation as in Section~\ref{sec:sig_strength}.

\begin{figure}[!tbp]
\centering
\includegraphics[width=.82\textwidth]{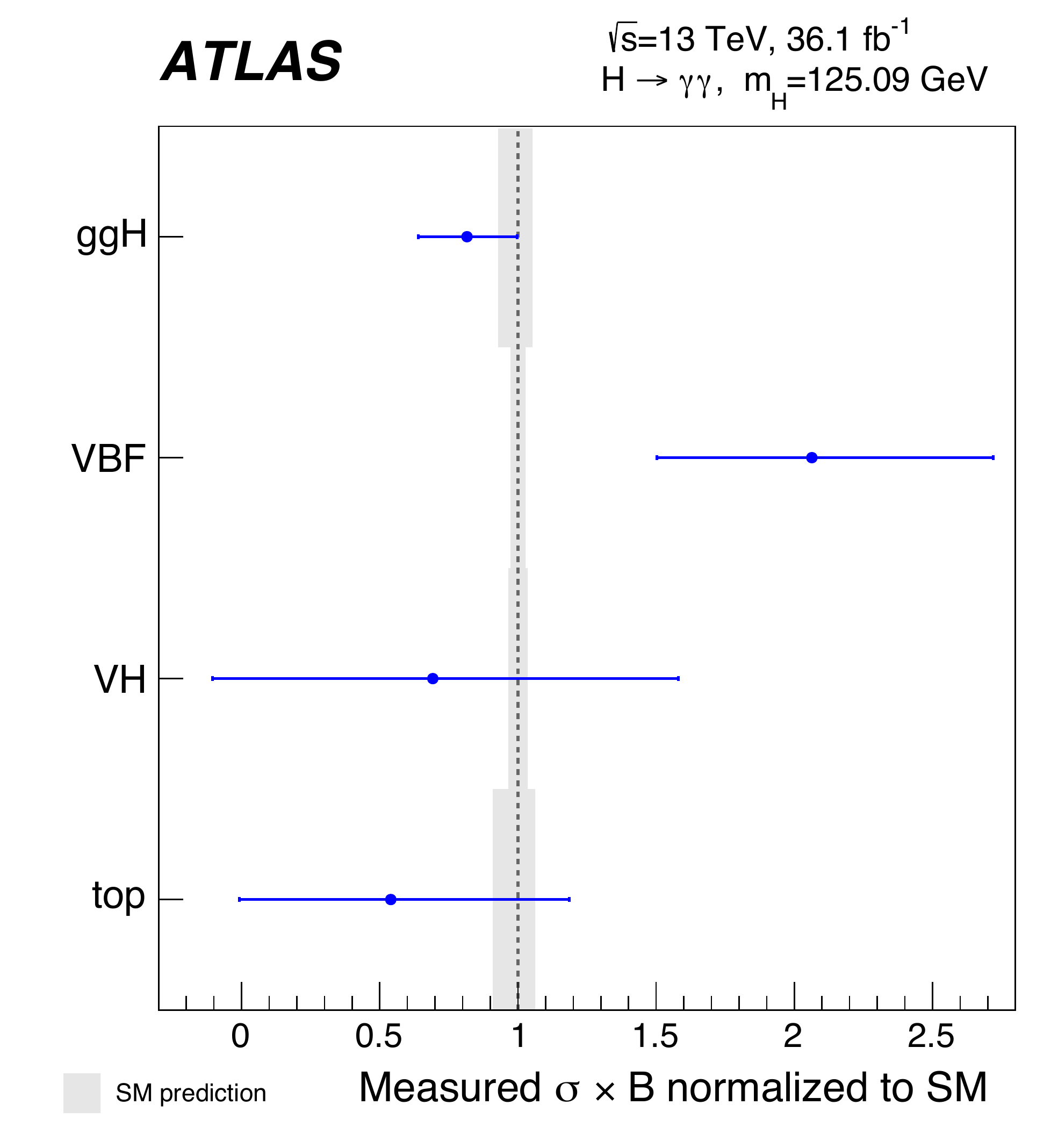}
\caption{Summary plot of the measured production-mode cross sections times the Higgs to diphoton branching ratio. 
For illustration purposes the central values have been divided by their SM expectations
but no additional theory uncertainties have been added to the uncertainty of the ratio. The uncertainties 
in the predicted SM cross sections are shown in gray bands in the plot.
The fitted value of $\sigma_\mathrm{top}$ corresponds to the sum of \ttH, \tHqb, and \tHW\ production-mode cross sections
under the assumption that their relative ratios are as predicted by the SM. The \VH\ production mode cross-section values are
determined under the assumption that the ratio of the \WH\ and \ZH\ production-mode cross sections is as predicted by the SM
and includes production from both the quark and gluon initial states. 
The \bbH\ contributions are merged with \ggH.
}
\label{fig:XSResults}
\end{figure}

The production-mode cross sections are summarized in Figure~\ref{fig:XSResults} and Table~\ref{tab:fourxs_results}.

\begin{table}[!tp]
\caption{
Best-fit values and uncertainties of the production-mode cross sections times branching ratio. 
The SM predictions~\cite{deFlorian:2016spz} with their uncertainties are shown for each production process. Uncertainties smaller than 0.05 are displayed as 0.0.
} \begin{center}
\renewcommand{\arraystretch}{1.3}
\begin{tabular}{l|clr@{}lcr@{}l|c}
\hline \hline
Process   & Result &\multicolumn{6}{c|}{Uncertainty [fb] } & SM prediction \\
 ($|y_H|<2.5$)& [fb] & Total & & Stat. & Exp. & Theo. & & [fb] \\\hline \hline
\ggH  &  82   & \asym{}{19}{18}   & \lg & \err{16}          & \asym{}{7}{6}     & \asym{}{5}{4}     & \rg & \asym{102}{5}{7} \\
\VBF  &  16   & \asym{}{5}{4}     & \lg & \err{4}           & \err{2}           & \asym{}{3}{2}     & \rg & \small $8.0 \pm 0.2$ \\
\VH   &   3   & \err{4}           & \lg & \asym{}{4}{3}     & \err{1}           & \asym{}{1}{0}     & \rg & \small $4.5 \pm 0.2$ \\
Top   &  0.7  & \asym{}{0.9}{0.7} & \lg & \asym{}{0.8}{0.7} & \asym{}{0.2}{0.1} & \asym{}{0.2}{0.0} & \rg & \small $1.3 \pm 0.1$ \\
\hline \hline
\end{tabular}
\end{center}
\label{tab:fourxs_results}
\end{table}

\begin{figure}[!tbp]
\centering
\includegraphics[width=.82\textwidth]{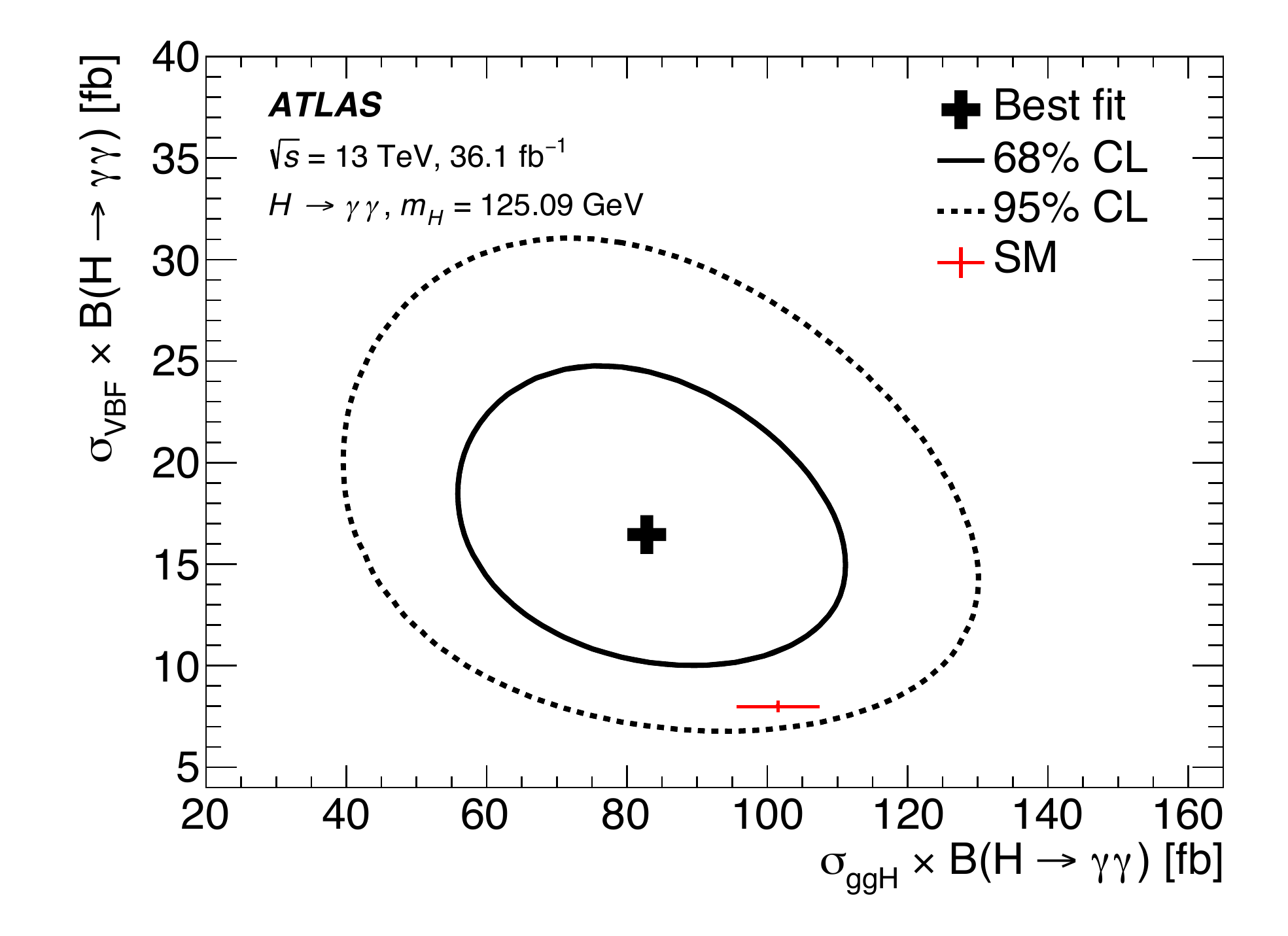}
\caption{Likelihood contours in the $(\sigma_\mathrm{ggH}\tbfhyy, \sigma_\mathrm{VBF}\tbfhyy)$ plane, compared to the Standard Model prediction (red cross) for a Higgs boson mass $m_H = 125.09\GeV$.}
\label{fig:ggH_VBF}
\end{figure}

The 68\% and 95\% CL two-dimensional contours of $\sigma_\mathrm{ggH}\tbfhyy$ and $\sigma_\mathrm{VBF}\tbfhyy$ are shown
in Figure~\ref{fig:ggH_VBF}, profiling $\sigma_\mathrm{VH}\tbfhyy$ and $\sigma_\mathrm{top}\tbfhyy$ in the fits. The SM
expectation of $\sigma_\mathrm{ggH}\tbfhyy$ vs $\sigma_\mathrm{VBF}\tbfhyy$ is within the 95\% CL contour of this
measurement.

To remove the impact of possible deviations in the $H \to
\gamma\gamma$ branching ratio, ratios of the production-mode cross sections
to the \ggH\ cross section are also extracted.
Such ratios, normalized for convenience of presentation
to the central values of their SM predictions, are\footnote{The quoted theory uncertainty only accounts for the uncertainty in the acceptance. The production cross-section uncertainties are not included in the uncertainty budget. Uncertainties smaller than 0.05 are displayed as 0.0.}
\begin{align*}
\frac{\sigma_\mathrm{VBF}/\sigma_\mathrm{ggH}}{(\sigma_\mathrm{VBF}/\sigma_\mathrm{ggH})^\mathrm{SM}} &= 2.5\ ^{+1.3}_{-0.9} = 2.5\ ^{+1.1}_{-0.8}\,\mathrm{(stat.)}\ ^{+0.5}_{-0.3}\,\mathrm{(exp.)}\ ^{+0.5}_{-0.3}\,\mathrm{(theo.)}\\
\frac{\sigma_\mathrm{VH} / \sigma_\mathrm{ggH}}{(\sigma_\mathrm{VH} / \sigma_\mathrm{ggH})^\mathrm{SM}} &= 0.9\ ^{+1.3}_{-1.0} = 0.9\ ^{+1.2}_{-0.9}\,\mathrm{(stat.)}\ ^{+0.3}_{-0.3}\,\mathrm{(exp.)}\ ^{+0.2}_{-0.1}\,\mathrm{(theo.)}\\
\frac{\sigma_\mathrm{top} / \sigma_\mathrm{ggH}}{(\sigma_\mathrm{top} / \sigma_\mathrm{ggH})^\mathrm{SM}} &= 0.7\ ^{+0.8}_{-0.7} = 0.7\ ^{+0.8}_{-0.7}\,\mathrm{(stat.)}\ ^{+0.2}_{-0.1}\,\mathrm{(exp.)}\ ^{+0.2}_{-0.0}\,\mathrm{(theo.)}\\
\end{align*}

The ratios are also given in Table~\ref{tab:fourxs_results_ratios}, along with their statistical, experimental and theoretical uncertainties without the normalization to the central values of the SM predictions. Both the measurements of the \ggH\ and \VBF\ production modes and the evaluations of the \VH\ and top production modes agree within $1$--$2~\sigma$ with the SM expectations. Appendix~\ref{app:summary_coup:xs} provides the expected uncertainties for the production mode cross sections.

\begin{table}[!tbp]
\caption{
Ratios of the production-mode cross sections with respect to the \ggH\ cross section and uncertainties 
are shown. The SM predictions~\cite{deFlorian:2016spz} with their uncertainties are shown for each production process.
} \begin{center}
\renewcommand{\arraystretch}{1.3}
\begin{tabular}{l|clr@{}lcr@{}l|c}
\hline \hline
Process   & Result &\multicolumn{6}{c|}{Uncertainty } & SM prediction \\
 ($|y_H|<2.5$)&  & Total & & Stat. & Exp. & Theo. & &  \\\hline \hline
$\sigma_\mathrm{VBF}/\sigma_\mathrm{ggH}$  &  0.20   & \asym{}{0.10}{0.07}   & \lg & \asym{}{0.09}{0.06}         & \asym{}{0.04}{0.02}      & \asym{}{0.04}{0.02}     & \rg & \asym{0.078}{0.005}{0.006} \\
$\sigma_\mathrm{VH}/\sigma_\mathrm{ggH}$  &  0.04   & \asym{}{0.06}{0.05}     & \lg & \asym{}{0.06}{0.04}         & \asym{}{0.01}{0.01}     & \asym{}{0.01}{0.01}     & \rg & \asym{0.045}{0.004}{0.005} \\
$\sigma_\mathrm{top}/\sigma_\mathrm{ggH}$   &   0.009   & \asym{}{0.010}{0.009}   & \lg &  \asym{}{0.010}{0.009}   &  \asym{}{0.002}{0.001}    &  \asym{}{0.002}{0.001}     & \rg & \asym{0.012}{0.001}{0.002} \\
\hline \hline
\end{tabular}
\end{center}
\label{tab:fourxs_results_ratios}
\end{table}

\subsubsection{Simplified template cross sections}
\label{sec:resultsSTRONG}

As the current data are not yet sensitive to all of the 31 regions with $|y_H|<\nobreak2.5$ (assuming SM acceptance)
 of the ``stage-1'' scheme of the simplified template
cross-section framework, simplified template cross sections are reported for 10 phase space regions obtained from merging
the initial 31 as described in Section~\ref{sec:intro_stxs} and Table~\ref{tab:STXS}. To retain
sensitivity to BSM Higgs boson production, the $\pT^H>\nobreak200\,\GeV$ gluon--gluon fusion and
$\pT^j>\nobreak200\,\GeV$ VBF regions are not merged with other regions. 
This scheme has been chosen to reduce strong anti-correlations between the measured 
cross sections and to keep measurements near or below 100\% total uncertainty.
In the likelihood, the signal yield $N_\mathrm{sig}^i$ in each category $i$ is the sum over the yields
$N_{\mathrm{sig},r}^i$ expected from each of the 9 regions $r$ of phase space, where $N_{\mathrm{sig},r}^i = \int L\,\mathrm{d}t \times\sigma_r^\mathrm{SM} \times \bfhyySM\times \epsilon_r^i$ and the additional
region corresponds to the difference of the cross sections for the $\pT^H>\nobreak200\,\GeV$ gluon--gluon fusion and
$\pT^j>\nobreak200\,\GeV$ VBF regions. The observed cross sections are reported in Table~\ref{tab:stxs_results}.
These measurements have been defined to minimize theoretical uncertainties and are strongly dominated by experimental uncertainty,
hence only the total systematic uncertainty is reported.

\begin{table}[!tp]
\caption{
Best-fit values and uncertainties of the simplified template cross sections times branching ratio.
The SM predictions~\cite{deFlorian:2016spz} are shown for each region. 
}
\begin{center}
\renewcommand{\arraystretch}{1.2}
\begin{tabular}{l|rlr@{}lr@{}l@{ }l|r@{ }l}
\hline \hline
Measurement region & \multirow{2}{*}{Result} & \multicolumn{6}{c|}{ Uncertainty } & \multicolumn{2}{c}{\multirow{2}{*}{SM prediction}} \\
($|y_H| < 2.5$)    & & Total & & Stat. & Syst. &&&& \\
\hline \hline
$\ggH, \mathrm{0~jet}$   &  37 & \asym{}{16}{15} & \lg & \err{14} & \asym{}{6}{5} & \rg & fb & $63 \pm 5$ & fb \tspp \\
$\ggH, \mathrm{1~jet}, \pT^{H} < 60$ \GeV  &  13 & \asym{}{13}{12} & \lg & \err{12} & \asym{}{5}{4} & \rg & fb & $15 \pm 2$ & fb \tspp \\
$\ggH, \mathrm{1~jet}, 60 \leq \pT^{H} < 120$ \GeV & 5 & \err{6} & \lg & \err{6} & \asym{}{2}{1} & \rg & fb & $10 \pm 2$ & fb \tspp \\
$\ggH, \mathrm{1~jet}, 120 \leq \pT^{H} < 200$ \GeV & 2.8 & \asym{}{1.7}{1.6} & \lg & \asym{}{1.6}{1.5} & \asym{}{0.7}{0.5} & \rg & fb & $1.7 \pm 0.3$ & fb \tspp \\
$\ggH, \geq 2~\mathrm{jet}$ & 20 & \asym{}{9}{8} & \lg & \err{8} & \asym{}{4}{3} & \rg & fb & $11 \pm 2$ & fb \tspp \\
$qq \rightarrow Hqq, \pT^{j} < 200$ \GeV & 15 & \asym{}{6}{5} & \lg & \err{5} & \asym{}{3}{2} & \rg & fb & $10 \pm 0.5$ & fb \tspp \\
$\ggH + qq \rightarrow Hqq, \mathrm{BSM-like}$ & 2.0 & \err{1.4} & \lg & \err{1.3} & \err{0.6} & \rg & fb & $1.8 \pm 0.4$ & fb \tspp \\
VH, leptonic & 0.7 & \asym{}{1.4}{1.3} & \lg & \asym{}{1.4}{1.2} & \asym{}{0.4}{0.3} & \rg & fb & $1.4 \pm 0.1$ & fb \tspp \\
Top          & 0.7 & \asym{}{0.8}{0.7} & \lg & \asym{}{0.8}{0.7} & \asym{}{0.2}{0.1} & \rg & fb & $1.3 \pm 0.1$ & fb \tspp \\
\hline \hline
\end{tabular}
\end{center}
\label{tab:stxs_results}
\end{table}

The evaluated cross sections including their correlations are summarized in Figures~\ref{fig:stxsResults} and~\ref{fig:stxsCorr}.
The expected Standard Model correlations can be found in Appendix~\ref{app:correlation}.
All observed cross sections are in agreement with the Standard Model values. The Standard Model prediction is determined using the generators in Section~\ref{sec:mc} and the theory uncertainties due to missing higher-order corrections and due to the chosen PDF set are constructed as described in Section~\ref{sec:theo}. The largest deviation ($1.7~\sigma$) from the SM prediction is found in the $\ggH, \mathrm{0~jet}$ bin. 
The difference of the cross sections for the $\pT^H>\nobreak200\,\GeV$ \ggH\ and
$\pT^j>\nobreak200\,\GeV$ \VBF\ regions is found to be $4.8^{+2.9}_{-2.7}$~\fb.

\begin{figure}[!tbp]
\centering
\includegraphics[width=.82\textwidth]{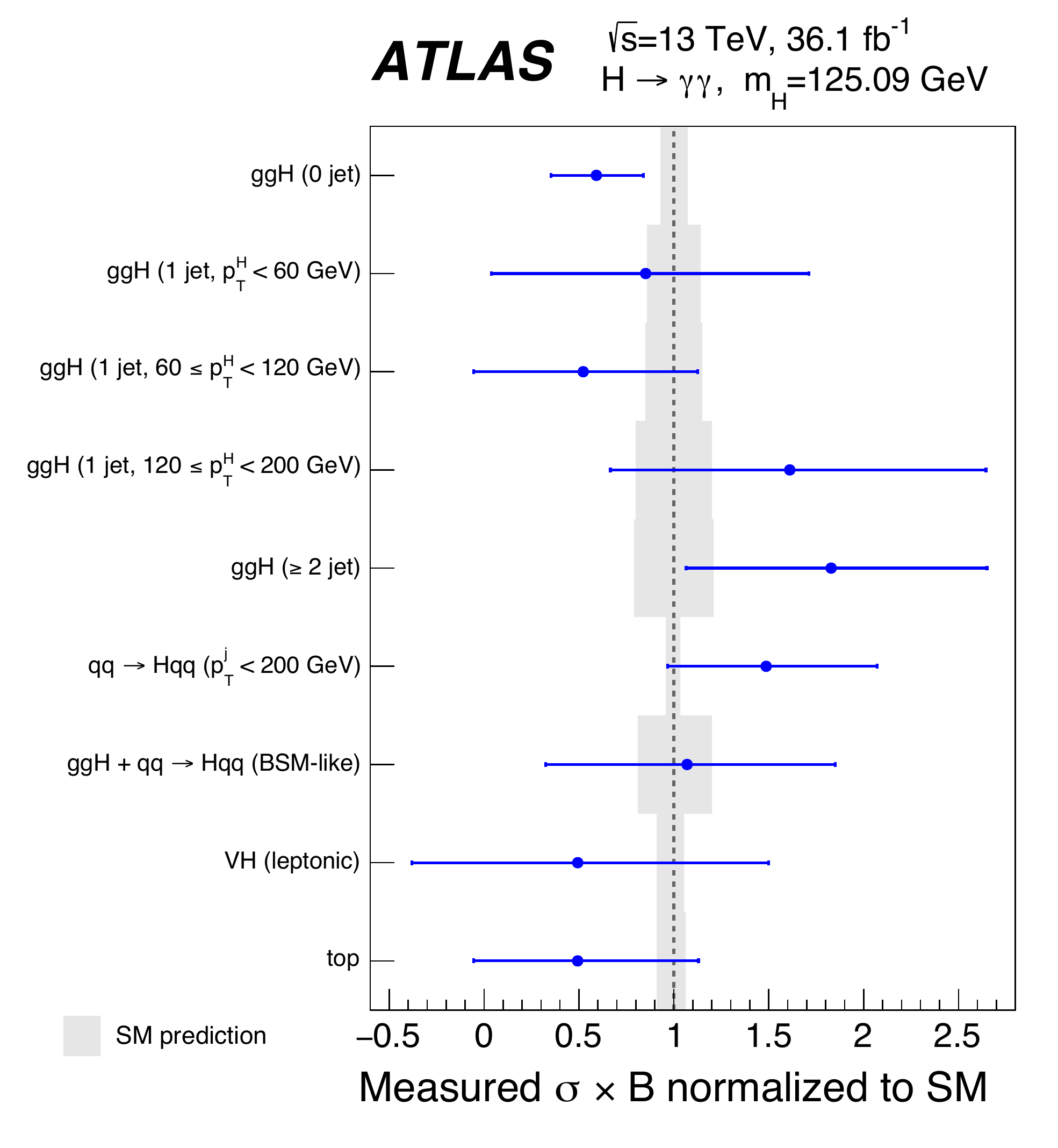}
\caption{Summary plot of the measured simplified template cross sections times
the Higgs to diphoton branching ratio. For illustration purposes the central values have been divided by their SM
expectations but no additional theory uncertainties have been included in the uncertainty of the ratio due to this. The uncertainties in the predicted SM cross sections are shown in gray in the plot. The definition of the measured regions can be found in Table~\ref{tab:STXS}. 
The fitted value of $\sigma (\mathrm{top})$ corresponds to the sum of \ttH and \tH\ production-mode 
cross sections under the assumption that their relative ratios are as predicted by the SM. The $\sigma (\mathrm{VH, leptonic})$ cross-section values are determined under the assumption that the ratio of the \WH\ and \ZH\ production mode cross
sections is as predicted by the SM and includes production from both the quark and gluon initial states. The \bbH\
contributions are merged with \ggH.
}
\label{fig:stxsResults}
\end{figure}

\begin{figure}[!tbp]
\centering
\includegraphics[width=.82\textwidth]{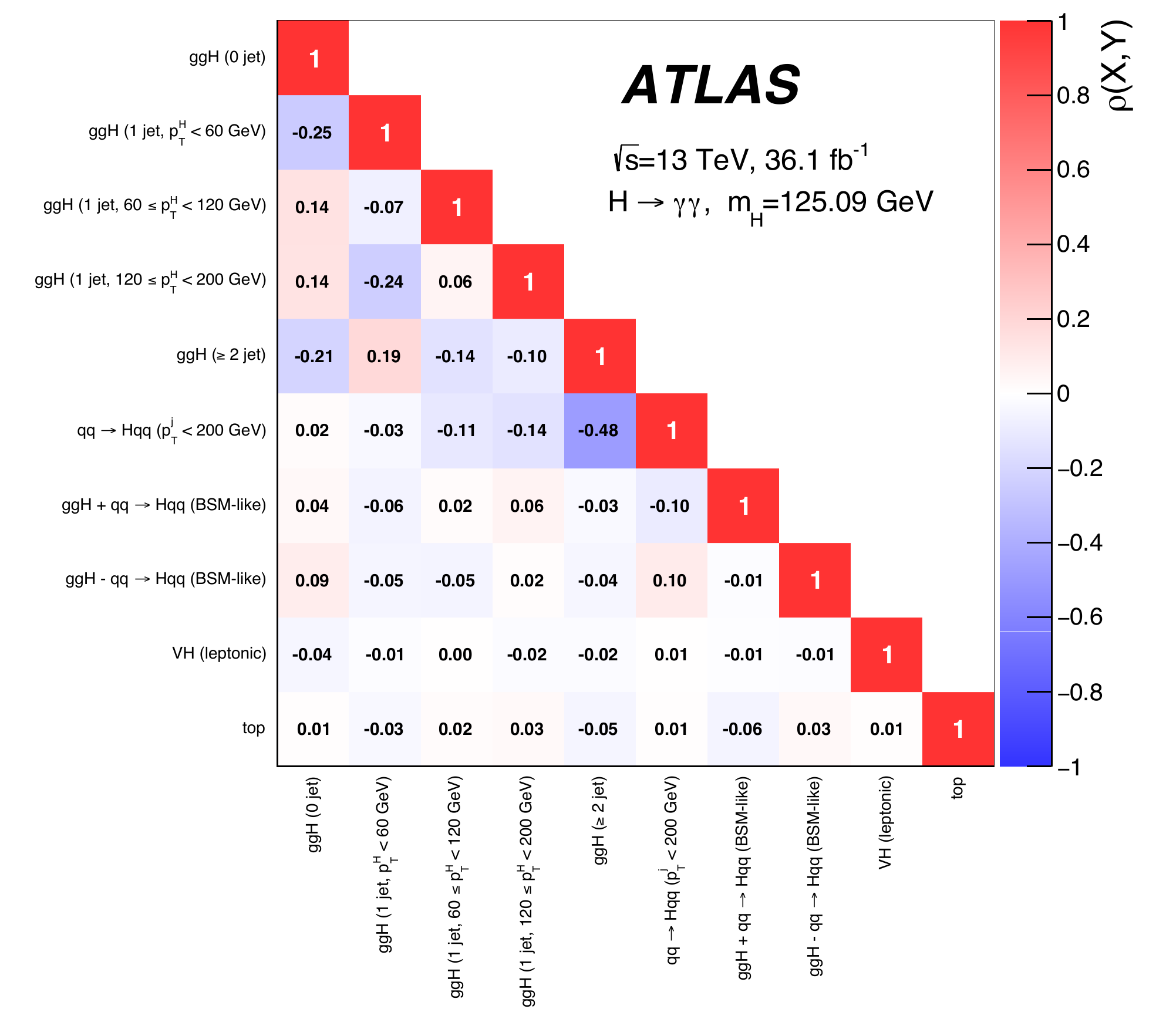}
\caption{Observed correlations between the measured simplified template cross sections, including both the statistical and systematic uncertainties. The color indicates the size of the correlation. }
\label{fig:stxsCorr}
\end{figure} 

Limits at 95\% CL on the $\ggH + qq \to Hqq$ BSM-like ($p_{\textrm T}^j > 200\,\GeV$) bin are set, profiling all other parameters, as shown in Table~\ref{tab:BSM_limits}. Appendix~\ref{app:summary_coup:stxs} provides the expected uncertainties for all quoted simplified template cross sections.

In Appendix~\ref{sec:appendix_MINSTXS} additional measurements are reported for a ``minimally merged'' set of 15 cross sections of kinematic regions defined by the requirement that the fits to expected event yields be stable even in the presence of large uncertainties or correlations.

\begin{table}[!tp]
\caption{Observed and expected upper limits at 95\% CL on the simplified template cross section times the Higgs 
  to diphoton branching ratio in the BSM sensitive phase space with $p_{\textrm T}^j > 200\,\GeV$. 
  The median expected limits are given for either the case when the
  true value of the cross section under study is SM-like ($\sigma = \sigma_{\rm SM}$)
  or zero. The $\pm 1~\sigma$ and $\pm 2~\sigma$ intervals for the
  expected upper limit ($\sigma=0\, \fb$) are also reported.
}
\label{tab:BSM_limits}
\vspace{1ex}
\renewcommand{\arraystretch}{1.2}
\begin{tabular}{c|c|c|c|c|c|c|c}
 \hline
 \hline
  Measurement          &   Observed &   Expected Limit    &   Expected Limit  &   $+2\sigma$ &   $+1\sigma$ &   $-1\sigma$ &   $-2\sigma$ \\
                       &            &   ($\sigma = \sigma_{\rm SM}$) & ($\sigma = 0\, \fb$) &              &              &              &              \\
 \hline \hline
 $ggH + qq \to Hqq$,  
                       &   4.4\,\fb &       4.3\,\fb  &      2.7\,\fb &     5.3\,\fb &     3.8\,\fb &     2.0\,\fb &     1.5\,\fb \\
                       
 $\mathrm{BSM-like}$ & & & & & & &  \\
 \hline
 \hline  
\end{tabular}
\end{table}

\clearpage

\subsubsection{Coupling-strength fits}

Following the tree-level-motivated framework and benchmark models recommended in Ref.~\cite{Heinemeyer:2013tqa}, measurements of Higgs boson coupling-strength modifiers $\kappa_j$ are implemented.
In the narrow width approximation for the Higgs boson, the cross section $\sigma(i \rightarrow H \rightarrow \gamma\gamma)$ can be parameterized as
\begin{equation*}
\sigma(i \rightarrow H \rightarrow \gamma\gamma) = \frac{\sigma_i(\vec\kappa)\,\Gamma^{\gamma\gamma}(\vec\kappa)}{\Gamma_H},
\end{equation*}

where $\Gamma_H$ is the total width of the Higgs boson and $\Gamma^{\gamma\gamma}$ is the partial decay width to two photons. 
A set of coupling-strength modifiers, $\vec\kappa$, is introduced to parameterize possible deviations from the SM predictions of the Higgs 
boson coupling to SM bosons and fermions. For a given production process or decay mode~$j$, a coupling-strength modifier $\kappa_j$ is defined such that:
\begin{equation*}
\kappa_j^2 = \sigma_j/\sigma_{j,\mathrm{SM}}\ \,\, \mathrm{or}\ \,\, \kappa_{\gamma}^2 = \Gamma^{\gamma\gamma}/\Gamma^{\gamma\gamma}_\mathrm{SM},
\end{equation*}

where all $\kappa_j$ values equal unity in the SM. Here, by construction, the SM cross sections and branching ratio include the 
best available higher-order QCD and EW corrections. This higher-order accuracy is not necessarily preserved for $\kappa_j$ values 
different from unity, but the dominant higher-order QCD corrections factorize to a large extent from any rescaling of the 
coupling strengths and are therefore assumed to remain valid over the entire range of $\kappa_j$ values considered.

Individual coupling-strength modifiers corresponding to tree-level Higgs boson couplings to different particles are introduced 
as well as two effective coupling-strength modifiers, $\kappa_g$ and $\kappa_\gamma$, which describe the loop processes for \ggH\ production 
and \hyy\ decay. This is possible because BSM particles that might be present in these loops are not expected to appreciably change 
the kinematics of the corresponding process. The \hgg\ and \hyy\ loop processes can thus be studied through these effective 
coupling-strength modifiers, providing sensitivity to potential BSM particles in the loops. 
In contrast, the \ggZH\ process, which occurs at LO through box and triangular loop diagrams, is always taken into account by
resolving the loop in terms of the corresponding coupling-strength modifiers ($\kappa_Z$ and $\kappa_t$). 
No decays to particles other than those predicted in the SM are assumed to take place. These considerations and the limited sensitivity of the data available in this analysis lead to introducing two distinct models. 

\begin{figure*}[!tbp]
  \centering
  \subfloat[] {\includegraphics[width=0.48\columnwidth]{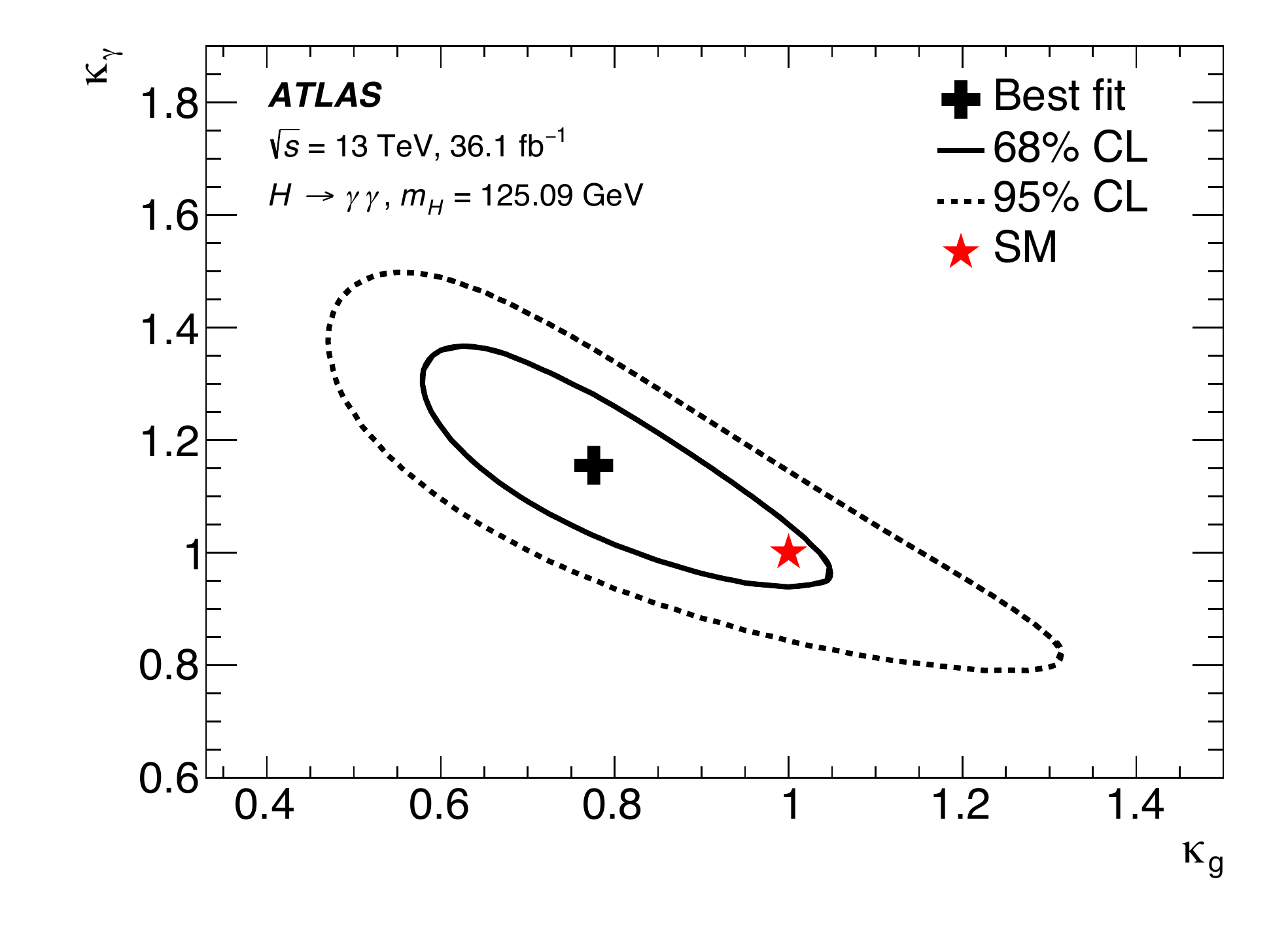} \label{fig:cg_cy}}
  \subfloat[] {\includegraphics[width=0.48\columnwidth]{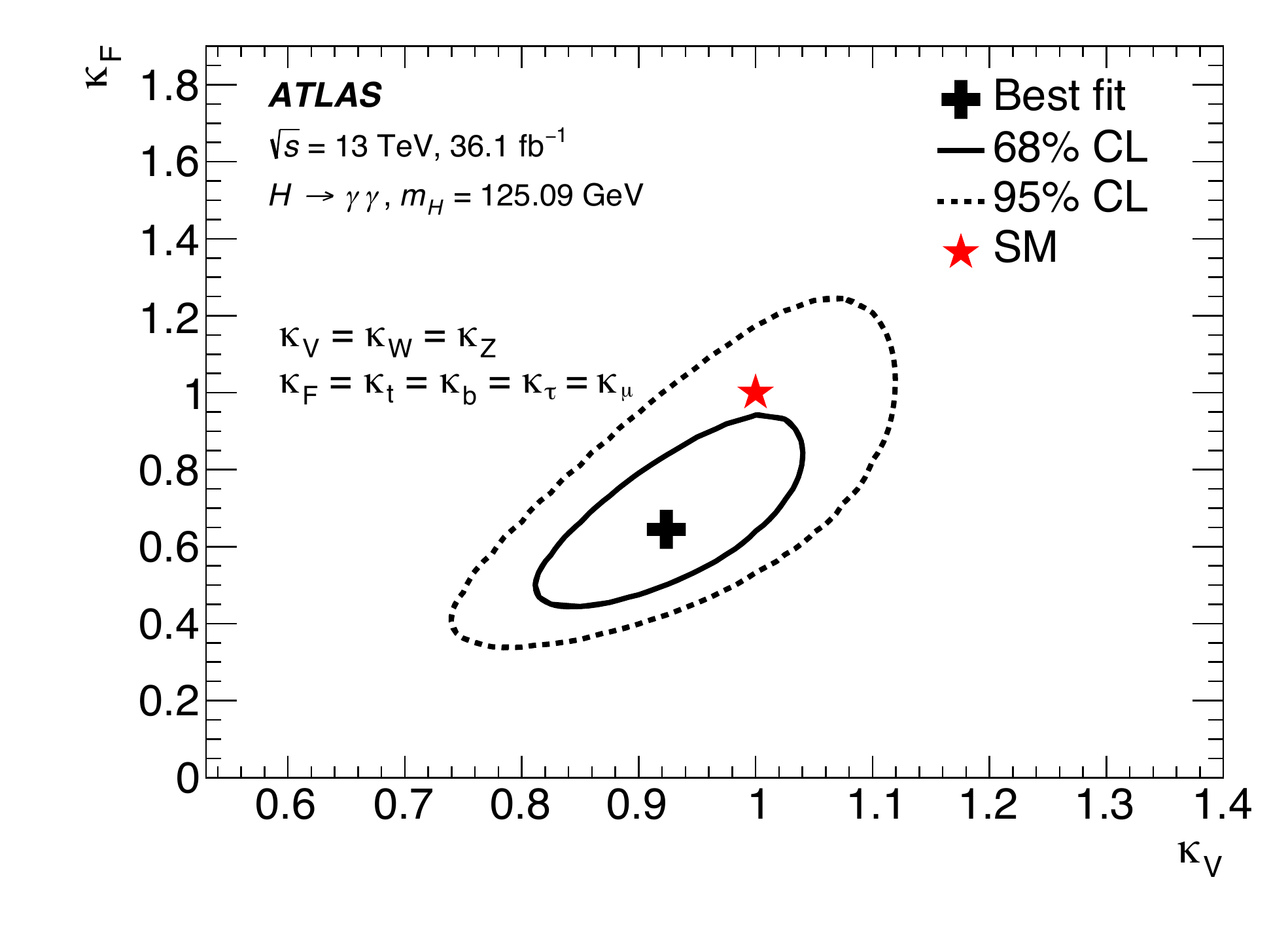} \label{fig:cf_cv}}
  \caption{
    Likelihood contours in (a) the $(\kappa_g, \kappa_\gamma)$ plane, and (b) the $(\kappa_V, \kappa_F)$ plane, compared to the Standard Model prediction (red star) for a Higgs boson mass $m_H = 125.09 \, \GeV$. In (a), all coupling-strength modifiers other than $\kappa_g$ and $\kappa_\gamma$ are fixed to their SM value. In (b), the \hgg\ and \hyy\ loops are resolved in terms of two universal coupling-strength modifiers $\kappa_F$ and $\kappa_V$, under the assumption that $\kappa_V = \kappa_W = \kappa_Z$ and $\kappa_F = \kappa_t = \kappa_b = \kappa_\tau = \kappa_\mu$.
  } 
  \label{fig:kappas}
\end{figure*}

In the first model, the two parameters $\kappa_g$ and $\kappa_\gamma$ introduced above are tested assuming that all other couplings are as in the SM. The 68\% and 95\% CL two-dimensional contours of both effective couplings are shown in Figure~\ref{fig:cg_cy} and the best fit values and uncertainties are $\kappa_g = 0.76^{+0.17}_{-0.14}$ and $\kappa_\gamma = 1.16^{+0.14}_{-0.14}$. 

In a second model, universal coupling-strength modifiers, $\kappa_F$ (for all fermions) and $\kappa_V$ (for all bosons), are defined that resolve the \hgg\ and \hyy\ loops:
\begin{equation*}
\begin{split}
\kappa_F &= \kappa_t = \kappa_b = \kappa_\tau = \kappa_\mu \, , \\
\kappa_V &= \kappa_W = \kappa_Z \, .
\end{split}
\end{equation*}
The 68\% and 95\% CL two-dimensional contours of both parameters are shown in Figure~\ref{fig:cf_cv} and the best fit values and uncertainties are $\kappa_F = 0.64^{+0.18}_{-0.14}$ and $\kappa_V = 0.92^{+0.08}_{-0.07}$. Due to the very limited sensitivity to $\kappa_b, \kappa_\tau$ and $\kappa_\mu$, the shown CLs would not change if these coupling-strength modifiers would be fixed to the SM expectation. 

The SM prediction is found within the 68\% CL contour for the first model and within the 95\% CL contour for the second model.

\begin{figure}[!tbp]
\centering
\includegraphics[width=.82\textwidth]{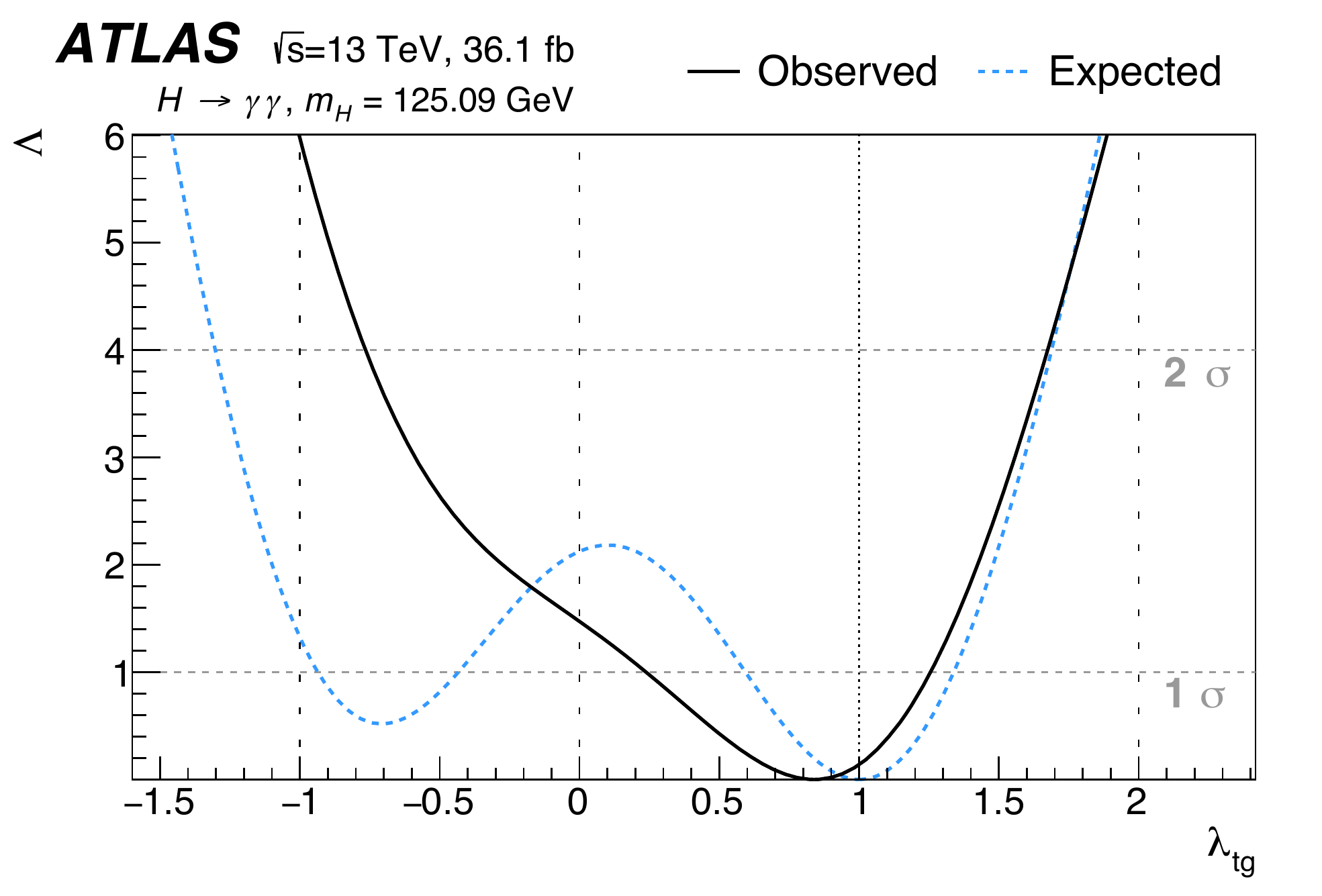}
\caption{The profile of negative log-likelihood $\Lambda$ of the observed and expected coupling-strength modifier ratio $\lambda_{tg} = \kappa_t / \kappa_g$. The parameters $\kappa_{g\gamma}$ and $\lambda_{Vg}$ are also profiled within the fit. 
The intersections of the solid and dashed curves with the horizontal dashed line at $\Lambda = 1$ and $\Lambda = 4$ indicate the 1 and 2 $\sigma$ confidence intervals of the observed and expected results, respectively.
  }
  
\label{fig:nll_comp_RFg}
\end{figure}

Finally, a set of three ratios is constructed to probe the loop vertices ($\kappa_g$, $\kappa_\gamma$), 
total width ($\kappa_H$), and the vector and top couplings ($\kappa_t$ and $\kappa_V$ respectively): 
$\kappa_{g\gamma} = \kappa_g \kappa_\gamma / \kappa_H$, $\lambda_{Vg} = \kappa_V / \kappa_g$, and $\lambda_{tg} = \kappa_t / \kappa_g$.
The parameter $\lambda_{tg}$ is allowed to be negative to exploit the sensitivity to the relative sign from the 
\tH\ and \ggZH\ processes. The expected and observed sensitivities to the relative sign are illustrated in
Figure~\ref{fig:nll_comp_RFg}. 
The bottom quark Yukawa coupling strength is kept fixed to the top quark Yukawa coupling strength ($\lambda_{bg} = \lambda_{tg}$);
this contribution is irrelevant to the $\lambda_{tg}$ measurement as there is no sensitivity to \bbH\ in the analysis.
All other parameters are assumed to be positive without losing generality. 
The inclusion of $\kappa_H$ in the parameterization allows for non-SM decays of the Higgs boson, 
but this parameter is not determined directly. The best fit values of these 
coupling ratios are summarized in Table~\ref{tab:lambda_results}. 

\begin{table}[!tp]
\caption{Best-fit values and uncertainties of $\kappa_{g\gamma}$, $\lambda_{Vg}$, and $\lambda_{tg}$. }
\label{tab:lambda_results}
\begin{center}
{\renewcommand{\arraystretch}{1.3}
\begin{tabular}{l|rlr@{}lcr@{}l} 
\hline
\hline
\multirow{2}{*}{Parameter} & \multirow{2}{*}{Result} & \multicolumn{6}{c}{Uncertainty} \\
                           &  & Total & & Stat. & Exp. & Theo. & \\
\hline \hline
$\kappa_{g\gamma}$ & 0.90 & \err{0.10}          & \lg & \err{0.09}          & \err{0.04}          & \asym{}{0.04}{0.03} & \rg \\
$\lambda_{Vg}$     & 1.41 & \asym{}{0.31}{0.26} & \lg & \asym{}{0.28}{0.23} & \asym{}{0.10}{0.07} & \asym{}{0.04}{0.03} & \rg \\
$\lambda_{tg}$     & 0.8  & \asym{}{0.4}{0.6}   & \lg & \asym{}{0.4}{0.6}   & \err{0.1}           & \asym{}{0.1}{0.0}   & \rg \\
\hline
\hline
\end{tabular}}
\end{center}
\end{table}

\clearpage


\clearpage

\section{Measurement of fiducial integrated and differential cross sections} \label{sec:methods_fid}

The measurement of fiducial integrated and differential cross sections provides an alternative way to study the properties of the Higgs boson and to search for physics beyond the Standard Model. The fiducial volumes are defined to closely mimic the detector-level photon and object selections described in Section~\ref{sec:evsel}. This reduces the model-dependence of the quoted cross sections in contrast to the per production mode simplified template cross-section measurements of Section~\ref{sec:resultsSTRONG}. The cross sections are determined by correcting measured signal yields for experimental inefficiencies and resolution effects, and by taking into account the integrated luminosity of the data. Rather than separating individual production modes, fiducial regions are defined such that they are enriched with a given production mode: Fiducial cross sections are measured in a variety of phase space regions, sensitive to for instance gluon--gluon fusion Higgs boson production, vector-boson fusion production, but also to production of the Higgs boson in association with charged leptons, top quarks and neutrinos. Differential and double-differential cross sections are reported for variables related to the diphoton kinematics and the jet activity produced in the Higgs boson events. The observed signal yields are corrected for detector effects resulting in cross sections measured at the particle level. The full statistical and systematic correlations between measured distributions are determined and are available in \hepdata\ along with the central values of the measured fiducial and differential cross sections to allow future comparisons and interpretations.

\subsection{Particle-level fiducial definition of the Higgs boson diphoton cross sections}\label{sec:fiddef}

The fiducial volume at particle level is defined using particles with a mean lifetime $c\tau>10$\,mm. Only photons and
leptons which do not originate from the decay of hadrons are considered.\footnote{Leptons originating from the decay of $\tau$ leptons are only considered if the $\tau$ lepton itself did not originate from the decay of hadrons.}
The two highest-\pt\ photons with $|\eta|<2.37$ -- excluding $1.37<|\eta|<1.52$ -- are selected as the diphoton system.
The leading (subleading) photon is required to satisfy $\pt / \mgg > 0.35\ (0.25)$, where $\mgg = \mH = 125.09 \, \GeV$.
Furthermore, for each photon the scalar \pt-sum of charged particles with $\pt>1$\,\GeV\ within a cone of $\Delta R=0.2$ around the photon is required to be less than 5\% of the photon \pt. The lepton four-momentum is defined as the combination of an electron (or muon) and all
nearby photons within $\Delta R< 0.1$ that do not originate from the decay of a hadron. Muons are required to have
$\pt>15$\,\GeV\ and $|\eta|<2.7$. Electrons are required to have $\pt>15$\,\GeV\ and $|\eta|<2.47$, excluding the region
$1.37<|\eta|<1.52$, and are rejected if the distance $\Delta R$ to a photon with $\pt>$ 15\,\GeV\ is less than 0.4. Jets are
reconstructed from all particles, excluding muons and neutrinos, using the \antikt\ algorithm with a radius parameter of
0.4. Unless stated otherwise, jets are required to have $\pt>30$~\GeV, $|y|<4.4$ and to be well separated from photons with $\pt>$ 15\,\GeV\
($\Delta R > 0.4$) and electrons ($\Delta R> 0.2$). The acceptance for the \VBF-enhanced fiducial region (introduced in Section~\ref{sec:fid_def}) is increased by loosening the \pT cut to 25 \GeV. Jets are considered to originate from a $b$-hadron if there is a $b$-hadron with $\pt > 5$~\GeV\ within a cone of size $\Delta R = 0.4$ around the jet.

The missing transverse momentum is
defined as the vector sum of neutrino transverse momenta, for neutrinos that do not originate from the decay of a hadron. 
The particle-level fiducial definition is summarized in Table~\ref{tab:criteria}.

\begin{table}[!tp]
  \caption{Summary of the particle-level definitions of the five fiducial integrated regions described in the text. The photon isolation $p_\text{T}^\text{iso,0.2}$ is defined analogously to the reconstructed-level track isolation as the transverse momentum of the system of charged particles within $\Delta R < 0.2$ of the photon.
  }
  \label{tab:criteria}
  \centering
\begin{tabular}{ l l }
  \hline\hline
  Objects & Definition \\
  \hline \hline
  Photons & $|\eta| < 1.37$ or $1.52 < |\eta| < 2.37$, \, $p_\text{T}^\text{iso,0.2}/p_\text{T}^\gamma < 0.05$ \\
  Jets & anti-$k_t$, $R = 0.4$, \, $p_{\rm T} > 30$\,\GeV{}, \, $|y|<4.4$ \\
  Leptons, $\ell$ & $e$ or $\mu$, \, $p_{\rm T} > 15$\,\GeV{}, \, $|\eta|<2.47$ for $e$ (excluding $1.37 < |\eta| < 1.52$) and $|\eta|<2.7$ for $\mu$   \\
  \hline \hline
  Fiducial region & Definition\\ \hline\hline
  Diphoton fiducial & $N_\gamma \geq 2$, \, $p_{\rm T}^{\gamma_1} > 0.35 \, \mgg = 43.8$ GeV, \, $p_{\rm T}^{\gamma_2} > 0.25 \, \mgg =  31.3$ GeV\\
  \VBF-enhanced & Diphoton fiducial, \, $N_j\geq2$ with $\pt^\mathrm{jet} > 25$ GeV, \, \\
  &  $\mjj > 400$~\GeV, \, $\deltayjj>2.8$, \, $\dphiggjj>2.6$\\
  $N_{\rm lepton} \ge 1$   & Diphoton fiducial, \, $N_\ell\geq 1$ \\
    High $E_\text{T}^\text{miss}$ & Diphoton fiducial, \, $E_\text{T}^\text{miss} > 80$~\GeV, \, $\ptgg > 80$~\GeV \\
  $t \bar t H$-enhanced  & Diphoton fiducial, \, $\left(N_j\geq 4, \, N_{b\text{-jets}}\geq 1\right)$ or $\left(N_j\geq 3, \, N_{b\text{-jets}}\geq 1, N_\ell \geq 1\right)$ \\  
  \hline\hline
\end{tabular}
\end{table}

\subsection{Fiducial integrated and differential cross sections}\label{sec:fid_def}

The cross section ($\sigma_i$) in a fiducial integrated region, and the differential cross section (${\rm d}\sigma_i/{\rm d}x$) in
a bin of variable $x$, are given by 
\begin{equation*}
\sigma_i = \frac{ N_i^{\mathrm{sig}}}{ c_i \int L\, \mathrm{d}t} \quad\quad\quad\text{and}\quad\quad\quad \frac{\mathrm{d}\sigma_i}{\mathrm{d} x} = \frac{N_i^{\mathrm{sig}}}{ c_i \, \Delta x_i \int L \, \mathrm{d}t}, 
\end{equation*} 
where $N_i^{\mathrm{sig}}$ is the number of signal events as introduced in Section~\ref{sec:statmod}, $ \int L\,
\mathrm{d}t$ is the integrated luminosity of the data set, $c_i$ is a correction factor that accounts for
detector inefficiency and resolution, and $\Delta x_i$ is the bin width. The correction factors are determined using the
simulated samples discussed in Section~\ref{sec:mc}. This bin-by-bin method showed similar performance to that of the non-regularized inversion of the full migration matrix and of regularized methods~\cite{DAgostini:1994zf,Hocker:1995kb,Malaescu:2009dm} within the current statistical accuracy and systematic uncertainties.

The correction factor is $0.75 \pm 0.03$ in the diphoton fiducial region, defined to unfold all signal events to the
fiducial definition of Section~\ref{sec:fiddef}, which is dominated by the photon identification and isolation efficiency. 
The correction factor also accounts for migrations caused by detector energy resolution and migration in and out of the fiducial phase space due to detector effects. In addition, the correction factor removes a small fraction (0.5\% for the diphoton fiducial region) of reconstructed $H \to ff \gamma$ Dalitz decays.\footnote{Here $f$ denotes any fermion but the top quark.} 

The correction factor is different in fiducial regions defined by associated jet activity, for example, taking values of $0.66$ and $0.87$ for the \ttH\ and \VBF\ fiducial regions defined in the next section, respectively. 
For the diphoton fiducial region the uncertainty in the correction factor is dominated by the theoretical modeling uncertainty.
For the \ttH\ and \VBF\ fiducial regions the uncertainties in the correction factors are dominated by uncertainties associated 
with the knowledge of the jet energy scale and energy resolution, as well as the theoretical modeling. A more complete breakdown 
is given in Section~\ref{sec:fid_syst} and Table~\ref{tab:syst-summary}.

\begin{figure}[!tbp]
  \centering
  \includegraphics[width=0.85\columnwidth]{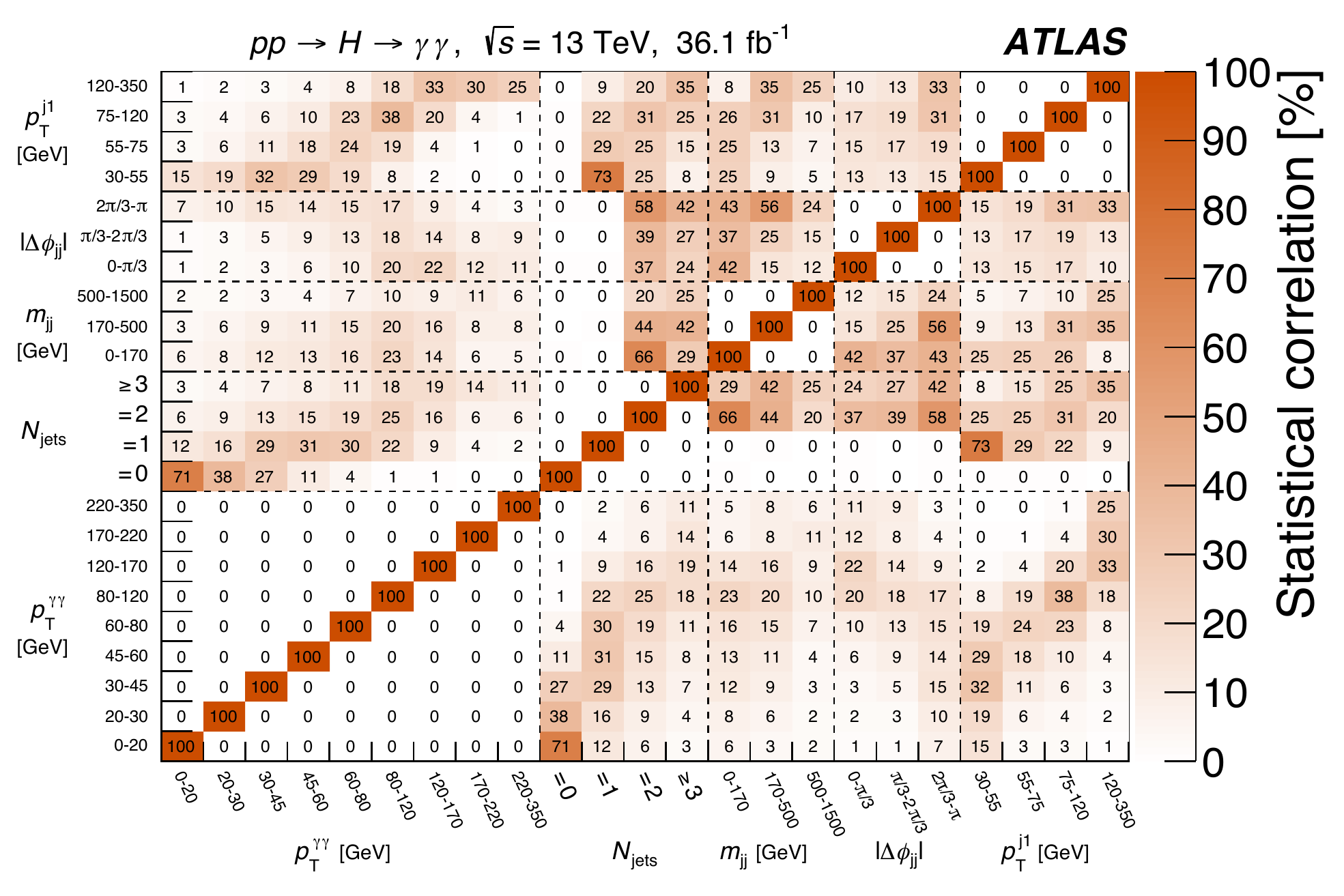}
  \caption{
    The observed statistical correlations between \ptgg, \njet, \mjj, \dphijjabs, and \ptjl\ are shown. These correlations were determined from an ensemble of 100,000 bootstrapped data sets which are each reanalyzed using an unbinned maximum-likelihood fit of the diphoton invariant mass spectrum to extract the correlations.
   } 
  \label{fig:stat_corr}
\end{figure}

The measured differential cross sections in different observables are partially statistically correlated, since they correspond to the same data set in a given fiducial region. These correlations are obtained using a random sampling with replacement method on the detector-level data, 
often referred to as 'bootstrapping'~\cite{PhysRevD.39.274}. Bootstrapped event samples are constructed from the data by assigning each event
a weight pulled from a Poisson distribution with unit mean. All measured differential distributions are then
reconstructed using the weighted events, and the signal yields in each bin of a differential distribution are determined
using an unbinned maximum-likelihood fit of the diphoton invariant mass spectrum. The procedure is repeated with
statistically independent weights and the correlation between two bins of different distributions is determined from the
obtained cross sections. Figure~\ref{fig:stat_corr} shows as an illustration the determined correlations 
between \ptgg, \njet, \mjj, \dphijjabs, and \ptjl: the lowest \ptgg\ bin, reconstructing events with a Higgs boson \pt\ between 0 and 20~\GeV, is highly correlated with the zero-jet bin. The lowest \ptjl\ bin, reconstructing events with a jet \pt\ between 30 and 55~\GeV, is strongly correlated with the one-jet bin.
And the lowest \mjj\ bin, reconstructing events with at least two jets and a dijet mass between 0 and 170~\GeV, is strongly correlated with the two jet bin.
The systematic correlations are obtained by fully correlating identical error sources described in
Section~\ref{sec:syst} across bins and observables to construct the corresponding systematic covariance matrix. 
Knowledge of these correlations allows to simultaneously analyze all fiducial regions, differential and double
differential cross sections. This is illustrated later in Section~\ref{sec:eft_res} with a simultaneous fit of the shown
\neftvars\ variables of Figure~\ref{fig:stat_corr} to set limits on new physics contributions. 

\subsection{Measurements of cross sections of fiducial integrated regions}\label{sec:details_fidreg}

Cross sections in five fiducial integrated regions are measured that target either specific Higgs boson production mechanisms or are sensitive to the presence of physics beyond the Standard Model. The selection criteria defining these regions are summarized in Table~\ref{tab:criteria} and a description of each region follows:

\begin{enumerate}

\item \underline{Diphoton fiducial:} This region unfolds all signal events after the selection presented in Section~\ref{sec:evsel}.

\item \underline{\VBF-enhanced:} This region retains all events with at least two jets and with an invariant dijet mass $m_{jj}$ of at least 400~\GeV, a large rapidity separation $\left| \Delta y_{jj} \right| > 2.8$, and an azimuthal difference between the Higgs boson and the dijet pair of $\dphiggjj > 2.6$. All variables are computed using the two highest-$\pt$ jets in the event with $\pt > 25 \, \GeV$ with matching detector-level cuts.

\item \underline{$N_{\rm lepton} \ge 1$:} This region retains events that contain at least
one electron or one muon with $\pt > 15$~\GeV. For electrons the pseudo-rapidity needs to satisfy $\left| \eta \right| < 2.47$ (excluding $1.37 < |\eta| < 1.52$) and for muons $\left| \eta \right| < 2.7$ is required. Such events are enriched in Higgs bosons produced in association with a vector boson. 

\item \underline{High $E_{\rm T}^{\rm miss}$:} This region retains events with missing transverse momentum $\met > 80$ 
\GeV\ and $\ptgg > 80$~\GeV\ is defined to study \VH\ production and possible contributions of Higgs boson production 
with dark matter particles. The simultaneous requirement that the Higgs boson system balances the missing transverse momentum reduces the fraction of selected events at detector level without particle-level $\met > 80$~\GeV.

\item \underline{\ttH-enhanced:} This region retains events with either at least one lepton and three jets or no leptons and four jets to study Higgs boson production in association with top quarks. In addition, one of the jets needs to be identified as originating from a bottom quark.

\end{enumerate}

\begin{figure*}[!tbp]
\begin{center}
\subfloat[] {\includegraphics[width=.75\textwidth]{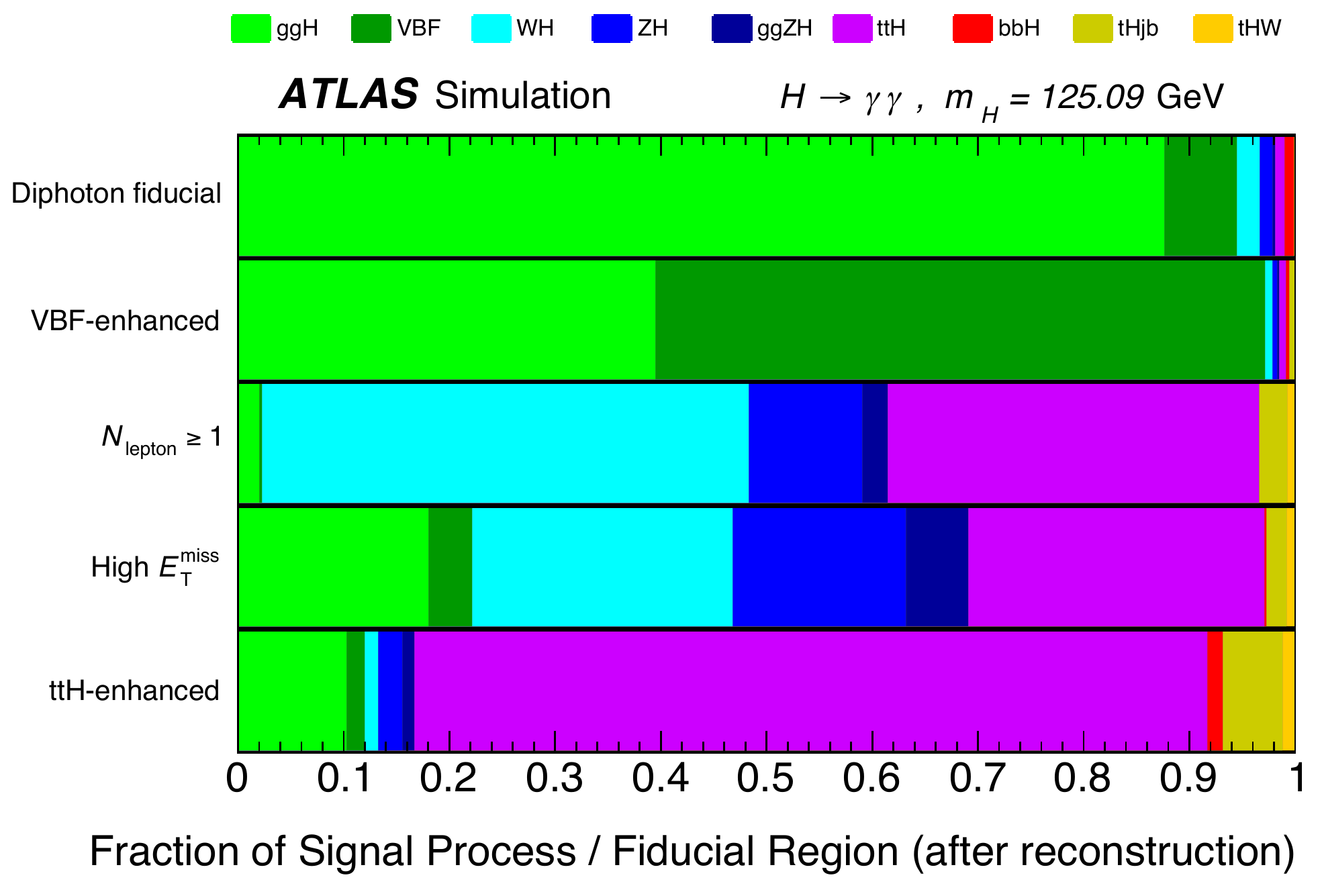}} \\
\subfloat[] {\includegraphics[width=.75\textwidth]{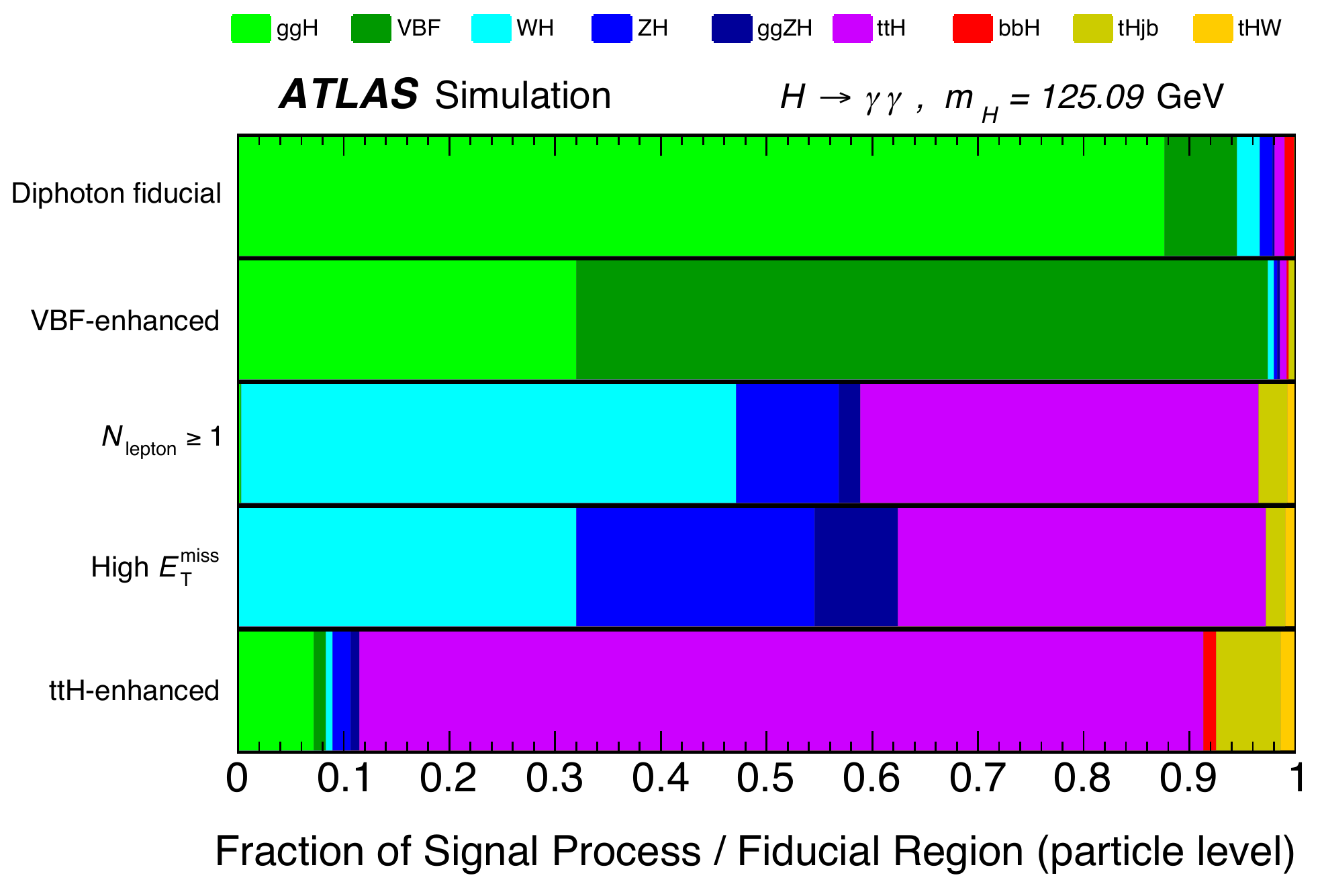}}
\caption{The expected composition of Higgs boson events in each fiducial region (a) after the reconstruction
and (b) at particle-level. Details about the reconstruction can be found in Section~\ref{sec:evsel} and the
definition of the particle-level fiducial volume is given in Section~\ref{sec:fiddef}. }
\label{fig:purity_fid}
\end{center}
\end{figure*}

The expected composition of Higgs boson events in the Standard Model after reconstruction and at particle level is
summarized in Figure~\ref{fig:purity_fid}. At particle level the \VBF-enhanced fiducial region contains about 65\% \VBF\ and 32\% \ggH\
events. The particle-level $N_{\rm lepton} \ge 1$ region is dominated by \WH\ (47\%), \ttH\ (37\%) and \ZH\ (13\%) production. The
particle-level high $E_{\rm T}^{\rm miss}$ region is populated by about equal amounts of \WH, \ZH, and \ttH\ (32\%, 30\%, and 35\%).
Finally, the particle-level \ttH-enhanced region contains about 80\% \ttH\ events. 

\begin{figure*}[!tbp]
  \centering
	{\includegraphics[width=0.8\columnwidth]{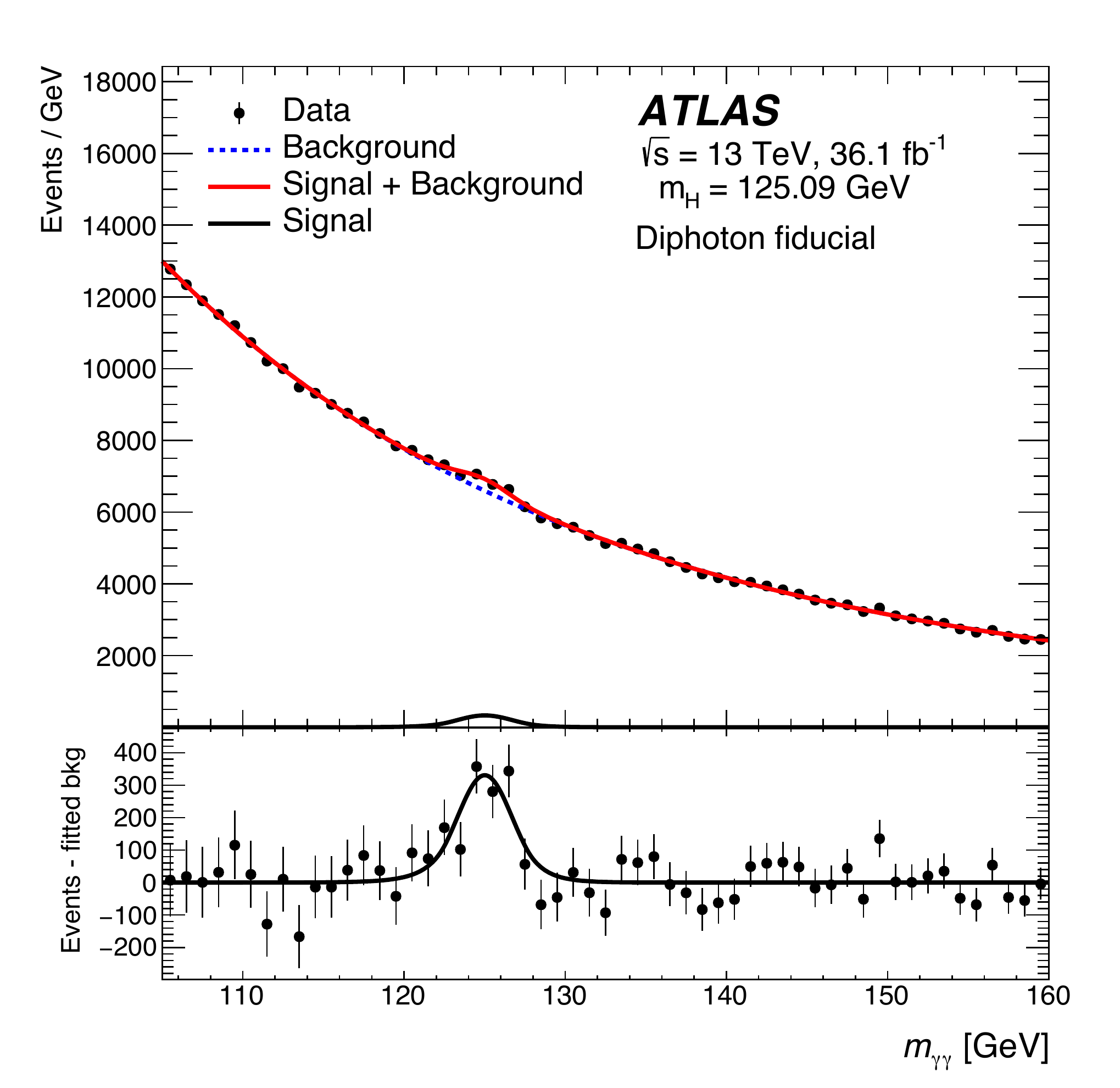} \label{fid_base}}
  \caption{
   Diphoton invariant mass \mgg\ spectrum observed in the 2015 and 2016 data at $\sqrt{s} = $ 13~\TeV\ for events in the diphoton
   fiducial region. The solid red curve shows the fitted signal-plus-background model when the Higgs boson mass is
   constrained to be $125.09\pm 0.24$~\GeV.  
   The background component of the fit is shown with the dotted blue curve. The signal component of the
   fit is shown with the solid black curve. The bottom plot shows the residuals between the data and the background component of the fitted model.
  } 
  \label{fig:fidmassspectra_incl}
\end{figure*}

\begin{figure*}[!tbp]
  \centering
	\subfloat[\VBF-enhanced] {\includegraphics[width=0.50\columnwidth]{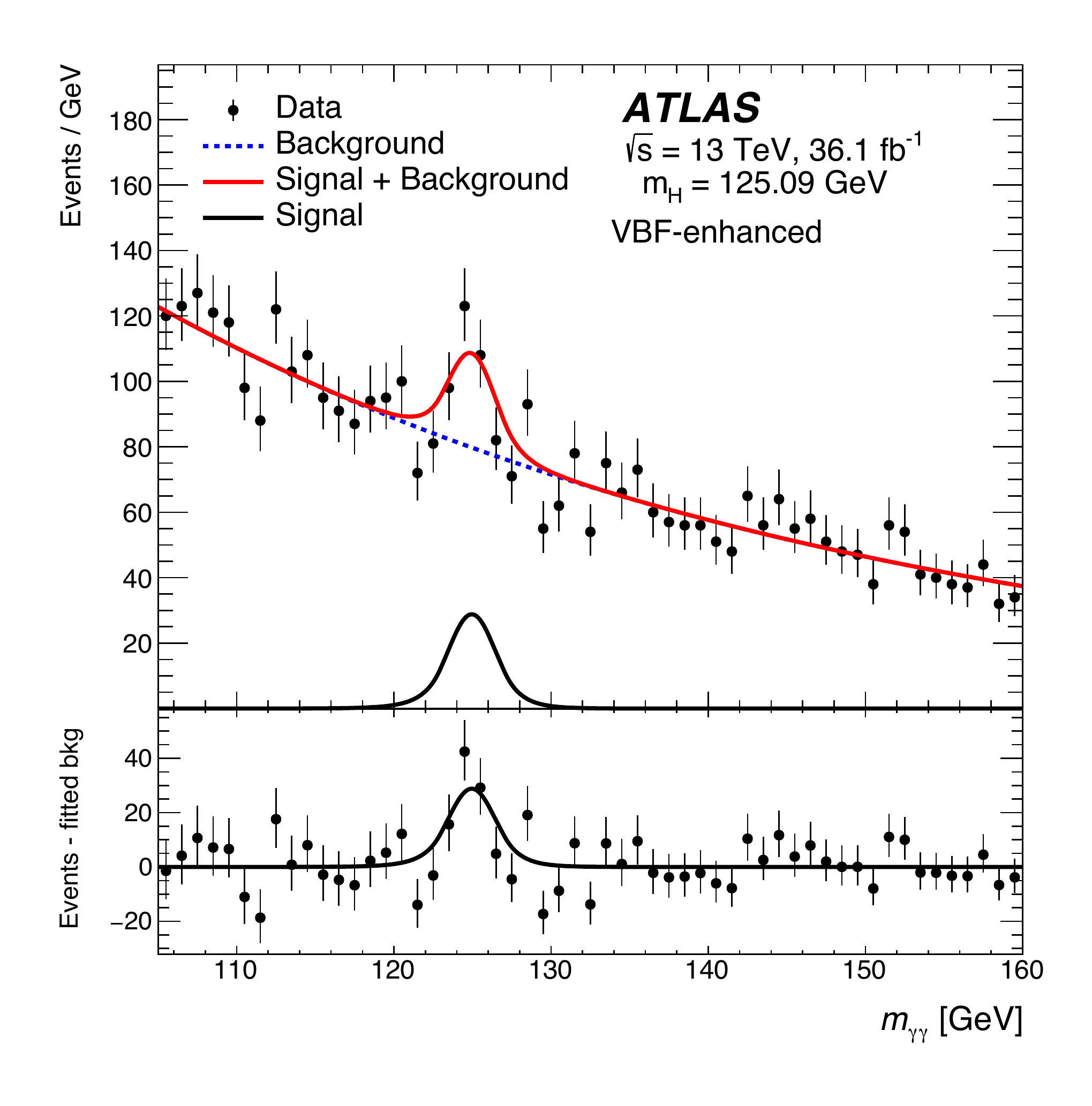} \label{fid_vbf}} 
	\subfloat[$N_{\rm lepton} \ge 1$] {\includegraphics[width=0.50\columnwidth]{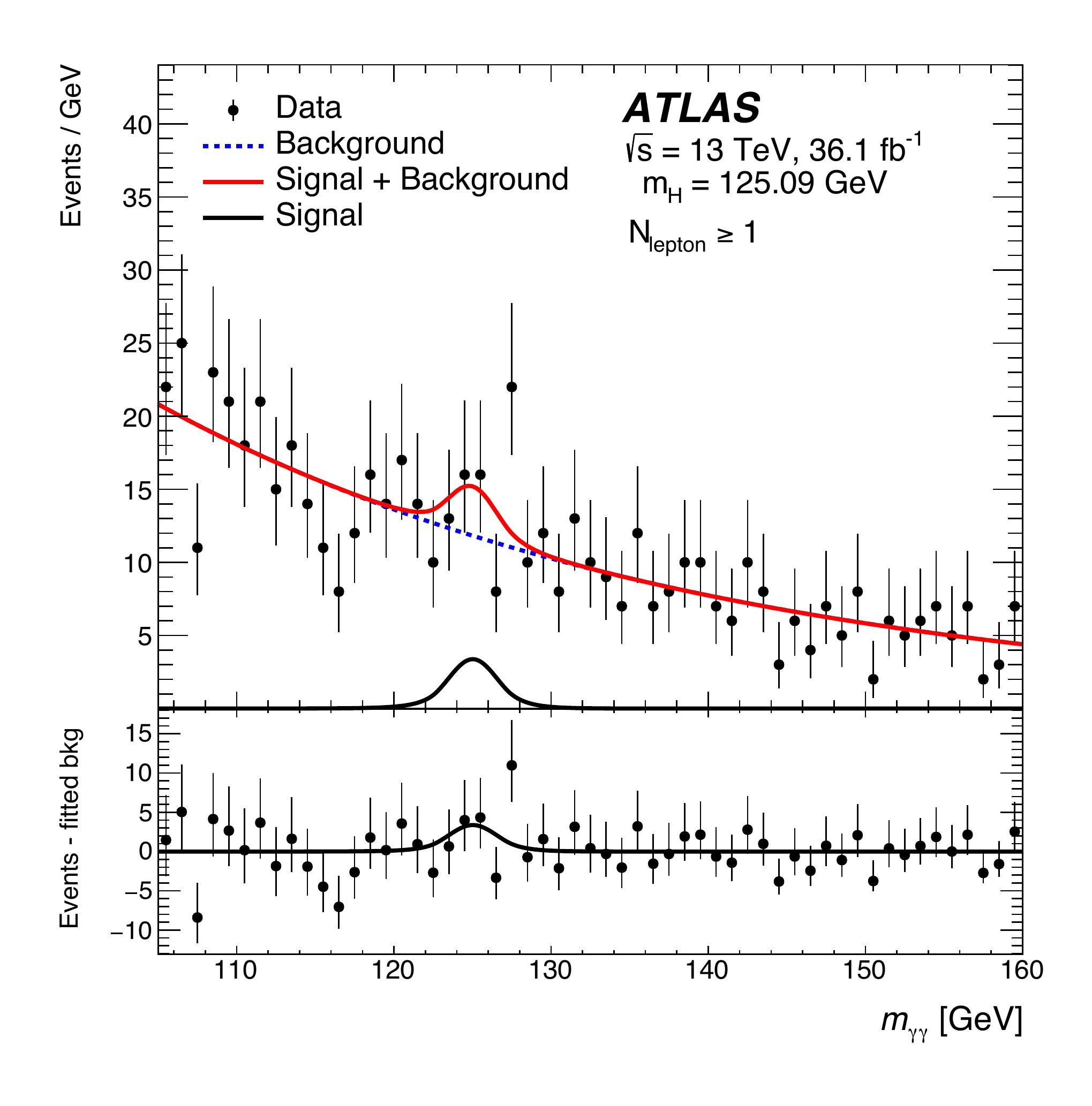}  \label{fid_lep}} \\
	\subfloat[High $E_{\rm T}^{\rm miss}$] {\includegraphics[width=0.50\columnwidth]{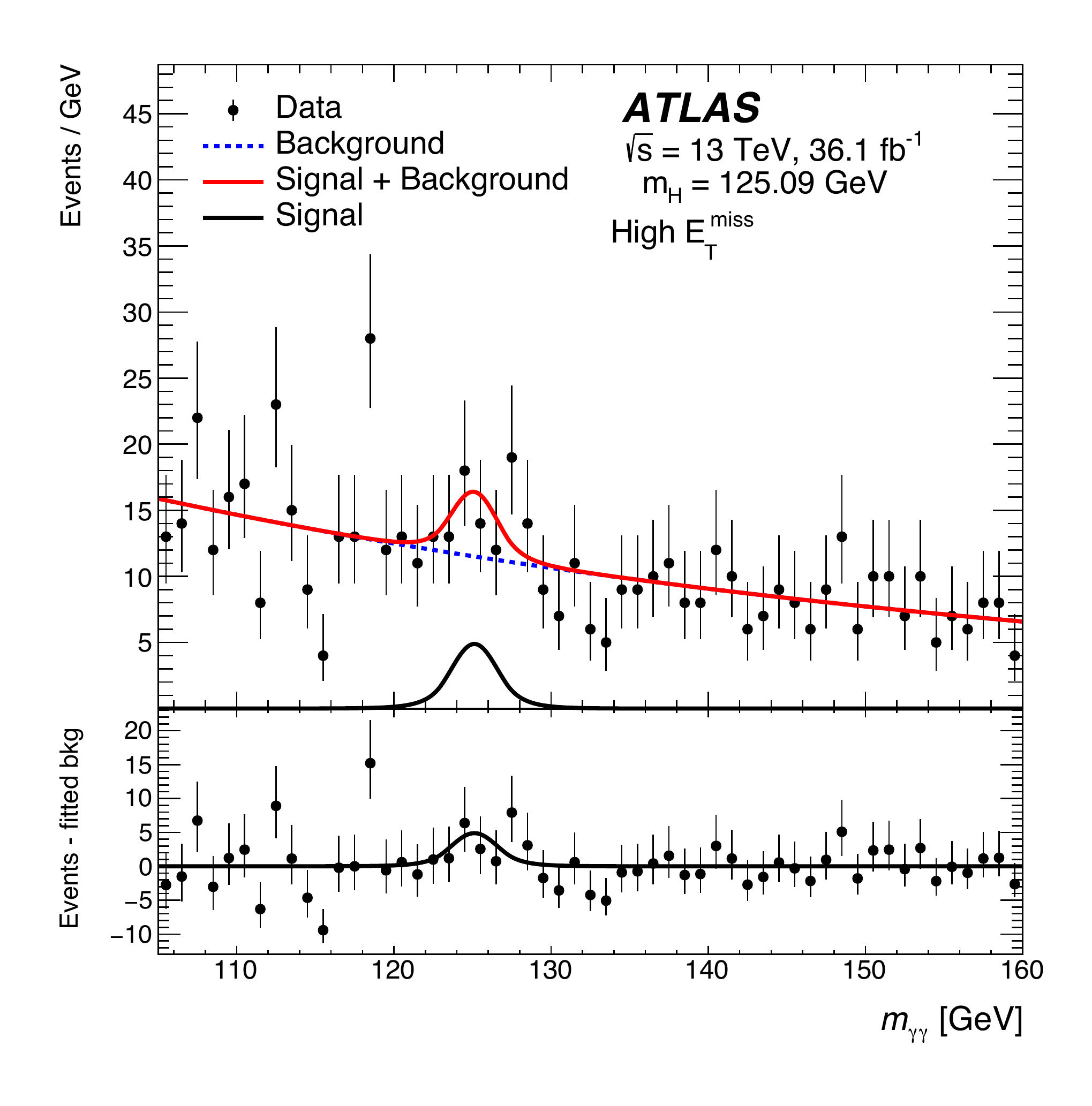} \label{fid_met}}
	\subfloat[$t \bar t H$-enhanced] {\includegraphics[width=0.50\columnwidth]{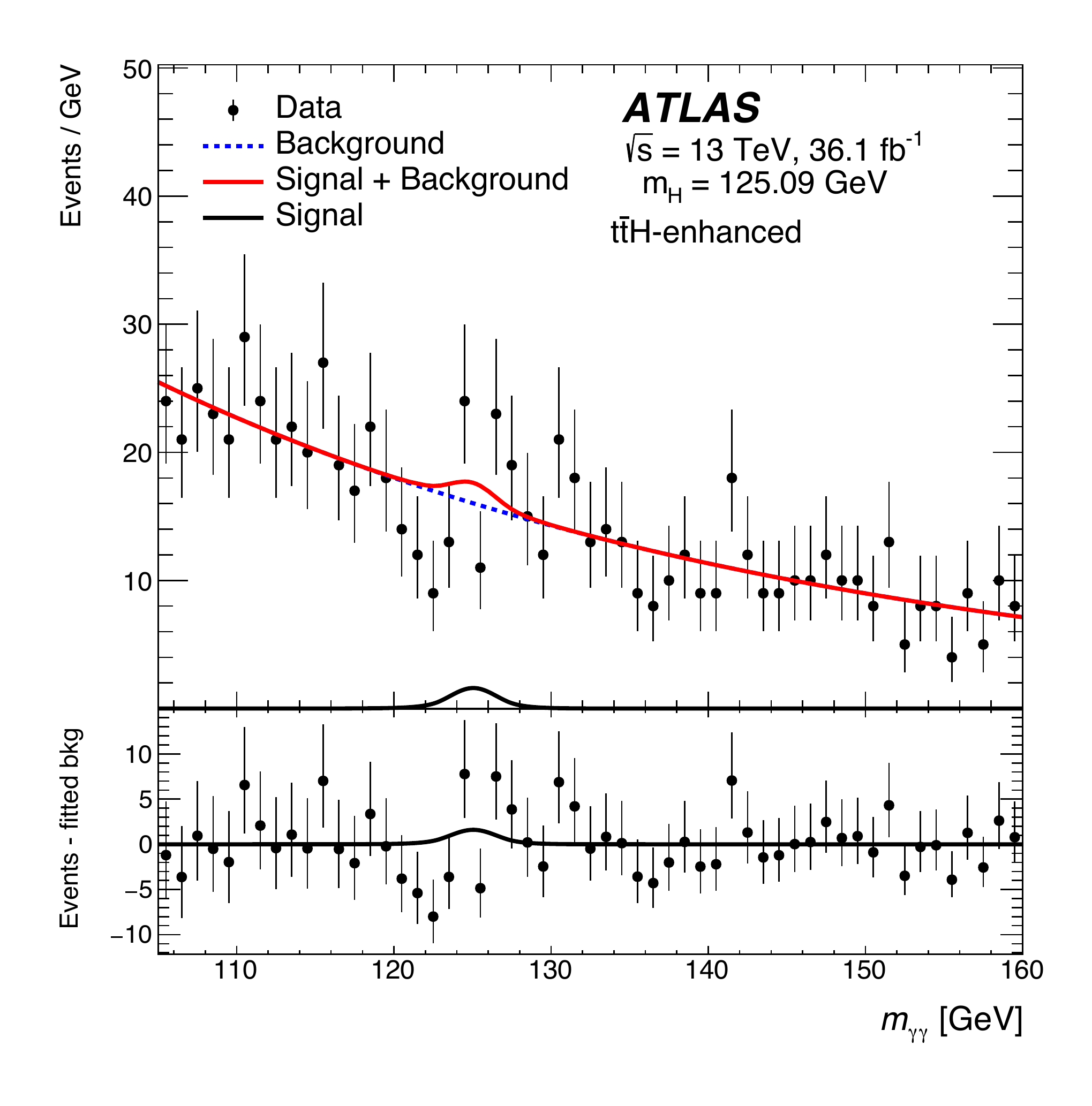} \label{fid_ttH}} 
  \caption{
   Diphoton invariant mass \mgg\ spectra observed in the 2015 and 2016 data at $\sqrt{s} = $ 13~\TeV\ for events in the
   (a) \VBF-enhanced, (b) $N_{\rm lepton} \ge 1$, (c) high $E_{\rm T}^{\rm miss}$, and (d) $t \bar t H$-enhanced fiducial regions.
   The solid red curve shows the fitted signal-plus-background model when the Higgs boson mass is
   constrained to be $125.09\pm 0.24$~\GeV.  
   The background component of the fit is shown with the dotted blue curve. The signal component of the
   fit is shown with the solid black curve. The bottom plot shows the residuals between the data and the background component of the fitted model.   
  } 
  \label{fig:fidmassspectra}
\end{figure*}

The fitted invariant mass spectra for all regions are shown in Figures~\ref{fig:fidmassspectra_incl} 
and~\ref{fig:fidmassspectra}. The results of signal-plus-background fits to these spectra is shown, displaying both the total sum and the
background-only component as well as the residuals between the data and the background.
In the diphoton fiducial region, the Higgs boson signal is clearly visible on the falling non-resonant background. In total, $1491 \pm 248 \, \text{(stat.)} \pm 64  \, \text{(syst.)}$ Higgs boson signal events are extracted. Clear evidence for Higgs boson production is observed in the \VBF-enhanced region with $117 \pm  26  \, \text{(stat.)} \pm 4  \, \text{(syst.)}$ signal events, corresponding to an observed significance of 4.2 standard deviations.

The remaining three regions all show positive signal yields with large, predominantly statistical, uncertainties: $14 \pm 11$, $19 \pm 11$, $6 \pm 15$ for the $N_{\rm lepton} \ge 1$, high $E_{\rm T}^{\rm miss}$, and \ttH-enhanced fiducial regions, respectively, and the error corresponds to the sum of the statistical and systematic uncertainties.

The cross section for $pp \to H \to \gamma\gamma$ measured in the diphoton fiducial region is 
\begin{align*}
 \sigma_{\rm fid} =  \fidxs \, ,
\end{align*}
which is to be compared with the Standard Model prediction of \fidxsSM. 
The gluon--gluon fusion contribution to the Standard Model prediction and its uncertainty are taken to be the N${}^{3}$LO QCD and NLO EW prediction of
Refs.~\cite{Anastasiou:2015ema, Anastasiou:2016cez, Actis:2008ug, Anastasiou:2008tj, Butterworth:2015oua,deFlorian:2016spz} 
corrected for the $H \to \gamma\gamma$ branching ratio and the fiducial acceptance. The fiducial acceptance is defined using the \nnlops\ prediction for gluon--gluon fusion~\cite{Hamilton:2013fea}. 
The contributions to the Standard Model prediction from the \VBF, \VH, \bbH\ and \ttH\ production mechanisms are determined using the
particle-level predictions normalized with theoretical calculations as discussed in Section~\ref{sec:mc},
and are collectively referred to as $XH$. The measured cross section is compatible with the Standard Model prediction and the observed \ggH\ coupling strength measured in Section~\ref{sec:methods_coup}, as the diphoton fiducial region is dominated by gluon--gluon fusion production.

The cross section of the \VBF-enhanced region is measured to be
\begin{align*}
 \sigma_{\rm VBF-enhanced} = \fidxsVBF\, ,
\end{align*}
which is to be compared with the Standard Model prediction of \fidxsSMVBF. The gluon--gluon fusion part of the SM prediction is constructed from the \nnlops\ prediction for gluon--gluon fusion normalized with the
N${}^{3}$LO in QCD and NLO EW prediction of Refs.~\cite{Anastasiou:2015ema,Anastasiou:2016cez,Actis:2008ug, Anastasiou:2008tj,
Butterworth:2015oua, deFlorian:2016spz}. This prediction is labeled as ``default MC'' in the following and includes all theory uncertainties related to gluon--gluon fusion as discussed in Section~\ref{sec:theo}. 

For the $N_{\rm lepton} \ge 1$, high $E_{\rm T}^{\rm miss}$, and \ttH-enhanced fiducial regions, limits on the cross sections are reported at the 95\% CL.\footnote{The quoted CL values were obtained using the unfolded cross sections and their corresponding uncertainties assuming Gaussian errors.} 

\begin{figure}[!tbp] 
\begin{center}
 \includegraphics[width=.8\textwidth]{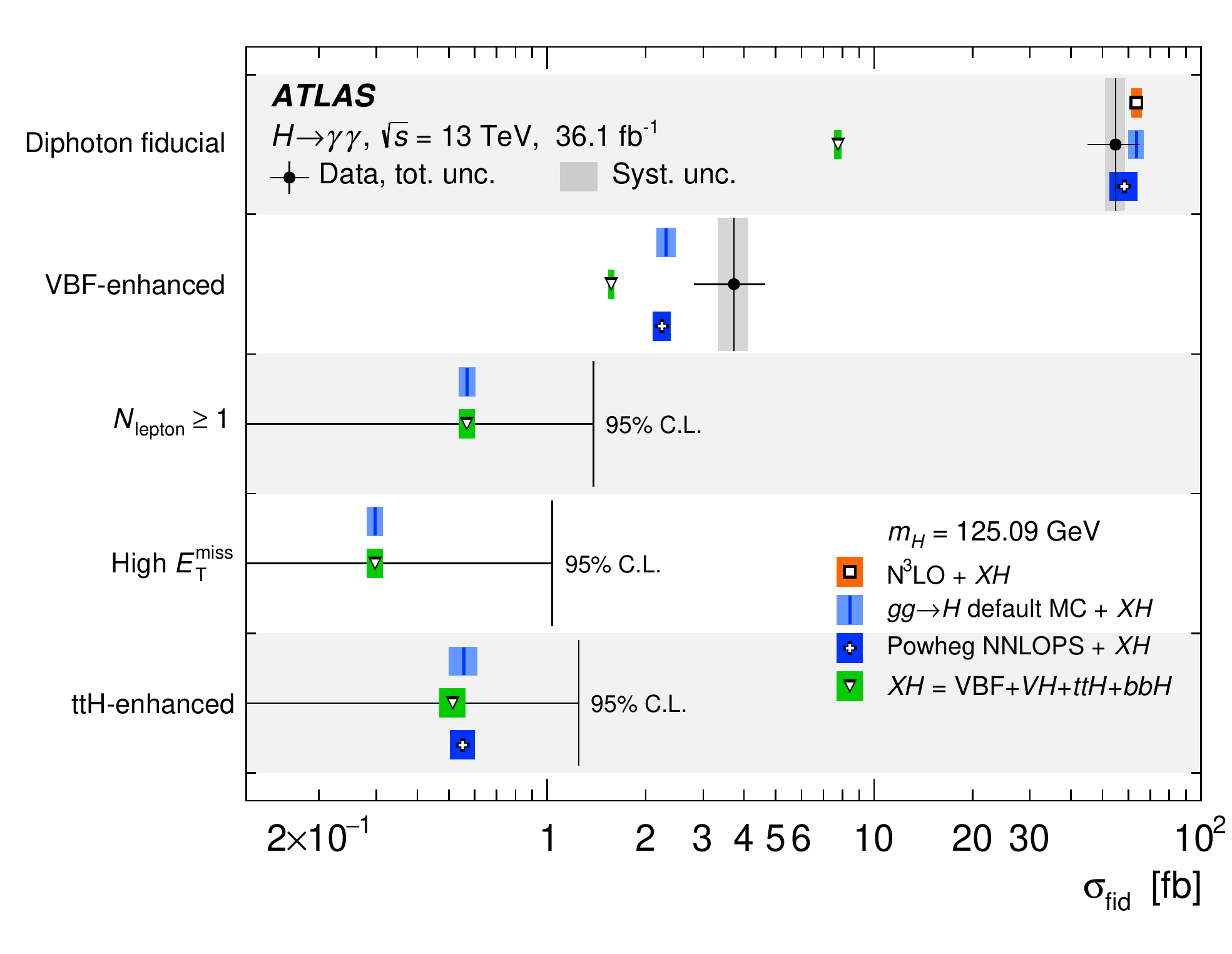}
\caption{The measured cross sections or cross-section upper limits of the diphoton, \VBF-enhanced, 
$N_{\rm lepton} \ge 1$, high $E_{\rm T}^{\rm miss}$, and $t \bar t H$-enhanced fiducial regions are shown. The intervals on the vertical axis each represent one of these fiducial regions. The data are shown as filled (black) circles. The error bar on each measured cross section represents the total uncertainty in the measurement, with the systematic uncertainty shown as a dark gray rectangle. Each cross section limit is shown at the 95\% confidence level. The measured cross sections are compared to a range of predictions and a detailed description of each prediction can be found in the text. All comparisons include the SM predictions arising from \VBF, \VH, \ttH, and \bbH, which are collectively labeled as $XH$.
}
\label{fig:res_fid_regions}
\end{center}
\end{figure}

Figure~\ref{fig:res_fid_regions} and Table~\ref{tab:xs} summarize measured cross sections of the fiducial regions and limits, and compare both to the Standard Model expectations, constructed as outlined
above. The \nnlops\ prediction, without any additional corrections, is also shown. The uncertainty band is estimated using  a set of scale variations and includes PDF uncertainties from eigenvector variations. The Standard Model predictions of all fiducial regions are in agreement with the corresponding measured cross sections.

\begin{table}[!tp]
  \caption{
  The measured cross sections in the diphoton, \VBF-enhanced, $N_{\rm lepton} \ge 1$, high $E_{\rm T}^{\rm miss}$,
  and \ttH-enhanced fiducial regions. The gluon--gluon fusion contribution to the Standard Model prediction of the diphoton
  fiducial region is taken to be the N${}^{3}$LO prediction of Refs.~\cite{Anastasiou:2015ema, Anastasiou:2016cez,Actis:2008ug, Anastasiou:2008tj, Butterworth:2015oua, deFlorian:2016spz} 
  corrected for the $H \to \gamma\gamma$ branching ratio and the fiducial acceptance. The gluon--gluon fusion contribution to the Standard Model for all the other
  regions is taken from the \nnlops\ prediction normalized with the N${}^{3}$LO prediction and includes all theory uncertainties related to gluon--gluon fusion as discussed in Section~\ref{sec:theo}. The contributions to the Standard Model prediction from
  \VBF, \VH\, \ttH\, and \bbH\ production mechanisms are determined using the particle-level predictions described in
  Section~\ref{sec:mc} normalized with theoretical calculations.
 }
\label{tab:xs}
   \centering
     \begin{tabular}{lccc}
       \hline\hline
         Fiducial region & Measured cross section & \multicolumn{2}{c}{SM prediction } \\
	   \hline\hline
	     Diphoton fiducial & \fidxs  & \fidxsSM & [N$^{3}$LO + $XH$] \\
	     VBF-enhanced  & \fidxsVBF  & \fidxsSMVBF & [default MC + $XH$]  \\
	     $N_{\rm lepton} \ge 1$ & $\leq$  1.39 fb \@ 95\% CL  &  $0.57 \pm 0.03 $ fb & [default MC + $XH$] \\
	     High $E_{\rm T}^{\rm miss}$ & $\leq$ 1.00 fb \@ 95\% CL &  $0.30 \pm 0.02$ fb & [default MC + $XH$] \\
	     \ttH-enhanced & $\leq$ 1.27 fb \@ 95\% CL  & $0.55 \pm 0.06 $ fb & [default MC + $XH$] \\
	     \hline\hline
         \end{tabular}
\end{table}

\clearpage

\subsection{Measurements of cross sections of inclusive and exclusive jet multiplicities}\label{sec:details_fidregjets}

The production of Higgs bosons in association with jets is sensitive to
the theoretical modeling in QCD and to the contribution of different Higgs boson production mechanisms. In the SM, events with zero or one jet are dominated by gluon--gluon fusion production. In events with two jets the contributions from
\VBF\ and \VH\ production modes become more important. Higgs boson production in association with top quarks (\ttH) can be probed in events with
the highest jet multiplicities. In BSM scenarios, the jet multiplicity distribution is sensitive to new heavy particles coupling to the
Higgs boson and vector bosons. For the measurements presented here, jet multiplicity bins with zero, one, two, and at least three jets with \pt\ larger than 30~\GeV\ and absolute rapidity $\left| y \right| < 4.4$ are defined. In addition, jet multiplicity bins with a \pt\ larger than 50~\GeV\ are defined for zero, one, or at least two jets. The measured cross sections are compared to a range of predictions of gluon--gluon fusion production:

\begin{itemize}

\item The parton-level N${}^{3}$LO QCD and NLO EW prediction of Refs.~\cite{Anastasiou:2015ema,Anastasiou:2016cez,Actis:2008ug,
Anastasiou:2008tj,Butterworth:2015oua, deFlorian:2016spz}. This prediction is shown for the inclusive zero-jet cross section. 

\item The parton-level JVE$+$N$^{3}$LO prediction of Ref.~\cite{Banfi:2015pju}, which includes NNLL resummation in QCD of the \pt\ of the leading jet which is matched to the N$^{3}$LO total cross section. This prediction is shown for the inclusive one-jet cross section. 

\item The parton-level \stwzblptw\ predictions of Refs.~\cite{Stewart:2013faa,Boughezal:2013oha}, which include NNLL$'$+NNLO resummation for the
\pt\ of the leading jet in QCD, combined with a NLL$'$+NLO resummation in QCD for the subleading jet.\footnote{The prime indicates that the leading contributions from N$^3$LL (resp. NNLL) are included along with the full NNLL (resp. NLL) corrections.} The numerical predictions for
$\sqrt{s}=13\,\TeV$ are taken from Ref.~\cite{deFlorian:2016spz}. This prediction is shown for the inclusive zero-, one- and two-jet cross sections as well as for the exclusive zero- and one-jet cross sections. 

\item The parton-level \nnlojet\ prediction of Refs.~\cite{Chen:2014gva,Chen:2016zka} is a fixed-order NNLO prediction in QCD for inclusive $H+$one-jet production.  This prediction is shown for the inclusive one-, two-jet, and three-jet cross sections as well as for the exclusive one- and two-jet cross sections.

\item The parton-level \gosam\ prediction of Refs.~\cite{Cullen:2011ac,Cullen:2014yla}, which provides the fixed-order loop contributions accurate at 
NLO in QCD in the inclusive $H$ + zero-jet, $H$ + one-jet, $H$ + two-jet, and $H$ + three-jet regions. The real-emission contributions at fixed order in QCD are provided by \sherpa~\cite{Gleisberg:2008ta}. This prediction is shown for the inclusive one-, two-jet, and three-jet cross sections as well as for the exclusive one- and two-jet cross sections.

\item The default MC prediction (\nnlops\ normalized with the N${}^{3}$LO in QCD and NLO EW cross section) introduced in Section~\ref{sec:details_fidreg}.
This prediction is shown for all measured inclusive and exclusive jet cross sections. 

\item The \nnlops\ prediction which is already described in Section~\ref{sec:mc}. This prediction is shown for all measured inclusive and exclusive jet  cross sections. 

\item The \sherpa\ (\meps) prediction of Refs.~\cite{Gleisberg:2008ta,Greiner:2015jha,Hoeche:2011fd,Hoeche:2012yf,Buschmann:2014sia,Bothmann:2016nao,
Gleisberg:2008fv,Krauss:2001iv,Cullen:2014yla,Schumann:2007mg,Hoche:2012wh,Hoeche:2014lxa} is accurate to NLO in QCD in the inclusive $H$ + zero-jet, $H$ + one-jet, $H$ + two-jet, and $H$ + three-jet regions and includes top-quark mass effects. The one-loop corrections are incorporated from GoSam~\cite{Cullen:2011ac,Cullen:2014yla} and the different jet multiplicity regions are merged using the \meps\ multijet merging technique. This prediction is shown for all measured inclusive and exclusive jet cross sections. 

\item The \amc\ prediction of Refs.~\cite{Alwall:2014hca, Frederix:2016cnl}, which includes up to two jets at NLO accuracy using the
\fxfx\ merging scheme~\cite{Frederix:2012ps}. The central merging scale is taken to be 30~\GeV. The generated events are passed to \pythia~\cite{Sjostrand:2007gs} to provide parton showering and hadronization to create the full final state (without underlying event). This prediction is shown for all measured inclusive and exclusive jet cross sections. 

\end{itemize}

All predictions but \nnlojet\ and \sherpa\ (\meps) use the NNLO PDF set following the \pdflhc\ recommendations. The
\nnlojet\ prediction uses the \ctften\ NNLO PDF set~\cite{Dulat:2015mca} and \sherpa\ (\meps) uses the
\nnpdfthree PDF set~\cite{Ball:2014uwa}. \gosam, \sherpa\ (\meps), and \nnlojet\ apply the kinematic selection on the final-state photons. 
For all other predictions, the fiducial acceptance  is determined
using \nnlops. The cross sections of all parton-level predictions are multiplied with isolation correction factors to account for the efficiency of the fiducial photon isolation criterion. The additional uncertainties in the isolation correction are determined by studying multiple event generators and/or event generator tunes, and are included in the uncertainty bands of the parton-level predictions. No correction factors nor additional uncertainties to account for the impact of hadronization and the underlying event activity are applied, so the theory uncertainties in the parton-level predictions may be incomplete, but example values for such corrections and their uncertainties can be found in Table~\ref{tab:NP} in Appendix~\ref{app:np_iso_factors}. All other acceptance and correction factors along with their associated uncertainties can also be found in Appendix~\ref{app:np_iso_factors}.

No $K$-factors are applied to the predictions and the contributions from $XH$ are also included in the comparison using the corresponding generators and cross sections described in Section~\ref{sec:mc}.

\begin{figure*}[!tbp]
  \centering
	\subfloat[] {\includegraphics[width=0.70\columnwidth]{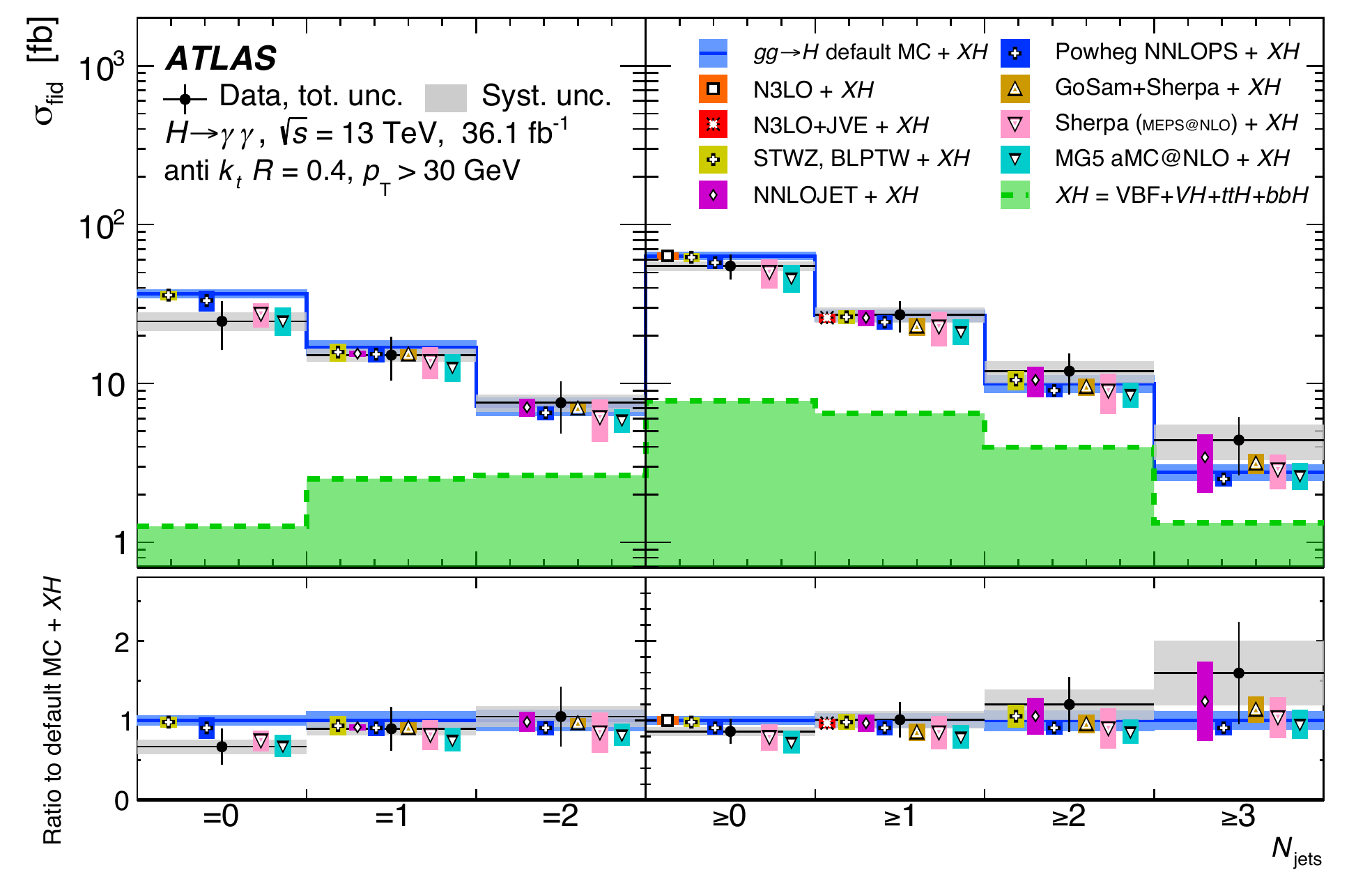}} \\
	\subfloat[] {\includegraphics[width=0.70\columnwidth]{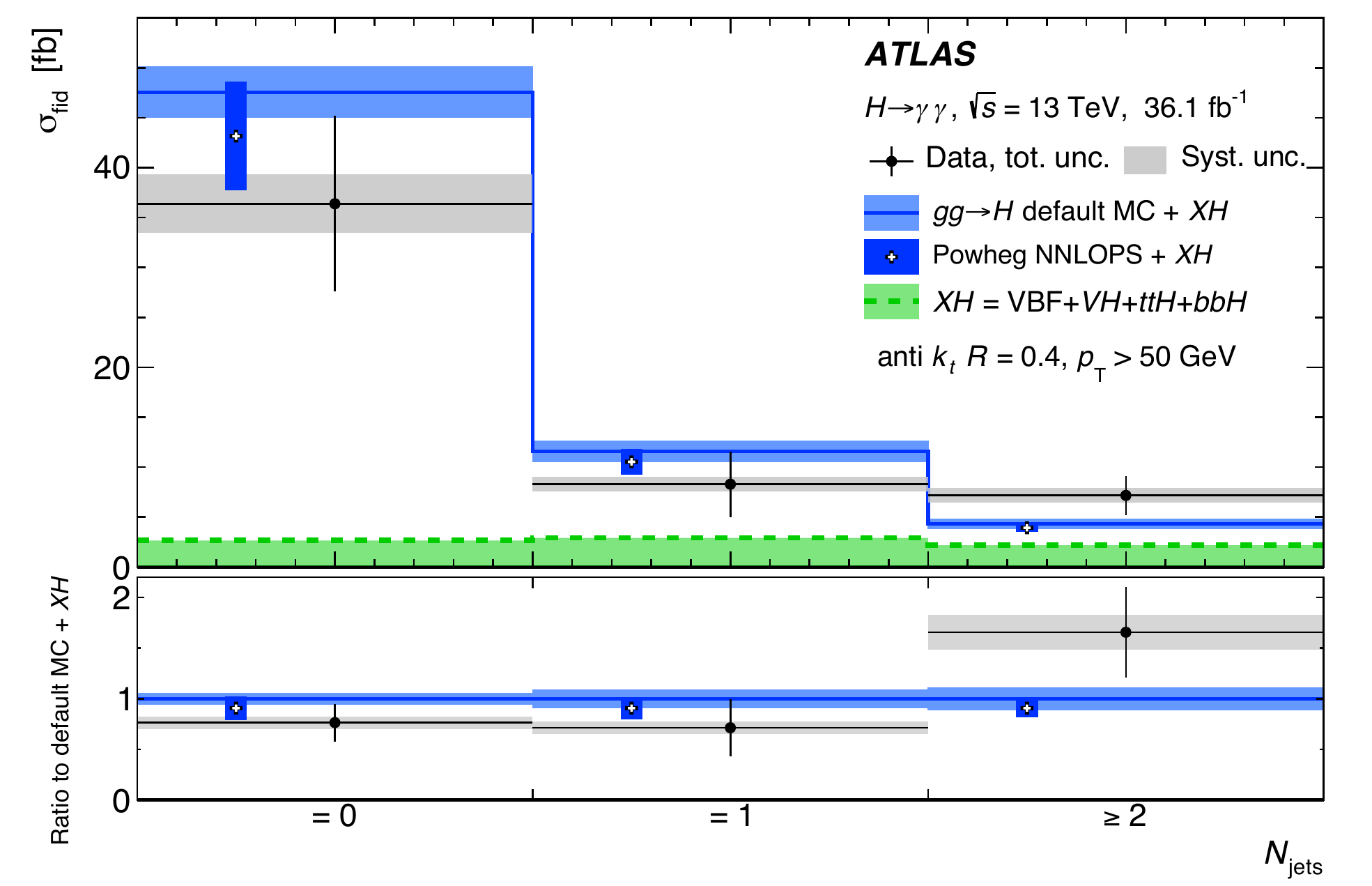}}
  \caption{
  Cross sections for $pp \to H \to \gamma \gamma$ as a function of inclusive and exclusive jet multiplicities for jets with (a) $\pt > 30~\GeV$ and (b) $\pT > 50~\GeV$. The data are shown as filled (black) circles. The vertical error bar on
   each data point represents the total uncertainty in the measured cross section and the shaded (gray) band is the
   systematic component. The measured differential cross sections are compared to a range of predictions and details can be found in the text. The width of the bands of each prediction reflects the total theoretical uncertainty. The small contribution from \VBF, \VH, \ttH, and \bbH\ is also shown as a (green) histogram and denoted by $XH$.
  }
  \label{fig:njets_incl_excl}
\end{figure*}

Figure~\ref{fig:njets_incl_excl}(a) shows exclusive and inclusive zero-, one- and two-jet cross sections
and the inclusive three-jet cross section for jets defined with $\pt > 30~\GeV$. 
Figure~\ref{fig:njets_incl_excl}(b) shows the exclusive zero- and one- and the inclusive two-jet cross section with $\pt > 50~\GeV$. The measured cross sections are in agreement with the Standard Model predictions, although there is a 1.5 $\sigma$ deficit
in the \njet = 0 cross section for jets defined with $\pt > 30~\GeV$. As shown in Figure~\ref{fig:stat_corr}, there is a sizeable positive correlation between zero-jet and low-\ptgg\ events, and a similar deficit is observed there (cf. Section~\ref{sec:diff_prodkin}).

\subsection{Measurements of differential and double-differential cross sections }\label{sec:details_diff}

Eleven fiducial differential cross sections are measured that characterize the Higgs boson 
production kinematics, the kinematics of jets produced in association with the Higgs boson, the spin and CP quantum numbers of the Higgs boson and variables sensitive to the VBF production mechanism. In addition, two double-differential cross sections are reported. The measurement of seven additional variables  can be found in Appendix~\ref{app:add_fid_meas}.

\subsubsection{Measurements of cross sections probing the Higgs boson production kinematics}\label{sec:diff_prodkin}

\begin{figure*}[!tbp]
  \centering
	\subfloat[] {\includegraphics[width=0.50\columnwidth]{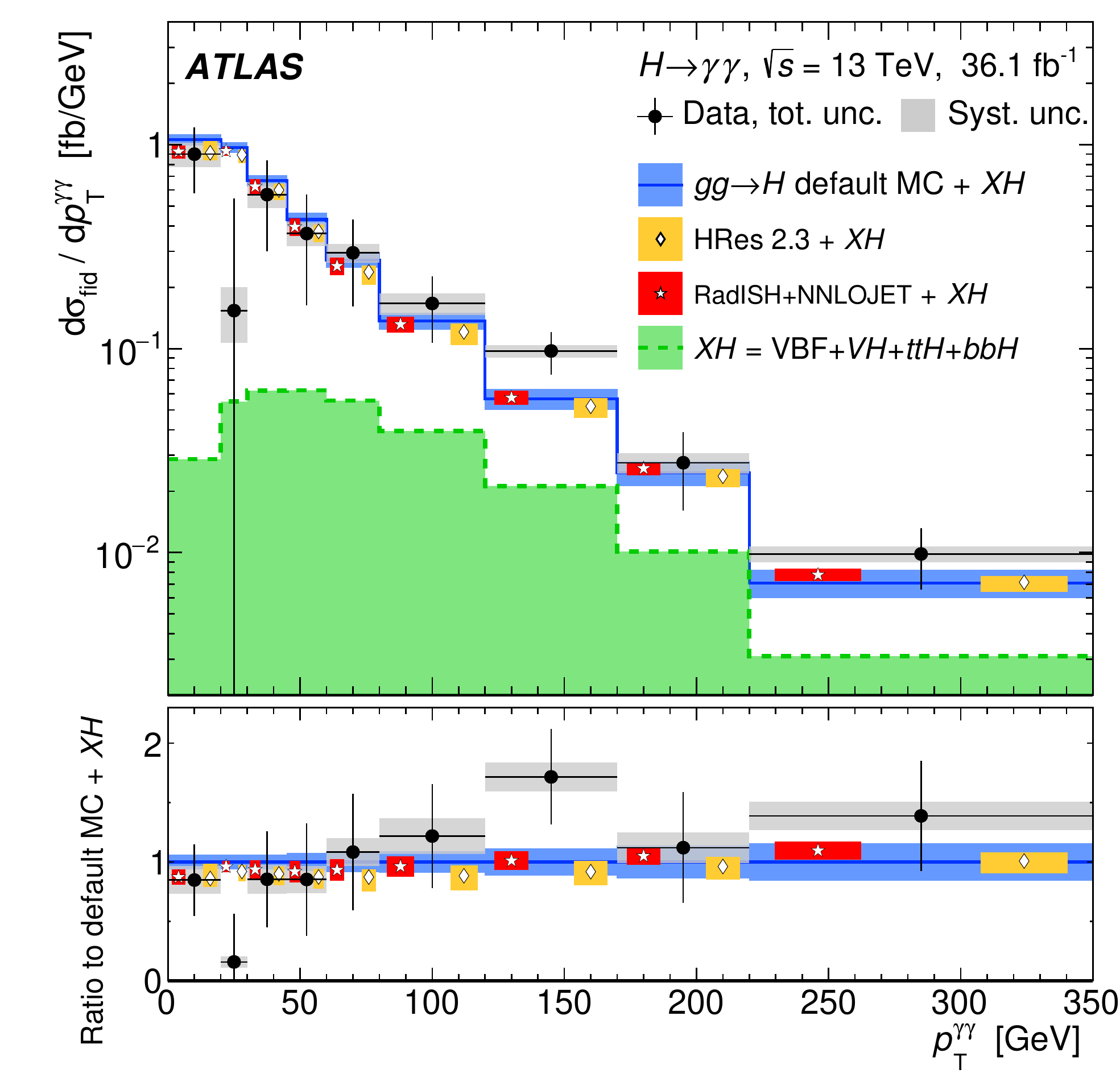}}
	\subfloat[] {\includegraphics[width=0.50\columnwidth]{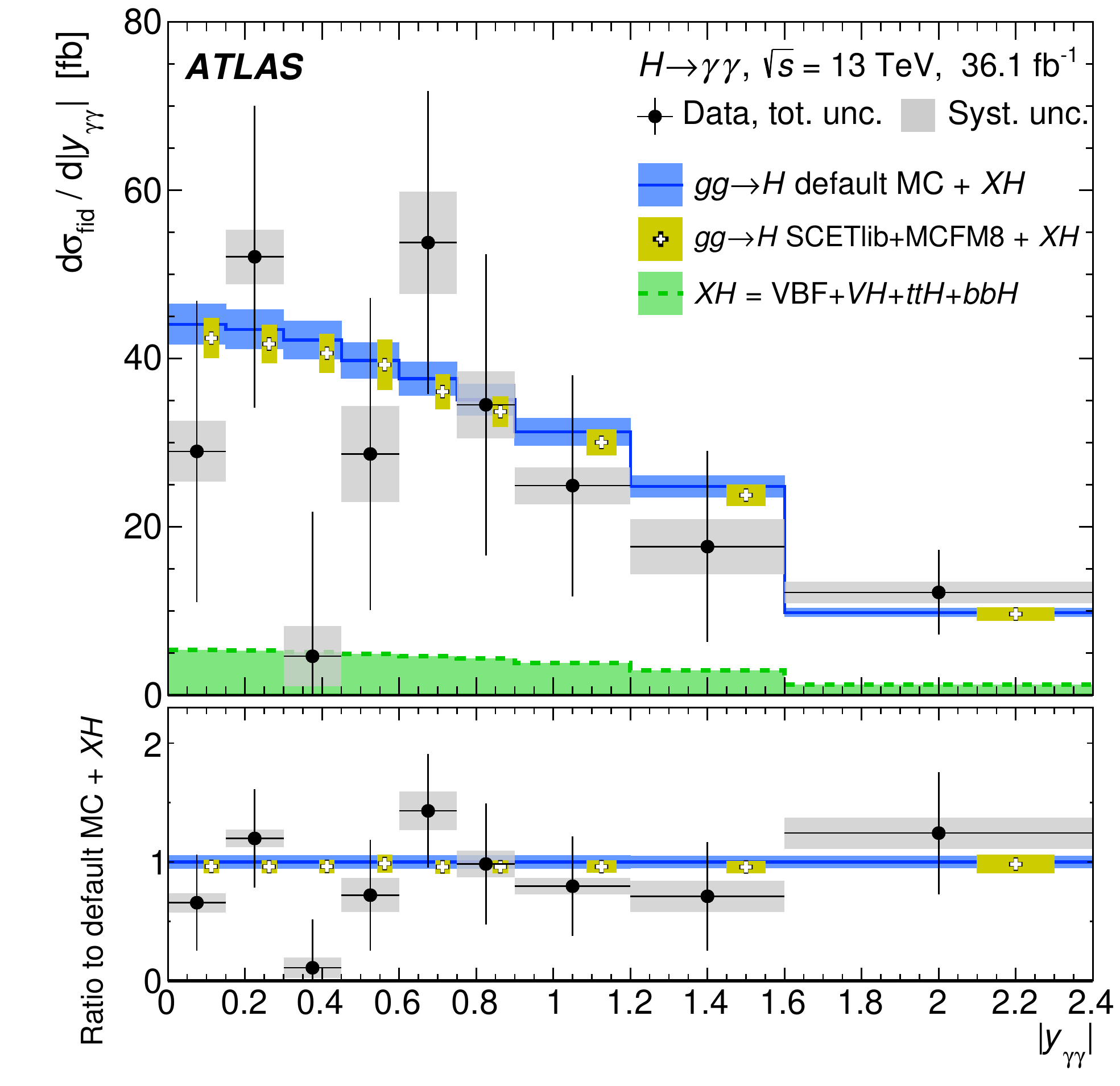}} \\
  \caption{
   The differential cross sections for $pp \to H \to \gamma \gamma$ as a function of (a) \ptgg\ and (b) \ygg\
   are shown and compared to the SM expectations.
  } 
  \label{fig:diff_pth_rap}
\end{figure*}

Measuring the transverse momentum of the diphoton system, \ptgg\ , probes the perturbative QCD modeling of the \ggH\ production mechanism which is mildly sensitive to the bottom- and charm-quark Yukawa couplings~\cite{Bishara:2016jga}. The distribution at high transverse momentum is sensitive to new heavy particles coupling to the Higgs boson and to the top-quark Yukawa coupling. The rapidity distribution of the diphoton system, \ygg, is also sensitive to the modeling of the \ggH\ production mechanism.
The differential cross sections for $pp \to H \to \gamma \gamma$ as a function of \ptgg\ and \ygg\ are shown in Figure~\ref{fig:diff_pth_rap}. The chosen bin widths are a compromise between retaining a sufficiently significant signal and providing spectra with good granularity. Each bin is chosen such that it retains an expected significance of at least two standard deviations, estimated using the \nnlops\ and additional predictions described in Section~\ref{sec:mc} as well as using a fit to $m_{\gamma\gamma}$ sidebands. The measurements are compared to several predictions of gluon--gluon fusion: 

\begin{itemize}
\item The default MC prediction (\nnlops\ normalized with the N${}^{3}$LO in QCD and NLO EW cross section) introduced in Section~\ref{sec:details_fidreg}.

\item \hres~\cite{deFlorian:2012mx, Grazzini:2013mca}, which provides predictions at NNLO with $p_\mathrm{T}^H$ resummation at NNLL and differentially in \ptgg\. Finite top-, bottom-, and charm-quark masses are included at NLO accuracy. The renormalization and factorization scales are chosen to be $\frac 1 2 \sqrt{m_H^2 + (p_\mathrm{T}^H)^2}$, and the two resummation scales are chosen to be $m_H/2$ and $2 m_b$.

\item \radish~\cite{Monni:2016ktx}, which provides predictions using a $p_\mathrm{T}^H$ resummation to NNLL and matching
to the one-jet NNLO differential spectrum from \nnlojet ~\cite{Chen:2014gva,Chen:2016zka}. The shown \ptgg\ \radish\ prediction does
include corrections from the finite top and bottom quark masses.

\item \scetlib\, which provides predictions at NNLO+NNLL$'$${}_{\varphi}$ accuracy derived by applying a resummation of the virtual corrections to the gluon form factor~\cite{Ebert:2017uel, SCETlib} and differentially in \ygg\ and \costhetastar.\footnote{The subscript $\varphi$ refers to the fact that the applied resummation is to the gluon form factor.} The underlying
NNLO predictions are obtained using \mcfme\ with zero-jettiness subtractions~\cite{Boughezal:2016wmq, Gaunt:2015pea}.

\end{itemize}

No additional $K$-factors are applied to the predictions, which all use the NNLO PDF set following the \pdflhc\
recommendations, and the fiducial acceptance for \radish\ is determined using \nnlops.
The \scetlib\ and \hres\ predictions include the kinematic acceptance and are corrected and apply correction factors accounting for the photon isolation efficiency as described in Section~\ref{sec:details_fidregjets}. As also mentioned in Section~\ref{sec:details_fidregjets}, no correction factors to  account for the impact of hadronization and the underlying-event activity are used.
The SM prediction shows a slight excess at low transverse momentum and low rapidity, and shows a slight deficit at large transverse momentum. The slightly harder Higgs boson transverse momentum shown in Figure~\ref{fig:diff_pth_rap} is consistent with the ATLAS Run 1 measurements in both the $H \to \gamma \gamma$ and $H \to ZZ^* \to 4 \ell$
decay channels~\cite{Aad:2014lwa,Aad:2014tca} and the measured zero-jet cross section. The Standard Model prediction is in agreement with the measured distributions.

\subsubsection{Measurements of cross sections probing the jet kinematics}\label{sec:diff_jet_kin}

\begin{figure*}[!tbp]
  \centering
	\subfloat[] {\includegraphics[width=0.50\columnwidth]{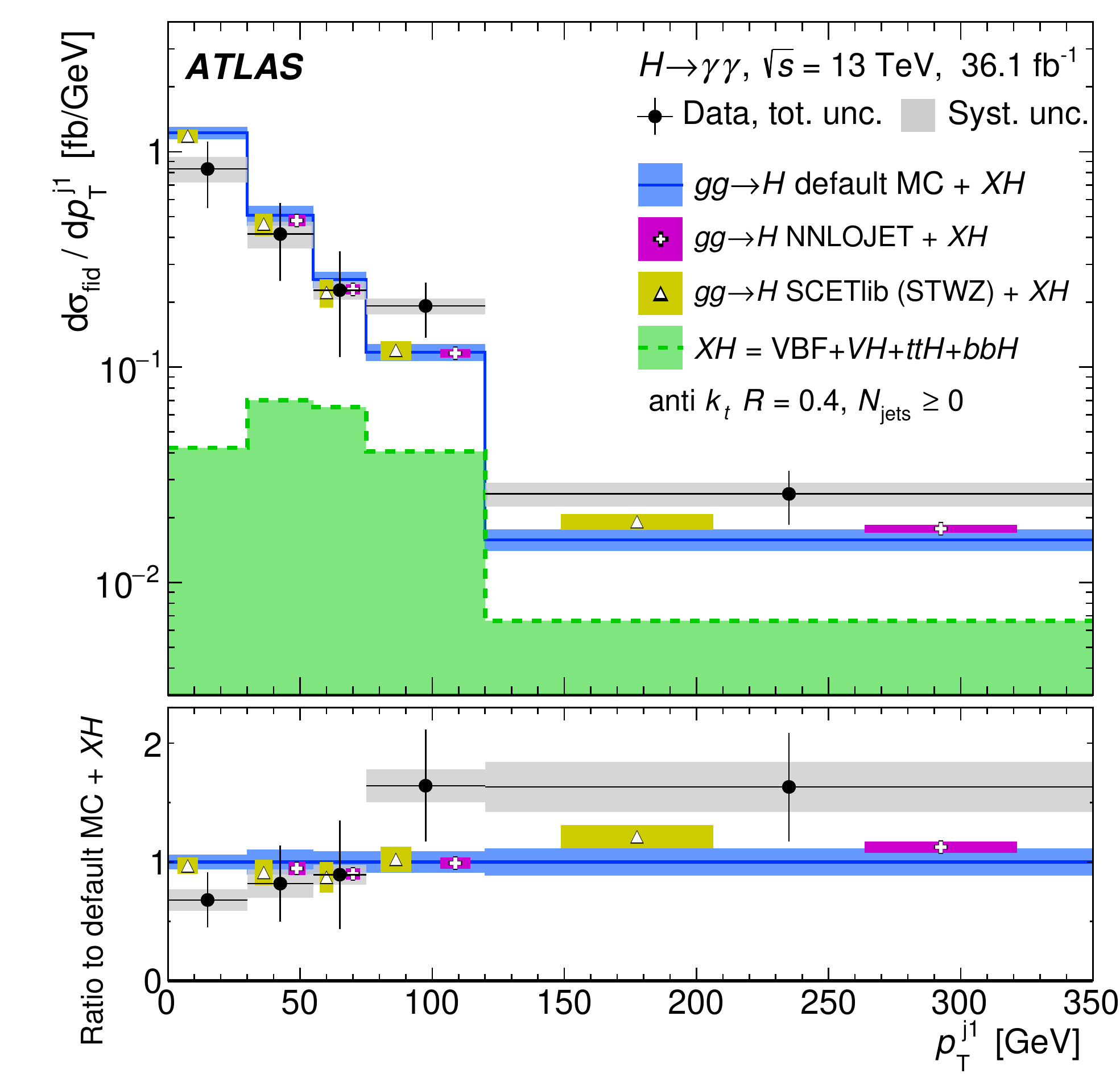}}
	\subfloat[] {\includegraphics[width=0.50\columnwidth]{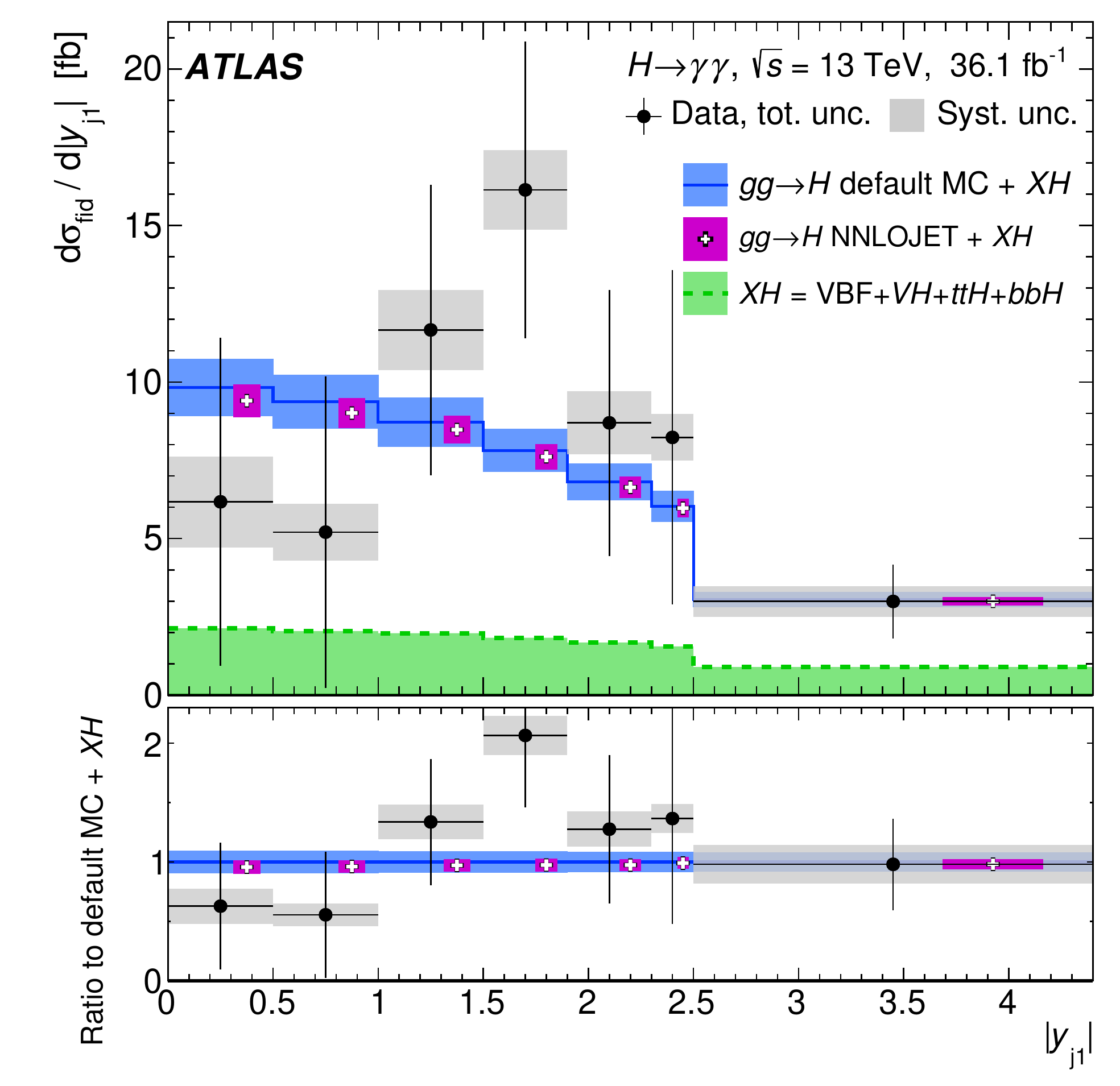}} \\
	\subfloat[] {\includegraphics[width=0.50\columnwidth]{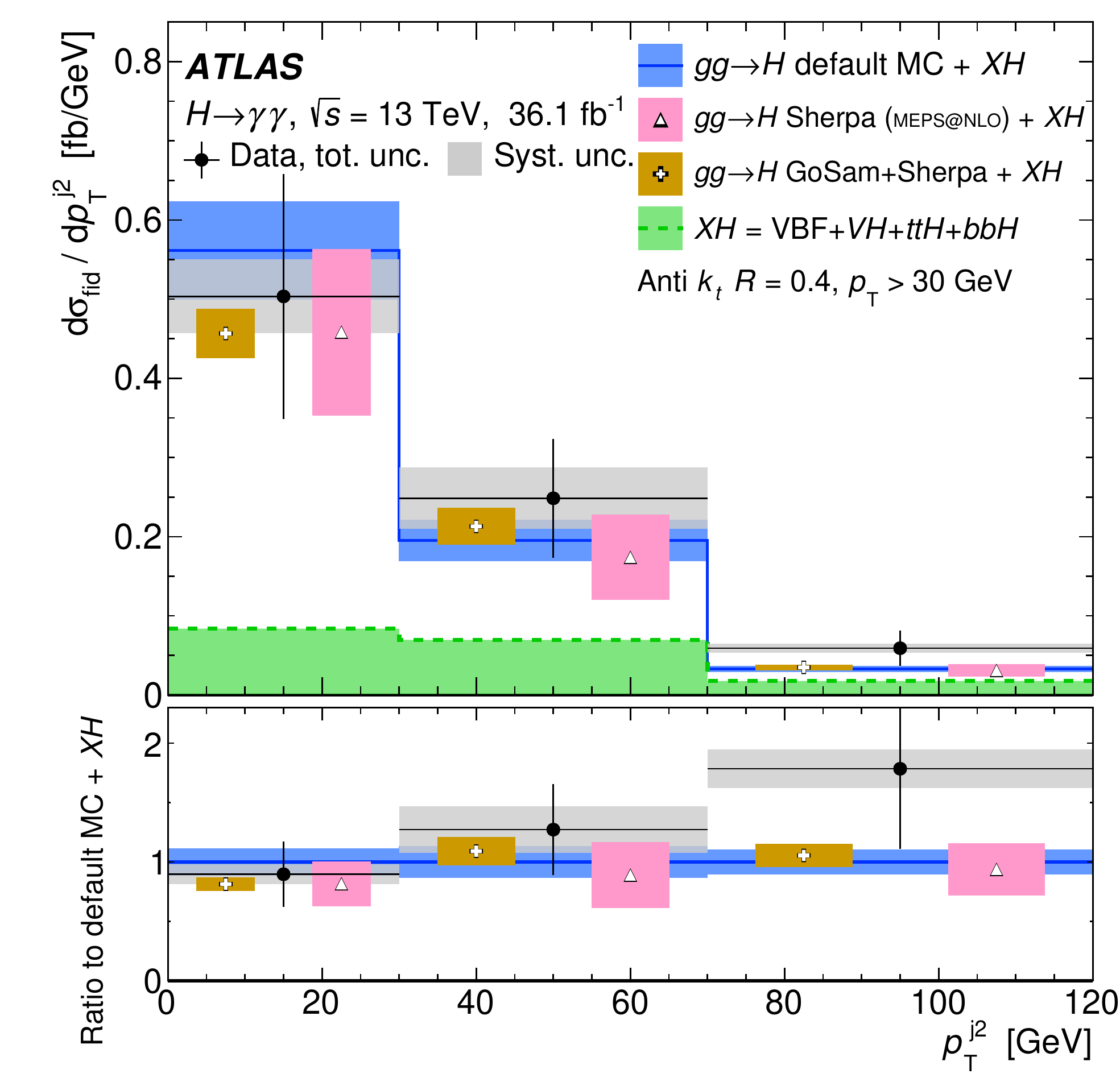}}
	\subfloat[] {\includegraphics[width=0.50\columnwidth]{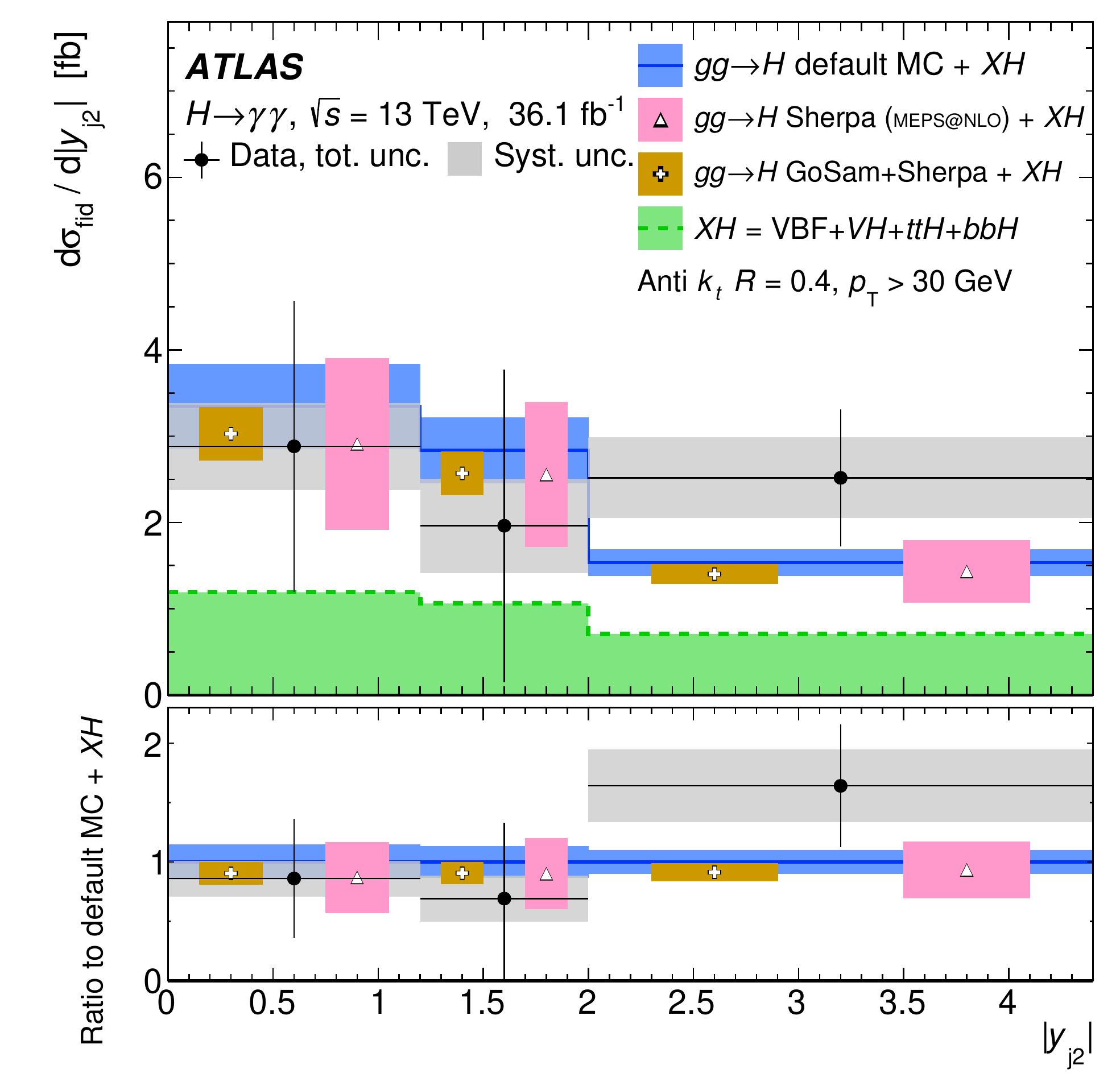}} \\
  \caption{
   The differential cross sections for $pp \to H \to \gamma \gamma$ as a function of (a) \ptjl, (b) \yjl, (c) \ptjsl, and (d) \yjsl\
   are shown and compared to the SM expectations. The data and theoretical predictions are presented in the same way as in Figure~\ref{fig:diff_pth_rap}.
   In addition, the \nnlojet\ and \scetlstwz\ predictions, the \nnlojet\ prediction, and the \sherpa\ (\meps) and \gosam\ predictions,
   described in the text, are displayed in (a), (b) and (c+d), respectively. 
   } 
  \label{fig:diff_pt_rap_jets}
\end{figure*}

The transverse momentum and absolute rapidity of the leading jet, \ptjl and \yjl, as well as the
transverse momentum and absolute rapidity of the subleading jet, \ptjsl\ and \yjsl,
are sensitive to the theoretical modeling and to the relative contributions of the different Higgs boson production mechanisms. 
The transverse momentum distribution of the leading jet probes the emission of energetic quarks and gluons.
In events with two jets, the contributions of \VBF\ and \VH\ productions become
more important. The differential cross sections for $pp \to H \to \gamma \gamma$ as a function of \ptjl, \yjl, \ptjsl, and \yjsl\ are
shown in Figure~\ref{fig:diff_pt_rap_jets}. The chosen bin widths are a compromise between keeping migrations between
  bins small whilst retaining enough statistical power to measure the differential spectra. The measured \ptjl\ spectrum
shown in Figure~\ref{fig:diff_pt_rap_jets}(a) is compared to the default MC prediction as introduced in the previous section as well as to the \nnlojet\ and \scetlstwz~\cite{SCETlib,Stewart:2013faa} predictions. Both the \nnlojet\ and \scetl\ predictions are
corrected using isolation correction factors to account for the impact of the isolation efficiency. In addition, the \nnlojet\ prediction is corrected for the kinematic acceptance and the uncertainties in these corrections is included in the uncertainty bands of both \nnlojet\ and \scetl. The first bin of the leading jet \pt\ spectrum represents zero-jet events that do not contain any jet with \pt > 30~\GeV. 
The predicted \pt\ distributions slightly exceed the measured distribution at low transverse momentum and all show a slight deficit at large
transverse momentum. Both are compatible with the observed slightly harder Higgs boson transverse momentum distribution.
The measured \yjl\ distribution shown in Figure~\ref{fig:diff_pt_rap_jets}(b) is compared to the default MC and the \nnlojet\ predictions: Both show a slight excess at
low rapidity. In Figure~\ref{fig:diff_pt_rap_jets}(c) the measured subleading jet \pt\ distribution is shown. The first bin of \ptjsl\ represents
one-jet events that do not contain two or more jets with \pt > 30~\GeV. The measured distribution is compared to the
default MC, \sherpa\ (\meps), and \gosam\ predictions, as introduced in Section~\ref{sec:details_fidregjets}. Finally, in Figure~\ref{fig:diff_pt_rap_jets}(d) the subleading jet rapidity distribution, \yjsl,
is shown and compared to the expectation from the default MC, \sherpa\ (\meps), and \gosam\ predictions. The SM predictions are in agreement with the measured distributions and no significant deviations are seen. 

\subsubsection{Measurements of cross sections probing spin and CP}\label{sec:diff_spin_cp}

The absolute value of the cosine of the angle between the beam axis and the photons in the
Collins--Soper frame~\cite{Collins:1984kg} of the Higgs boson, \costhetastar, can be used to study the spin
of the Higgs boson. The azimuthal angle between the two leading jets, \dphijj,\footnote{
To preserve the sign information, the azimuthal angles of the jets are ordered according to the jet with the highest rapidity. This definition of \dphijj\ is invariant under a redefinition of the ordering by choosing the opposite beam axis, as explained in Ref.~\cite{Klamke:2007cu}.} in events containing two or more jets is sensitive to the charge
conjugation and parity properties of the Higgs boson interactions with gluons and weak bosons in the gluon--gluon fusion and the VBF production channels,
respectively~\cite{Plehn:2001nj,Klamke:2007cu,Andersen:2010zx,Dolan:2014upa}.

\begin{figure*}[!tbp]
  \centering
	\subfloat[] {\includegraphics[width=0.50\columnwidth]{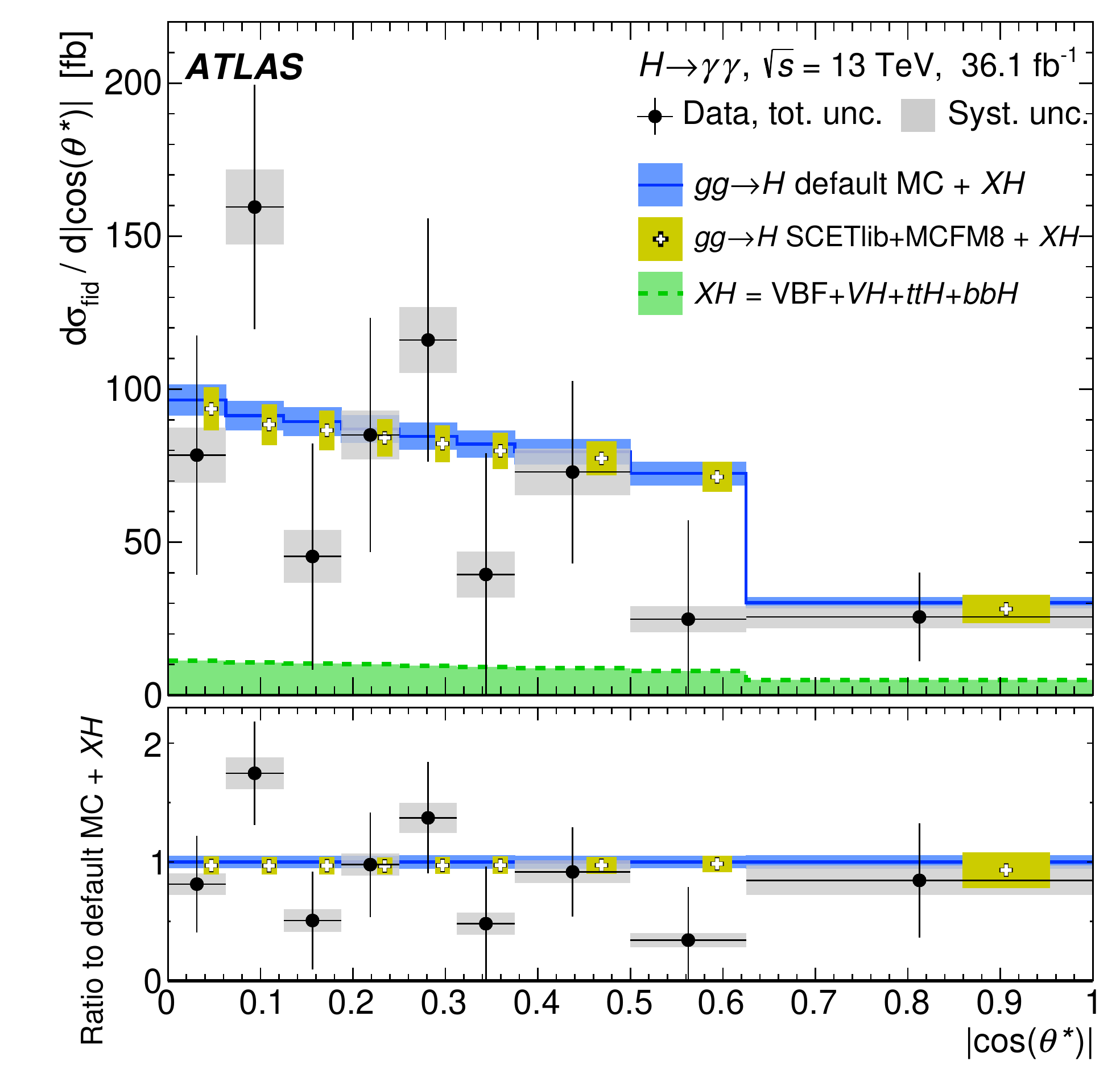}}
	\subfloat[] {\includegraphics[width=0.50\columnwidth]{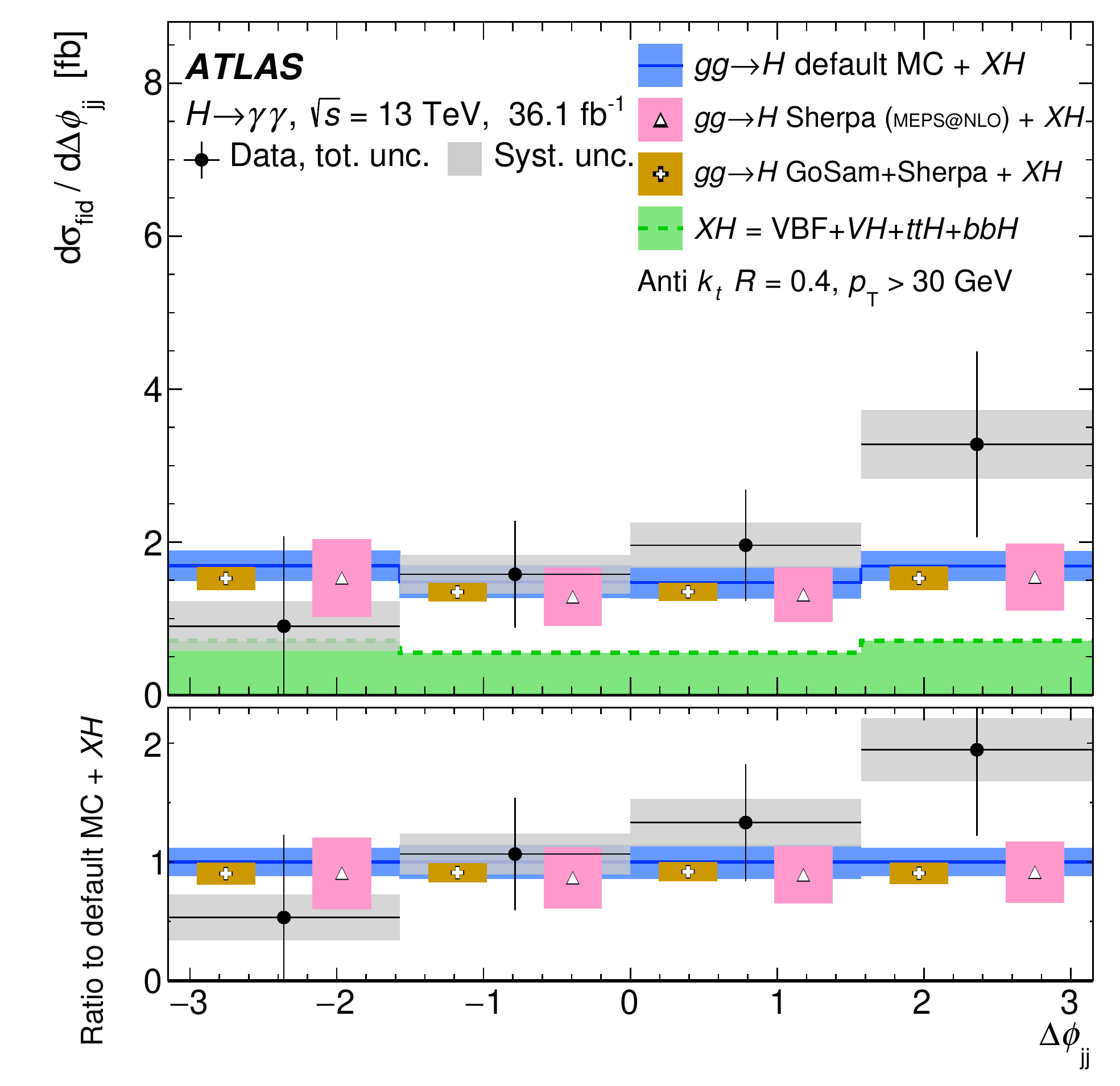}} \\
  \caption{
   The differential cross sections for $pp \to H \to \gamma \gamma$ as a function of (a) \costhetastar\ and (b) \dphijj\ are shown and 
   compared to the SM expectations. The data and theoretical predictions are presented in the same way as in Figure~\ref{fig:diff_pth_rap}.\
   In addition, the \scetlib\ prediction and the \sherpa\ (\meps) and \gosam\ predictions, described in the text, are displayed in (a) and (b), respectively. 
   } 
  \label{fig:diff_spin_cp}
\end{figure*}

The differential cross sections for $pp \to H \to \gamma \gamma$ as a function of \costhetastar\ and \dphijj\ are shown in Figure~\ref{fig:diff_spin_cp}. For a scalar particle \costhetastar, shows a strong drop around 0.6 due to the fiducial requirement on the photon system, whereas for a spin-2 particle, an enhancement would be present in precisely this region. 
The charge conjugation and parity properties of the Higgs boson are encoded in the azimuthal angle between the jets: For example, in gluon--gluon fusion, its distribution for a CP-even coupling has a dip at $\pm \frac \pi 2$ and present peaks at 0 and $\pm \pi$,
whereas for a purely CP-odd coupling it would present as peaks at $\pm \frac{\pi}{2}$ and dips at 0 and $\pm \pi$. For \VBF\,
the SM prediction for \dphijj\ is approximately constant with a slight rise towards \dphijj $= \pm\pi$. Any additional
anomalous CP-even or CP-odd contribution to the interaction between the Higgs boson and weak bosons would manifest itself
as an additional oscillatory component, and any interference between the SM and anomalous couplings can produce
distributions peaked at either \dphijj $ = 0$ or \dphijj $ = \pm
\pi$~\cite{Klamke:2007cu,Andersen:2010zx,Dolan:2014upa}. The shape of the distribution is therefore sensitive to the
relative contribution of gluon--gluon fusion and vector-boson fusion, as well as to the tensor structure of the interactions
between the Higgs boson and gluons or weak bosons. This is exploited in Section~\ref{sec:eft_res} to set limits on 
new physics contributions. To quantify the structure of the azimuthal angle between the two
jets, a ratio is defined as
\begin{align*}
 A_{\dphijjabs} = \frac{ \sigma( \dphijjabs < \frac{\pi}{3} ) - \sigma( \frac{\pi}{3} < \dphijjabs < \frac{2 \pi}{3} ) + \sigma( \dphijjabs > \frac{2 \pi}{3} ) }{ \sigma( \dphijjabs < \frac{\pi}{3} ) + \sigma( \frac{\pi}{3} < \dphijjabs < \frac{2 \pi}{3} ) + \sigma( \dphijjabs > \frac{2 \pi}{3} ) } \, ,
\end{align*}
which is motivated by a similar ratio presented in Ref.~\cite{Andersen:2010zx}. The measured ratio in data as determined by measuring \dphijjabs\ in three bins is
\begin{align*}
 A_{\dphijjabs}^{\rm meas} = 0.45^{+0.18}_{-0.24} \, {\rm(stat.)} \, {}^{+0.10}_{-0.11} \, {\rm(syst.)} \, .
\end{align*}
This value can be compared to the SM prediction from the default MC simulation. The predicted value is
$ A_{\dphijjabs}^{\rm SM} = 0.44 \pm 0.01$, consistent with the measured ratio.

In summary, the measured \costhetastar\ and \dphijj\ distributions are consistent with Standard Model predictions for a CP-even scalar particle.

\subsubsection{Cross sections probing the \VBF\ production mode}\label{sec:diff_vbf}

The distribution of the dijet rapidity separation, \deltayjj, the azimuthal angle between the dijet and diphoton systems, \dphiggjj, 
and the invariant mass of the leading and subleading jets, \mjj\, for events with two or more jets are sensitive to the differences between the gluon--gluon fusion
and VBF production mechanisms. In vector-boson fusion, the $t$-channel exchange of a $W/Z$ boson typically results in
two moderate-$\pt$ jets that are well separated in rapidity. Furthermore, quark/gluon radiation in the
rapidity interval between the two jets is suppressed in the VBF process when compared to the gluon--gluon fusion process,
because there is no color flow between the two jets. The \dphiggjj\ distribution for VBF production is therefore expected to be steeper and more
peaked towards $\dphiggjj=\pi$ than for gluon--gluon fusion.

\begin{figure*}[!tbp]
  \centering
	\subfloat[] {\includegraphics[width=0.50\columnwidth]{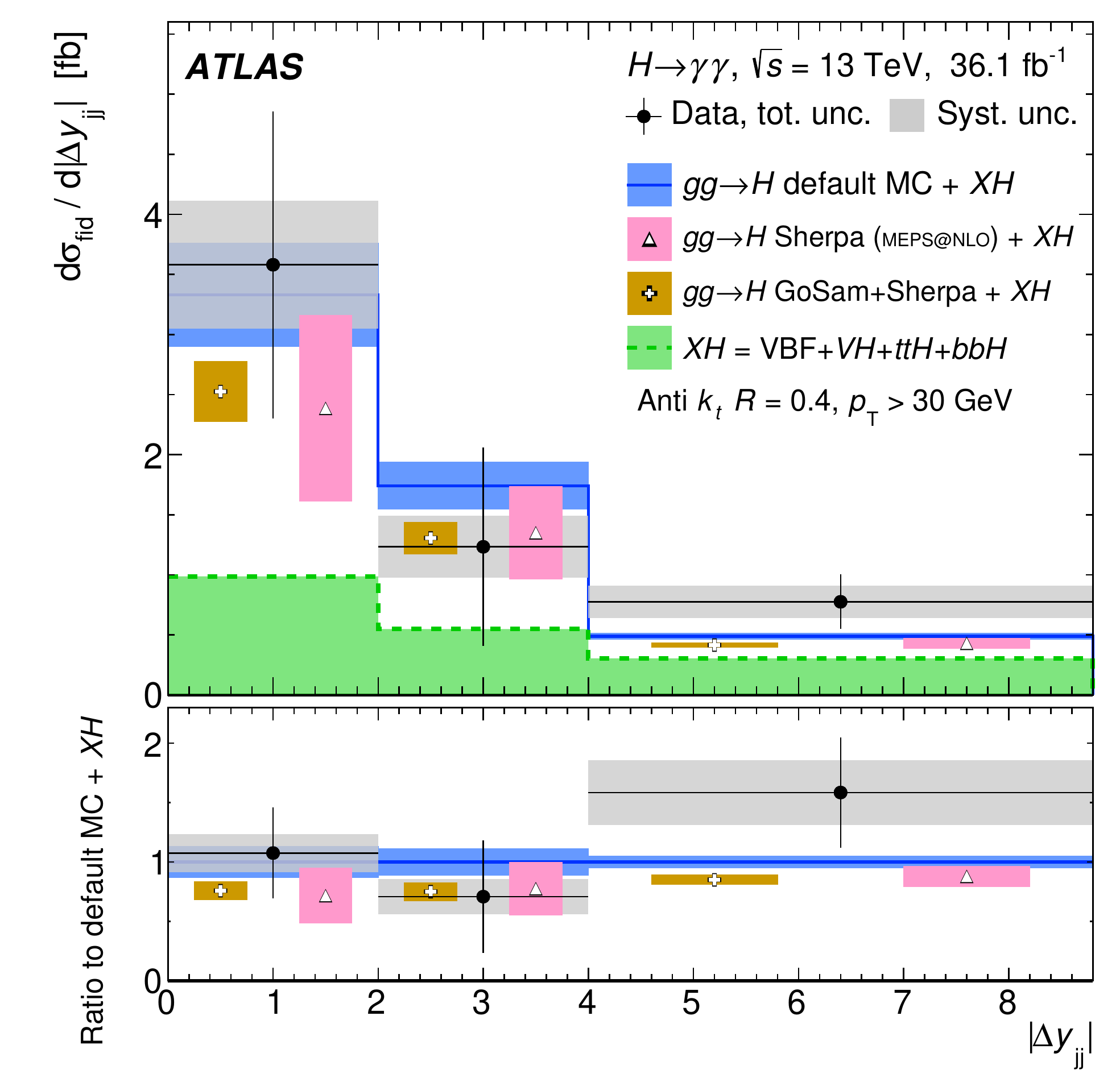}}
	\subfloat[] {\includegraphics[width=0.50\columnwidth]{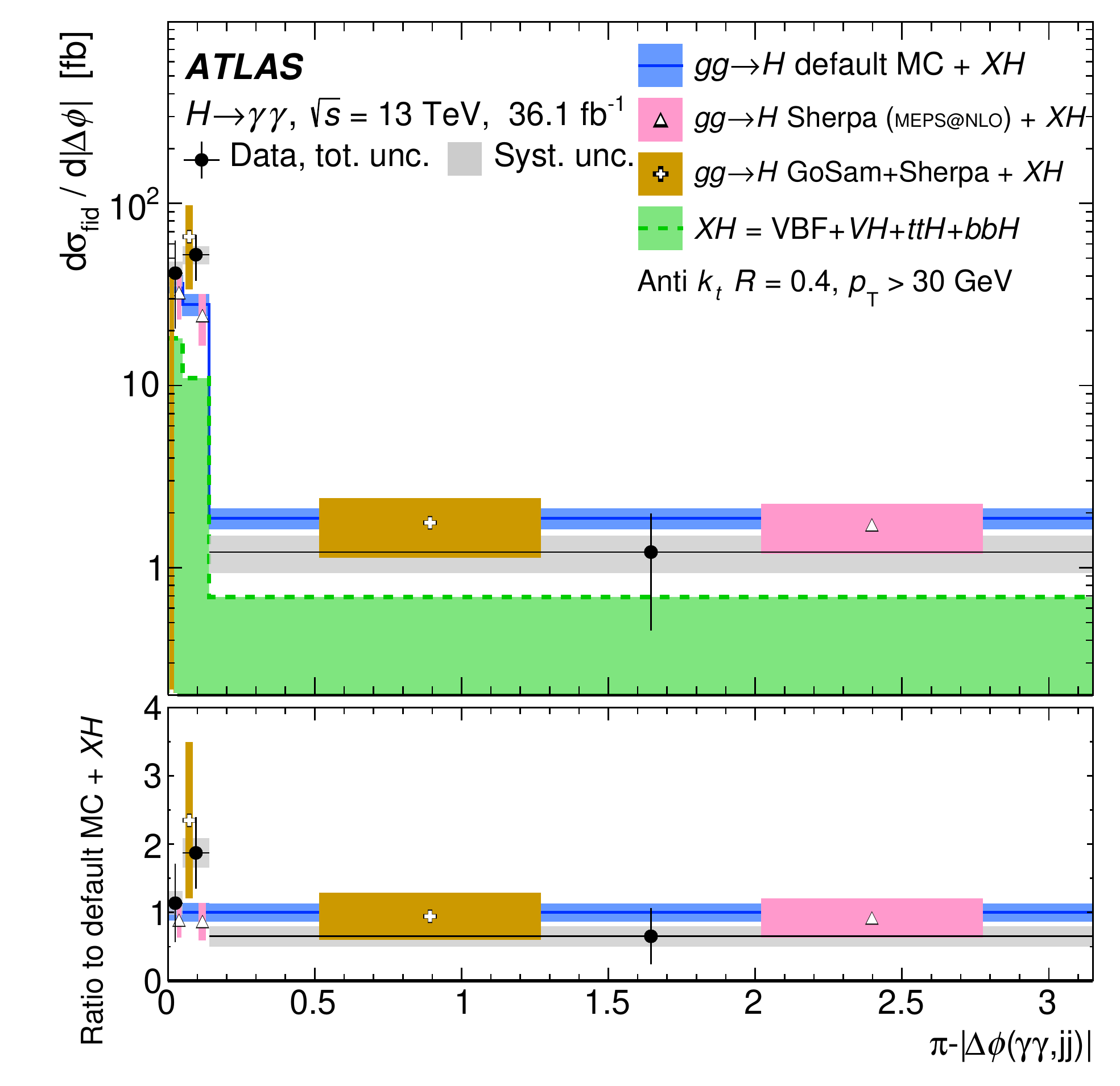}} \\
	\subfloat[] {\includegraphics[width=0.50\columnwidth]{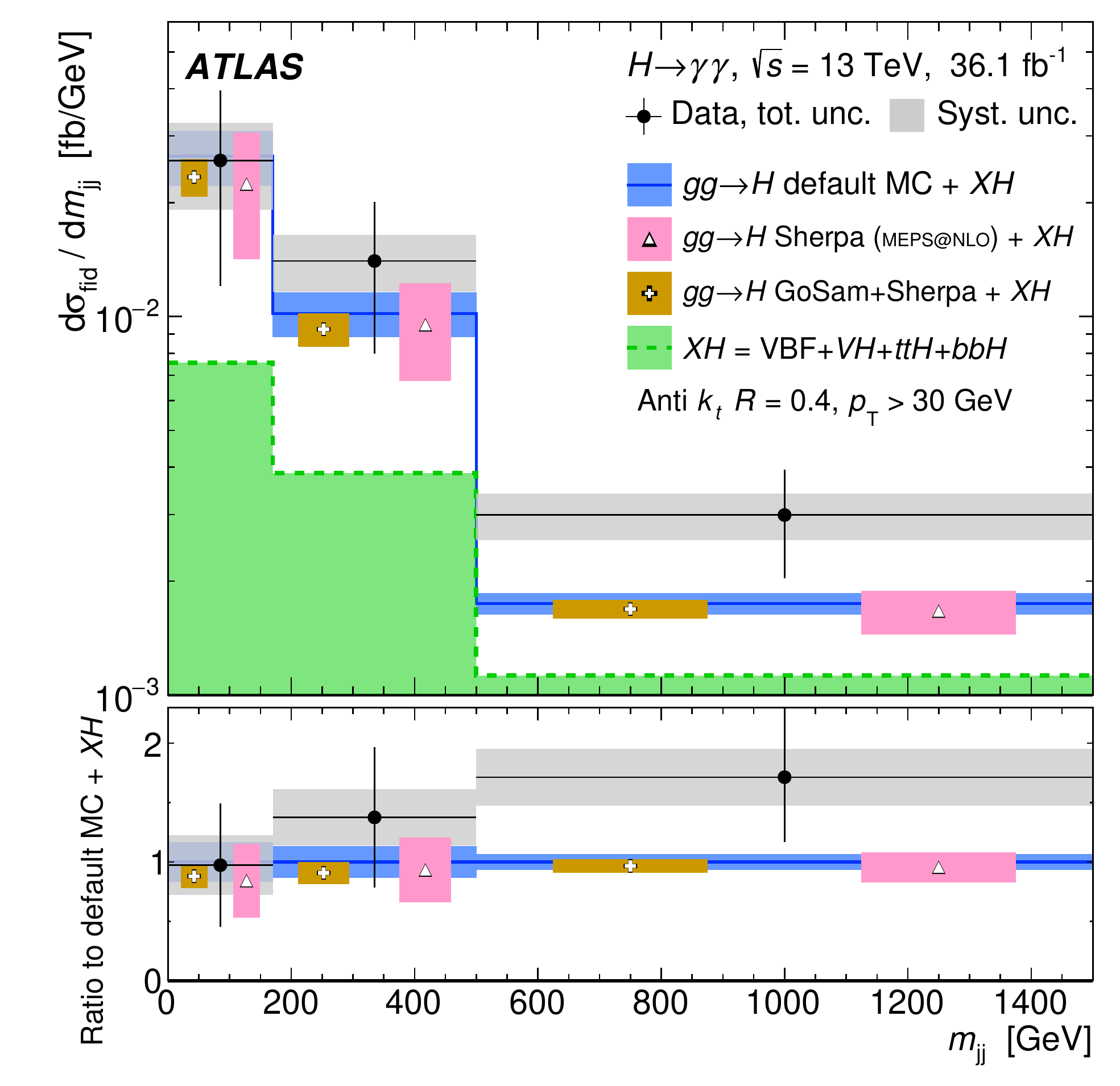}} 
  \caption{
   The differential cross sections for $pp \to H \to \gamma \gamma$ as a function of (a) \deltayjj, (b) $\pi -$ \dphiggjj, and (c) \mjj\ are shown and 
   compared to the SM expectations. The data and theoretical predictions are presented in the same way as in Figure~\ref{fig:diff_pth_rap}.
   In addition, the \sherpa\ (\meps) and \gosam\ predictions are shown for all three cross sections. 
   } 
  \label{fig:diff_VBF}
\end{figure*}

The differential cross sections for $pp \to H \to \gamma \gamma$ as a function of \deltayjj, \dphiggjj, and \mjj\ are shown 
for events with at least two jets with \pt\ > 30~\GeV\ in Figure~\ref{fig:diff_VBF}. These variables are used to 
discriminate between gluon--gluon fusion and the \VBF\ production of the Higgs boson and enter the multivariate classifier introduced
in Section~\ref{sec:vbf_category} that defines the categories used for the simplified template cross-section and coupling measurements.
The measured distributions are in agreement to the default MC, \sherpa\ (\meps), 
and the \gosam\ predictions. The accuracy of the fixed-order parton-level prediction from \gosam\ breaks down in the lowest
bin of $\pi - \dphiggjj$ and the measured cross section moderately exceeds the SM predictions at high \mjj\ values.

\subsubsection{Double-differential cross sections}\label{sec:diff_double}

\begin{figure*}[!tbp]
  \centering
	\subfloat[] {\includegraphics[width=0.5\columnwidth]{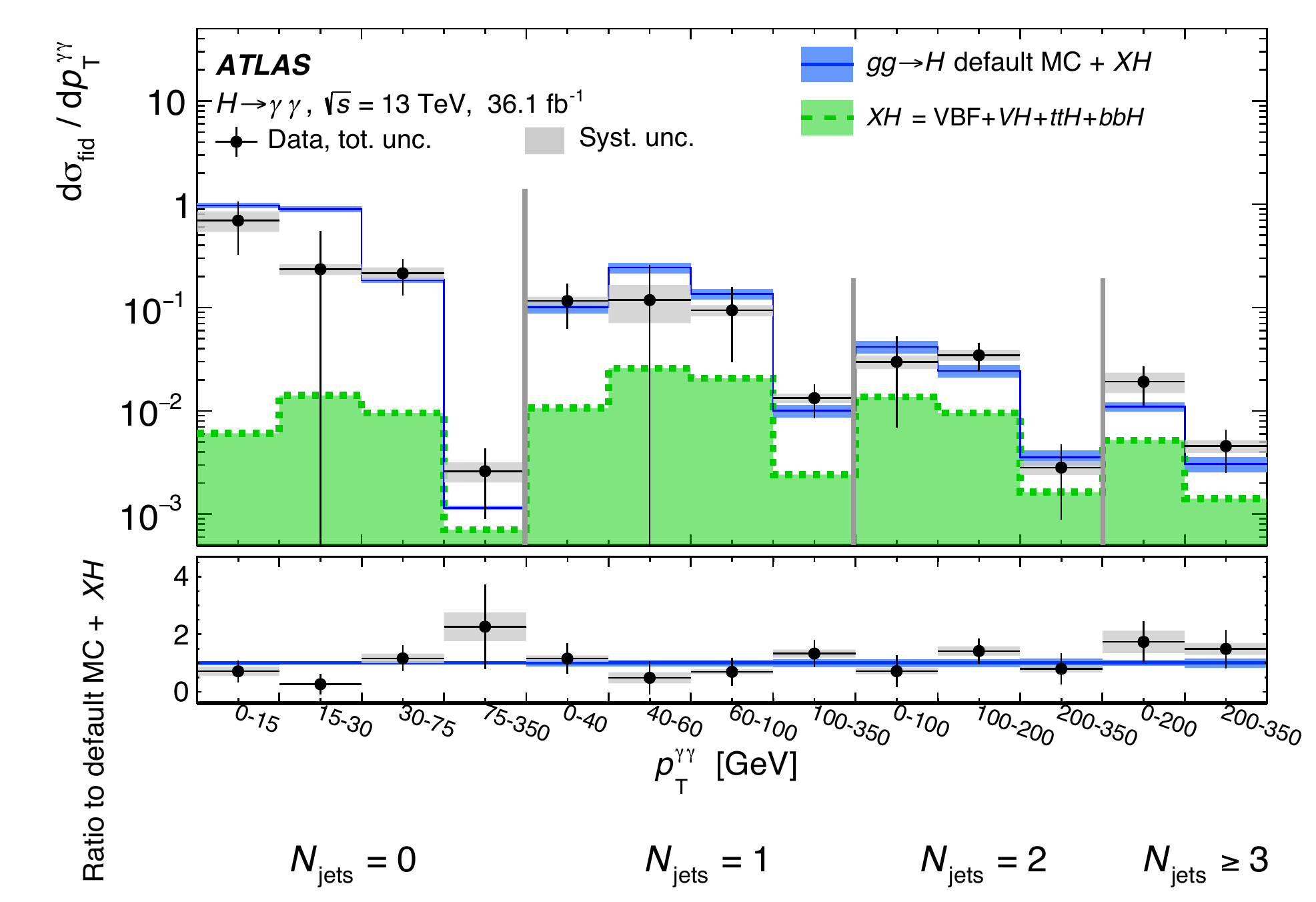}}
	\subfloat[] {\includegraphics[width=0.5\columnwidth]{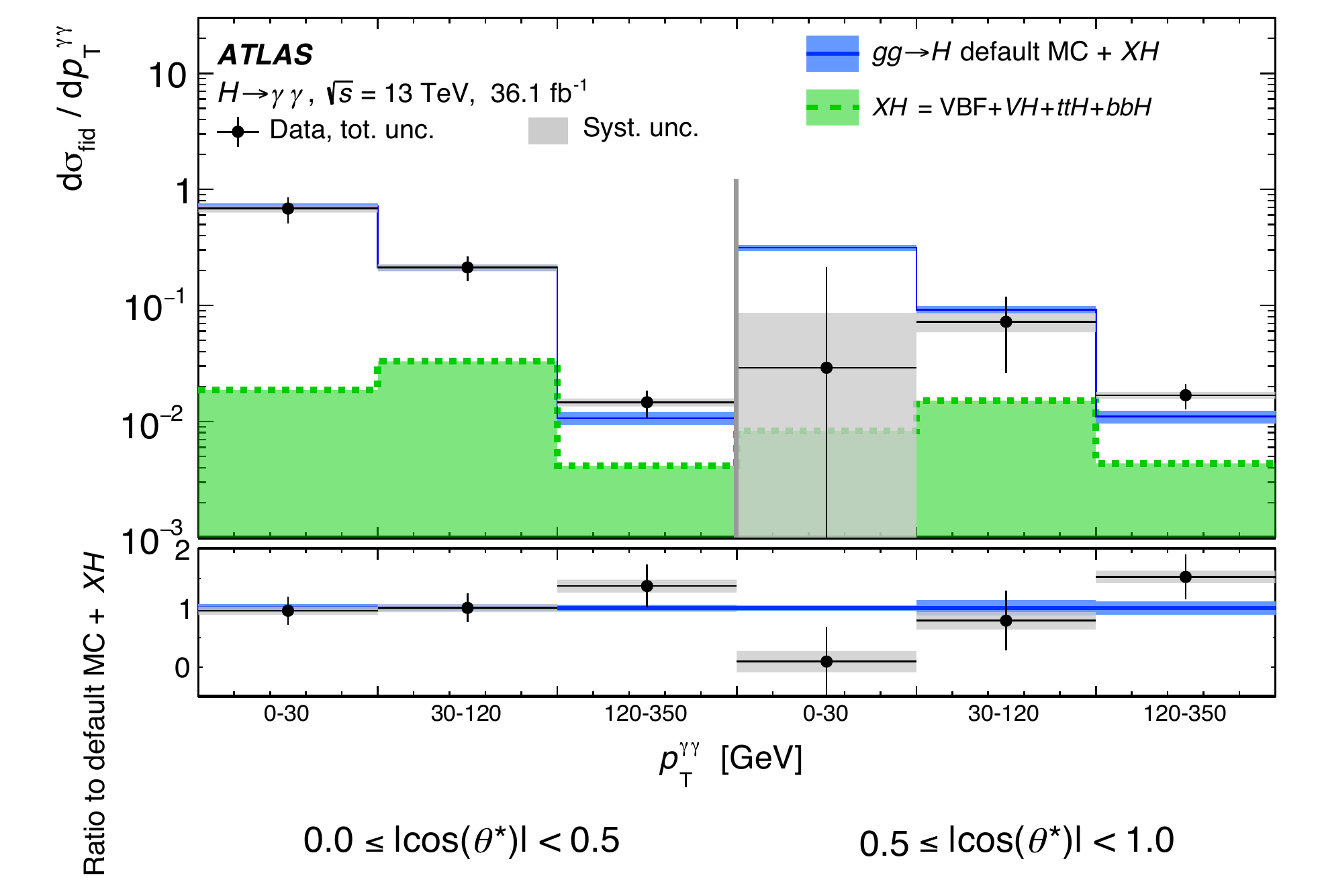}} 
  \caption{
   The double-differential cross section for $pp \to H \to \gamma \gamma$ as a function of (a) \ptgg\ and \njet, for jets with $\pt > 30~\GeV$, and (b) \ptgg\ and \costhetastar\ separating the
   two regions of $\costhetastar < 0.5$ and $\costhetastar > 0.5$ from each other.  The data and theoretical predictions are presented in the same way as in Figure~\ref{fig:diff_pth_rap}.
   } 
  \label{fig:diff_double}
\end{figure*}

The double-differential cross section for $pp \to H \to \gamma\gamma$ as a function of \ptgg\ and \njet, for jets with \mbox{$\pt >
30$~\GeV}, and \ptgg\ and \costhetastar\ are shown in Figure~\ref{fig:diff_double}. These cross sections are sensitive to the modeling of the Higgs boson kinematic, its production mechanisms, and its spin-CP properties. Both double-differential cross sections
are in agreement with the Standard Model expectation.

\subsubsection{Impact of systematic uncertainties on results}\label{sec:fid_syst}

\begin{figure}[!tbp]
  \begin{center}
  	\subfloat[] {\includegraphics[width=.5\textwidth]{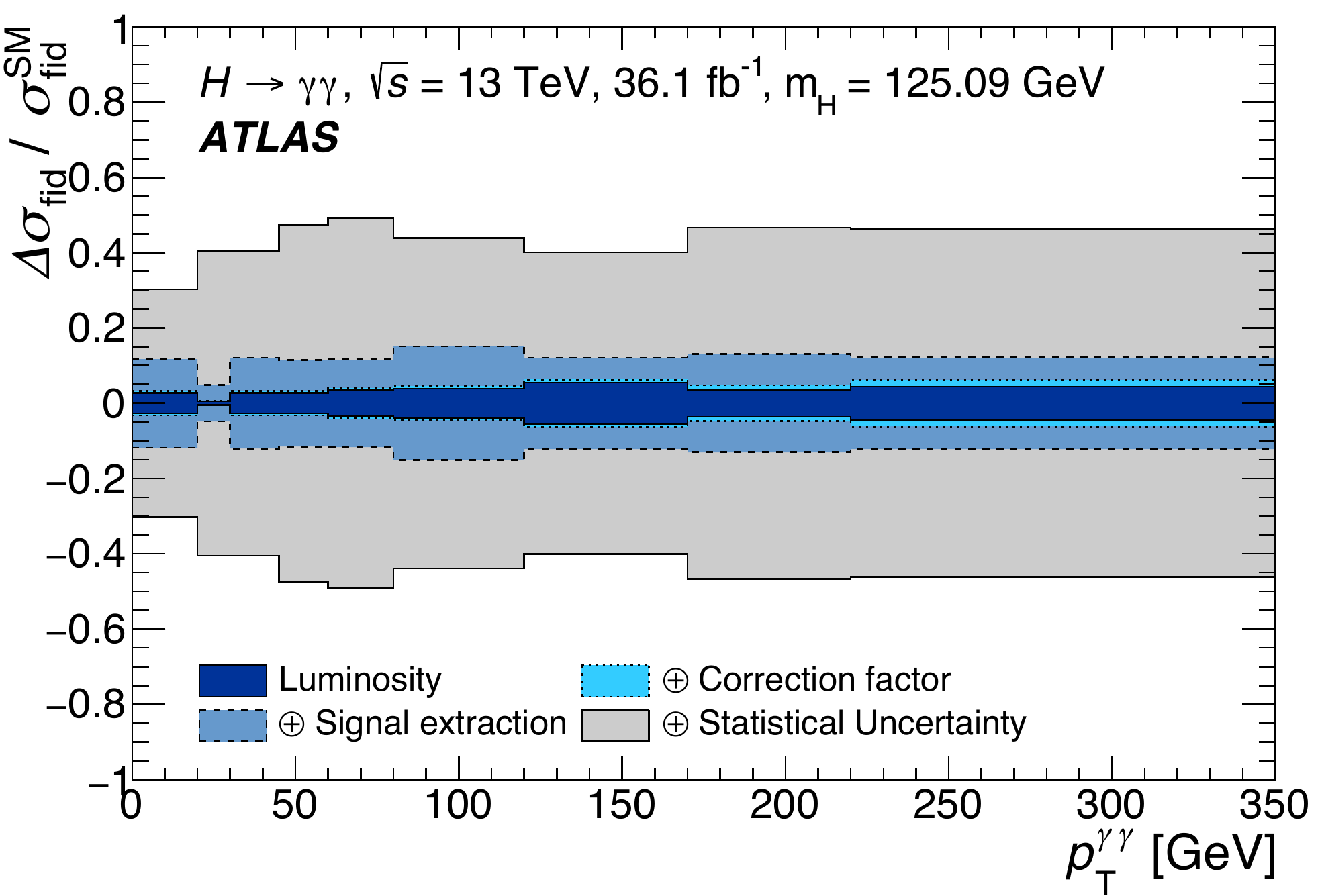}}
  	\subfloat[] {\includegraphics[width=.5\textwidth]{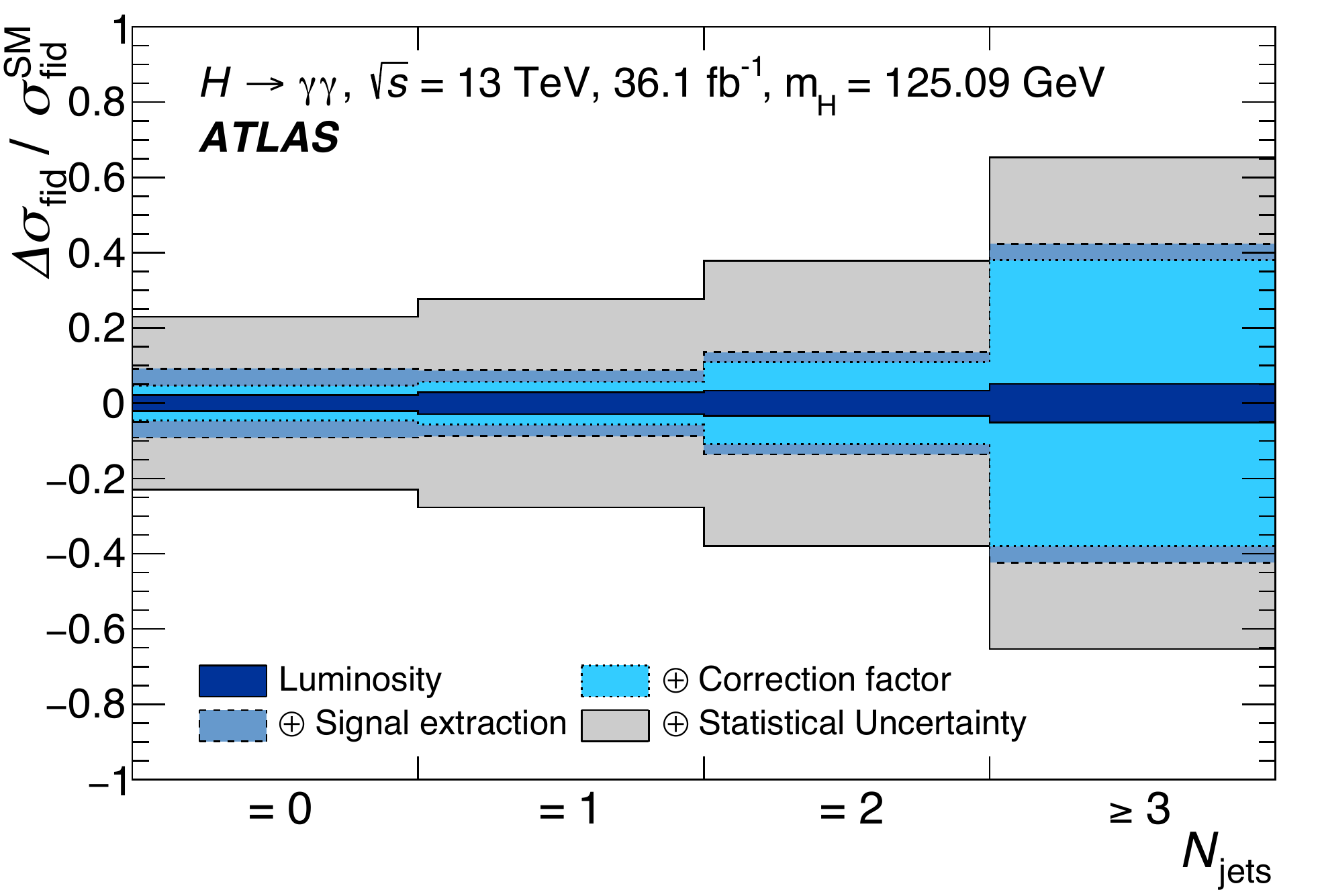}} \\
  	\subfloat[] {\includegraphics[width=.5\textwidth]{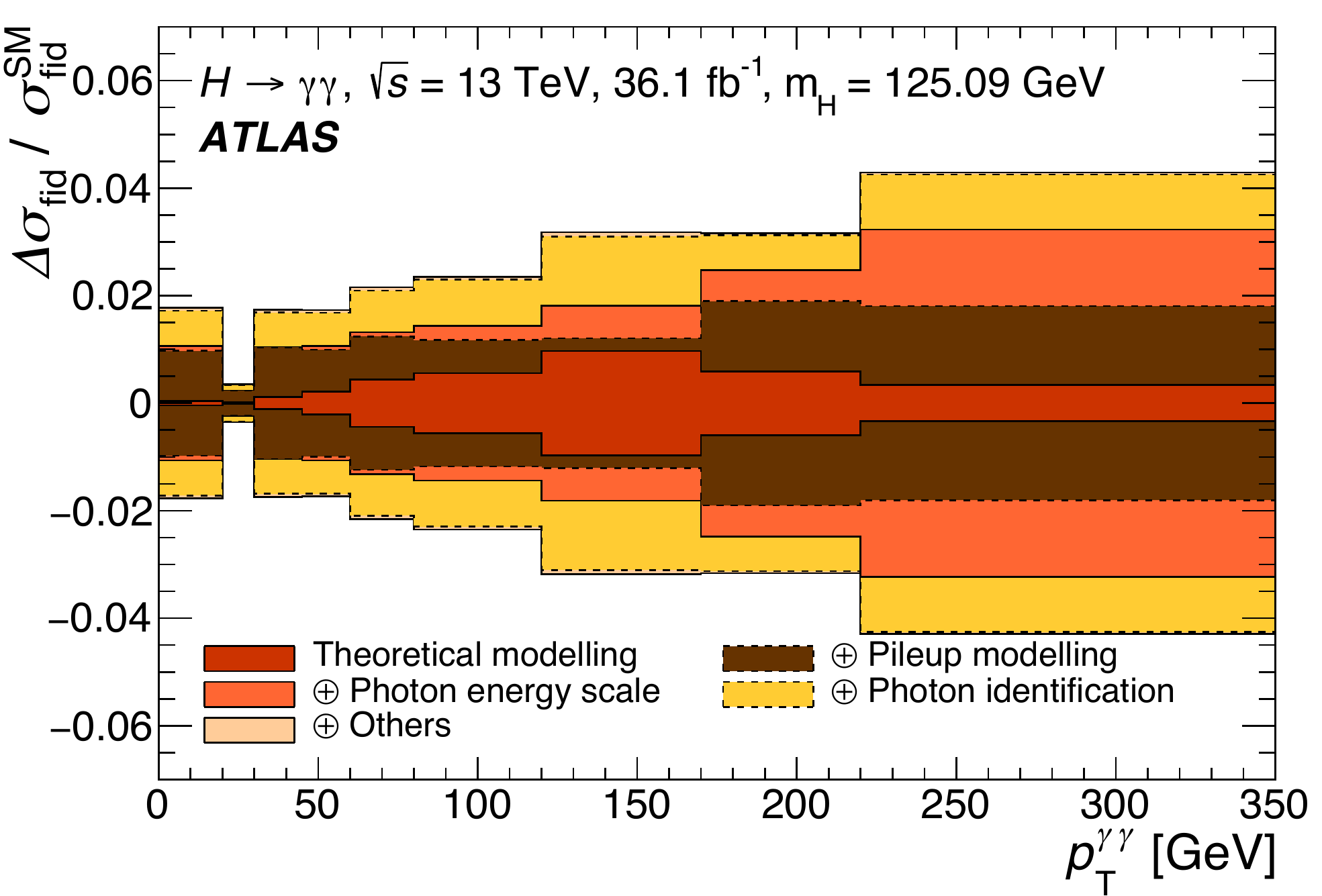}} 
  	\subfloat[] {\includegraphics[width=.5\textwidth]{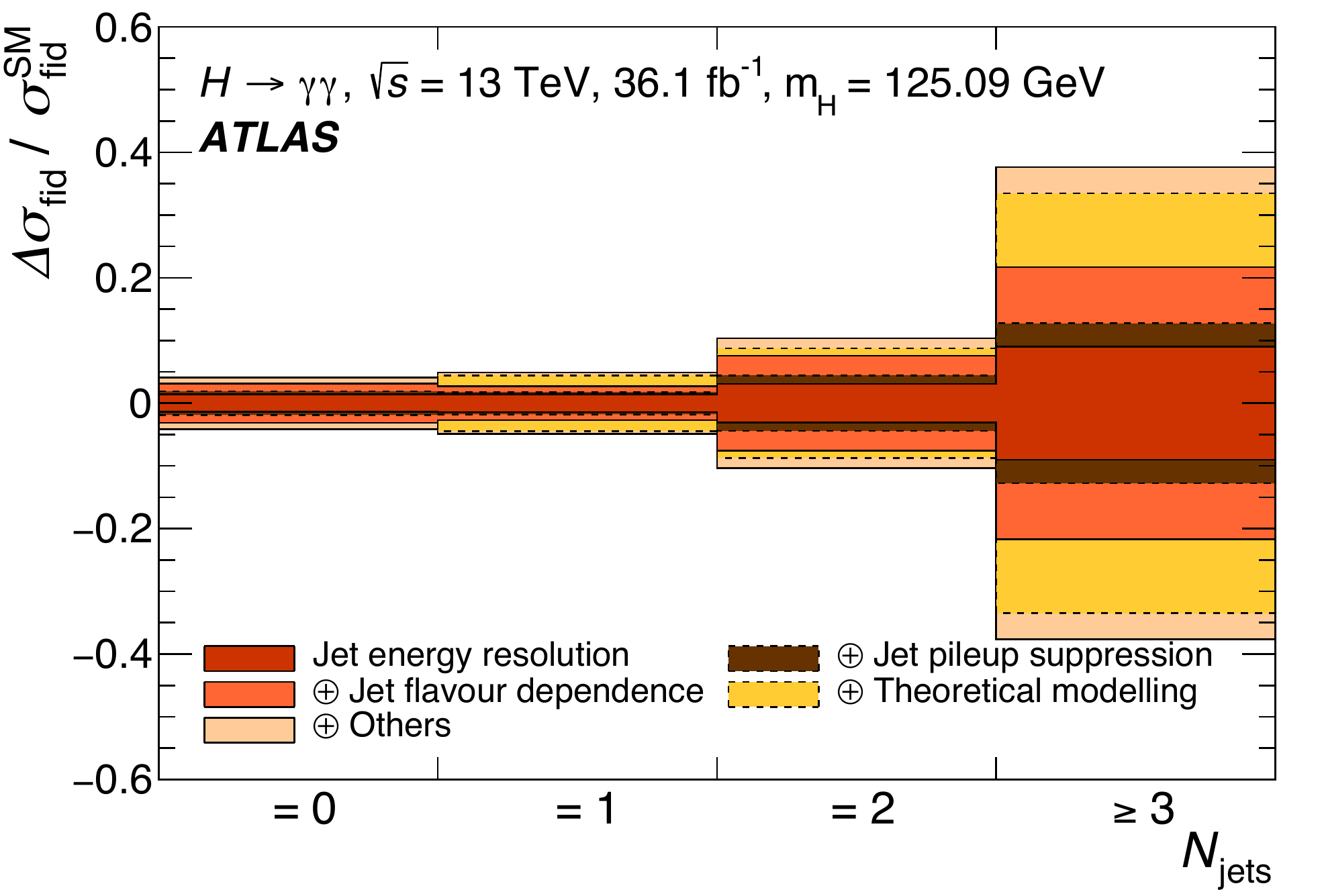}}
       \caption[]{
       The relative size of systematic uncertainties associated with the signal extraction, the correction factors
       (experimental and theoretical modeling) and the luminosity on the differential cross sections are shown as a function of 
       (a) \pttgg\ and (b) \njet. 
       The statistical uncertainty associated with the signal extraction is also shown
       as a gray band. For completeness, the relevant components of the uncertainties in the correction factors are shown as a function of (c) \pttgg\ and (d) \njet.}     
    \label{fig:syst-breakdown}
  \end{center}
\end{figure}

A summary of the uncertainties in the measured cross sections of the fiducial regions are shown in
Table~\ref{tab:syst-summary}. As an example concerning the differential measurements, a breakdown of the systematic uncertainties in the differential cross sections as a function of \ptgg\ and \njet\ is shown in Figure~\ref{fig:syst-breakdown}. The measurements are dominated by the
statistical uncertainties. For the systematic uncertainties, the uncertainty in the fitted signal yield, due to the background modeling and the photon energy resolution, is typically more important than the uncertainty in the correction factor due to the
theoretical modeling. The jet energy scale and resolution uncertainties become increasingly
important for high-jet multiplicities and in the \ttH- and \VBF-enhanced phase space.

\begin{table}[!tp]
  \caption{The expected uncertainties, expressed in percent, in the cross sections measured in the diphoton fiducial, VBF-enhanced,
  $N_{\rm lepton} \ge 1$, $t \bar t H$-enhanced, and high $E_{\rm T}^{\rm miss}$ regions. 
  The fit systematic uncertainty includes the effect of the photon energy
  scale and resolution, and the impact of the background modeling on the signal yield. The theoretical modeling uncertainty is defined
  as the envelope of the signal composition, the modeling of Higgs boson transverse momentum and rapidity distribution, and the uncertainty of parton shower and the underlying event (labeled as ``UE/PS'') as described in Section~\ref{sec:theo_fid}. }
  \label{tab:syst-summary}
\centering
  \begin{tabular}{lccccc}
  \hline\hline
  Source  & \multicolumn{5}{c}{Uncertainty in fiducial cross section} \\
  & Diphoton & VBF-enhanced &  $N_{\rm lepton} \ge 1$ & $t \bar t H$-enhanced & High $E_{\rm T}^{\rm miss}$ \\
  \hline
  \hline
  Fit (stat.) & 17\% & 22\% & 72\%  & 176\% &  53\% \\
  Fit (syst.) & 6\% & 9\% & 27\% & 138\% & 13\% \\   
  $\quad$ Photon energy scale \& resolution         & 4.3\% & 3.5\% & 3.1\% & 10\% & 4.1\% \\
  $\quad$ Background modeling & 4.2\% & 7.8\% & 26.7\% & 138\%  & 12.2\%  \\
  \hline
  Photon efficiency & 1.8\% & 1.8\% & 1.8\% & 1.8\% & 1.9\%\\
  Jet energy scale/resolution & - & 8.9\% & - & 4.5\% & 6.9\%\\
  $b$-jet flavor tagging & - & - & - & 3\% & -\\
  Lepton selection & - & - & 0.7\% & 0.2\% & - \\
  Pileup  & 1.1\% & 2.9\% & 1.3\% & 2.5\% & 2.5\%\\
  Theoretical modeling & 0.1\% & 4.5\% & 4.0\% & 8.1\% & 31\% \\
  $\quad$ Signal composition         & 0.1\% & 4.5\% & 3.1\% & 8.1\% & 25\% \\
  $\quad$ Higgs boson $\pt^H$ \& $|y_H|$ & 0.1\% & 0.9\% & 0.2\% & 0.7\%  & 0.1\%  \\
  $\quad$ UE/PS                         & -   & 0.3\% & 0.7\% & 1.1\% & 31\% \\
  Luminosity     & 3.2\% & 3.2\% & 3.2\% & 3.2\% & 3.2\%\\ \hline
  Total & 18\% & 26\% & 77\% & 224\% & 63\% \\
  \hline\hline
  \end{tabular}
\end{table}

\subsubsection{Compatibility of measured distributions with the Standard Model}

\begin{table}[!tp]
\centering
\caption{Probabilities from a $\chi^{2}$ test for the comparison between data and the default SM prediction.}
\begin{tabular}{cc}
\hline \hline
Distribution & Default MC Prediction \\
\hline \hline
\ptgg & 51\% \\
\ygg & 57\% \\
\ptjl & 32\% \\
\yjl & 66\% \\
\ptjsl & 61\% \\
\yjsl & 56\% \\
\costhetastar & 47\% \\
\dphijj & 64\% \\
\deltayjj & 53\% \\
\dphiggjj & 43\% \\
\mjj & 54\% \\
\njet ($\pt > 30$~\GeV) & 56\% \\
\njet ($\pt > 50$~\GeV)& 19\% \\
\hline \hline
\end{tabular} 
\label{tab:prob_vars}
\end{table}

\begin{figure*}[!tbp]
 \centering
\subfloat[] {\includegraphics[width=0.50\columnwidth]{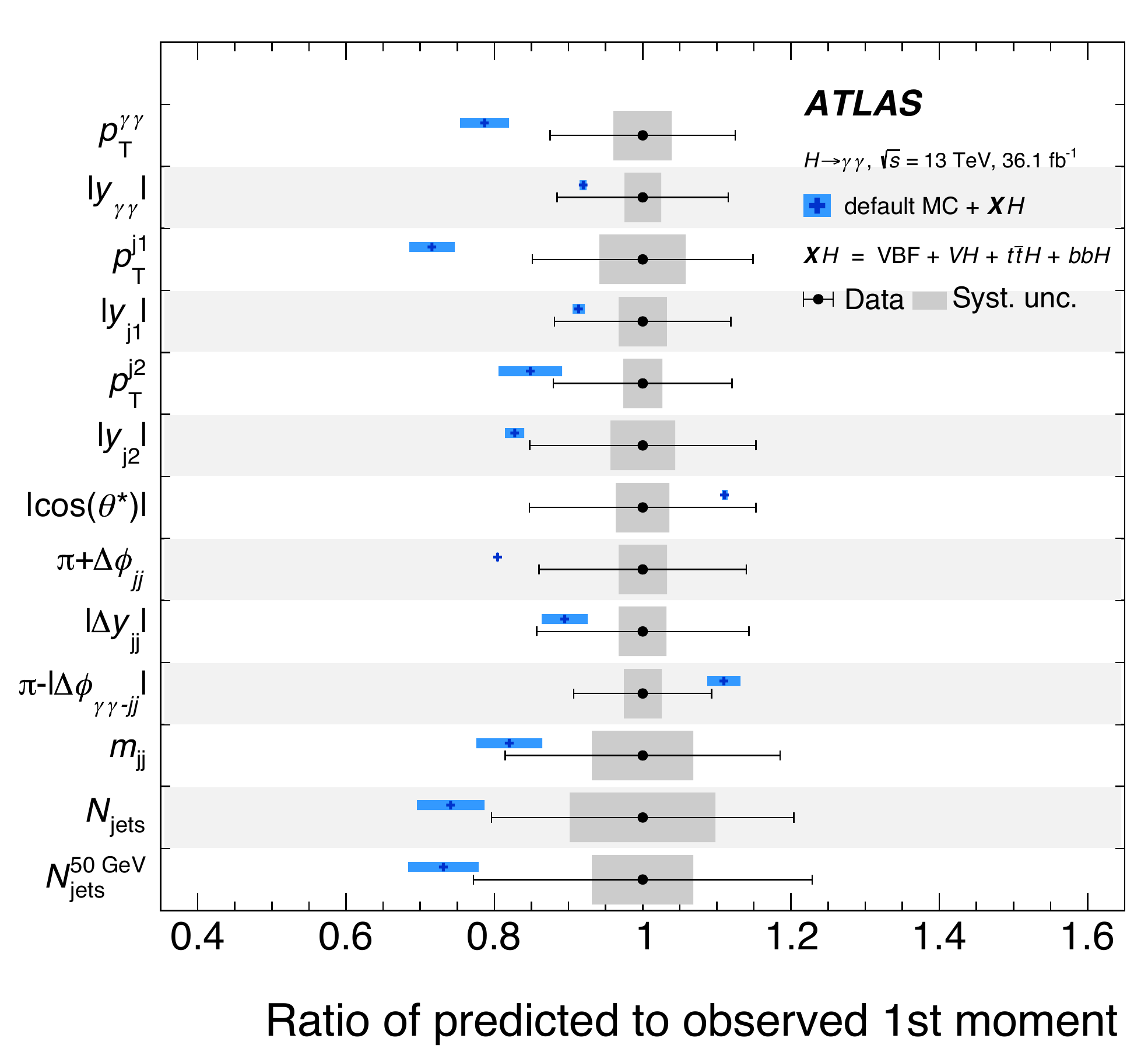}}
\subfloat[] {\includegraphics[width=0.50\columnwidth]{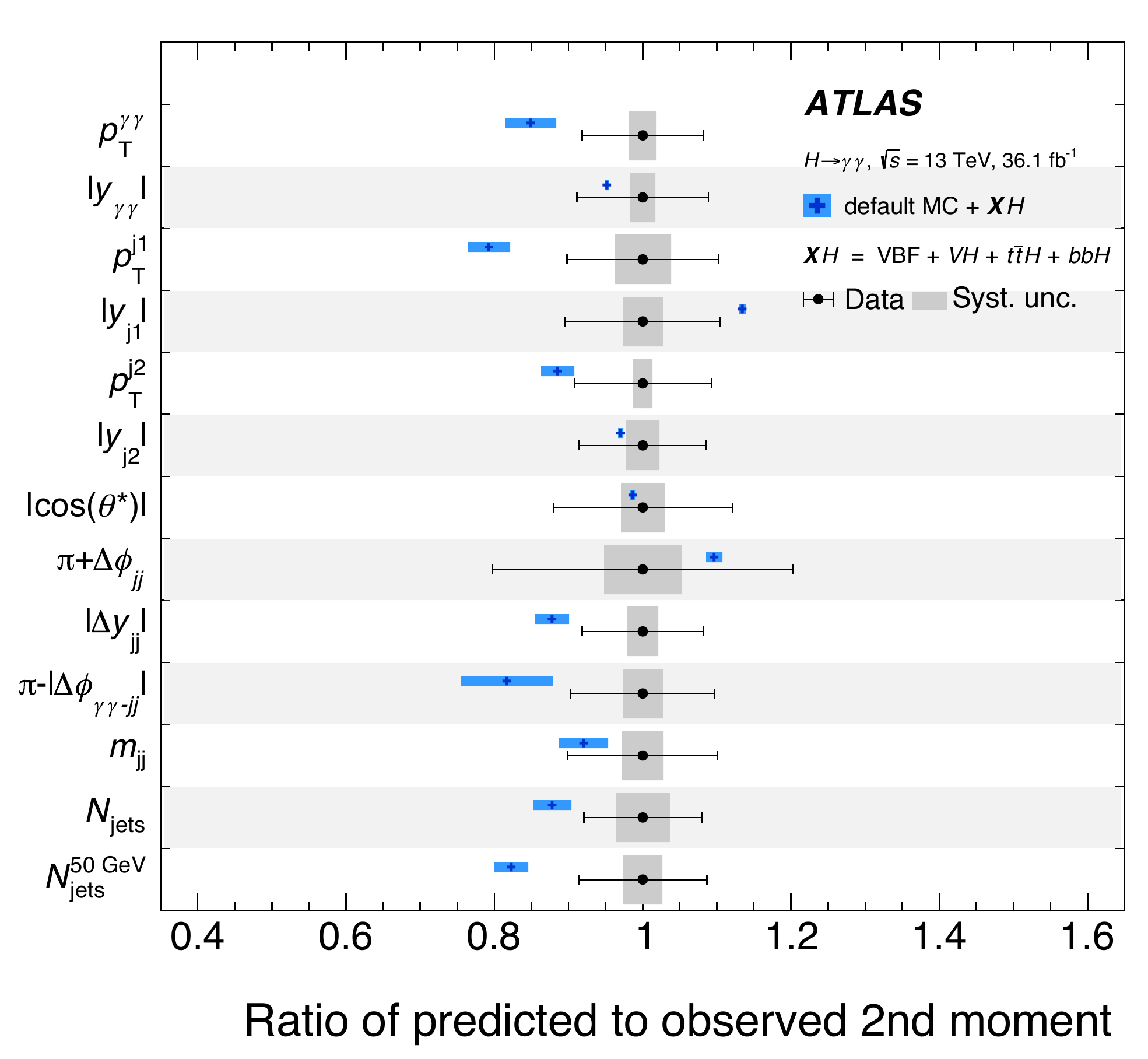}} \\
 \caption{
  (a)  The ratio of the first moment (mean) of each differential distribution predicted by the Standard Model to that
  observed in the data. The SM moment is calculated by using the default MC distributions for gluon--gluon fusion and the other production mechanisms. 
  (b) The ratio of the second moment (RMS) of each differential distribution predicted by the Standard Model to that observed in the data.
  The intervals on the vertical axes each represent one of the differential distributions. The band for
  the theoretical prediction represents the corresponding uncertainty in that prediction (see text for
  details). The error bar on the data represents the total uncertainty in the measurement, with the
  gray band representing only the systematic uncertainty.
 } 
 \label{fig:diff_moments}
\end{figure*}

The compatibility between the measured distributions and the Standard Model is tested by comparing the first and second
moments of the measured distributions. Figure~\ref{fig:diff_moments} shows the first and second moments (mean and RMS) of the distributions and compares them to the moments of the default MC prediction, as calculated from the measured and predicted cross-section bins. The theory uncertainties are constructed as outlined in Section~\ref{sec:theo}. The measured Higgs boson transverse momentum has somewhat higher first and second moments than the Standard Model prediction, which is consistent with the previous observations~\cite{Aad:2014lwa,Aad:2014tca}. The leading-jet \pt\ spectrum shows the same feature. In addition a $\chi^2$ test is carried out for all distributions reported in Section~\ref{sec:methods_fid}: The resulting $p$-values are reported in Table~\ref{tab:prob_vars}, which confirms that within the current uncertainties the data are in agreement with the SM predictions.

\clearpage

\subsubsection{Search for anomalous Higgs-boson interactions using an effective field theory approach}\label{sec:eft_res}

The strength and tensor structure of the Higgs-boson interactions can be investigated using an effective field theory
approach, which introduces additional CP-even and CP-odd interactions that change the event rates and the kinematic properties of the Higgs boson and associated jet spectra from those in the Standard Model. The parameters of the
effective field theory are probed using a fit to \neftvars\ of the most relevant differential cross sections. The effective Lagrangian of
Ref.~\cite{Contino:2013kra} is used which adds dimension-six operators of the Strongly Interacting Light Higgs formulation~\cite{Giudice:2007fh} to the Standard Model interactions. The $H \to \gamma\gamma$ differential cross sections are mainly sensitive to the operators that affect the Higgs boson interactions with gauge bosons and the relevant terms in the Lagrangian can be specified by 
\begin{eqnarray*}\label{eq:effT} 
{\cal L}_{\rm eff} =  &  \bar{c}_{g} {\cal O}_{g} + \bar{c}_{HW} {\cal O}_{HW} + \bar{c}_{HB} {\cal O}_{HB} \\
                      & \quad + \,\,  \tilde{c}_{g} \tilde{\cal O}_{g} + \tilde{c}_{HW} \tilde{\cal O}_{HW} + \tilde{c}_{HB} \tilde{\cal O}_{HB}, 
\end{eqnarray*} 
where $\bar{c}_i$ and $\tilde c_i$ are dimensionless Wilson coefficients specifying the strength of the new CP-even and CP-odd
interactions, respectively, and the dimension-six operators ${\cal O}_i$ and $\tilde{{\cal O}_i}$ are those described in
Refs.~\cite{Contino:2013kra,Alloul:2013naa}. In the SM, all of the Wilson coefficients are equal to zero. 
The ${\cal O}_g$ and $\tilde{\cal O}_g$ operators introduce new interactions between the Higgs boson and two gluons and can
be probed through the gluon--gluon fusion Higgs production mechanism. The ${\cal O}_{HW}$ and $\tilde{\cal O}_{HW}$ operators
introduce new $HWW$, $HZZ$ and $HZ\gamma$ interactions. The  $HZZ$ and $HZ\gamma$ interactions are also impacted by
${\cal O}_{HB}$ and $\tilde{\cal O}_{HB}$.
The ${\cal O}_{HW}$, $\tilde{\cal O}_{HW}$, ${\cal O}_{HB}$ and $\tilde{\cal O}_{HB}$ operators can be probed through
vector-boson fusion and associated production. Other operators in the full effective Lagrangian of Ref.~\cite{Contino:2013kra} can also modify Higgs-boson interactions but are not considered here due to the lack of sensitivity of the $H \to \gamma\gamma$ decay channel. Combinations of some of the CP-even operators have been constrained using global fits to experimental data from LEP and the LHC~\cite{Contino:2013kra,Pomarol:2013zra,Ellis:2014jta}.

The effective Lagrangian has been implemented in FeynRules \cite{Alloul:2013naa}.\footnote{The implementation in
Ref.~\cite{Alloul:2013naa} involves a redefinition of the gauge boson propagators that results in unphysical amplitudes
unless certain physical constants are also redefined. The original implementation did not include the redefinition of
these physical constants. However, the impact of redefining the physical constants is found to be negligible on the
predicted cross sections across the range of Wilson coefficients studied. The relative change in the predicted Higgs
boson cross sections as functions of the different Wilson coefficients is also found to agree with that predicted by the
Higgs characterization framework~\cite{Artoisenet:2013puc}, with less than 2\% variation across the parameter ranges
studied.} Parton-level event samples are produced for specific values of Wilson coefficients by interfacing the
universal file output from FeynRules to the \amc event generator~\cite{Alwall:2014hca}. 
Higgs bosons are produced via gluon--gluon fusion with up to two additional partons in the final state using leading-order matrix elements. 

The generated events are passed to \pythia~\cite{Sjostrand:2007gs} to provide parton showering, hadronization and
underlying event and the zero-, one- and two-parton events are merged using the MLM matching scheme~\cite{Mangano:2006rw} 
to create the full final state. Event samples containing a Higgs boson produced either in association with a vector boson or via vector-boson fusion are produced using leading-order matrix
elements and passed through the \pythia\ generator. For each production mode, the Higgs boson mass is set to 125~\GeV\ and
events are generated using the \nnpdflo\ PDF set~\cite{Ball:2014uwa} and the A14 parameter
set~\cite{ATL-PHYS-PUB-2014-021}. All other Higgs boson production modes are assumed to occur as predicted by the SM.

Event samples are produced for different values of a given Wilson coefficient.  The particle-level differential cross
sections are produced using {\sc Rivet}~\cite{Buckley:2010ar}. The {\sc Professor} method~\cite{Buckley:2009bj} is used
to interpolate between these samples, for each bin of each distribution, to provide a parameterization of the
effective Lagrangian prediction. The parameterization function is determined using 11 samples when studying a single
Wilson coefficient, whereas 25 samples are used when studying two Wilson coefficients simultaneously. As the Wilson
coefficients enter the effective Lagrangian in a linear fashion, second-order polynomials are used to predict the cross
sections in each bin. The method was validated by comparing the differential cross sections obtained with the
parameterization function to the predictions obtained with dedicated event samples generated at the specific point in
parameter space. 

The model implemented in FeynRules fixes the Higgs boson width to be that of the SM, $\Gamma_{H} = 4.07$~\MeV
~\cite{Heinemeyer:2013tqa}. The cross sections are scaled by $\Gamma_{H} / (\Gamma_{H} + \Delta \Gamma)$, where $\Delta
\Gamma$ is the change in partial widths due to a specific choice of Wilson coefficient. The change in partial widths is
determined for each Higgs coupling using the partial-width calculator in \amc and normalized to reproduce the SM
prediction from {\sc Hdecay} \cite{Djouadi:1997yw}. 

The leading-order predictions obtained from \amc are reweighted to account for higher-order QCD and electroweak
corrections to the SM process, assuming that these corrections factorize from the new physics effects. The differential
cross section as a function of variable $X$ for a specific choice of Wilson coefficient, $c_i$, is given by
\begin{equation*} \label{eq:sigma}
\frac{{\rm d} \sigma}{{\rm d}X} = \sum_j \left( \frac{{\rm d}\sigma_j}{{\rm d}X} \right)^{\rm ref} \cdot \left(
\frac{{\rm d}\sigma_j}{{\rm d}X} \right)_{c_i}^{\rm MG5} / \left( \frac{{\rm d}\sigma_j}{{\rm d}X} \right)_{c_i=0}^{\rm
MG5},
\end{equation*}
where the summation $j$ is over the different Higgs boson production mechanisms, `MG5' labels the interpolated
\amc prediction and `ref' labels a reference sample for SM Higgs boson production. For the reference 
sample the default MC simulation is used. 

\begin{figure*}[!tbp]
\centering
 \subfloat[] {\includegraphics[width=0.75\columnwidth]{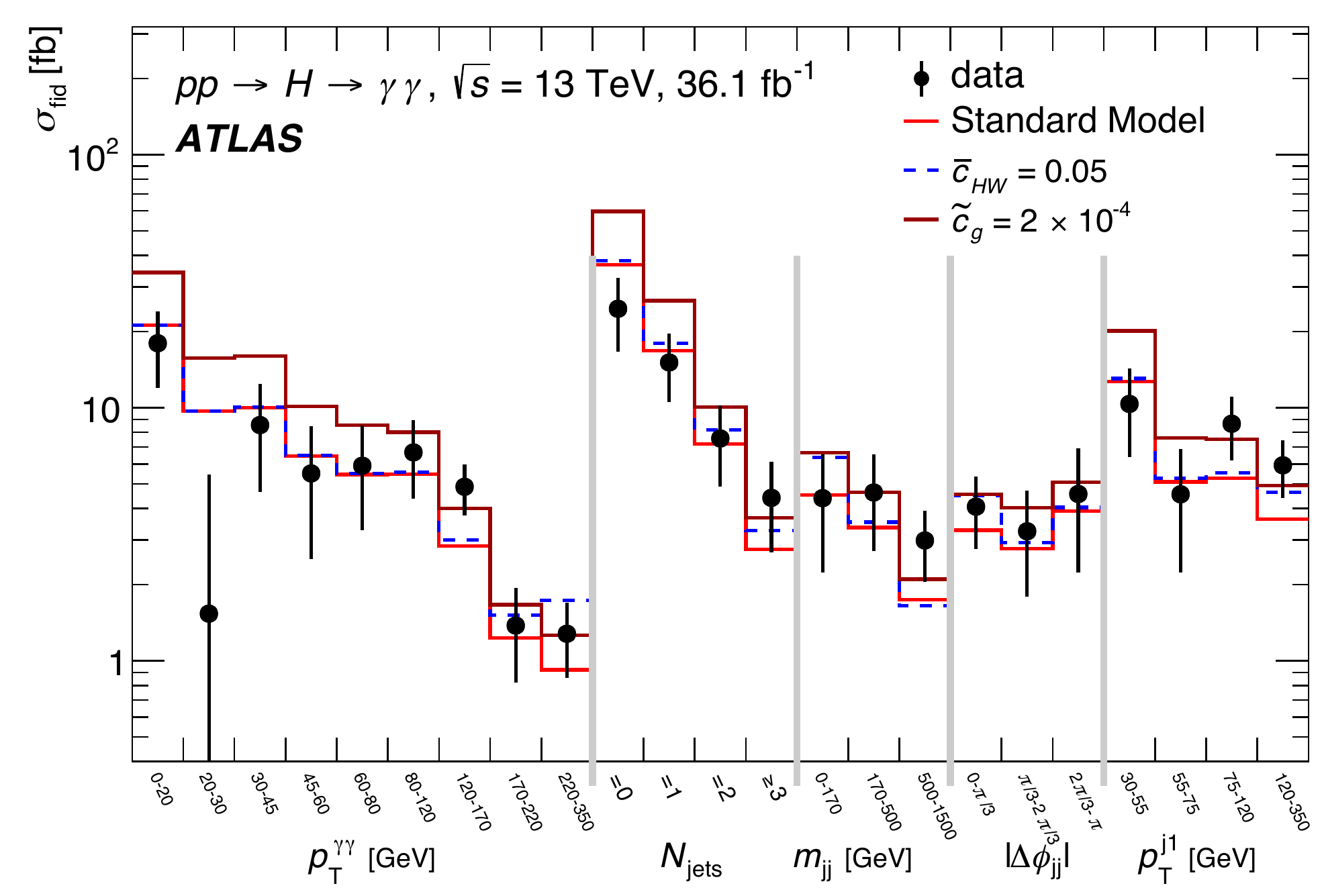}} \\
 \subfloat[] {\includegraphics[width=0.75\columnwidth]{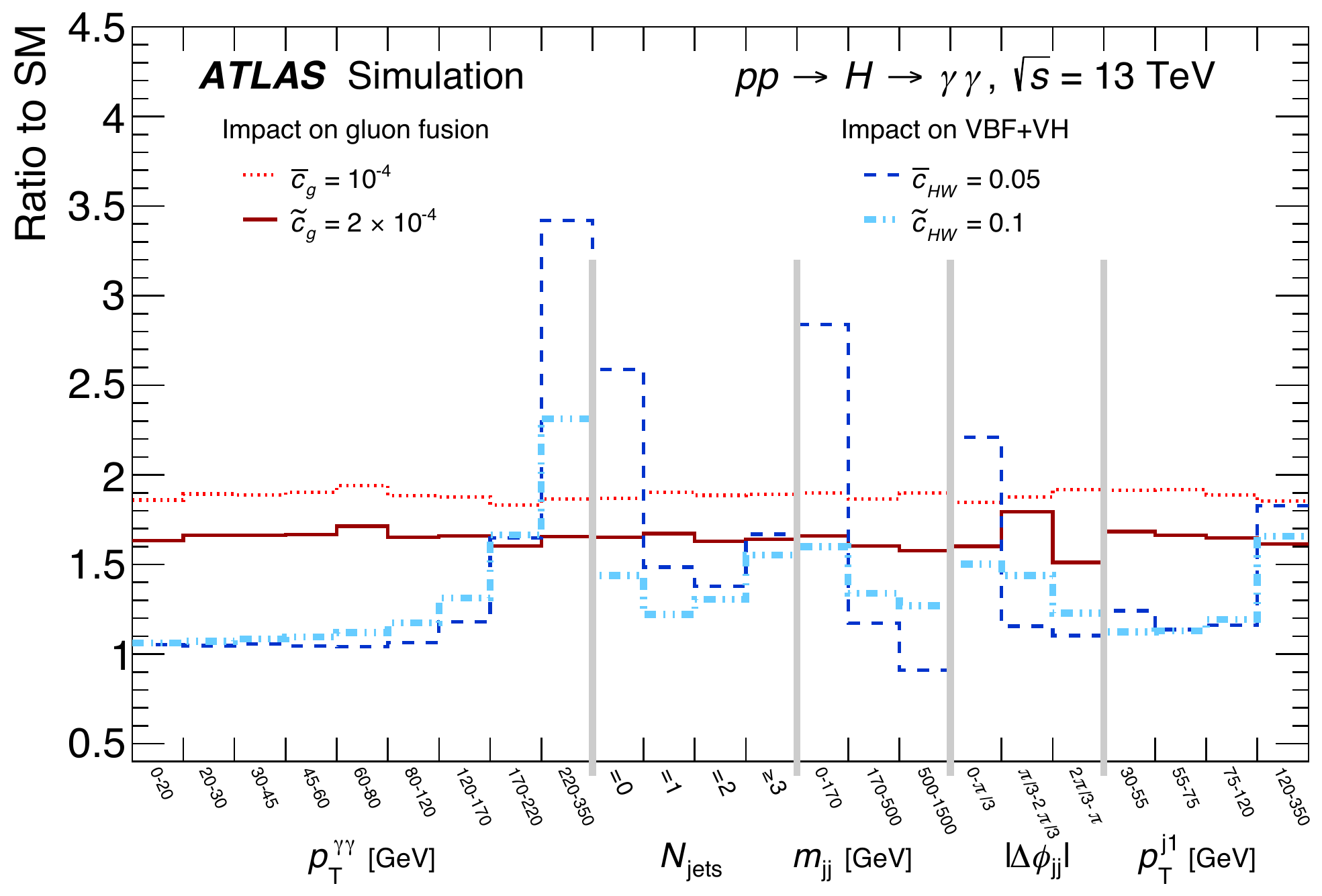}}
\caption{ 
(a) The measured differential cross sections as a function of \ptgg, \njet, \mjj, \dphijjabs, and \ptjl\ are compared to the SM
hypothesis and two non-SM hypotheses with $\bar{c}_{g}=1\times10^{-4}$ and $\bar{c}_{HW}=0.05$, respectively.
(b) Ratios of differential cross sections, as predicted for specific choices of Wilson coefficient, to the differential cross
sections predicted by the SM: the impact of non-zero $\bar c_g$ and $\tilde c_g$ is shown relative to the SM \ggH\ prediction,
while the impact of non-zero $\bar c_{HW}$ and $\tilde c_{HW}$ is shown relative to the SM \VBF+\VH\ prediction.
}
\label{fig:super-plot}
\end{figure*}

The measured differential cross sections of \ptgg, \njet, \mjj, \dphijjabs, and \ptjl\ are compared in
Figure~\ref{fig:super-plot}(a) to the SM hypothesis and to two non-SM hypotheses, specified by
$\bar{c}_{g}=2\times10^{-4}$ and $\bar{c}_{HW}=0.05$, respectively. The new CP-odd gluon--gluon fusion 
operator results in a large increase in rate and the additional CP-even $WH$ operator leads to a larger number of Higgs boson with sizeable \pT\ and an increased number of zero-jet events. 

The ratios of the expected differential cross sections to the SM predictions for some representative values of the Wilson
coefficients are shown in Figure~\ref{fig:super-plot}(b). The impact of the $\bar{c}_{g}$ and $\tilde{c}_{g}$ coefficients are
presented for the gluon--gluon fusion production: it displays a large change in the overall cross-section normalization.
The $\tilde{c}_{g}$ coefficient also changes the shape of the $\Delta\phi_{jj}$ distribution, which is expected from
consideration of the tensor structure of CP-even and CP-odd interactions~\cite{Klamke:2007cu, Andersen:2010zx}. In contrast, the
impact of the $\bar{c}_{HW}$ and $\tilde{c}_{HW}$ coefficients are presented specifically for the VBF$+$VH production channel: one expects large 
shape changes in all of the studied distributions and the $\Delta\phi_{jj}$ distribution is known to
discriminate between CP-even and CP-odd interactions in the \VBF\-production channel~\cite{Plehn:2001nj}.

Limits on Wilson coefficients are set by constructing a likelihood function 
\begin{align*}
 \mathcal{L} = \frac{1}{\sqrt{ \left(2\pi\right)^k | C |}} \exp \left( - \frac 1 2 \left( \vec \sigma_{\rm data} - \vec \sigma_{\rm pred} \right)^T C^{-1} \left(  \vec \sigma_{\rm data} - \vec \sigma_{\rm pred}  \right) \right) \, ,
\end{align*} 
where $\vec \sigma_{\rm data}$ and $\vec \sigma_{\rm pred}$ are $k$-dimensional vectors from the measured and predicted differential cross sections
of the \neftvars\ analyzed observables, $C = C_{\rm stat} + C_{\rm syst} + C_{\rm theo}$ is the $k\times k$ total covariance
matrix defined by the sum of the statistical, systematic and theoretical covariances, and $|C|$ denotes its
determinant. The theory covariance is constructed as outlined in Section~\ref{sec:theo} and includes no additional uncertainty to account for
the factorization assumption in Eq.~\ref{eq:sigma}. Based on this likelihood, one can construct a $\chi^2$ test to test the compatibility of the \neftvars\ distributions with the SM and 
a probability of 93\% is found. In what follows, the likelihood function is numerically maximized to determine $\mathcal{L}_{\rm max}$ and
confidence limits for one or several Wilson coefficients are determined via 
\begin{align*} 
 1 - \text{CL} = \int_{-2 \ln \mathcal{L}(c_i) + 2 \ln \mathcal{L}_{\rm max}  }^{\infty} \, \text{d} x \, f(x) \, , 
\end{align*} 
with $\mathcal{L}(c_i)$ denoting the likelihood value evaluated for a given Wilson coefficient value $c_i$, and $f(x)$
denoting the distribution of the test statistic. The coverage of the confidence limit is determined using ensembles of pseudo-experiments. Form factors are sometimes used to regularize the change of the cross section above a momentum scale $\Lambda_{\rm FF}$. This was investigated by reweighting the VBF$+$VH samples using form-factor predictions from {\sc VBFNLO~\cite{Arnold:2008rz}}. The impact on the $\bar c_{HW}$ and
$\tilde{c}_{HW}$ limits is negligible for $\Lambda_{\rm FF}>$1~\TeV.

In Table~\ref{tab:1D_scans}, the observed and expected 95\% CL limits for four Wilson coefficient fits are given. The limit for
$\bar c_{g}$ is derived by fixing all other Wilson coefficients to zero. This additional interaction can interfere
with the corresponding SM interaction and destructive interference causes the gluon--gluon fusion production-mode cross section to
be zero at $\bar c_{g} \sim -2.2 \times 10^{-4}$. The $\tilde c_{g}$ coefficient is also derived after setting all
Wilson coefficients to zero. Due to the CP conjugate structure of the interaction, no interference with the SM process is
possible. The 95\% CL limit for $\bar c_{HW}$ is obtained after setting $\bar c_{HB} = \bar c_{HW}$ to ensure that the partial
width for $H \to Z \, \gamma$ is unchanged from the SM prediction (Values of $\left| \bar c_{HW} - \bar c_{HB}
\right| > 0.03$ lead to a very large decay rate for the $H \to Z \gamma$ process that is contradicted by the
experimental constraints reported by ATLAS~\cite{Aad:2014fia,Aaboud:2017uhw}) and setting all other Wilson coefficients to zero.
Finally, the 95\% CL limit for $\tilde c_{HW}$ is given after setting $\tilde c_{HB} = \tilde c_{HW}$ to ensure a SM decay
rate for $H \to Z \, \gamma$  and all other Wilson coefficients to zero. The observed limits are improved by about a factor
of two compared to the Run 1 analysis of Ref.~\cite{Aad:2015tna}.

\begin{figure*}[!tbp]
 \centering
 \includegraphics[width=0.85\columnwidth]{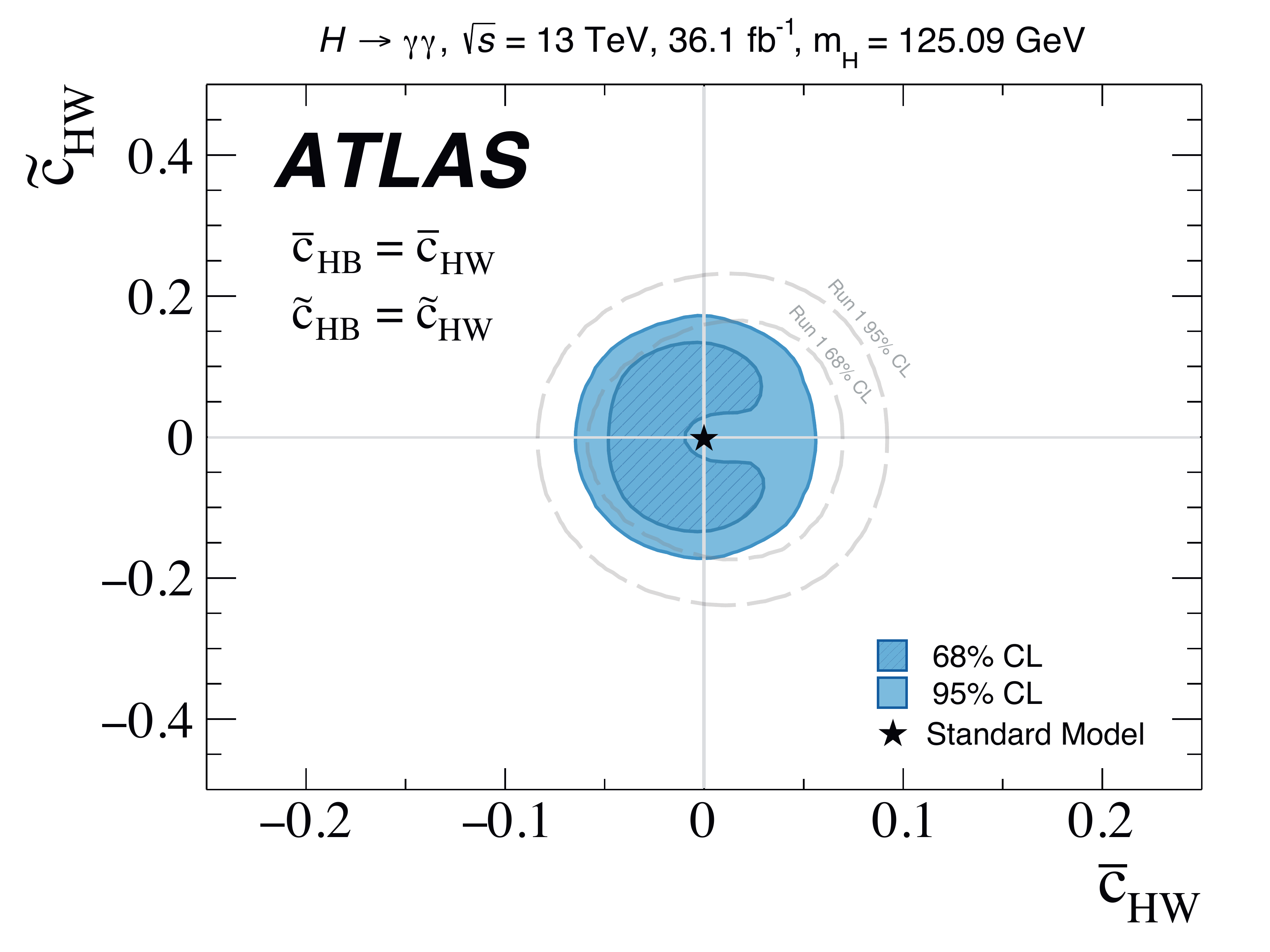}
 \caption{
 The observed 68\% (dark) and 95\% (light) confidence level regions from the simultaneous fit to the $\bar c_{HW}$ and $\tilde c_{HW}$ Wilson coefficients. The values of $\bar c_{HB}$ and $\tilde c_{HB}$ are set to be equal to $\bar c_{HW}$ and $\tilde c_{HW}$, respectively, and all other Wilson coefficients are set to zero, except for $\bar c_{HB}$ and
 $\tilde c_{HB}$ which are set to be equal to $\bar c_{HW}$ and $\tilde c_{HW}$, respectively. 
 The SM expectation at $(0,0)$ is also shown, together with the Run-1 confidence regions reported in Ref.~\cite{Aad:2015tna}.}
 \label{fig:heft_scans}
\end{figure*}

Figure~\ref{fig:heft_scans} shows the 68\% and 95\% confidence regions obtained from scanning $\bar c_{HW}$ and $\tilde
c_{HW}$ simultaneously, with the other two Wilson coefficients set to $\bar c_{HB} = \bar c_{HW}$ and $\tilde c_{HB} = \tilde c_{HW}$. All other Wilson coefficients are fixed at zero. The $\bar c_{HW}$ and $\tilde c_{HW}$ Wilson coefficients produce large shape changes in all distributions, as shown in Figure~\ref{fig:super-plot}, and the obtained limits are strongest when fitting all five distributions simultaneously. The shape of the observed 68\% confidence regions thus results from both shape and yield differences between data and expectations: 
the operators proportional to $\bar c_{HW}$ can destructively interfere with the SM contributions, a negative value of $\bar c_{HW}$ reduces the overall predicted cross section in the zero-jet and the lowest $m_{jj}$ bins, where deficits are observed in the data. The operators proportional to $\tilde c_{HW}$ can only increase the cross section from its SM value and can increase the predicted cross sections in the higher jet bins and the tails of the distributions (cf. Figure~\ref{fig:super-plot}). If only shape information is used to constrain the Wilson coefficients, the reported limits on $\bar c_{HW}$ and $\tilde c_{HW}$ weaken by about 20\% and 50\%, respectively. As also shown in Figure~\ref{fig:heft_scans}, these results display significant improvements on similar limits obtained from the Run-1 analysis~\cite{Aad:2015tna}. All reported results assume that QCD effects and new physics effects factorise. This assumption cannot be avoided with the current state-of-the-art implementation of the effective Lagrangian of Ref.~\cite{Contino:2013kra}. The full statistical and systematic correlations between measured distributions and all measured fiducial and differential cross sections are available in \hepdata\ to allow future interpretations with better models.

\begin{table}[!tp]
\caption{ Observed allowed ranges at 95\% CL for the $\bar c_g$ and $\bar c_{HW}$ Wilson coefficients and the
CP-conjugate coefficients. Limits on $\bar c_g$ and $\tilde c_g$ are each derived with all other Wilson coefficients set to
zero. Limits on $\bar c_{HW}$ and $\tilde c_{HW}$ are derived with $\bar c_{HB} = \bar c_{HW}$ and  $\tilde c_{HB} =
\tilde c_{HW}$, respectively. 
\label{tab:1D_scans} }
\centering
\vspace{2ex}
\begin{tabular}{ccc}
\hline\hline
 Coefficient & Observed 95\% CL limit & Expected 95\% CL limit  \\
\hline\hline
$\bar c_{g}$  & $[ -0.8, 0.1  ] \times 10^{-4}$ $\cup$ $[ -4.6, -3.8 ] \times 10^{-4}$  & $[ -0.4, 0.5  ] \times 10^{-4}$  $\cup$ $[ -4.9, -4.1 ] \times 10^{-4}$  \\
$\tilde c_{g}$ & $[ -1.0, 0.9 ] \times 10^{-4}$ & $[ -1.4, 1.3 ] \times 10^{-4}$  \\
$\bar c_{HW}$ & $[-5.7, 5.1] \times 10^{-2}$  & $[-5.0,5.0] \times 10^{-2}$ \\
$\tilde c_{HW}$ & $[-0.16,0.16]$ & $[-0.14,0.14]$ \\
\hline\hline
\end{tabular}

\end{table}

\clearpage

\section{Summary and conclusions} \label{sec:conclusion}

Measurements of Higgs boson cross sections in the Higgs boson to diphoton decay channel are performed using $pp$
collision data recorded by the ATLAS experiment at the LHC. The data were taken at a center-of-mass energy of $\sqrt{s} =
13 \, \TeV$ and correspond to an integrated luminosity of \lumi\,fb$^{-1}$. All measurements assume a Higgs boson mass 
of $125.09 \pm 0.24$~\GeV. The measured signal strength relative to the Standard Model expectation is
found to be:
\begin{align*}
\mu &= 0.99\ ^{+0.15}_{-0.14} \ .
\end{align*}
Signal strengths of the main production modes are measured separately via event reconstruction categories that are designed to be sensitive to the specific production modes. They are found to be:
\begin{align*}
\mu_\mathrm{ggH} = 0.81\ ^{+0.19}_{-0.18} \, , \quad
\mu_\mathrm{VBF} = 2.0\ ^{+0.6}_{-0.5}  \, , \quad
\mu_\mathrm{VH}  = 0.7\ ^{+0.9}_{-0.8}  \, , \quad \mathrm{and} \quad
\mu_\mathrm{top} = 0.5\ ^{+0.6}_{-0.6}  \, . 
\end{align*}
The total uncertainties of both the global and the production mode signal strengths is dominated by their 
respective statistical uncertainties. The global signal strength measurement improves on the precision of 
the previous ATLAS measurement in the diphoton channel by a factor of two~\cite{HIGG-2013-08}. The ggH (VBF) signal strength is 
measured to be $1~\sigma$ below ($2~\sigma$ above) the Standard Model expectation. The precision of the 
coupling-strength measurements involving top quarks improves by about a factor of three
compared to the previous ATLAS measurement in the diphoton channel~\cite{HIGG-2013-08}. These improvements result from a combination of the larger Higgs boson sample collected at $\sqrt{s} = 13$~\TeV, from the use of multivariate techniques to target the \VBF, \VH, and top-quark associated production modes more efficiently, from the improved precision of the \ggH\ Standard Model theory predictions, and from a significant reduction of some of the experimental uncertainties such as the photon energy resolution.
Production mode cross-section measurements for a Higgs boson of rapidity $|y_H| < 2.5$ for gluon--gluon fusion, vector-boson fusion, and Higgs boson production in association with vector bosons or a
top quark pair are reported:
\begin{align*}
\sigma_\mathrm{ggH} = 82 \asym{}{19}{18} \, \mathrm{fb}  \, , \quad
\sigma_\mathrm{VBF} = 16 \asym{}{5}{4} \, \mathrm{fb}  \, , \quad
\sigma_\mathrm{VH}  = 3 \err{4} \, \mathrm{fb} \, , \quad \mathrm{and} \quad
\sigma_\mathrm{top} = 0.7\asym{}{0.9}{0.7}\, \mathrm{fb}  \, .
\end{align*}
These values can be compared to the Standard Model expectations of 
\begin{align*}
\sigma_\mathrm{ggH}^\mathrm{SM} = \asym{102}{5}{7} \, \mathrm{fb}  \, , \quad
\sigma_\mathrm{VBF}^\mathrm{SM} = 8 \pm 0.2\, \mathrm{fb}  \, , \quad
\sigma_\mathrm{\VH}^\mathrm{SM} = 5 \pm 0.2\, \mathrm{fb}  \, , \quad
\sigma_\mathrm{top}^\mathrm{SM} = 1.3 \pm 0.1\, \mathrm{fb}  \, , \quad
\end{align*}
and show a similar level of agreement as that obtained with the coupling-strength measurements.

Nine measurements of so-called simplified template cross sections, designed to measure the different Higgs boson production processes in specific regions of phase space, are reported:
\begin{align*}
\sigma(\mathrm{ggH}, \mathrm{0~jet}) &= 37  \asym{}{16}{15} \, \mathrm{fb} \, , \\
\sigma(\mathrm{ggH}, \mathrm{1~jet}, \pT^{H} < 60\ \GeV) &= 13  \asym{}{13}{12} \, \mathrm{fb} \, , \\
\sigma(\mathrm{ggH}, \mathrm{1~jet}, 60 \leq \pT^{H} < 120\ \GeV) &= 5  \err{6}  \, \mathrm{fb}\, , \\
\sigma(\mathrm{ggH}, \mathrm{1~jet}, 120 \leq \pT^{H} < 200\ \GeV) &= 2.8  \asym{}{1.7}{1.6} \, \mathrm{fb} \, , \\
\sigma(\mathrm{ggH}, \geq 2~\mathrm{jet}) &= 20  \asym{}{9}{8} \, \mathrm{fb} \, , \\
\sigma(qq \rightarrow Hqq, \pT^{j} < 200~\GeV) &= 15  \asym{}{6}{5} \, \mathrm{fb} \, , \\
\sigma(ggH + qq \rightarrow Hqq, \mathrm{BSM-like}) &= 2.0  \err{1.4} \, \mathrm{fb} \, , \\
\sigma(\mathrm{VH}, \mathrm{leptonic}) &= 0.7  \asym{}{1.4}{1.3} \, \mathrm{fb} \, , \\
\sigma(\mathrm{top}) &= 0.7  \asym{}{0.8}{0.7} \, \mathrm{fb} \, .
\end{align*}
All reported results show agreement with the Standard Model expectation.

Higgs boson coupling-strength modifiers are reported and two models are investigated: the first one reports results on effective coupling-strength modifiers for Higgs boson production in gluon--gluon fusion and decay, $\kappa_g$ and $\kappa_\gamma$, respectively. They are found to be:
\begin{align*}
 \kappa_g      = 0.76^{+0.17}_{-0.14} \, , \quad \mathrm{and} \quad
 \kappa_\gamma = 1.16^{+0.14}_{-0.14} \, .
\end{align*}
The second model resolves the Higgs boson production and decay loops in terms of the more fundamental fermionic and vector boson couplings under the assumption of universal coupling-strength modifiers for all fermions and vector bosons, namely $\kappa_V$ and $\kappa_F$, respectively.
They are found to be:
\begin{align*}
 \kappa_V = 0.92^{+0.08}_{-0.07} \, , \quad \mathrm{and} \quad
 \kappa_F = 0.64^{+0.18}_{-0.14}  \, .
\end{align*}

Fiducial cross-section measurements are reported for a Higgs boson decaying into two isolated photons with
transverse momentum greater than 35\% and 25\% of the diphoton invariant mass (corresponding to a photon $p_T$ of $43.8$ \GeV\ and $31.3$ \GeV), and with $|\eta|<2.37$, excluding the region of $1.37 < |\eta|<1.52$. The total fiducial cross section is measured to be
\begin{align*}
 \sigma_{\rm fid} = \fidxs \, , 
 \end{align*}
and is in agreement with the Standard Model expectation of \fidxsSM. Additional cross sections in
fiducial regions probing Higgs boson production from vector-boson fusion or associated with large missing transverse
momentum, leptons or top quarks are reported. The cross section for the \VBF-enhanced region is measured to be
\begin{align*}
 \sigma_{\rm VBF-enhanced} = \fidxsVBF\, ,
\end{align*}
which is to be compared with the Standard Model prediction of \fidxsSMVBF. The larger measured cross section is 
consistent with the \VBF\ signal-strength measurement reported above, if one scales the expected SM contributions from \VBF\ (about 65\%) and \ggH\ (about 35\%) in this fiducial region with the corresponding measured signal strengths. For the remaining fiducial regions, limits at 95\% CL are reported
\begin{align*}
 \sigma_{N_{\rm lepton} \ge 1}  < 1.39 \,  \mathrm{fb} \,\, , \quad
 \sigma_{{\rm High} \, E_{\rm T}^{\rm miss}}  < 1.00 \, \mathrm{fb} \, , \quad \mathrm{and} \quad
 \sigma_{\ttH{\rm -enhanced}}  < 1.27 \, \mathrm{fb} \, ,
\end{align*}
which can be compared with the Standard Model expectations of $0.57 \pm 0.03 \, \mathrm{fb}$, $0.30 \pm 0.02 \, \mathrm{fb}$, 
and $0.55 \pm 0.06 \, \mathrm{fb}$, respectively.

The fiducial cross sections for different jet multiplicities is reported and compared to several theoretical predictions.
Eleven differential cross sections and two double-differential cross sections are reported for events belonging to the
inclusive diphoton fiducial region, as a function of kinematic variables of the diphoton system or of
jets produced in association with the Higgs boson. The reported cross sections are sensitive to the Higgs boson
production kinematics, the jet kinematics, the spin and CP quantum numbers of the Higgs boson, and the \VBF\
production mechanism. 
All measured differential cross sections are compared to predictions and no significant deviation from 
the Standard Model expectation is observed. The full statistical and systematic correlations between measured distributions
are determined and are available in \hepdata\ along with the central values of the measured fiducial and differential
cross sections to allow future comparisons and interpretations. 

The strength and tensor structure of the Higgs boson interactions is investigated using \neftvars\ differential variables and an effective Lagrangian, which introduces additional CP-even and CP-odd interactions. No significant new physics contributions are observed and the reported 68\% and 95\% limits on such contributions have improved by a factor of two in comparison to the previous ATLAS measurement. 

The measurements presented in this paper lay the foundation for further studies. All reported results are statistically limited and their precision will further improve with the full data set to be recorded during Run~2 of the LHC.

\clearpage


\section*{Acknowledgments}


We thank CERN for the very successful operation of the LHC, as well as the
support staff from our institutions without whom ATLAS could not be
operated efficiently.

We acknowledge the support of ANPCyT, Argentina; YerPhI, Armenia; ARC, Australia; BMWFW and FWF, Austria; ANAS, Azerbaijan; SSTC, Belarus; CNPq and FAPESP, Brazil; NSERC, NRC and CFI, Canada; CERN; CONICYT, Chile; CAS, MOST and NSFC, China; COLCIENCIAS, Colombia; MSMT CR, MPO CR and VSC CR, Czech Republic; DNRF and DNSRC, Denmark; IN2P3-CNRS, CEA-DRF/IRFU, France; SRNSFG, Georgia; BMBF, HGF, and MPG, Germany; GSRT, Greece; RGC, Hong Kong SAR, China; ISF, I-CORE and Benoziyo Center, Israel; INFN, Italy; MEXT and JSPS, Japan; CNRST, Morocco; NWO, Netherlands; RCN, Norway; MNiSW and NCN, Poland; FCT, Portugal; MNE/IFA, Romania; MES of Russia and NRC KI, Russian Federation; JINR; MESTD, Serbia; MSSR, Slovakia; ARRS and MIZ\v{S}, Slovenia; DST/NRF, South Africa; MINECO, Spain; SRC and Wallenberg Foundation, Sweden; SERI, SNSF and Cantons of Bern and Geneva, Switzerland; MOST, Taiwan; TAEK, Turkey; STFC, United Kingdom; DOE and NSF, United States of America. In addition, individual groups and members have received support from BCKDF, the Canada Council, CANARIE, CRC, Compute Canada, FQRNT, and the Ontario Innovation Trust, Canada; EPLANET, ERC, ERDF, FP7, Horizon 2020 and Marie Sk{\l}odowska-Curie Actions, European Union; Investissements d'Avenir Labex and Idex, ANR, R{\'e}gion Auvergne and Fondation Partager le Savoir, France; DFG and AvH Foundation, Germany; Herakleitos, Thales and Aristeia programmes co-financed by EU-ESF and the Greek NSRF; BSF, GIF and Minerva, Israel; BRF, Norway; CERCA Programme Generalitat de Catalunya, Generalitat Valenciana, Spain; the Royal Society and Leverhulme Trust, United Kingdom.

The crucial computing support from all WLCG partners is acknowledged gratefully, in particular from CERN, the ATLAS Tier-1 facilities at TRIUMF (Canada), NDGF (Denmark, Norway, Sweden), CC-IN2P3 (France), KIT/GridKA (Germany), INFN-CNAF (Italy), NL-T1 (Netherlands), PIC (Spain), ASGC (Taiwan), RAL (UK) and BNL (USA), the Tier-2 facilities worldwide and large non-WLCG resource providers. Major contributors of computing resources are listed in Ref.~\cite{ATL-GEN-PUB-2016-002}.


\clearpage

\clearpage


\appendix
\part*{Appendix}\label{sec:appendix}
\addcontentsline{toc}{part}{Appendix}


\section{Simplified template cross-section framework}
\label{sec:appendix_STXS}

As introduced in Section~\ref{sec:intro_stxs}, this paper includes
cross-section measurements using the so called ``stage-1'' of the 
simplified template cross-section framework~\cite{LesHouches,deFlorian:2016spz}.
In the full stage-1 proposal, template cross sections are defined
in 31 regions of phase space with $|y_H|<\nobreak2.5$. These regions 
have been chosen to minimize the dependence on theoretical
uncertainties and isolate possible BSM effects, while maximizing 
the experimental sensitivity. 

The 31 regions of particle-level phase space corresponding to the stage 1 of the
template cross-section approach are the following~\cite{LesHouches,deFlorian:2016spz}:
\begin{itemize}
\item Gluon--gluon fusion (11 regions). Gluon--gluon fusion events, together
  with $gg\to ZH$ events followed by hadronic decays of the $Z$ boson,
  are split according to the number of jets in the event in 0, 1, and $\ge 2$-jet events.
  Jets are reconstructed from all stable particles\footnote{The Higgs boson is treated as stable and consequently its decay products are removed from the jet finding.} 
  with lifetime greater than 10~ps using the anti-$k_t$ algorithm~\cite{Cacciari:2008gp} with a jet
  radius parameter $R=0.4$  and must have $\pT>30~\GeV$. The region containing two or more jets is
  split into two, with one of the two subregions (``VBF-like'') 
  containing events
  with a topology similar to vector-boson fusion events
  (invariant mass of the leading-\pT\ jet pair $\mjj > 400~\GeV$, and
  rapidity separation between the two jets $\deltayjj > 2.8$).
  The one-jet and non-VBF-like two-jet regions are further split according to 
  the transverse momentum of the Higgs boson in ``low'' (0--60~\GeV),
  ``medium'' (60--120~\GeV), ``high'' (120--200~\GeV) and ``BSM''
  ($>200~\GeV$) regions.
  The VBF-like events are further split into exclusive two-jet-like
  and inclusive three-jet-like events through a requirement on the transverse momentum
  \pTHjj\ of the system formed by the two photons and the two
  leading-\pT\ jets ($\pTHjj < 25~\GeV$ or $\pTHjj > 25~\GeV$, respectively).
  The separation between events with zero, one, or two or more jets
  probes perturbative QCD predictions.
  Events containing a very high transverse momentum Higgs-boson of more than 200~\GeV\ are sensitive
  to BSM contributions, such as those from loop-induced amplitudes mediated by hypothetical particles heavier than the top-quark.
   
\item Vector-boson fusion (5 regions). Vector-boson fusion events, and $VH$ events followed by hadronic
  $V$-boson decays, are first split according to the $\pT$ of the leading jet.
  Events that contain at least one jet with a transverse momentum greater than 200~\GeV,
  which are sensitive to BSM contributions, are measured
  separately in a ``VBF BSM'' category.
  The remaining events are separated into \VBF-like events, \VH-like events, and events that have a \ggH-like topology (referred to as ``Rest'').
  VBF-like events satisfy the same $\mjj$ and $\deltayjj$ requirements as for the gluon--gluon fusion VBF-like category and are
  similarly split into ``two-jet'' and ``$\ge 3$-jet'' events by requiring $\pTHjj < 25$~\GeV\ or $\pTHjj > 25$~\GeV,
  respectively.
  \VH-like events are selected by requiring that they have at least two jets and an invariant mass of $60~\GeV < \mjj < 120~\GeV$.

\item Associated production with vector bosons decaying to leptons (11 regions).
  \VH\ events are first split according to their production mode ($q\bar{q}' \to WH$, $q\bar{q}\to ZH$, or $gg \to ZH$).
  Events are separated further into regions of the vector boson transverse momentum $\pt^V$, and of jet multiplicity.
  For $gg \to ZH$, two regions are defined with $\pT^V $ (``low'': 0--150~\GeV, and ``high'': > 150~\GeV). The ``high-$\pT^V$'' $gg\to ZH$ region
  is further split into zero-jet and $\ge 1$-jet regions.
  Regions sensitive to BSM contributions with $\pt^V > 250$~\GeV\ are defined for the $q\bar{q}\to VH$ production modes and 
  two further $\pT^V$ regions are defined (``low'': 0--150~\GeV, and ``high'': 150--250~\GeV).
  The ``high-$\pT^V$'' $q\bar{q}\to VH$ region is finally split into zero-jet and $\ge 1$-jet regions.
  
\item Associated production with top and bottom quarks (4 regions). \ttH, $t$-channel \tH,
  $W$-associated \tH, and \bbH\ events are classified according to their production mode,
  with no further separation into specific regions of phase space. 
\end{itemize}

Table~\ref{tab:SMacc} summarizes the acceptances for each of the stage-1 STXS \hgg\ regions, and for five \qqH\ regions, split into their \VBF, \WH, and \ZH\ respective contributions. The table also lists the summed acceptance for the 11 \VH\ leptonic regions, separately for the $gg \to ZH$, $q\bar{q}' \to WH$ and $q\bar{q} \to ZH$ processes. Finally, the acceptances are shown for the rarer production processes: \ttH, $t$-channel \tH, $W$-associated \tH, and \bbH. All STXS regions require $|y_H| < 2.5$ and are determined using the samples summarized in Table~\ref{mc_table}.

\begin{table}[!tp]
\caption{
The SM acceptances of stage-1 STXS regions useful to the results presented in this paper.  For the \hgg\ regions each 
acceptance is relative to inclusive \hgg\ production; for all other regions, each acceptance is relative to 
the inclusive process shown at the top of the column.  All regions require $|y_H| < 2.5$. 
}
\begin{center}
\renewcommand{\arraystretch}{1.2}
\begin{tabular}{lrrr}
\hline \hline
\hgg\ regions & 0-jet & 1-jet & $\geq 2$-jet \\
\hline\hline
$p_{\textrm T}^H < 60$ \GeV & 0.562 & 0.134 & 0.025 \\
$60\,\GeV \leq p_{\textrm T}^H <120$ \GeV & - & 0.093 & 0.038 \\
$120\,\GeV \leq p_{\textrm T}^H <200$ \GeV & - & 0.015 & 0.020 \\
$p_{\textrm T}^H \geq 200$ \GeV & - & 0.003 & 0.009 \\  
VBF-like & & & \\
~~$p_{\textrm T}^{Hjj} <25$ \GeV & - & - & 0.006 \\
~~$p_{\textrm T}^{Hjj} \geq 25$ \GeV & - & - & 0.007 \\
\hline \hline
\qqH\ regions & VBF & $q\bar{q}' \to WH$ & $q\bar{q} \to ZH$ \\
\hline \hline
$p_{\textrm T}^j \geq 200$ \GeV  & 0.043 & 0.027 & 0.029 \\
$p_{\textrm T}^j < 200$ \GeV     & & & \\
~~VH-like                           & 0.023 & 0.189 & 0.224 \\
~~Rest                              & 0.556 & 0.368 & 0.363 \\
~~VBF-like & & & \\
~~~~$p_{\textrm T}^{Hjj} <25$ \GeV     & 0.235 & 0.002 & 0.002 \\
~~~~$p_{\textrm T}^{Hjj} \geq 25$ \GeV & 0.074 & 0.007 & 0.008 \\
\hline \hline
VH, leptonic region & $gg \to ZH$ & $q\bar{q}' \to WH$ & $q\bar{q} \to ZH$ \\
\hline \hline
 & 0.289 & 0.286 & 0.265 \\
\hline \hline
Top region & \ttH & $t$-ch. \tH & $W$-assoc. \tH \\
\hline \hline
 & 0.987 & 0.921 & 0.989 \\
\hline \hline
Beauty region & \bbH \\
\hline \hline
  & 0.945  \\

\hline \hline
\label{tab:SMacc}
\end{tabular}
\end{center}
\end{table}

\clearpage


\section{Minimally merged simplified template cross sections}
\label{sec:appendix_MINSTXS}

\begin{figure}[ht!]
\centering
\includegraphics[width=.82\textwidth]{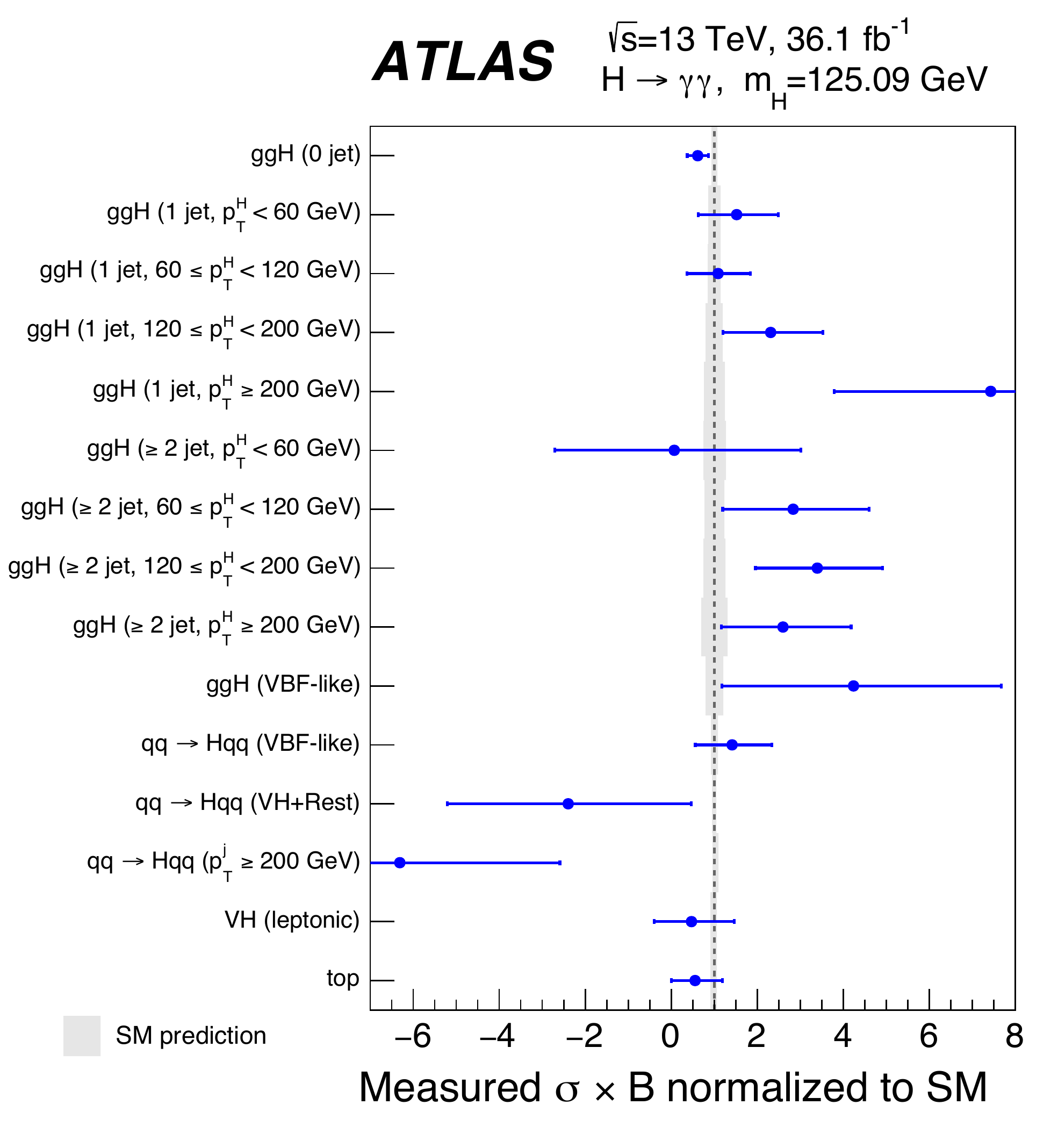}
\caption{Summary plot of the measured simplified template cross sections times
 the Higgs to diphoton branching ratio, as defined in Table~\ref{tab:STXS_WEAK}. 
For illustration purposes, the central values have been divided by their SM 
expectations but no additional SM uncertainties have been folded into the measurement. 
The uncertainties in the SM predicted cross sections are shown in gray in the plot. 
The fitted value of $\sigma (\mathrm{top})$ corresponds to the sum of the \ttH, \tHqb, and \tHW\
production-mode cross sections under the assumption that their relative ratios are as predicted by the SM. The $\sigma (\mathrm{VH, leptonic})$ cross-section values are determined under the assumption that the ratio of the \WH\ and \ZH\ production mode cross
sections is as predicted by the SM and includes production from both the quark and gluon initial states. The \bbH\
contributions are merged with \ggH.
}
\label{fig:stxsResults_WEAK}
\end{figure}

In this appendix, the measurement for a minimally merged set of fifteen simplified template cross section 
regions is presented. The merged regions are defined in Table~\ref{tab:STXS_WEAK} and the extracted
cross sections are summarized in Table~\ref{tab:stxsWEAK_results} and Figure~\ref{fig:stxsResults_WEAK}.

\begin{table}[!tp]
  \small
  \caption{The kinematic regions of the stage 1 of the simplified template cross sections,
    along with the intermediate (minimally merged set of) regions 
    used for the measurements presented in this appendix.
    The $VH$-like, VBF-like, and rest regions are defined as in Table~\ref{tab:STXS} and Appendix~\ref{sec:appendix_STXS}.  All regions require $|y_H|<2.5$.
    The leading jet transverse momentum is denoted by $\pT^j$.
    In total, the cross sections for fifteen kinematic regions are measured.}
\begin{center}
\begin{tabular}{lll}
\hline \hline
Process & Measurement region & Particle-level stage 1 region \\
\hline\hline
\ggH\ + $gg\to Z(\to qq)H$ & 0-jet & 0-jet \\
                        & 1-jet, $p_{\textrm T}^H < 60$ \GeV & 1-jet, $p_{\textrm T}^H < 60$ \GeV \\
                        & 1-jet, $60 \leq p_{\textrm T}^H <120$ \GeV & 1-jet, $60 \leq p_{\textrm T}^H <120$ \GeV \\
                        & 1-jet, $120 \leq p_{\textrm T}^H <200$ \GeV& 1-jet, $120 \leq p_{\textrm T}^H <200$ \GeV \\
                        & 1-jet, $p_{\textrm T}^H >200$ \GeV& 1-jet, $p_{\textrm T}^H >200$ \GeV \\
                        & $\geq 2$-jet, $p_{\textrm T}^H < 60$ \GeV & $\geq 2$-jet, $p_{\textrm T}^H < 60$ \GeV \\
                        & $\geq 2$-jet, $60 \leq p_{\textrm T}^H <120$ \GeV & $\geq 2$-jet, $60 \leq p_{\textrm T}^H <120$ \GeV \\
                        & $\geq 2$-jet, $120 \leq p_{\textrm T}^H <200$ \GeV & $\geq 2$-jet, $120 \leq p_{\textrm T}^H <200$ \GeV \\
                        & $\geq 2$-jet, $p_{\textrm T}^H >200$ \GeV & $\geq 2$-jet, $p_{\textrm T}^H >200$\, \GeV \\
                        & VBF-like & VBF-like, $p_{\textrm T}^{Hjj} <25$ \GeV \\
                        & & VBF-like, $p_{\textrm T}^{Hjj} \geq 25$ \GeV \\
\hline
$qq' \to H qq'$ (VBF + $VH$) & $p_{\textrm T}^j < 200$ \GeV, VBF-like & $p_{\textrm T}^j < 200$ \GeV, VBF-like, $p_{\textrm T}^{Hjj} <25$ \GeV  \\
                             & & $p_{\textrm T}^j < 200$ \GeV, VBF-like, $p_{\textrm T}^{Hjj} \geq 25$ \GeV \\
                             & $p_{\textrm T}^j < 200$ \GeV, VH+Rest & $p_{\textrm T}^j < 200$ \GeV, VH-like                      \\
                             & & $p_{\textrm T}^j < 200$ \GeV, Rest                         \\
                             & $p_{\textrm T}^j > 200$ \GeV, BSM-like & $p_{\textrm T}^j > 200$ \GeV \\ 
\hline
$VH$ (leptonic decays) & $VH$ leptonic & $q\bar{q}\to ZH$, $\pT^Z<150$ \GeV \\
                       &               & $q\bar{q}\to ZH$, $150 <\pT^Z<250$ \GeV, 0-jet \\
                       &               & $q\bar{q}\to ZH$, $150 <\pT^Z<250$ \GeV, $\ge 1$-jet \\
                       &               & $q\bar{q}\to ZH$, $\pT^Z>250$ \GeV\\
                       &               & $q\bar{q}\to WH$, $\pT^W<150$ \GeV \\
                       &               & $q\bar{q}\to WH$, $150 <\pT^W<250$ \GeV, 0-jet \\
                       &               & $q\bar{q}\to WH$, $150 <\pT^W<250$ \GeV, $\ge 1$-jet \\
                       &               & $q\bar{q}\to WH$, $\pT^W>250$ \GeV\\
                       &               & $gg\to ZH$, $\pT^Z<150$ \GeV \\
                       &               & $gg\to ZH$, $\pT^Z>150$ \GeV, 0-jet \\
                       &               & $gg\to ZH$, $\pT^Z>150$ \GeV, $\ge 1$-jet \\
\hline
Top-associated production & top          & \ttH \\
                          &              & $\tHW$ \\
                          &              & $\tHqb$ \\
\hline
\bbH                      & merged w/ \ggH & \bbH \\
\hline \hline
\label{tab:STXS_WEAK}
\end{tabular}
\end{center}
\end{table}

\begin{table}[!tp]
\caption{
  Best-fit values and uncertainties of the simplified template cross sections times branching ratio, as defined in Table~\ref{tab:STXS_WEAK}. The SM predictions~\cite{deFlorian:2016spz} are shown for each region. 
}
\begin{center}
\renewcommand{\arraystretch}{1.2}
\begin{tabular}{lrlr@{}lr@{}l@{ }lr@{ }l}
\hline \hline
Measurement region & \multirow{2}{*}{Result} & \multicolumn{6}{c}{ Uncertainty } & \multicolumn{2}{c}{\multirow{2}{*}{SM prediction}} \\
($|y_H| < 2.5$)    & & Total & & Stat. & Syst. &&&& \\
\hline\hline
$ggH, \mathrm{0~jet}$   &  38 & \asym{}{16}{15} & \lg & \err{14} & \asym{}{6}{5} & \rg & fb & $63 \pm 5$ & fb \tspp \\
$ggH, \mathrm{1~jet}, \pT^{H} < 60$ \GeV  &  23 & \asym{}{14}{13} & \lg & \err{13} & \asym{}{5}{4} & \rg & fb & $15 \pm 2$ & fb \tspp \\
$ggH, \mathrm{1~jet}, 60 \leq \pT^{H} < 120$ \GeV & 11 & \err{8} & \lg & \err{7} & \asym{}{3}{2} & \rg & fb & $10 \pm 2$ & fb \tspp \\
$ggH, \mathrm{1~jet}, 120 \leq \pT^{H} < 200$ \GeV & 4.0 & \asym{}{2.1}{1.9} & \lg & \err{1.8} & \asym{}{0.9}{0.6} & \rg & fb & $1.7 \pm 0.3$ & fb \tspp \\
$ggH, \mathrm{1~jet}, \pT^{H} \geq 200$ \GeV & 2.6 & \asym{}{1.6}{1.2} & \lg & \asym{}{1.3}{1.1} & \asym{}{0.8}{0.5} & \rg & fb & $0.4 \pm 0.1$ & fb \tspp \\
$ggH, \geq 2~\mathrm{jet}, \pT^{H} < 60$ \GeV & 0 & \err{8} & \lg & \err{8} & \asym{}{3}{2} & \rg & fb & $3 \pm 1$ & fb \tspp \\
$ggH, \geq 2~\mathrm{jet}, 60 \leq \pT^{H} < 120$ \GeV & 12 & \asym{}{8}{7} & \lg & \err{7} & \asym{}{3}{2} & \rg & fb & $4 \pm 1$ & fb \tspp \\
$ggH, \geq 2~\mathrm{jet}, 120 \leq \pT^{H} < 200$ \GeV & 7.9 & \asym{}{3.5}{3.4} & \lg & \err{3.3} & \asym{}{1.1}{0.9} & \rg & fb & $2.3 \pm 0.6$ & fb \tspp \\
$ggH, \geq 2~\mathrm{jet}, \pT^{H} \geq 200$ \GeV & 2.6 & \asym{}{1.6}{1.4} & \lg & \asym{}{1.5}{1.4} & \asym{}{0.6}{0.5} & \rg & fb & $1.0 \pm 0.3$ & fb \tspp \\
$ggH, \mathrm{VBF-like}$ & 6.2 & \asym{}{5.0}{4.5} & \lg & \err{4.1} & \err{1.2} & \rg & fb & $1.5 \pm 0.3$ & fb \tspp \\
$qq \rightarrow Hqq, \mathrm{VBF-like}$ & 3.8 & \asym{}{2.5}{2.3} & \lg & \asym{}{2.2}{2.0} & \err{1.2} & \rg & fb & $2.7 \pm 0.2$ & fb \tspp \\
$qq \rightarrow Hqq, \mathrm{VH+Rest}$ & -19 & \err{22} & \lg & \asym{}{21}{20} & \asym{}{6}{7} & \rg & fb & $7.7 \pm 0.4$ & fb \tspp \\
$qq \rightarrow Hqq, \pT^{j} > 200$ \GeV & -3.2 & \asym{}{1.9}{2.0} & \lg & \err{1.7} & \asym{}{0.7}{0.9} & \rg & fb & $0.5 \pm 0.1$ & fb \tspp \\
VH, leptonic & 0.7 & \asym{}{1.4}{1.2} & \lg & \asym{}{1.4}{1.2} & \asym{}{0.4}{0.3} & \rg & fb & $1.4 \pm 0.1$ & fb \tspp \\
Top          & 0.7 & \asym{}{0.8}{0.7} & \lg & \asym{}{0.8}{0.7} & \asym{}{0.2}{0.1} & \rg & fb & $1.3 \pm 0.1$ & fb \tspp \\
\hline \hline
\end{tabular}
\end{center}
\label{tab:stxsWEAK_results}
\end{table}

\clearpage


\section{ Additional unfolded differential cross sections}\label{app:add_fid_meas}

This appendix presents additional measurements and comparisons to theoretical predictions to those discussed in Section~\ref{sec:details_diff}.

\begin{figure*}[!htbp]
  \centering
	\subfloat[] {\includegraphics[width=0.50\columnwidth]{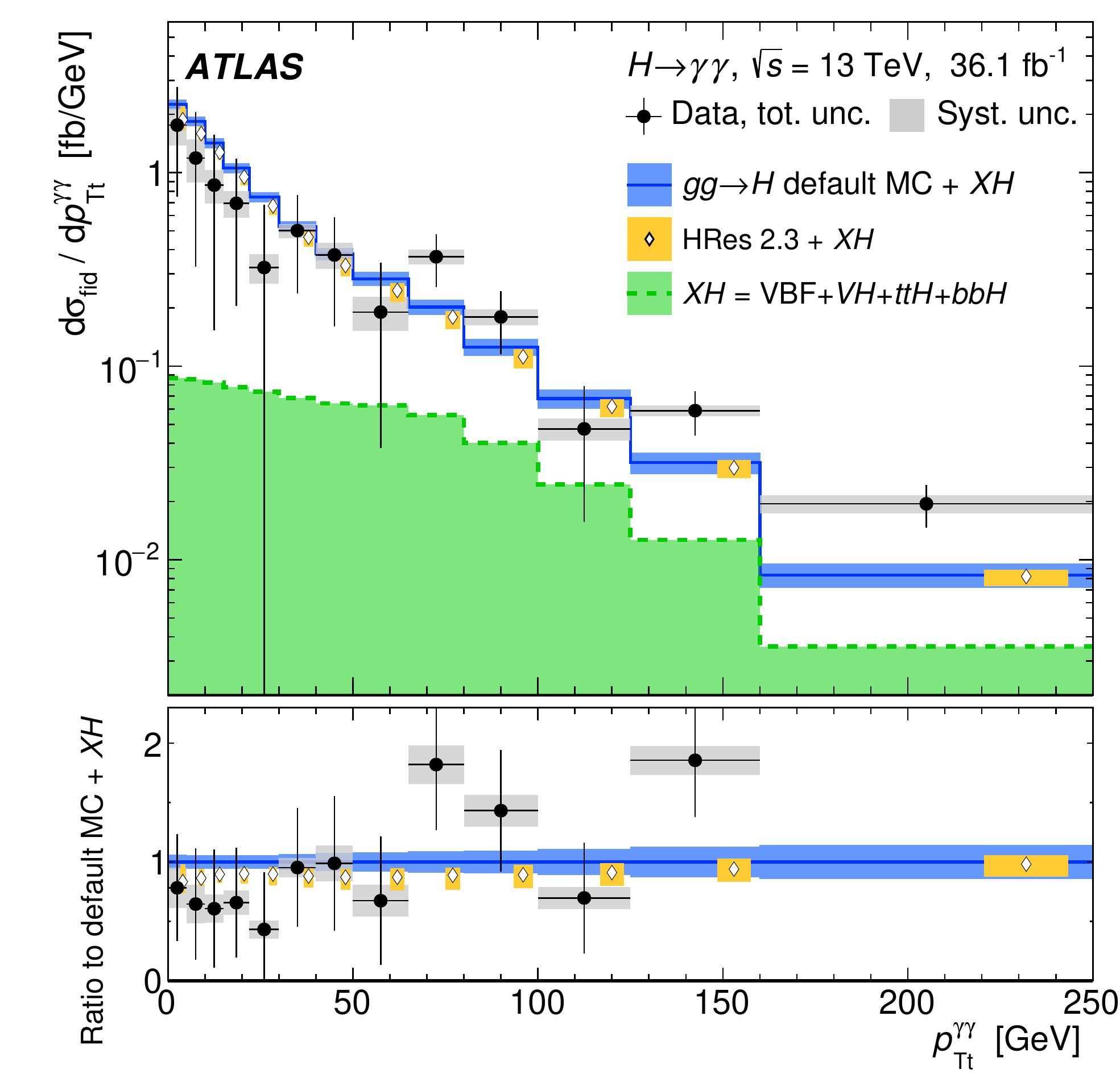}}
	\subfloat[] {\includegraphics[width=0.50\columnwidth]{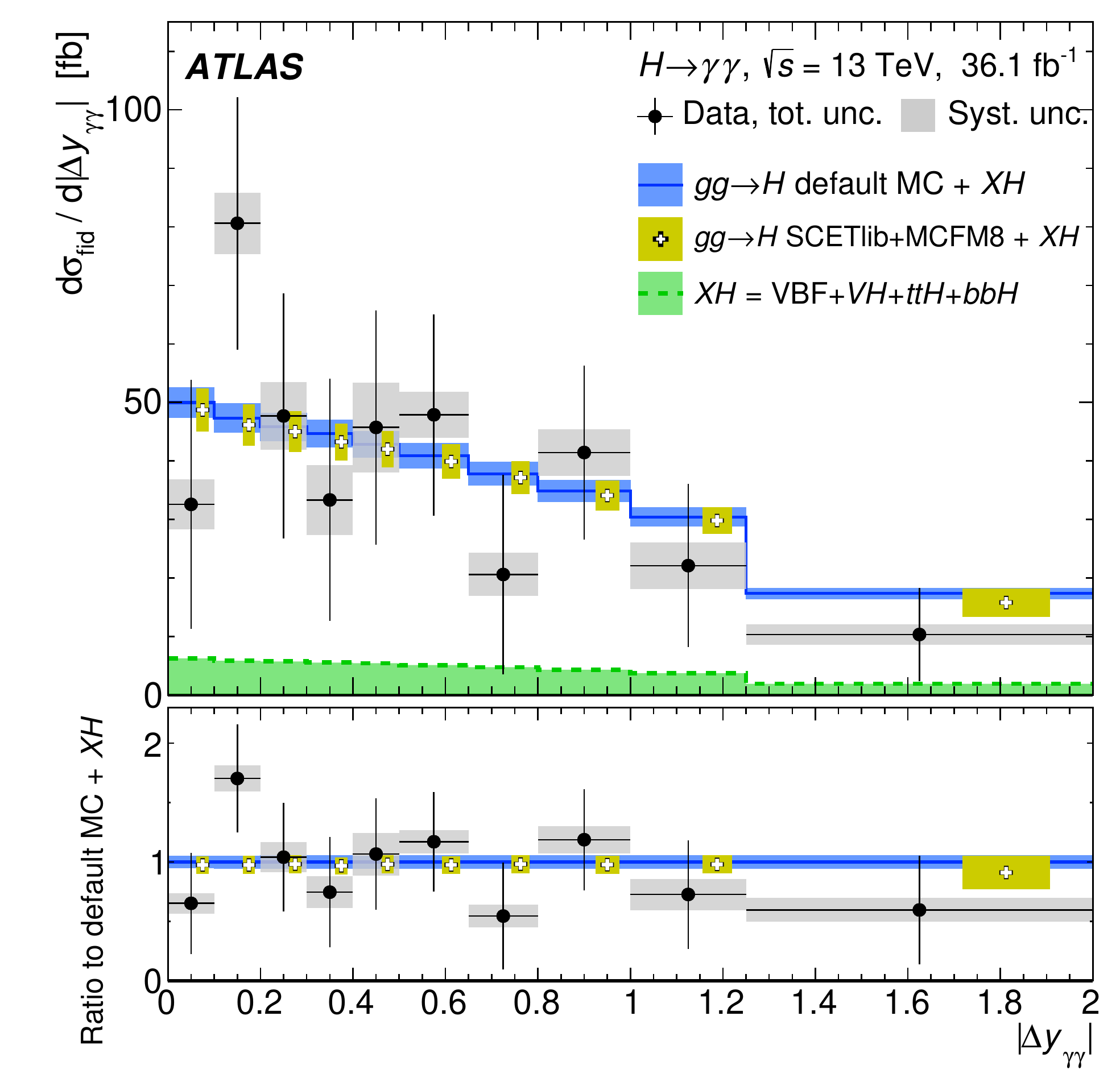}} 
  \caption{
   The differential cross sections for $pp \to H \to \gamma \gamma$ as a function of (a) \pttgg\ and (b) \deltaygg\
   are shown and compared to the SM expectations. The data are shown as filled (black) circles. The vertical error bar on
   each data point represents the total uncertainty in the measured cross section and the shaded (gray) band is the
   systematic component. The SM prediction, defined using the \nnlops\ prediction for gluon--gluon fusion and the default MC
   samples for the other production mechanisms, is presented as a hatched (blue) band, with the width of the band
   reflecting the total theoretical uncertainty (see text for details). The small contribution from \VBF, \VH\, \ttH\,
   and \bbH\ is also shown as a (green) histogram and denoted by $XH$. The default MC has been normalized with the
   N${}^{3}$LO prediction of Refs.~\cite{Anastasiou:2015ema,Anastasiou:2016cez,Actis:2008ug, Anastasiou:2008tj,Butterworth:2015oua, deFlorian:2016spz}.
   In addition, the \hres\ and \scetlib\ predictions, described in Section~\ref{sec:details_diff}, are displayed in (a) and (b), respectively. 
   } 
  \label{fig:diff_ptt_dyy}
\end{figure*}

Figure~\ref{fig:diff_ptt_dyy} shows differential cross sections as a function of \pttgg, the orthogonal component of the diphoton momentum when projected on the axis given by the difference of the 3-momenta of the two photons, as well as \deltaygg, the rapidity separation of the two photons. 

\begin{figure*}[!htbp]
  \centering
	\subfloat[] {\includegraphics[width=0.50\columnwidth]{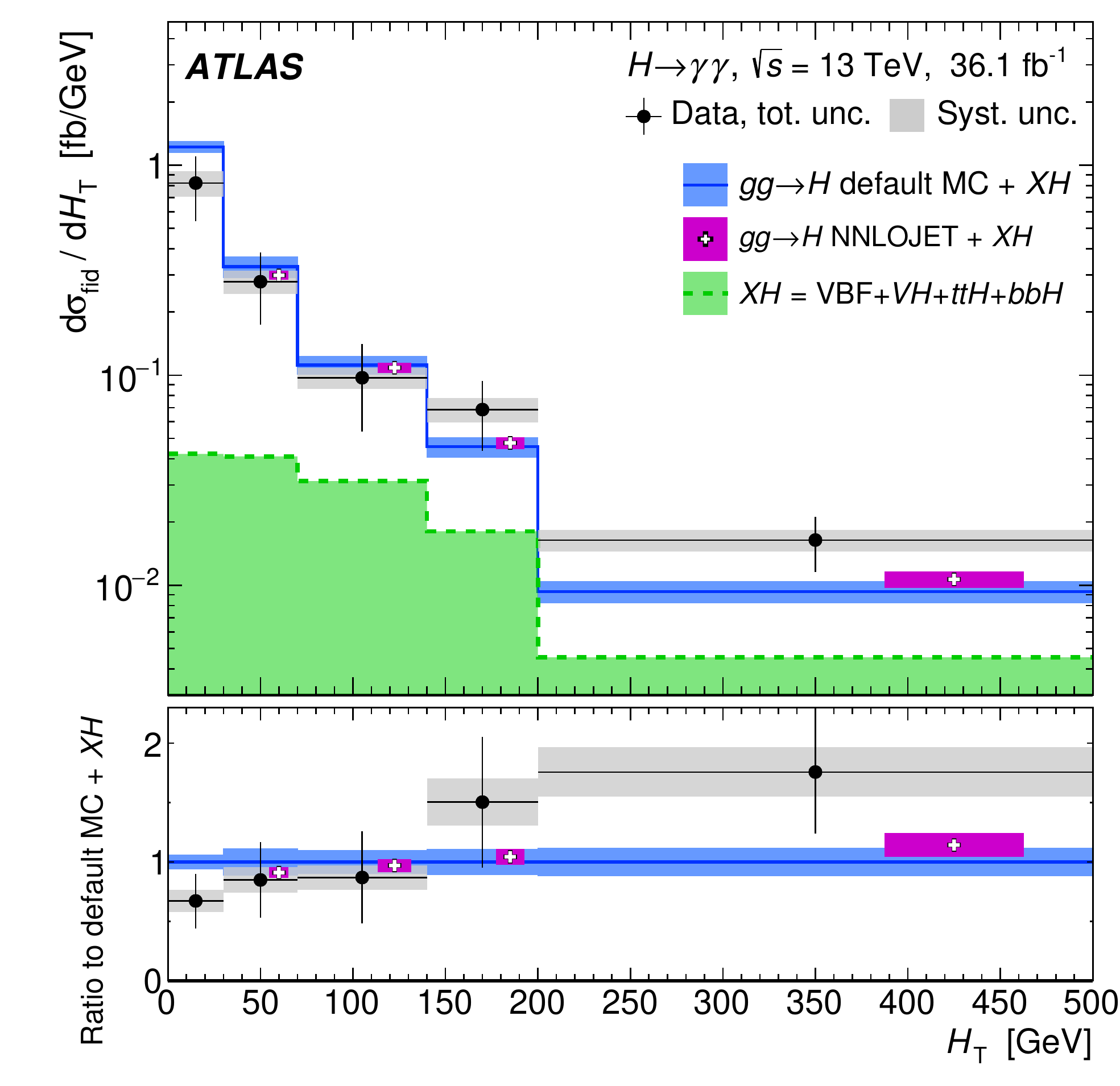}}
	\subfloat[] {\includegraphics[width=0.50\columnwidth]{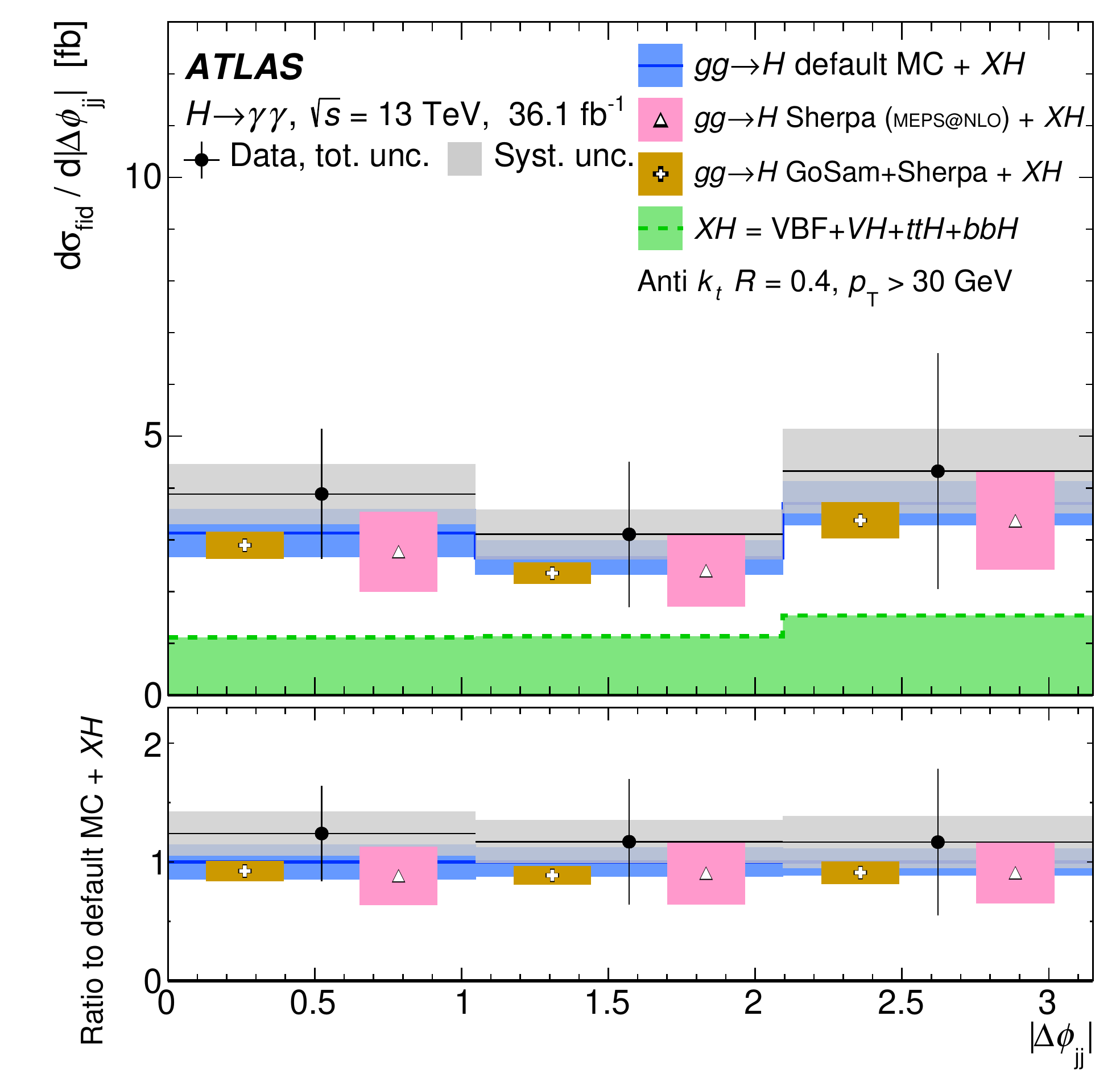}} \\
	\subfloat[] {\includegraphics[width=0.50\columnwidth]{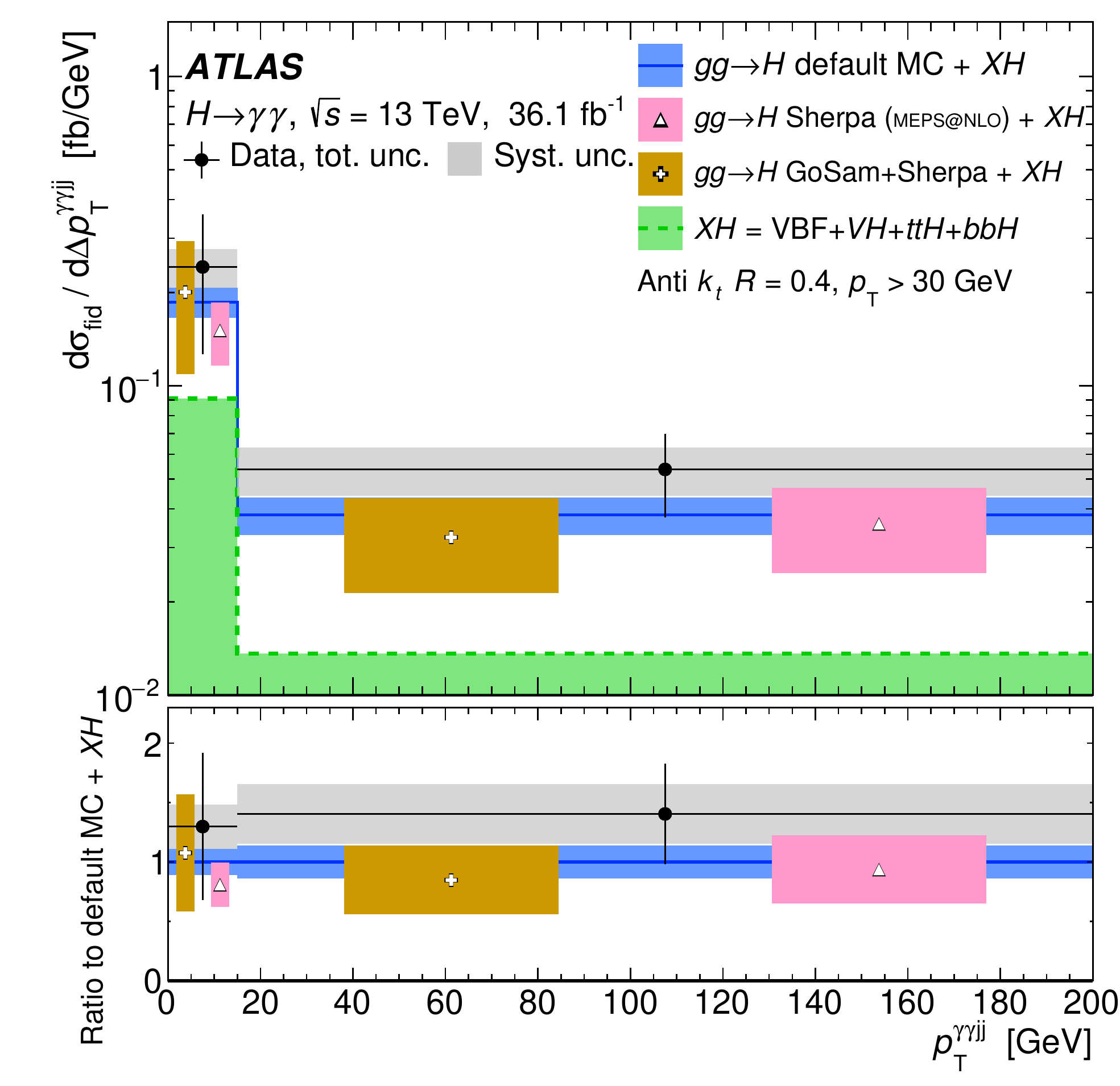}} 
  \caption{
   The differential cross sections for $pp \to H \to \gamma \gamma$ as a function of (a) \HT, (b) \dphijjabs, and (c) \ptggjj\
   are shown and compared to the SM expectations. The data and theoretical predictions are presented in the same way as in Figure~\ref{fig:diff_ptt_dyy}.
   In addition, the \nnlojet\ prediction is displayed in (a), and the \sherpa\ and \gosam\ predictions are displayed in (b) and (c). More details
   of these predictions can be found in Section~\ref{sec:diff_prodkin}.
   }
  \label{fig:diff_HT_vbf}
\end{figure*}

Figure~\ref{fig:diff_HT_vbf} shows differential cross sections as a function of \HT, the scalar sum of all reconstructed jets in a given event with $\pt > 30$~\GeV, the absolute value of the azimuthal difference \dphijjabs\ between the leading and subleading jet in events with at least two jets, and the vectorial sum of the transverse momentum of the diphoton system and the leading and subleading jet system, \ptggjj, in events with at least two jets.

\begin{figure*}[!htbp]
  \centering
	\subfloat[] {\includegraphics[width=0.50\columnwidth]{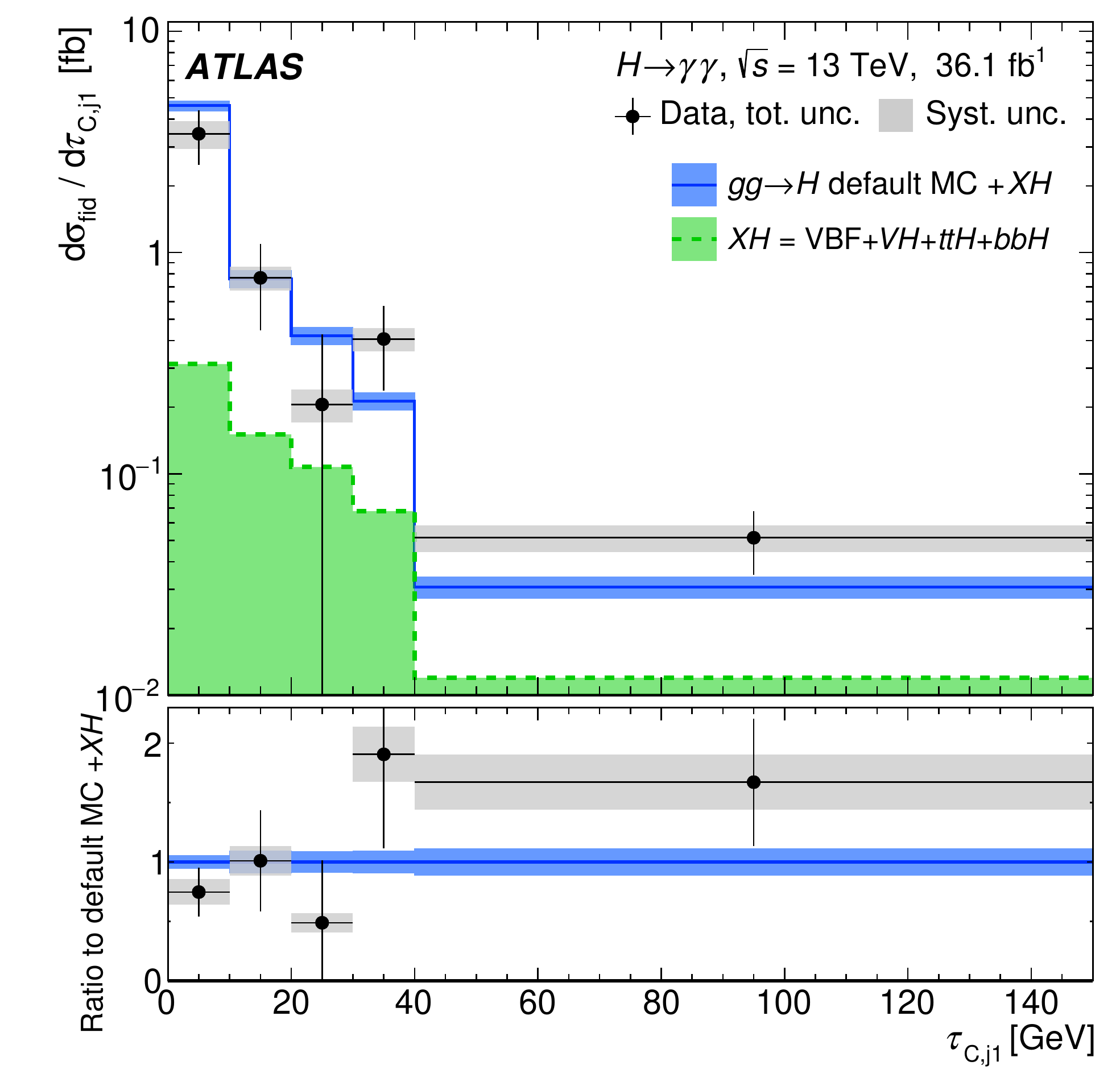}}
	\subfloat[] {\includegraphics[width=0.50\columnwidth]{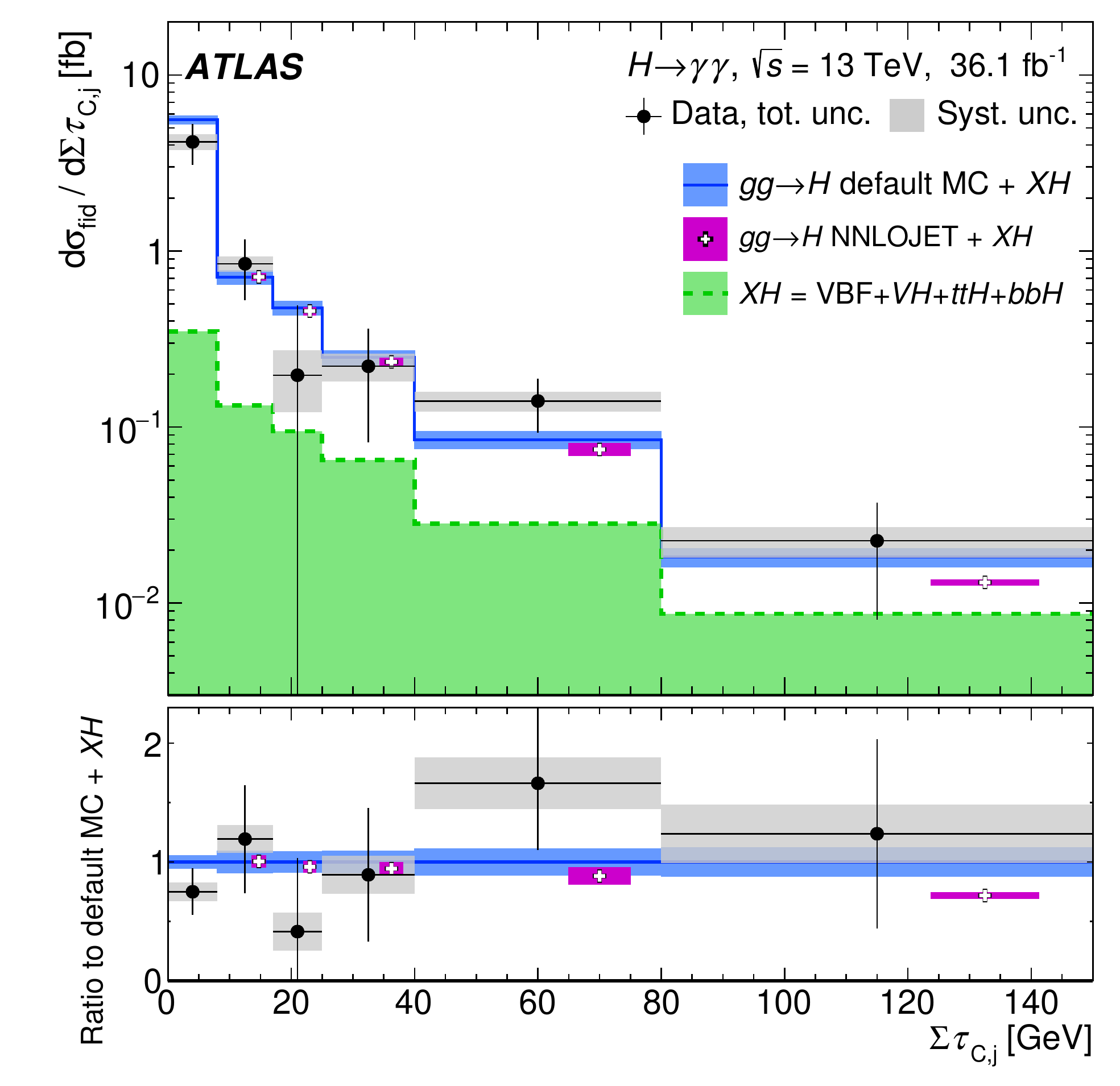}} 
  \caption{
   The differential cross sections for $pp \to H \to \gamma \gamma$ as a function of (a) \taujet\ and (b) \sumtaujet\
   are shown and compared to the SM expectations. The data and theoretical predictions are presented in the same way as in Figure~\ref{fig:diff_ptt_dyy}. In addition, the \nnlojet\ prediction is displayed in (b). 
   }
  \label{fig:diff_tj_stj}
\end{figure*}

Figure~\ref{fig:diff_tj_stj} displays measurements of the beam-thrust-like variables \taujet\ and \sumtaujet. For a given jet, $\tau$ is defined by
\begin{align}
 \tau = \frac{m_\mathrm{T}}{2 \cosh y^*} \, , \quad y^* = y - y_{\gamma\gamma} \, , \qquad m_\mathrm{T} = \sqrt{ \pT^2 + m^2 } \, ,
\end{align}
where $y$ is the jet rapidity and $m$ is the jet mass. The variable \taujet\ refers to the highest-$\tau$ jet, and
\sumtaujet\ is the scalar sum of $\tau$ for all jets with $\tau > 8$~\GeV. For large jet rapidities, $\tau$ corresponds to
the small light-cone component of the jet, $p_{\rm jet}^+ = E_{\rm jet} - \left| p_{z,{\rm jet}} \right|$, while the sum
is closely related to the beam-thrust global event shape~\cite{Stewart:2009yx}, as measured in the diphoton rest frame.

\begin{figure*}[!htbp]
 \centering
\subfloat[] {\includegraphics[width=0.50\columnwidth]{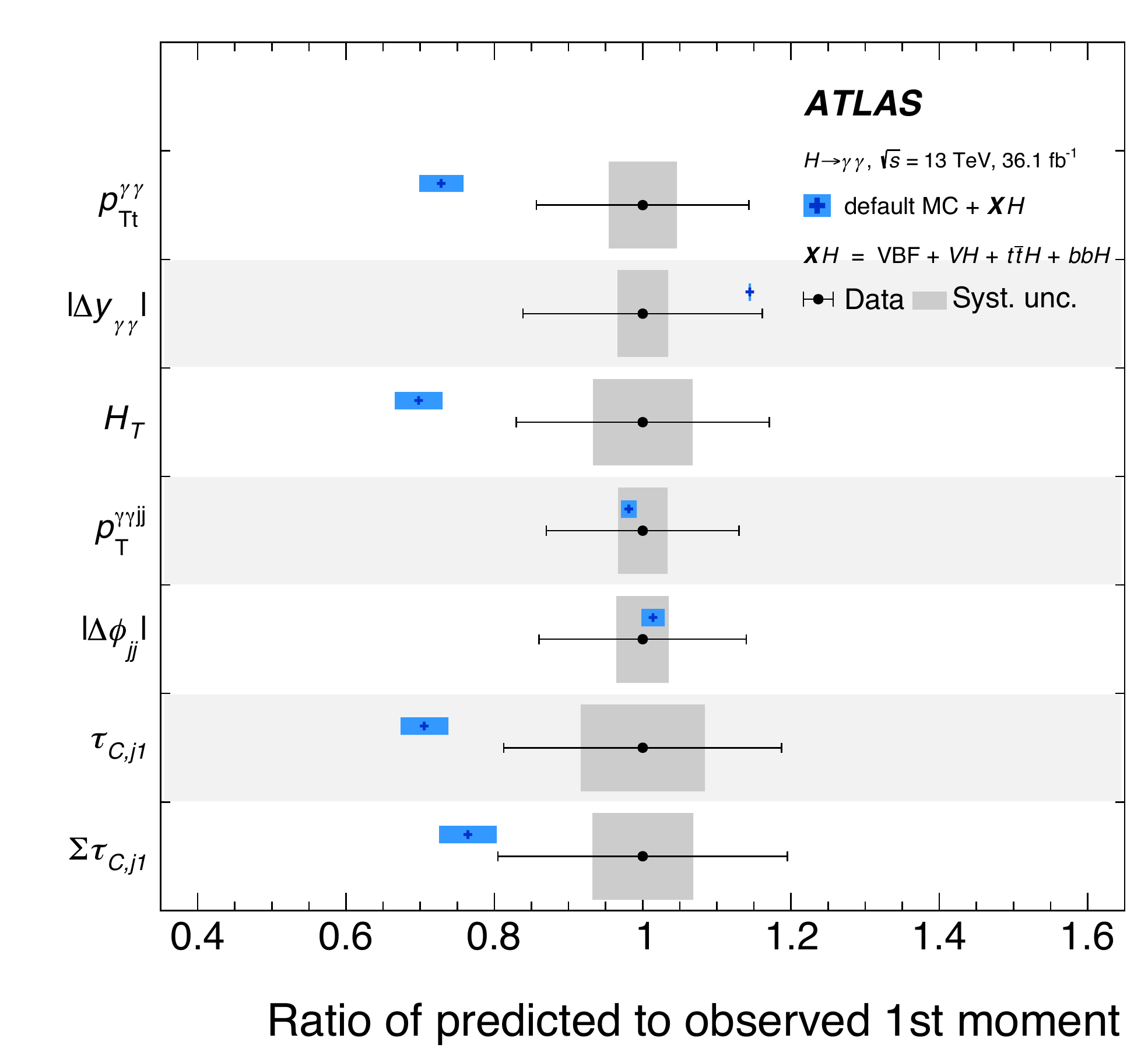}}
\subfloat[] {\includegraphics[width=0.50\columnwidth]{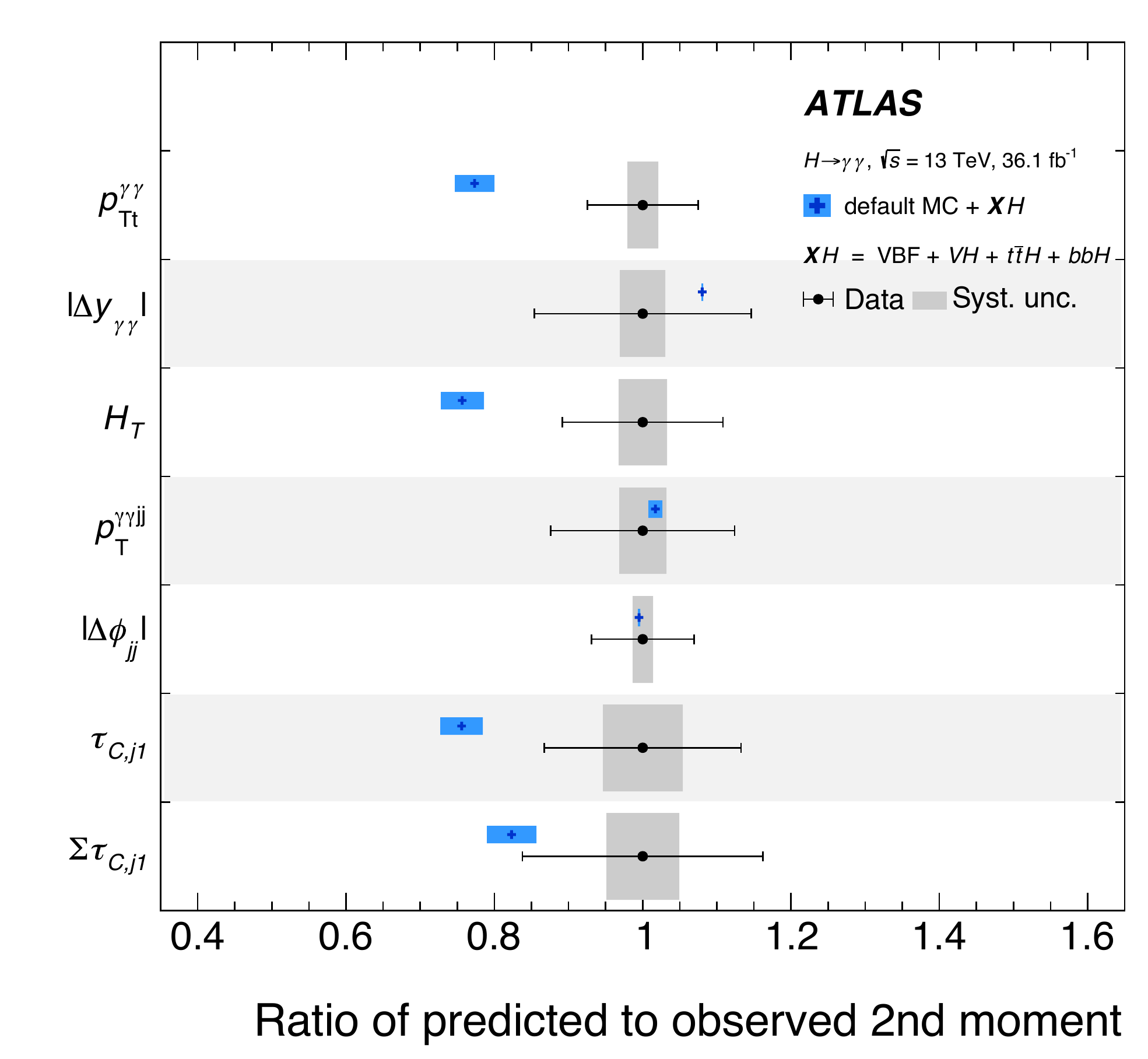}} \\
 \caption{
  (a)  The ratio of the first moment (mean) of each differential distribution predicted by the Standard Model to that
  observed in the data. The SM moment is calculated by using the \nnlops\ prediction for gluon--gluon fusion and the default
  MC samples for the other production mechanisms. (b) The ratio of the second moment (RMS)
  of each differential distribution predicted by the Standard Model to that observed in the data.
  The intervals on the vertical axes each represent one of the differential distributions. The band for
  the theoretical prediction represents the corresponding uncertainty in that prediction (see text for
  details). The error bar on the data represents the total uncertainty in the measurement, with the
  gray band representing only the systematic uncertainty.
 } 
 \label{fig:diff_moments_aux}
\end{figure*}

Figure~\ref{fig:diff_moments_aux} shows the first and second moments of each of the additional differential distributions. The
data are compared to a variety of theoretical predictions. In general, the SM predictions are in agreement with the measured distributions.

\section{Diphoton acceptance, photon isolation and non-perturbative correction factors for parton-level gluon--gluon fusion calculations}\label{app:np_iso_factors}

This appendix presents the diphoton acceptance factors that are applied to parton-level calculations of Higgs
production via gluon--gluon fusion, in order to correctly account for the diphoton selection criteria applied to the Higgs-boson decay products, are shown in Table~\ref{tab:acc} for the fiducial and differential
cross sections presented in Section~\ref{sec:details_diff} and Appendix~\ref{app:add_fid_meas}. Multiplicative isolation efficiency and non-perturbative correction factors that account for the efficiency of the photon isolation
criterion and the impact of hadronization and underlying-event activity are presented in
Tables~\ref{tab:iso} and ~\ref{tab:NP}, respectively. The isolation efficiency is defined as the fraction of selected
diphoton events (i.e. within the kinematic acceptance) that also satisfy the isolation criteria,
and is determined using samples before including hadronization and the underlying-event activity. The
non-perturbative correction factors are defined as the ratios of cross sections produced with
and without hadronization and the underlying event. The default non-perturbative correction is taken as the central value of an envelope formed
from multiple event generators and/or event generator tunes, with the uncertainty taken to 
be the maximal deviation observed in the envelope. Table~\ref{tab:NPiso} also provides the combined non-perturbative and isolation correction with a total uncertainty that takes into account the correlations between the uncertainties of both factors. Note though that no non-perturbative correction factors are applied to the SM predictions presented in this paper.

A summary of the binning of all differential variables is given in Table~\ref{tab:Bins}.

\begin{sidewaystable}[htb!]
\centering
\scalebox{0.7}{
\begin{tabular}{lrrrrrrrrrrrrrr
}
\hline\hline
& Bin 1 & Bin 2 & Bin 3 & Bin 4 & Bin 5 & Bin 6 & Bin 7 & Bin 8 & Bin 9 & Bin 10 & Bin 11 & Bin 12 & Bin 13 \\
\hline\hline
Diphoton fiducial & $51.9 \pm 0.4$ & -- &-- &-- &-- &-- &-- &-- &-- &-- &-- &-- &-- &-- \\  
\ptgg &52.3 $\pm 0.5$ & 52.9 $\pm 0.5$ & 52.0 $\pm 0.4$ & 50.8 $\pm 0.4$ & 50.0 $\pm 0.3$ & 48.6 $\pm 0.3$ & 51.0 $\pm 0.3$ & 57.7 $\pm 0.3$ & 65.2 $\pm 0.3$ & -- &-- &-- &-- &-- \\  
\ygg &74.0 $\pm 0.2$ & 73.6 $\pm 0.2$ & 72.2 $\pm 0.2$ & 69.3 $\pm 0.1$ & 67.1 $\pm 0.1$ & 64.8 $\pm 0.1$ & 61.5 $\pm 0.1$ & 56.0 $\pm 0.2$ & 31.4 $\pm 0.1$ & -- &-- &-- &-- &-- \\  
\pttgg &49.5 $\pm 0.6$ & 49.4 $\pm 0.6$ & 49.2 $\pm 0.6$ & 48.8 $\pm 0.6$ & 49.0 $\pm 0.6$ & 49.8 $\pm 0.6$ & 52.8 $\pm 0.7$ & 59.7 $\pm 0.5$ & 69.5 $\pm 0.5$ & 73.2 $\pm 0.5$ & 75.7 $\pm 0.5$ & 78.3 $\pm 0.4$ & 82.9 $\pm 0.4$ & -- \\  
\costhetastar&78.8 $\pm 0.6$ & 74.9 $\pm 0.6$ & 73.3 $\pm 0.6$ & 71.7 $\pm 0.6$ & 69.6 $\pm 0.6$ & 67.6 $\pm 0.7$ & 65.8 $\pm 0.7$ & 60.4 $\pm 0.7$ & 23.6 $\pm 0.5$ & -- &-- &-- &-- &-- \\  
\deltaygg &80.0 $\pm 0.5$ & 76.1 $\pm 0.5$ & 74.9 $\pm 0.5$ & 73.7 $\pm 0.6$ & 72.4 $\pm 0.6$ & 70.6 $\pm 0.6$ & 68.6 $\pm 0.6$ & 67.4 $\pm 0.6$ & 65.3 $\pm 0.7$ & 50.8 $\pm 0.7$ & -- &-- &-- &-- \\  
\njet, \pt $>$ 30~\GeV &51.8 $\pm 0.5$ & 51.8 $\pm 0.4$ & 52.3 $\pm 0.4$ & 54.2 $\pm 0.7$ & -- &-- &-- &-- &-- &-- &-- &-- &-- &-- \\  
\njet, \pt $>$ 50~\GeV &51.2 $\pm 0.5$ & 54.2 $\pm 0.4$ & 55.8 $\pm 0.7$ & -- &-- &-- &-- &-- &-- &-- &-- &-- &-- &-- \\  
\ptjl &51.8 $\pm 0.5$ & 49.8 $\pm 0.4$ & 56.1 $\pm 0.4$ & 52.7 $\pm 0.4$ & 55.5 $\pm 0.6$ & -- &-- &-- &-- &-- &-- &-- &-- &-- \\  
\ptjsl &51.8 $\pm 0.4$ & 51.9 $\pm 0.4$ & 57.1 $\pm 0.8$ & -- &-- &-- &-- &-- &-- &-- &-- &-- &-- &-- \\  
\HT &51.8 $\pm 0.5$ & 51.1 $\pm 0.4$ & 52.1 $\pm 0.4$ & 53.8 $\pm 0.5$ & 57.3 $\pm 0.9$ & -- &-- &-- &-- &-- &-- &-- &-- &-- \\  
\yjl &55.2 $\pm 0.4$ & 54.6 $\pm 0.4$ & 53.1 $\pm 0.4$ & 53.0 $\pm 0.4$ & 52.3 $\pm 0.4$ & 50.4 $\pm 0.4$ & 48.2 $\pm 0.4$ & -- &-- &-- &-- &-- &-- &-- \\  
\yjsl &54.1 $\pm 0.5$ & 53.5 $\pm 0.5$ & 52.3 $\pm 0.5$ & -- &-- &-- &-- &-- &-- &-- &-- &-- &-- &-- \\  
\mjj &52.8 $\pm 0.5$ & 53.3 $\pm 0.6$ & 51.6 $\pm 0.6$ & -- &-- &-- &-- &-- &-- &-- &-- &-- &-- &-- \\  
\deltayjj &70.2 $\pm 3.7$ & 68.7 $\pm 3.4$ & 67.2 $\pm 3.4$ & -- &-- &-- &-- &-- &-- &-- &-- &-- &-- &-- \\  
\dphijjabs &53.8 $\pm 0.6$ & 53.3 $\pm 0.5$ & 51.6 $\pm 0.4$ & -- &-- &-- &-- &-- &-- &-- &-- &-- &-- &-- \\  
\dphijj    &51.9 $\pm 0.4$ & 54.1 $\pm 0.6$ & 53.4 $\pm 0.6$ & 52.0 $\pm 0.4$ & -- &-- &-- &-- &-- &-- &-- &-- &-- &-- \\  
\ptggjj    &47.6 $\pm 0.4$ & 54.4 $\pm 0.5$ & -- &-- &-- &-- &-- &-- &-- &-- &-- &-- &-- &-- \\  
\dphiggjj  &50.9 $\pm 0.6$ & 54.0 $\pm 0.4$ & 54.0 $\pm 0.4$ & -- &-- &-- &-- &-- &-- &-- &-- &-- &-- &-- \\  
\taujet &51.0 $\pm 0.5$ & 50.9 $\pm 0.4$ & 55.9 $\pm 0.4$ & 63.0 $\pm 0.5$ & 59.0 $\pm 0.7$ & -- &-- &-- &-- &-- &-- &-- &-- &-- \\  
\sumtaujet &51.0 $\pm 0.5$ & 50.5 $\pm 0.4$ & 52.3 $\pm 0.4$ & 59.5 $\pm 0.4$ & 57.6 $\pm 0.5$ & 61.7 $\pm 0.9$ & -- &-- &-- &-- &-- &-- &-- &-- \\  
\ptgg [\costhetastar $<$ 0.5] &72.5 $\pm 0.6$ & 71.1 $\pm 0.5$ & 54.0 $\pm 0.3$ & -- &-- &-- &-- &-- &-- &-- &-- &-- &-- &-- \\  
\ptgg [0.5 $\leq$ \costhetastar $<$ 1.0] &32.4 $\pm 0.4$ & 30.4 $\pm 0.3$ & 55.1 $\pm 0.6$ & -- &-- &-- &-- &-- &-- &-- &-- &-- &-- &-- \\  
\ptgg [\njet=0] &51.2 $\pm 0.5$ & 52.1 $\pm 0.5$ & 52.2 $\pm 0.5$ & 59.9 $\pm 1.9$ & -- &-- &-- &-- &-- &-- &-- &-- &-- &-- \\  
\ptgg [\njet=1] &58.4 $\pm 0.5$ & 49.6 $\pm 0.4$ & 48.8 $\pm 0.4$ & 51.6 $\pm 0.5$ & -- &-- &-- &-- &-- &-- &-- &-- &-- &-- \\  
\ptgg [\njet=2] &52.5 $\pm 0.4$ & 50.1 $\pm 0.4$ & 63.1 $\pm 0.3$ & -- &-- &-- &-- &-- &-- &-- &-- &-- &-- &-- \\  
\ptgg [\njet$\geq$3] &52.2 $\pm 0.4$ & 63.0 $\pm 0.3$ & -- &-- &-- &-- &-- &-- &-- &-- &-- &-- &-- &-- \\  
\hline\hline
\end{tabular}
}
\caption{ 
Diphoton kinematic acceptances in percent for gluon--gluon fusion for the diphoton fiducial region and all differential variable bins studied in this paper, defined as the probability to fulfill the diphoton kinematic criteria: $\pT/m_{\gamma\gamma} <$ 0.35 (0.25) for the leading (subleading) photon and $|\eta_\gamma| < 2.37$. The factors are
evaluated using the \nnlops event generator. Uncertainties are taken from PDF variations. QCD scale variations have a negligible impact on these factors. The range of
each bin is given in Table~\ref{tab:Bins}. 
}
\label{tab:acc}
\end{sidewaystable}

\begin{sidewaystable}[htb!]
\centering
\scalebox{0.7}{
\begin{tabular}{lrrrrrrrrrrrrrrr}
\hline\hline
& Bin 1 & Bin 2 & Bin 3 & Bin 4 & Bin 5 & Bin 6 & Bin 7 & Bin 8 & Bin 9 & Bin 10 & Bin 11 & Bin 12 & Bin 13 \\
\hline\hline
Diphoton fiducial &97.7 $\pm 0.1$ & -- &-- &-- &-- &-- &-- &-- &-- &-- &-- &-- &-- &-- \\  
\ptgg &98.4 $\pm 0.1$ & 97.6 $\pm 0.1$ & 97.2 $\pm 0.0$ & 96.9 $\pm 0.0$ & 97.0 $\pm 0.0$ & 97.7 $\pm 0.1$ & 98.5 $\pm 0.0$ & 98.7 $\pm 0.1$ & 98.9 $\pm 0.2$ & -- &-- &-- &-- &-- \\  
\ygg &97.6 $\pm 0.1$ & 97.6 $\pm 0.1$ & 97.7 $\pm 0.1$ & 97.6 $\pm 0.2$ & 97.7 $\pm 0.2$ & 97.7 $\pm 0.0$ & 97.7 $\pm 0.1$ & 97.8 $\pm 0.0$ & 98.0 $\pm 0.0$ & -- &-- &-- &-- &-- \\  
\pttgg &97.2 $\pm 0.2$ & 97.4 $\pm 0.0$ & 97.5 $\pm 0.1$ & 97.7 $\pm 0.0$ & 97.9 $\pm 0.0$ & 97.9 $\pm 0.0$ & 98.0 $\pm 0.1$ & 98.0 $\pm 0.1$ & 98.2 $\pm 0.1$ & 98.5 $\pm 0.1$ & 98.9 $\pm 0.0$ & 99.1 $\pm 0.0$ & 99.2 $\pm 0.2$ & -- \\  
\costhetastar &97.9 $\pm 0.0$ & 97.9 $\pm 0.1$ & 97.9 $\pm 0.0$ & 97.8 $\pm 0.1$ & 97.8 $\pm 0.0$ & 97.7 $\pm 0.2$ & 97.7 $\pm 0.1$ & 97.6 $\pm 0.1$ & 97.6 $\pm 0.1$ & -- &-- &-- &-- &-- \\  
\deltaygg &97.9 $\pm 0.1$ & 98.0 $\pm 0.1$ & 97.9 $\pm 0.1$ & 97.9 $\pm 0.1$ & 97.9 $\pm 0.2$ & 97.9 $\pm 0.1$ & 97.8 $\pm 0.0$ & 97.7 $\pm 0.1$ & 97.7 $\pm 0.1$ & 97.4 $\pm 0.1$ & -- &-- &-- &-- \\  
\njet, \pt $>$ 30~\GeV &97.7 $\pm 0.1$ & 97.8 $\pm 0.1$ & 97.8 $\pm 0.1$ & 97.8 $\pm 0.1$ & -- &-- &-- &-- &-- &-- &-- &-- &-- &-- \\  
\njet, \pt $>$ 50~\GeV &97.7 $\pm 0.1$ & 97.8 $\pm 0.1$ & 98.1 $\pm 0.0$ & -- &-- &-- &-- &-- &-- &-- &-- &-- &-- &-- \\  
\ptjl &97.7 $\pm 0.1$ & 97.6 $\pm 0.1$ & 97.8 $\pm 0.0$ & 97.8 $\pm 0.2$ & 98.3 $\pm 0.0$ & -- &-- &-- &-- &-- &-- &-- &-- &-- \\  
\ptjsl &97.8 $\pm 0.1$ & 97.7 $\pm 0.1$ & 98.3 $\pm 0.0$ & -- &-- &-- &-- &-- &-- &-- &-- &-- &-- &-- \\  
\HT &97.7 $\pm 0.1$ & 97.7 $\pm 0.1$ & 97.8 $\pm 0.1$ & 98.0 $\pm 0.2$ & 98.2 $\pm 0.0$ & -- &-- &-- &-- &-- &-- &-- &-- &-- \\  
\yjl &97.7 $\pm 0.1$ & 97.7 $\pm 0.1$ & 97.8 $\pm 0.1$ & 97.8 $\pm 0.1$ & 97.8 $\pm 0.1$ & 97.7 $\pm 0.4$ & 97.8 $\pm 0.1$ & -- &-- &-- &-- &-- &-- &-- \\  
\yjsl &97.8 $\pm 0.1$ & 97.8 $\pm 0.1$ & 97.8 $\pm 0.1$ & -- &-- &-- &-- &-- &-- &-- &-- &-- &-- &-- \\  
\mjj &97.7 $\pm 0.0$ & 97.8 $\pm 0.1$ & 98.2 $\pm 0.1$ & -- &-- &-- &-- &-- &-- &-- &-- &-- &-- &-- \\  
\deltayjj &97.8 $\pm 0.1$ & 97.8 $\pm 0.0$ & 97.8 $\pm 0.3$ & -- &-- &-- &-- &-- &-- &-- &-- &-- &-- &-- \\  
\dphijjabs &98.1 $\pm 0.1$ & 97.9 $\pm 0.1$ & 97.4 $\pm 0.1$ & -- &-- &-- &-- &-- &-- &-- &-- &-- &-- &-- \\  
\dphijj &97.6 $\pm 0.1$ & 98.0 $\pm 0.2$ & 98.1 $\pm 0.0$ & 97.5 $\pm 0.1$ & -- &-- &-- &-- &-- &-- &-- &-- &-- &-- \\  
\ptggjj &98.3 $\pm 0.2$ & 97.7 $\pm 0.1$ & -- &-- &-- &-- &-- &-- &-- &-- &-- &-- &-- &-- \\  
\dphiggjj &98.4 $\pm 0.1$ & 97.8 $\pm 0.2$ & 96.4 $\pm 0.2$ & -- &-- &-- &-- &-- &-- &-- &-- &-- &-- &-- \\  
\taujet &97.7 $\pm 0.1$ & 97.8 $\pm 0.1$ & 97.7 $\pm 0.1$ & 97.6 $\pm 0.1$ & 97.9 $\pm 0.0$ & 98.1 $\pm 0.2$ & -- &-- &-- &-- &-- &-- &-- &-- \\  
\sumtaujet &97.7 $\pm 0.2$ & 97.7 $\pm 0.1$ & 97.6 $\pm 0.0$ & 97.7 $\pm 0.1$ & 97.9 $\pm 0.2$ & 98.3 $\pm 0.1$ & -- &-- &-- &-- &-- &-- &-- &-- \\  
\ptgg [\costhetastar $<$ 0.5] &98.3 $\pm 0.1$ & 97.0 $\pm 0.0$ & 98.4 $\pm 0.2$ & 98.9 $\pm 0.4$ & -- &-- &-- &-- &-- &-- &-- &-- &-- &-- \\  
\ptgg [0.5 $\leq$ \costhetastar $<$ 1.0] &97.6 $\pm 0.1$ & 97.4 $\pm 0.1$ & 98.9 $\pm 0.0$ & 99.0 $\pm 1.0$ & -- &-- &-- &-- &-- &-- &-- &-- &-- &-- \\  
\ptgg [\njet=0] &98.6 $\pm 0.1$ & 97.8 $\pm 0.1$ & 96.4 $\pm 0.0$ & 91.2 $\pm 1.3$ & 100 & -- &-- &-- &-- &-- &-- &-- &-- &-- \\  
\ptgg [\njet=1] &97.5 $\pm 0.3$ & 97.6 $\pm 0.2$ & 97.7 $\pm 0.0$ & 98.4 $\pm 0.1$ & -- &-- &-- &-- &-- &-- &-- &-- &-- &-- \\  
\ptgg [\njet=2] &97.0 $\pm 0.3$ & 98.4 $\pm 0.1$ & 98.8 $\pm 0.1$ & -- &-- &-- &-- &-- &-- &-- &-- &-- &-- &-- \\  
\ptgg [\njet$\geq$3] &97.4 $\pm 0.2$ & 98.5 $\pm 0.7$ & -- &-- &-- &-- &-- &-- &-- &-- &-- &-- &-- &-- \\  
\hline\hline
\end{tabular}
}
\caption{
 Isolation efficiencies in percent for gluon--gluon fusion $H \to \gamma\gamma$ for the diphoton fiducial region and all differential variable bins in this analysis. The isolation
efficiency is defined as the probability for both photons to fulfill the isolation criteria (as described in Section~\ref{sec:fiddef}) for events that
satisfy the diphoton kinematic criteria. Regions of phase space where no reliable estimate could be obtained are listed as '100' without uncertainties. 
Uncertainties are assigned in the same way as for the non-perturbative correction factors: by varying the
fragmentation and underlying-event modeling. These factors can be multiplied by the kinematic acceptance factors (see Table~\ref{tab:acc}) to extrapolate an
inclusive gluon--gluon fusion Higgs prediction to the fiducial volume used in this analysis. The range of each bin is given in Table~\ref{tab:Bins}.
}
\label{tab:iso}
\end{sidewaystable}

\begin{sidewaystable}[htb!]
\centering
\scalebox{0.7}{
\begin{tabular}{lrrrrrrrrrrrrrrr}
\hline\hline
& Bin 1 & Bin 2 & Bin 3 & Bin 4 & Bin 5 & Bin 6 & Bin 7 & Bin 8 & Bin 9 & Bin 10 & Bin 11 & Bin 12 & Bin 13 \\
\hline\hline
Diphoton fiducial &100 $\pm 2$ & -- &-- &-- &-- &-- &-- &-- &-- &-- &-- &-- &-- &-- \\  
\ptgg &99.7 $\pm 0.4$ & 100.3 $\pm 0.3$ & 100.2 $\pm 0.4$ & 99.7 $\pm 0.6$ & 100.0 $\pm 0.7$ & 99.8 $\pm 0.5$ & 100.4 $\pm 0.6$ & 99.9 $\pm 0.9$ & 100.1 $\pm 0.6$ & -- &-- &-- &-- &-- \\
\ygg &99.8 $\pm 0.4$ & 99.9 $\pm 0.3$ & 100.1 $\pm 0.2$ & 100.0 $\pm 0.5$ & 100.0 $\pm 0.3$ & 100.2 $\pm 0.4$ & 99.9 $\pm 0.5$ & 100.2 $\pm 0.3$ & 100.0 $\pm 0.4$ & -- &-- &-- &-- &-- \\
\pttgg &100.0 $\pm 0.5$ & 100.1 $\pm 0.3$ & 100.0 $\pm 0.4$ & 100.3 $\pm 0.3$ & 100.0 $\pm 0.8$ & 100.3 $\pm 0.7$ & 99.8 $\pm 0.7$ & 100.0 $\pm 0.3$ & 99.9 $\pm 1.1$ & 99.8 $\pm 0.8$ & 99.8 $\pm 1.1$ & 100.6 $\pm 0.9$ & 100.4 $\pm 1.5$ & -- \\
\costhetastar &100.1 $\pm 0.5$ & 100.0 $\pm 0.5$ & 99.8 $\pm 0.5$ & 99.7 $\pm 0.4$ & 99.9 $\pm 0.4$ & 100.0 $\pm 0.4$ & 100.2 $\pm 0.5$ & 100.0 $\pm 0.4$ & 100.2 $\pm 0.2$ & -- &-- &-- &-- &-- \\
\deltaygg &99.9 $\pm 0.7$ & 100.0 $\pm 0.5$ & 100.0 $\pm 0.9$ & 99.7 $\pm 0.5$ & 99.7 $\pm 0.4$ & 100.1 $\pm 0.4$ & 99.9 $\pm 0.3$ & 100.0 $\pm 0.5$ & 100.1 $\pm 0.7$ & 100.1 $\pm 0.2$ & -- &-- &-- &-- \\
\njet, \pt $>$ 30~\GeV &102 $\pm 4$ & 99.2 $\pm 2.8$ & 98.2 $\pm 7.3$ & 94.4 $\pm 11.9$ & -- &-- &-- &-- &-- &-- &-- &-- &-- &-- \\
\njet, \pt $>$ 50~\GeV&100.7 $\pm 2.2$ & 98.7 $\pm 3.6$ & 96.4 $\pm 6.5$ & -- &-- &-- &-- &-- &-- &-- &-- &-- &-- &-- \\
\ptjl &102 $\pm 4$ & 98.7 $\pm 5.2$ & 98.0 $\pm 4.7$ & 98.1 $\pm 4.3$ & 97.4 $\pm 3.7$ & -- &-- &-- &-- &-- &-- &-- &-- &-- \\
\ptjsl &99.2 $\pm 2.8$ & 97.2 $\pm 9.2$ & 97.3 $\pm 6.2$ & -- &-- &-- &-- &-- &-- &-- &-- &-- &-- &-- \\
\HT &102 $\pm 4$ & 99.1 $\pm 3.8$ & 98.1 $\pm 5.3$ & 97.0 $\pm 6.7$ & 96.7 $\pm 6.5$ & -- &-- &-- &-- &-- &-- &-- &-- &-- \\
\yjl &98.5 $\pm 4.6$ & 98.9 $\pm 4.4$ & 98.5 $\pm 4.7$ & 98.0 $\pm 4.5$ & 98.4 $\pm 4.3$ & 99.3 $\pm 4.9$ & 97.9 $\pm 4.9$ & -- &-- &-- &-- &-- &-- &-- \\
\yjsl&96.7 $\pm 8.6$ & 97.6 $\pm 8.4$ & 97.0 $\pm 8.9$ & -- &-- &-- &-- &-- &-- &-- &-- &-- &-- &-- \\
\mjj &96.1 $\pm 9.4$ & 98.6 $\pm 8.0$ & 98.4 $\pm 8.3$ & -- &-- &-- &-- &-- &-- &-- &-- &-- &-- &-- \\
\deltayjj &96.5 $\pm 8.3$ & 98.5 $\pm 9.0$ & 96.5 $\pm 10.4$ & -- &-- &-- &-- &-- &-- &-- &-- &-- &-- &-- \\
\dphijjabs &95.2 $\pm 8.3$ & 97.0 $\pm 7.9$ & 100.3 $\pm 10.6$ & -- &-- &-- &-- &-- &-- &-- &-- &-- &-- &-- \\
\dphijj &99.0 $\pm 9.6$ & 95.8 $\pm 8.0$ & 95.5 $\pm 8.3$ & 98.7 $\pm 9.5$ & -- &-- &-- &-- &-- &-- &-- &-- &-- &-- \\
\ptggjj &94.3 $\pm 12.4$ & 97.8 $\pm 7.5$ & -- &-- &-- &-- &-- &-- &-- &-- &-- &-- &-- &-- \\
\dphiggjj &96.6 $\pm 7.1$ & 96.3 $\pm 7.7$ & 102.5 $\pm 14.3$ & -- &-- &-- &-- &-- &-- &-- &-- &-- &-- &-- \\
\taujet &102 $\pm 4$ & 98.1 $\pm 4.3$ & 98.0 $\pm 5.1$ & 99.2 $\pm 5.0$ & 99.9 $\pm 4.9$ & 98.8 $\pm 4.1$ & -- &-- &-- &-- &-- &-- &-- &-- \\
\sumtaujet&100.9 $\pm 2.9$ & 98.0 $\pm 4.0$ & 100.1 $\pm 4.2$ & 98.8 $\pm 5.3$ & 98.8 $\pm 6.0$ & 97.8 $\pm 5.3$ & -- &-- &-- &-- &-- &-- &-- &-- \\
\ptgg [\costhetastar $<$ 0.5] &100.1 $\pm 1.0$ & 100.2 $\pm 0.7$ & 100.0 $\pm 0.7$ & 100.8 $\pm 6.8$ & -- &-- &-- &-- &-- &-- &-- &-- &-- &-- \\
\ptgg [0.5 $\leq$ \costhetastar $<$ 1.0] &100.2 $\pm 0.9$ & 99.8 $\pm 0.8$ & 100.2 $\pm 1.3$ & 100.3 $\pm 4.1$ & -- &-- &-- &-- &-- &-- &-- &-- &-- &-- \\
\ptgg [\njet=0] &98.6 $\pm 2.3$ & 99.8 $\pm 0.7$ & 103.1 $\pm 4.7$ & 110.1 $\pm 23.7$ & -- & -- &-- &-- &-- &-- &-- &-- &-- &-- \\
\ptgg [\njet=1] &98.2 $\pm 9.4$ & 99.1 $\pm 0.9$ & 103.2 $\pm 5.8$ & 104.9 $\pm 8.1$ & -- &-- &-- &-- &-- &-- &-- &-- &-- &-- \\
\ptgg [\njet=2] &99.9 $\pm 5.2$ & 102.5 $\pm 7.9$ & 106.1 $\pm 12.3$ & -- &-- &-- &-- &-- &-- &-- &-- &-- &-- &-- \\
\ptgg [\njet$\geq$3] &100.1 $\pm 2.5$ & 102 $\pm 11$ & -- &-- &-- &-- &-- &-- &-- &-- &-- &-- &-- &-- \\
\hline\hline
\end{tabular}
}
\caption{ 
Non-perturbative correction factors in percent accounting for the impact of hadronization and the underlying-event
activity for the diphoton fiducial region and all differential variable bins.
Uncertainties are evaluated by deriving these factors using
different generators and tunes as described in the text. The range of each bin is given in Table~\ref{tab:Bins}.
}
\label{tab:NP}
\end{sidewaystable}

\begin{sidewaystable}[htb!]
\centering
\scalebox{0.7}{
\begin{tabular}{lrrrrrrrrrrrrrrr}
\hline\hline
& Bin 1 & Bin 2 & Bin 3 & Bin 4 & Bin 5 & Bin 6 & Bin 7 & Bin 8 & Bin 9 & Bin 10 & Bin 11 & Bin 12 & Bin 13 \\
\hline\hline
Diphoton fiducial &97.7 $\pm 2.4$ & -- &-- &-- &-- &-- &-- &-- &-- &-- &-- &-- &-- &-- \\  
\ptgg &97.8 $\pm 0.5$ & 97.7 $\pm 0.6$ & 97.2 $\pm 1.0$ & 96.3 $\pm 1.0$ & 96.5 $\pm 0.9$ & 96.9 $\pm 0.7$ & 98.4 $\pm 1.0$ & 98.3 $\pm 1.1$ & 99.1 $\pm 0.7$ & -- &-- &-- &-- &-- \\  
\ygg &97.2 $\pm 0.7$ & 97.3 $\pm 0.5$ & 97.4 $\pm 0.6$ & 97.4 $\pm 0.8$ & 97.5 $\pm 0.6$ & 97.5 $\pm 0.7$ & 97.5 $\pm 0.6$ & 97.5 $\pm 0.7$ & 97.8 $\pm 0.8$ & -- &-- &-- &-- &-- \\  
\pttgg &97.0 $\pm 0.6$ & 97.4 $\pm 0.7$ & 97.2 $\pm 0.6$ & 97.6 $\pm 0.7$ & 97.1 $\pm 0.9$ & 97.8 $\pm 1.0$ & 97.5 $\pm 0.9$ & 97.6 $\pm 0.6$ & 97.8 $\pm 1.4$ & 98.0 $\pm 0.8$ & 98.1 $\pm 1.2$ & 99.6 $\pm 1.0$ & 99.5 $\pm 1.6$ & -- \\  
\costhetastar &97.7 $\pm 0.8$ & 97.7 $\pm 0.7$ & 97.3 $\pm 0.8$ & 97.1 $\pm 0.7$ & 97.4 $\pm 0.6$ & 97.5 $\pm 0.5$ & 97.6 $\pm 0.7$ & 97.3 $\pm 0.6$ & 97.5 $\pm 0.7$ & -- &-- &-- &-- &-- \\  
\deltaygg &97.6 $\pm 0.9$ & 97.9 $\pm 0.5$ & 97.6 $\pm 1.1$ & 97.4 $\pm 0.7$ & 97.1 $\pm 0.6$ & 97.7 $\pm 0.7$ & 97.4 $\pm 0.7$ & 97.5 $\pm 0.9$ & 97.3 $\pm 0.6$ & 97.3 $\pm 0.8$ & -- &-- &-- &-- \\  
\njet, \pt $>$ 30~\GeV &99.3 $\pm 3.8$ & 96.5 $\pm 3.1$ & 95.5 $\pm 7.6$ & 91.9 $\pm 12.2$ & -- &-- &-- &-- &-- &-- &-- &-- &-- &-- \\  
\njet, \pt $>$ 50~\GeV &98.4 $\pm 1.5$ & 96.1 $\pm 3.9$ & 94.2 $\pm 6.9$ & -- &-- &-- &-- &-- &-- &-- &-- &-- &-- &-- \\  
\ptjl &99.3 $\pm 3.8$ & 95.9 $\pm 5.5$ & 95.3 $\pm 5.0$ & 95.5 $\pm 4.8$ & 95.1 $\pm 4.3$ & -- &-- &-- &-- &-- &-- &-- &-- &-- \\  
\ptjsl &96.5 $\pm 3.1$ & 94.4 $\pm 9.5$ & 95.1 $\pm 6.5$ & -- &-- &-- &-- &-- &-- &-- &-- &-- &-- &-- \\  
\HT &99.3 $\pm 3.8$ & 96.4 $\pm 4.1$ & 95.4 $\pm 5.7$ & 94.5 $\pm 7.0$ & 94.5 $\pm 6.8$ & -- &-- &-- &-- &-- &-- &-- &-- &-- \\  
\yjl &95.8 $\pm 5.0$ & 96.2 $\pm 4.7$ & 95.9 $\pm 5.0$ & 95.3 $\pm 4.9$ & 95.6 $\pm 4.7$ & 96.8 $\pm 5.4$ & 95.2 $\pm 5.3$ & -- &-- &-- &-- &-- &-- &-- \\  
\yjsl &94.1 $\pm 9.0$ & 94.8 $\pm 8.7$ & 94.3 $\pm 9.2$ & -- &-- &-- &-- &-- &-- &-- &-- &-- &-- &-- \\  
\mjj &93.3 $\pm 9.6$ & 95.9 $\pm 8.3$ & 96.0 $\pm 8.7$ & -- &-- &-- &-- &-- &-- &-- &-- &-- &-- &-- \\  
\deltayjj &93.8 $\pm 8.6$ & 95.8 $\pm 9.3$ & 93.8 $\pm 10.8$ & -- &-- &-- &-- &-- &-- &-- &-- &-- &-- &-- \\  
\dphijjabs &92.9 $\pm 8.7$ & 94.4 $\pm 8.3$ & 97.2 $\pm 11.0$ & -- &-- &-- &-- &-- &-- &-- &-- &-- &-- &-- \\  
\dphijj &96.0 $\pm 9.9$ & 93.4 $\pm 8.3$ & 93.0 $\pm 8.5$ & 95.7 $\pm 9.7$ & -- &-- &-- &-- &-- &-- &-- &-- &-- &-- \\  
\ptggjj  &92.3 $\pm 12.6$ & 94.9 $\pm 7.8$ & -- &-- &-- &-- &-- &-- &-- &-- &-- &-- &-- &-- \\  
\dphiggjj &94.6 $\pm 7.4$ & 93.6 $\pm 8.0$ & 97.9 $\pm 14.7$ & -- &-- &-- &-- &-- &-- &-- &-- &-- &-- &-- \\  
\taujet &99.3 $\pm 3.8$ & 95.4 $\pm 4.8$ & 95.2 $\pm 5.5$ & 96.3 $\pm 5.2$ & 97.1 $\pm 5.4$ & 96.5 $\pm 4.5$ & -- &-- &-- &-- &-- &-- &-- &-- \\  
\sumtaujet&98.7 $\pm 2.3$ & 95.3 $\pm 4.5$ & 97.2 $\pm 4.6$ & 96.0 $\pm 5.6$ & 96.3 $\pm 6.2$ & 95.7 $\pm 5.8$ & -- &-- &-- &-- &-- &-- &-- &-- \\  
\ptgg [\costhetastar $<$ 0.5] &98.1 $\pm 0.7$ & 97.0 $\pm 0.7$ & 98.2 $\pm 1.0$ & 99.8 $\pm 6.8$ & -- &-- &-- &-- &-- &-- &-- &-- &-- &-- \\  
\ptgg [0.5 $\leq$ \costhetastar $<$ 1.0] &97.6 $\pm 0.6$ & 96.9 $\pm 1.1$ & 98.8 $\pm 1.4$ & 99.7 $\pm 4.5$ & -- &-- &-- &-- &-- &-- &-- &-- &-- &-- \\  
\ptgg [\njet=0] &96.7 $\pm 2.4$ & 97.3 $\pm 1.1$ & 99.0 $\pm 4.6$ & 102.0 $\pm 24.5$ & -- & -- &-- &-- &-- &-- &-- &-- &-- &-- \\  
\ptgg [\njet=1] &95.4 $\pm 9.7$ & 96.3 $\pm 1.3$ & 100.4 $\pm 5.6$ & 103.2 $\pm 7.9$ & -- &-- &-- &-- &-- &-- &-- &-- &-- &-- \\  
\ptgg [\njet=2] &96.5 $\pm 5.4$ & 100.5 $\pm 7.7$ & 104.5 $\pm 12.2$ & -- &-- &-- &-- &-- &-- &-- &-- &-- &-- &-- \\  
\ptgg [\njet$\geq$3] &97.3 $\pm 3.1$ & 99.9 $\pm 10.6$ & -- &-- &-- &-- &-- &-- &-- &-- &-- &-- &-- &-- \\  
\hline\hline
\end{tabular}
}
\caption{ 
Combined non-perturbative (Table~\ref{tab:NP}) and particle-level isolation correction factors (Table~\ref{tab:iso}) in percent accounting for the impact of hadronization and the underlying-event
activity for the diphoton fiducial region and all differential variable bins.
The uncertainties in the combined values properly take into account the correlations between both multiplicative factors.
}
\label{tab:NPiso}
\end{sidewaystable}

\begin{sidewaystable}[htb!]
\centering
\scalebox{0.6}{
\begin{tabular}{lrrrrrrrrrrrrrrrrr} 
\hline\hline
Bin & 1 & 2 & 3 & 4 & 5 & 6 & 7 & 8 & 9 & 10 & 11 & 12 & 13 \\
\hline\hline
\ptgg $\,\,$ [GeV] & 0 -- 20 & 20 -- 30 & 30 -- 45 & 45 -- 60 & 60 -- 80 & 80 -- 120 & 120 -- 170 & 170 -- 220 & 220 -- 350 \\
\ygg &  0 -- 0.15 & 0.15 -- 0.3 & 0.3 -- 0.45 & 0.45 -- 0.6 & 0.6 -- 0.75 & 0.75 -- 0.9 & 0.9 -- 1.2 & 1.2 -- 1.6 & 1.6 -- 2.4 \\
\pttgg $\,\,$ [GeV] & 0 -- 5 & 5 -- 10 & 10 -- 15 & 15 -- 22 & 22 -- 30 & 30 -- 40 & 40 -- 50 & 50 -- 65 & 65 -- 80 & 80 --100 & 100 -- 125 & 125 -- 160 & 160 -- 250 \\
\costhetastar & 0 -- 0.0625 & 0.0625 -- 0.125 & 0.125 -- 0.1875 & 0.1875 -- 0.25 & 0.25 -- 0.3125 & 0.3125 -- 0.375 & 0.375 -- 0.5 & 0.5 -- 0.625 & 0.625 -- 1 \\
\deltaygg &  0 -- 0.1 & 0.1 -- 0.2 & 0.2 -- 0.3 & 0.3 -- 0.4 & 0.4 -- 0.5 & 0.5 -- 0.65 & 0.65 -- 0.8 & 0.8 -- 1 & 1 -- 1.25 & 1.25 -- 2 \\
\njet, \pt $>$ 30~\GeV & 0 & 1 & 2 & $\ge 3$ \\
\njet, \pt $>$ 50~\GeV & 0 & 1 & $\ge 2$ \\
\ptjl $\,\,$ [GeV] &  0 -- 30 & 30 -- 55 & 55 -- 75 & 75 -- 120 & 120 -- 350  \\
\ptjsl $\,\,$ [GeV] & 0 -- 30 & 30 -- 70 & 70 -- 120 \\
\HT $\,\,$ [GeV] & 0 -- 30 & 30 -- 70 & 70 -- 140 & 140 -- 200 & 200 -- 500 \\
\yjl &  0 -- 0.5 & 0.5 -- 1 & 1 -- 1.5 & 1.5 -- 1.9 & 1.9 -- 2.3 & 2.3 -- 2.5 & 2.5 -- 4.4 \\
\yjsl & 0 -- 1.2 & 1.2 -- 2 & 2 -- 4.4 \\
\mjj $\,\,$ [GeV] &  0 -- 170 & 170 -- 500 & 500 -- 1500 \\
\deltayjj &  0 -- 2 & 2 -- 4 & 4 -- 8.8 \\
\dphijjabs & 0 -- $\frac{\pi}{3}$ & $\frac{\pi}{3}$ -- $\frac{2 \pi}{3}$ & $\frac{2 \pi}{3}$ -- $\pi$ \\
\dphijj & -- $\pi$ -- $--\frac{\pi}{2}$ &  $--\frac{\pi}{2}$ -- 0 & 0 -- $\frac{\pi}{2}$ &  $\frac{\pi}{2}$ -- $\pi$ \\
\ptggjj $\,\,$ [GeV] &  0 -- 15 & 15 -- 200  \\
\dphiggjj & 0 -- 3.01 & 3.01 -- 3.1 & 3.1 -- $\pi$ \\
\taujet $\,\,$ [GeV] & 0 -- 10 & 10 -- 20 & 20 -- 30 & 30 -- 40 & 40 -- 150 \\
\sumtaujet $\,\,$ [GeV]  & 0 -- 8 & 8 -- 17 & 17 -- 25 & 25 -- 40 & 40 -- 80 & 80 -- 150 \\
\ptgg [\costhetastar $<$ 0.5] $\,\,$ [GeV] & 0 -- 30 & 30 -- 120 & 120 -- 350 \\
\ptgg [0.5 $\leq$ \costhetastar $<$ 1.0] $\,\,$ [GeV] & 0 -- 30 & 30 -- 120 & 120 -- 350 \\
\ptgg [\njet=0] $\,\,$ [GeV] & 0 -- 15 & 15 -- 30 & 30 -- 75 & 75 -- 350 \\
\ptgg [\njet=1] $\,\,$ [GeV] & 0 -- 40 & 40 -- 60 & 60 -- 100 & 100 -- 350 \\
\ptgg [\njet=2] $\,\,$ [GeV] & 0 -- 100 & 100 -- 200 & 200 -- 350  \\
\ptgg [\njet$\geq$3] $\,\,$ [GeV] & 0 -- 200 & 200 -- 350 \\
\hline\hline
\end{tabular}
}
\caption{ 
 Bin ranges for each of the studied variables.
}
\label{tab:Bins}
\end{sidewaystable}


\section{Supplement to event categorization}\label{app:suppmaterial}

Table~\ref{tab:SB90_tableExp} summarizes the number of expected signal events and measured background events in the
smallest interval expected to contain 90\% of the expected SM signal events, together with the expected signal purity
and local significance in the same interval, for each of the event reconstruction categories. The definition of the categories can be found in 
Table~\ref{tab:cat-summary} in Section~\ref{sec:cat_summary}.

Table~\ref{tab:purity_table} summarizes the fractions of signal events from the different production modes expected in each reconstruction category, as illustrated in Figure~\ref{fig:purity1D}.

Table~\ref{tab:bkg_fct} summarizes the chosen background function used in each reconstruction category. 

\begin{table}[!tp]
  \begin{center}
    \caption{The effective signal mass resolutions $\sigma_{68}$ ($\sigma_{90}$) in GeV defined as half the width containing 68\% (90\%) of the signal events for listed for each reconstructed category. Further, the numbers of background events $B_{90}$, measured by fits to the data, in the smallest interval expected to contain 90\% of the SM signal events $S_{90}$ are given, accompanied by the expected purities $f_{90} \equiv S_{90}/(S_{90}+B_{90})$ and expected significances $Z_{90} \equiv \sqrt{ 2((S_{90}+B_{90})\log(1+S_{90}/B_{90}) - S_{90}) }$. \label{tab:SB90_tableExp}   }
\begin{tabular}{l|cc|cccc}
   \hline\hline
            Category  & $\sigma_{68}$ [GeV] & $\sigma_{90}$ [GeV] & $S_{90}$  &  $B_{90}$  &  $f_{90}$  &  $Z_{90}$  \\
   \hline\hline 
                  ttH lep 0fwd  &  1.7  &  3.0  &   0.93  &      3.6  &      0.21  &      0.47  \\
                  ttH lep 1fwd  &  1.7  &  3.0  &   0.99  &      1.9  &      0.34  &      0.67  \\
                       ttH lep  &  1.6  &  2.9  &    2.1  &      2.7  &      0.44  &      1.16  \\
                  ttH had BDT1  &  1.6  &  2.8  &    1.3  &      2.0  &      0.40  &      0.85  \\
                  ttH had BDT2  &  1.6  &  2.9  &    1.6  &      3.9  &      0.29  &      0.75  \\
                  ttH had BDT3  &  1.6  &  2.9  &   0.54  &      2.3  &      0.19  &      0.35  \\
                  ttH had BDT4  &  1.6  &  2.9  &    2.2  &     14.0  &      0.14  &      0.58  \\
                   tH had 4j1b  &  1.7  &  3.0  &    2.3  &       48  &      0.05  &      0.32  \\
                   tH had 4j2b  &  1.7  &  3.1  &   0.56  &      6.8  &      0.08  &      0.21  \\
                      VH dilep  &  1.7  &  3.0  &   0.84  &      1.1  &      0.43  &      0.72  \\
                   VH lep High  &  1.5  &  2.8  &    1.4  &      2.4  &      0.37  &      0.82  \\
                    VH lep Low  &  1.8  &  3.3  &    5.8  &       52  &      0.10  &      0.79  \\
                   VH MET High  &  1.6  &  2.8  &    1.2  &      2.3  &      0.34  &      0.72  \\
                    VH MET Low  &  1.8  &  3.3  &   0.56  &      3.4  &      0.14  &      0.30  \\
                       jet BSM  &  1.4  &  2.6  &     24  &      280  &      0.08  &      1.41  \\
                  VH had tight  &  1.5  &  2.8  &     11  &       47  &      0.19  &      1.55  \\
                  VH had loose  &  1.7  &  3.1  &     15  &      220  &      0.06  &      0.98  \\
  VBF tight, high $p_{T}^{Hjj}$ &  1.7  &  2.8  &     18  &      120  &      0.13  &      1.62  \\
  VBF loose, high $p_{T}^{Hjj}$ &  1.8  &  3.1  &     15  &      250  &      0.06  &      0.93  \\
   VBF tight, low $p_{T}^{Hjj}$ &  1.6  &  2.9  &     12  &       12  &      0.50  &      3.12  \\
   VBF loose, low $p_{T}^{Hjj}$ &  1.8  &  3.3  &     17  &      110  &      0.14  &      1.62  \\
                    ggH 2J BSM  &  1.4  &  2.6  &    6.8  &       26  &      0.21  &      1.29  \\
                   ggH 2J High  &  1.6  &  2.9  &     26  &      280  &      0.08  &      1.53  \\
                    ggH 2J Med  &  1.8  &  3.2  &     65  &     1700  &      0.04  &      1.56  \\
                    ggH 2J Low  &  1.9  &  3.4  &     73  &     3100  &      0.02  &      1.30  \\
                    ggH 1J BSM  &  1.4  &  2.6  &    2.0  &      7.1  &      0.22  &      0.72  \\
                   ggH 1J High  &  1.6  &  2.9  &     28  &      240  &      0.11  &      1.80  \\
                    ggH 1J Med  &  1.8  &  3.2  &    140  &     2900  &      0.05  &      2.61  \\
                    ggH 1J Low  &  1.9  &  3.4  &    260  &     8000  &      0.03  &      2.89  \\
                    ggH 0J Fwd  &  2.1  &  3.8  &    520  &    21000  &      0.02  &      3.62  \\
                    ggH 0J Cen  &  1.6  &  2.7  &    300  &     5300  &      0.05  &      4.07  \\
\hline\hline                    
\end{tabular}
\end{center}
\end{table}

\begin{table}[!tp]
\begin{center}
\caption{
Composition of the selected Higgs boson events, in terms of the different production modes, as expected for each reconstructed category. The total expected numbers of Higgs boson events are given in the column labeled $N_H$.
\label{tab:purity_table}
}
\begin{tabular}{r|c|ccccccccc}
\hline\hline
 & & \multicolumn{9}{c}{Composition [\%]} \\
                        Category &  $N_H$ & $\ggH$ & $\VBF$ &  $\WH$ &  $\ZH$ &   ggZH & $\ttH$ & $\bbH$ & $\tHqb$& $\tHW$ \\
\hline\hline

                     tH lep 0fwd &    1.0 &    4.1 &    0.2 &    5.6 &    2.2 &    0.6 &   75.7 &    0.9 &    8.2 &    2.5 \\
                     tH lep 1fwd &    1.1 &    1.8 &    0.2 &    1.4 &    0.8 &    0.2 &   79.4 &    0.2 &   13.5 &    2.6 \\
                         ttH lep &    2.4 &    --- &    --- &    0.2 &    0.1 &    --- &   96.0 &    0.1 &    1.0 &    2.6 \\
                    ttH had BDT1 &    1.4 &    1.2 &    0.1 &    0.1 &    0.5 &    0.2 &   95.0 &    0.1 &    0.7 &    2.1 \\
                    ttH had BDT2 &    1.8 &    3.6 &    0.3 &    0.8 &    1.2 &    0.4 &   89.3 &    0.2 &    1.8 &    2.4 \\
                    ttH had BDT3 &    0.6 &    3.5 &    0.5 &    1.0 &    2.0 &    1.1 &   86.1 &    0.5 &    3.1 &    2.2 \\
                    ttH had BDT4 &    2.5 &    7.0 &    0.8 &    1.4 &    2.7 &    1.7 &   79.4 &    0.3 &    4.3 &    2.4 \\
                     tH had 4j1b &    2.5 &   35.4 &    4.0 &    4.3 &    5.7 &    2.2 &   36.4 &    2.2 &    8.5 &    1.3 \\
                     tH had 4j2b &   0.62 &   23.8 &    2.8 &    1.6 &    9.8 &    3.6 &   39.0 &    8.3 &   10.5 &    0.6 \\
                        VH dilep &   0.93 &    --- &    --- &    --- &   76.9 &   18.9 &    4.0 &    --- &    --- &    0.2 \\
                     VH lep High &    1.5 &    0.2 &    --- &   76.2 &    3.5 &    1.2 &   16.4 &    --- &    1.2 &    1.3 \\
                      VH lep Low &    6.4 &   11.4 &    1.1 &   68.0 &    6.8 &    1.3 &    8.5 &    0.9 &    1.6 &    0.4 \\
                     VH MET High &    1.3 &    1.3 &    0.1 &   22.4 &   48.1 &   18.5 &    8.3 &    --- &    0.6 &    0.7 \\
                      VH MET Low &   0.62 &   11.9 &    0.4 &   23.4 &   48.0 &   15.2 &    0.5 &    0.3 &    0.2 &    --- \\
                         jet BSM &     27 &   59.9 &   25.8 &    5.9 &    3.3 &    1.1 &    3.0 &    0.1 &    0.6 &    0.2 \\
                    VH had tight &     12 &   52.4 &    3.5 &   23.8 &   13.5 &    4.4 &    1.9 &    0.1 &    0.2 &    0.1 \\
                    VH had loose &     16 &   67.3 &    4.9 &   14.6 &    8.8 &    2.2 &    1.6 &    0.4 &    0.3 &    0.1 \\
   VBF tight, high $p_{T}^{Hjj}$ &     20 &   46.9 &   48.3 &    1.2 &    0.7 &    0.6 &    0.8 &    0.3 &    1.2 &    --- \\
   VBF loose, high $p_{T}^{Hjj}$ &     17 &   69.9 &   23.8 &    2.2 &    1.3 &    0.6 &    0.9 &    0.8 &    0.6 &    --- \\
    VBF tight, low $p_{T}^{Hjj}$ &     14 &   13.0 &   86.7 &    0.1 &    0.1 &    --- &    --- &    --- &    0.1 &    --- \\
    VBF loose, low $p_{T}^{Hjj}$ &     19 &   32.5 &   66.6 &    0.3 &    0.2 &    0.1 &    --- &    0.2 &    0.1 &    --- \\
                      ggH 2J BSM &    7.5 &   76.1 &   10.3 &    4.9 &    2.8 &    1.8 &    3.0 &    0.2 &    0.6 &    0.2 \\
                     ggH 2J High &     29 &   75.8 &   12.8 &    4.8 &    2.6 &    1.3 &    2.0 &    0.1 &    0.4 &    0.1 \\
                      ggH 2J Med &     72 &   77.6 &   12.2 &    4.4 &    2.6 &    0.6 &    1.5 &    0.7 &    0.4 &    --- \\
                      ggH 2J Low &     81 &   79.1 &    9.5 &    4.5 &    2.9 &    0.3 &    1.1 &    2.3 &    0.3 &    --- \\
                      ggH 1J BSM &    2.2 &   72.4 &   16.9 &    6.0 &    2.7 &    1.5 &    0.3 &    --- &    0.1 &    --- \\
                     ggH 1J High &     32 &   76.0 &   17.5 &    3.4 &    1.9 &    0.8 &    0.1 &    0.3 &    --- &    --- \\
                      ggH 1J Med &    160 &   83.6 &   11.7 &    2.6 &    1.5 &    0.2 &    --- &    0.4 &    --- &    --- \\
                      ggH 1J Low &    290 &   90.5 &    5.7 &    1.7 &    0.9 &    --- &    --- &    1.1 &    --- &    --- \\
                      ggH 0J Fwd &    580 &   97.0 &    1.2 &    0.5 &    0.4 &    --- &    --- &    0.9 &    --- &    --- \\
                      ggH 0J Cen &    330 &   97.3 &    1.1 &    0.4 &    0.3 &    --- &    --- &    0.9 &    --- &    --- \\
\hline\hline                      
\end{tabular}
\end{center}
\end{table}

\begin{table}[!tp]
  \begin{center}
    \caption{ The used background functions are listed: Either a power law ($m_{\gamma\gamma}^\alpha$), exponential function of a first order polynomial ($e^{ m_{\gamma\gamma} \alpha}$) or a second order polynomial ($e^{ m_{\gamma\gamma} \alpha + m_{\gamma\gamma}^2 \beta}$) are used to describe the non-resonant diphoton background. \label{tab:bkg_fct} }
\begin{tabular}{l|c}
   \hline\hline
            Category  & Background function \\
   \hline\hline 
                  ttH lep 0fwd  & Power law \\
                  ttH lep 1fwd  & Power law \\
                       ttH lep  & Power law \\
                  ttH had BDT1  & Exponential of a first order polynomial \\
                  ttH had BDT2  & Exponential of a first order polynomial \\
                  ttH had BDT3  & Exponential of a first order polynomial \\
                  ttH had BDT4  & Exponential of a first order polynomial \\
                   tH had 4j1b  & Power law \\
                   tH had 4j2b  & Power law \\
                      VH dilep  & Power law \\
                   VH lep High  & Exponential of a first order polynomial \\
                    VH lep Low  & Exponential of a first order polynomial \\
                   VH MET High  & Exponential of a first order polynomial \\
                    VH MET Low  & Exponential of a first order polynomial \\
                       jet BSM  & Exponential of a first order polynomial \\
                  VH had tight  & Exponential of a first order polynomial \\
                  VH had loose  & Exponential of a first order polynomial \\
  VBF tight, high $p_{T}^{Hjj}$ & Exponential of a first order polynomial \\
  VBF loose, high $p_{T}^{Hjj}$ & Exponential of a first order polynomial \\
   VBF tight, low $p_{T}^{Hjj}$ & Exponential of a first order polynomial \\
   VBF loose, low $p_{T}^{Hjj}$ & Exponential of a first order polynomial \\
                    ggH 2J BSM  & Power law \\
                   ggH 2J High  & Power law \\
                    ggH 2J Med  & Exponential of a second order polynomial  \\
                    ggH 2J Low  & Exponential of a second order polynomial \\
                    ggH 1J BSM  & Exponential of a first order polynomial \\
                   ggH 1J High  & Power law \\
                    ggH 1J Med  & Exponential of a second order polynomial \\
                    ggH 1J Low  & Exponential of a second order polynomial \\
                    ggH 0J Fwd  & Exponential of a second order polynomial \\
                    ggH 0J Cen  & Exponential of a second order polynomial \\
\hline\hline                    
\end{tabular}
\end{center}
\end{table}

\clearpage

\section{Limits on $\mu_{\mathrm{ZH}}$ and $\mu_{\mathrm{WH}}$ using pseudo-experiments}\label{app:mu_WH_n_mu_ZH}

As discussed in Section~\ref{sec:sig_strength}, Table~\ref{tab:muLegacy_limits_WH_vs_ZH} shows the observed and expected limits for $\mu_{\mathrm{VH}}$, and separately for $\mu_{\mathrm{ZH}}$ and $\mu_{\mathrm{WH}}$, as obtained using the asmptotic approximation and ensembles of pseudo-experiments.

\begin{table}[!tp]
\begin{center}
  \caption{
Observed and expected 95\% CL limits for the signal strengths of the VH associated production processes.
The observed asymptotic limit on $\mu_{\mathrm{VH}}$ is compared to that obtained using an ensemble of pseudo-experiments (PEs). Separate observed limits obtained from toys are reported for $\mu_{\mathrm{ZH}}$ and $\mu_{\mathrm{WH}}$. These are shown for the background-only case ($\mu_i=0$), together with the $\pm 1~\sigma$ and $\pm 2~\sigma$ intervals.
}
\label{tab:muLegacy_limits_WH_vs_ZH}
\begin{tabular}{cccccccc}
\hline
\hline  
  Measurement          &   Observed &   Expected Limit    &   Expected Limit &   $+2\sigma$ &   $+1\sigma$ &   $-1\sigma$ &   $-2\sigma$ \\
                       &            &   ($\mu = 1$)   &  ($\mu = 0$) &              &              &              &              \\
\hline\hline
  $\mu_{\mathrm{VH}}$  &        2.3 &             2.5 &           1.5 &          3.1 &          2.2 &          1.1 &          0.8 \\
\hline
  $\mu_{\mathrm{VH}}$ (PE)  &        2.2 &              &           1.5 &          3.1 &          2.2 &          1.1 &          1.0 \\
  $\mu_{\mathrm{ZH}}$ (PE)  &        2.3 &              &           3.1 &          6.2 &          4.4 &          2.2 &          1.9 \\
  $\mu_{\mathrm{WH}}$ (PE)  &        4.5 &              &           2.7 &          4.9 &          3.8 &          1.8 &          1.4 \\
\hline
\hline
\end{tabular}
\end{center}
\end{table}


\section{Summary of couplings results}\label{app:summary_coup}

In this Appendix the expected and observed central values and uncertainties of signal strength measurements, production mode cross section measurements, and simplified template cross section measurements from Section~\ref{sec:results_coup} and Appendix~\ref{sec:appendix_MINSTXS} are summarized.

\subsection{Signal strengths}\label{app:summary_coup:mu}

Table~\ref{tab:mu_obs_vs_exp} summarizes the observed and expected signal strengths for inclusive production and for
various production modes.

\begin{table}[!tp]
  \begin{center}
    \caption{Observed and expected signal strengths for inclusive production and for various production modes. Uncertainties smaller than 0.5 (0.05) are displayed as 0 (0.0).}
    \label{tab:mu_obs_vs_exp}
    {\renewcommand{\arraystretch}{1.3}
    \begin{tabular}{l|clcr|clcr}
      \hline
      \hline
      Production mode & \multicolumn{4}{c|}{Observed $\mu$} & \multicolumn{4}{c}{Expected $\mu$} \\ 
      & Result & Stat & Exp & Theo & Result & Stat & Exp & Theo \\
      \hline\hline
      Inclusive & $0.99$ & $\ ^{+0.12}_{-0.12}$ & $\ ^{+0.06}_{-0.05}$ & $\ ^{+0.07}_{-0.05}$
                & $1.00$ & $\ ^{+0.12}_{-0.12}$ & $\ ^{+0.07}_{-0.06}$ & $\ ^{+0.07}_{-0.05}$ \\
      ggH       & $0.81$ & $\ ^{+0.16}_{-0.16}$ & $\ ^{+0.07}_{-0.06}$ & $\ ^{+0.07}_{-0.05}$
                & $1.00$ & $\ ^{+0.16}_{-0.17}$ & $\ ^{+0.08}_{-0.06}$ & $\ ^{+0.08}_{-0.06}$ \\
      VBF       & $2.0$ & $\ ^{+0.5}_{-0.5}$ & $\ ^{+0.3}_{-0.2}$ & $\ ^{+0.3}_{-0.2}$
                & $1.0$ & $\ ^{+0.4}_{-0.4}$ & $\ ^{+0.2}_{-0.1}$ & $\ ^{+0.2}_{-0.1}$ \\
      $VH$      & $0.7$ & $\ ^{+0.8}_{-0.8}$ & $\ ^{+0.2}_{-0.2}$ & $\ ^{+0.2}_{-0.1}$
                & $1.0$ & $\ ^{+0.8}_{-0.7}$ & $\ ^{+0.2}_{-0.2}$ & $\ ^{+0.1}_{-0.1}$ \\
      $ttH+tH$  & $0.5$ & $\ ^{+0.6}_{-0.5}$ & $\ ^{+0.1}_{-0.1}$ & $\ ^{+0.1}_{-0.0}$
                & $1.0$ & $\ ^{+0.7}_{-0.6}$ & $\ ^{+0.1}_{-0.1}$ & $\ ^{+0.2}_{-0.0}$ \\
      \hline
      \hline
    \end{tabular}
    }
  \end{center}
\end{table}

\subsection{Production mode cross sections}\label{app:summary_coup:xs}

Table~\ref{tab:xs_obs_vs_exp} summarizes the observed and expected cross sections times diphoton branching ratio for various production modes, in the fiducial region $|y_H| < 2.5$.

\begin{table}[!tp]
  \caption{Observed and expected cross sections times diphoton branching ratio for various production modes, in the fiducial region $|y_H| < 2.5$. Uncertainties smaller than 0.5 (0.05) are displayed as 0 (0.0).}
  \label{tab:xs_obs_vs_exp}
  \begin{center}
    {\renewcommand{\arraystretch}{1.3}
    \begin{tabular}{l|clcr|clcr}
      \hline
      \hline
      Production mode & \multicolumn{4}{c|}{Observed $\sigma \tbfhyy$ [fb]} & \multicolumn{4}{c}{Expected $\sigma \tbfhyy$ [fb]}\\
      $(|y_H|<2.5)$ & Result & Stat & Exp & Theo & Result & Stat & Exp & Theo \\
      \hline\hline
      ggH & $82$  & $\ ^{+16}_{-16}$ & $\ ^{+7}_{-6}$ & $\ ^{+5}_{-4}$ 
          & $102$ & $\ ^{+17}_{-17}$ & $\ ^{+8}_{-6}$ & $\ ^{+5}_{-4}$\\
      VBF & $17$  & $\ ^{+5}_{-4}$ & $\ ^{+2}_{-2}$ & $\ ^{+3}_{-2}$
          & $8$   & $\ ^{+3}_{-3}$ & $\ ^{+1}_{-1}$ & $\ ^{+2}_{-1}$ \\
      $VH$& $\phantom{0}3$ & $\ ^{+4}_{-3}$ & $\ ^{+1}_{-1}$ & $\ ^{+1}_{-0}$
          & $\phantom{0}5$ & $\ ^{+4}_{-3}$ & $\ ^{+1}_{-1}$ & $\ ^{+0}_{-0}$\\
      $ttH+tH$ & $\phantom{0}0.7$ & $\ ^{+0.8}_{-0.7}$ & $\ ^{+0.2}_{-0.1}$ & $\ ^{+0.2}_{-0.0}$
               & $\phantom{0}1.3$ & $\ ^{+0.9}_{-0.8}$ & $\ ^{+0.2}_{-0.1}$ & $\ ^{+0.3}_{-0.1}$\\
      \hline
      \hline
    \end{tabular}
    }
  \end{center}
\end{table}

\subsection{Simplified template cross sections}\label{app:summary_coup:stxs}

Table~\ref{tab:stxs_obs_vs_exp} summarizes the observed and expected simplified template cross sections times diphoton branching ratio, in the fiducial region $|y_H| < 2.5$.

\begin{table}[!tp]
  \caption{Observed and expected simplified template cross sections times diphoton branching ratio, in the fiducial region $|y_H| < 2.5$.}
  \label{tab:stxs_obs_vs_exp}
  \begin{center}
    {\renewcommand{\arraystretch}{1.3}
    \begin{tabular}{l|ccc|ccc}
      \hline
      \hline
      Simplified fiducial region & \multicolumn{3}{c|}{Observed $\sigma \tbfhyy$ [fb]} & \multicolumn{3}{c}{Expected $\sigma \tbfhyy$ [fb]}\\
      $(|y_H|<2.5)$ & Result & Stat & Syst & Result & Stat & Syst \\
      \hline\hline
      $ggH, \mathrm{0~jet}$ & $37$ & $\ ^{+14}_{-14}$ & $\ ^{+6}_{-5}$
                            & $63$ & $\ ^{+15}_{-15}$ & $\ ^{+8}_{-6}$ \\
      $ggH, \mathrm{1~jet}, \pT^{H} < 60$ \GeV & $13$ & $\ ^{+12}_{-12}$ & $\ ^{+5}_{-4}$
                                                & $15$ & $\ ^{+12}_{-12}$ & $\ ^{+6}_{-4}$ \\      
      $ggH, \mathrm{1~jet}, 60 \leq \pT^{H} < 120$ \GeV & $5$ & $\ ^{+6}_{-6}$ & $\ ^{+2}_{-1}$
                                                         & $10$ & $\ ^{+6}_{-6}$ & $\ ^{+2}_{-1}$ \\      
      $ggH, \mathrm{1~jet}, 120 \leq \pT^{H} < 200$ \GeV & $2.8$ & $\ ^{+1.6}_{-1.5}$ & $\ ^{+0.7}_{-0.5}$
                                                          & $1.7$ & $\ ^{+1.6}_{-1.6}$ & $\ ^{+0.5}_{-0.4}$ \\      
      $ggH, \geq 2~\mathrm{jet}$ & $20$ & $\ ^{+8}_{-8}$ & $\ ^{+4}_{-3}$
                                 & $11$ & $\ ^{+8}_{-8}$ & $\ ^{+3}_{-2}$ \\      
      $qq \rightarrow Hqq, \pT^{j} < 200$ \GeV & $15$ & $\ ^{+5}_{-5}$ & $\ ^{+3}_{-2}$
                                               & $10$ & $\ ^{+5}_{-5}$ & $\ ^{+2}_{-1}$ \\      
      $ggH + qq \rightarrow Hqq, \mathrm{BSM-like}$ & $2.0$ & $\ ^{+1.3}_{-1.3}$ & $\ ^{+0.6}_{-0.6}$
                                                    & $1.8$ & $\ ^{+1.3}_{-1.3}$ & $\ ^{+0.5}_{-0.5}$ \\      
      $\mathrm{VH, leptonic}$ & $0.7$ & $\ ^{+1.4}_{-1.2}$ & $\ ^{+0.4}_{-0.3}$
                              & $1.4$ & $\ ^{+1.3}_{-1.2}$ & $\ ^{+0.3}_{-0.3}$ \\      
      $ttH+tH$ & $0.7$ & $\ ^{+0.8}_{-0.7}$ & $\ ^{+0.2}_{-0.1}$
               & $1.3$ & $\ ^{+0.9}_{-0.8}$ & $\ ^{+0.3}_{-0.1}$ \\
      \hline
      \hline
    \end{tabular}
    }
  \end{center}
\end{table}

\subsection{Minimally merged simplified template cross sections}\label{app:summary_coup:stxsmin}

Table~\ref{tab:stxsmin_obs_vs_exp} summarizes the observed and expected minimally merged simplified template cross sections times diphoton branching ratio, in the fiducial region $|y_H| < 2.5$.

\begin{table}[!tp]
  \caption{Observed and expected simplified template cross sections times diphoton branching ratio, in the fiducial region $|y_H| < 2.5$.}
  \label{tab:stxsmin_obs_vs_exp}
  \begin{center}
    {\renewcommand{\arraystretch}{1.3}
    \begin{tabular}{l|ccc|ccc}
      \hline
      \hline
      Simplified fiducial region & \multicolumn{3}{c|}{Observed $\sigma \tbfhyy$ [fb]} & \multicolumn{3}{c}{Expected $\sigma \tbfhyy$ [fb]}\\
      $(|y_H|<2.5)$ & Result & Stat & Syst & Result & Stat & Syst \\
      \hline\hline
      $ggH, \mathrm{0~jet}$ & $38$ & $\ ^{+14}_{-14}$ & $\ ^{+6}_{-5}$
                            & $63$ & $\ ^{+15}_{-15}$ & $\ ^{+8}_{-6}$ \\
      $ggH, \mathrm{1~jet}, \pT^{H} < 60$ \GeV & $23$ & $\ ^{+13}_{-13}$ & $\ ^{+5}_{-4}$
                                                & $15$ & $\ ^{+12}_{-13}$ & $\ ^{+6}_{-4}$ \\      
      $ggH, \mathrm{1~jet}, 60 \leq \pT^{H} < 120$ \GeV & $11$ & $\ ^{+7}_{-7}$ & $\ ^{+3}_{-2}$
                                                         & $10$ & $\ ^{+8}_{-8}$ & $\ ^{+2}_{-2}$ \\      
      $ggH, \mathrm{1~jet}, 120 \leq \pT^{H} < 200$\GeV & $4.0$ & $\ ^{+1.8}_{-1.8}$ & $\ ^{+0.9}_{-0.6}$
                                                          & $1.7$ & $\ ^{+1.9}_{-1.8}$ & $\ ^{+0.6}_{-0.4}$ \\      
      $ggH, \mathrm{1~jet}, \pT^{H} \geq 200$ \GeV & $2.6$ & $\ ^{+1.3}_{-1.1}$ & $\ ^{+0.8}_{-0.5}$
                                                    & $0.4$ & $\ ^{+1.1}_{-0.9}$ & $\ ^{+0.5}_{-0.4}$ \\      
      $ggH, \geq 2~\mathrm{jet}, \pT^{H} < 60$ \GeV & $0$ & $\ ^{+8}_{-8}$ & $\ ^{+3}_{-2}$
                                                     & $3$ & $\ ^{+8}_{-8}$ & $\ ^{+4}_{-2}$ \\      
      $ggH, \geq 2~\mathrm{jet}, 60 \leq \pT^{H} < 120$ \GeV & $12$ & $\ ^{+7}_{-7}$ & $\ ^{+3}_{-2}$
                                                              &  $4$ & $\ ^{+7}_{-7}$ & $\ ^{+2}_{-1}$ \\      
      $ggH, \geq 2~\mathrm{jet}, 120 \leq \pT^{H} < 200$ \GeV & $7.9$ & $\ ^{+3.3}_{-3.3}$ & $\ ^{+1.1}_{-0.9}$
                                                               & $2.3$ & $\ ^{+3.4}_{-3.3}$ & $\ ^{+0.8}_{-0.7}$ \\      
      $ggH, \geq 2~\mathrm{jet}, \pT^{H} \geq 200$ \GeV & $2.6$ & $\ ^{+1.5}_{-1.4}$ & $\ ^{+0.6}_{-0.5}$
                                                         & $1.0$ & $\ ^{+1.4}_{-1.3}$ & $\ ^{+0.5}_{-0.4}$ \\      
      $ggH,\mathrm{VBF-like}$ & $6.2$ & $\ ^{+4.1}_{-4.1}$ & $\ ^{+1.2}_{-1.2}$
                              & $1.5$ & $\ ^{+3.9}_{-3.8}$ & $\ ^{+1.4}_{-1.0}$ \\      
      $qq \rightarrow Hqq, \mathrm{VBF-like}$ & $3.8$ & $\ ^{+2.2}_{-2.0}$ & $\ ^{+1.2}_{-1.2}$
                                              & $2.7$ & $\ ^{+2.0}_{-1.8}$ & $\ ^{+0.8}_{-0.5}$ \\      
      $qq \rightarrow Hqq, \mathrm{VH+Rest}$ & $-19$ & $\ ^{+21}_{-20}$ & $\ ^{+6}_{-7}$
                                             &   $8$ & $\ ^{+22}_{-21}$ & $\ ^{+6}_{-5}$ \\      
      $qq \rightarrow Hqq, \pt^{j} > 200$ \GeV & $-3.2$ & $\ ^{+1.7}_{-1.7}$ & $\ ^{+0.7}_{-0.9}$
                                               &  $0.5$ & $\ ^{+1.7}_{-1.7}$ & $\ ^{+0.6}_{-0.6}$ \\      
      $\VH\,\mathrm{leptonic}$ & $0.7$ & $\ ^{+1.4}_{-1.2}$ & $\ ^{+0.4}_{-0.3}$
                              & $1.4$ & $\ ^{+1.3}_{-1.2}$ & $\ ^{+0.3}_{-0.3}$ \\      
      $ttH+tH$ & $0.7$ & $\ ^{+0.8}_{-0.7}$ & $\ ^{+0.2}_{-0.1}$
               & $1.3$ & $\ ^{+0.9}_{-0.8}$ & $\ ^{+0.3}_{-0.1}$ \\
      \hline
      \hline
    \end{tabular}
    }
  \end{center}
\end{table}

\clearpage


\section{Observed and expected correlation maps}
\label{app:correlation}

\begin{figure}[ht!]
\centering
\includegraphics[width=.48\textwidth]{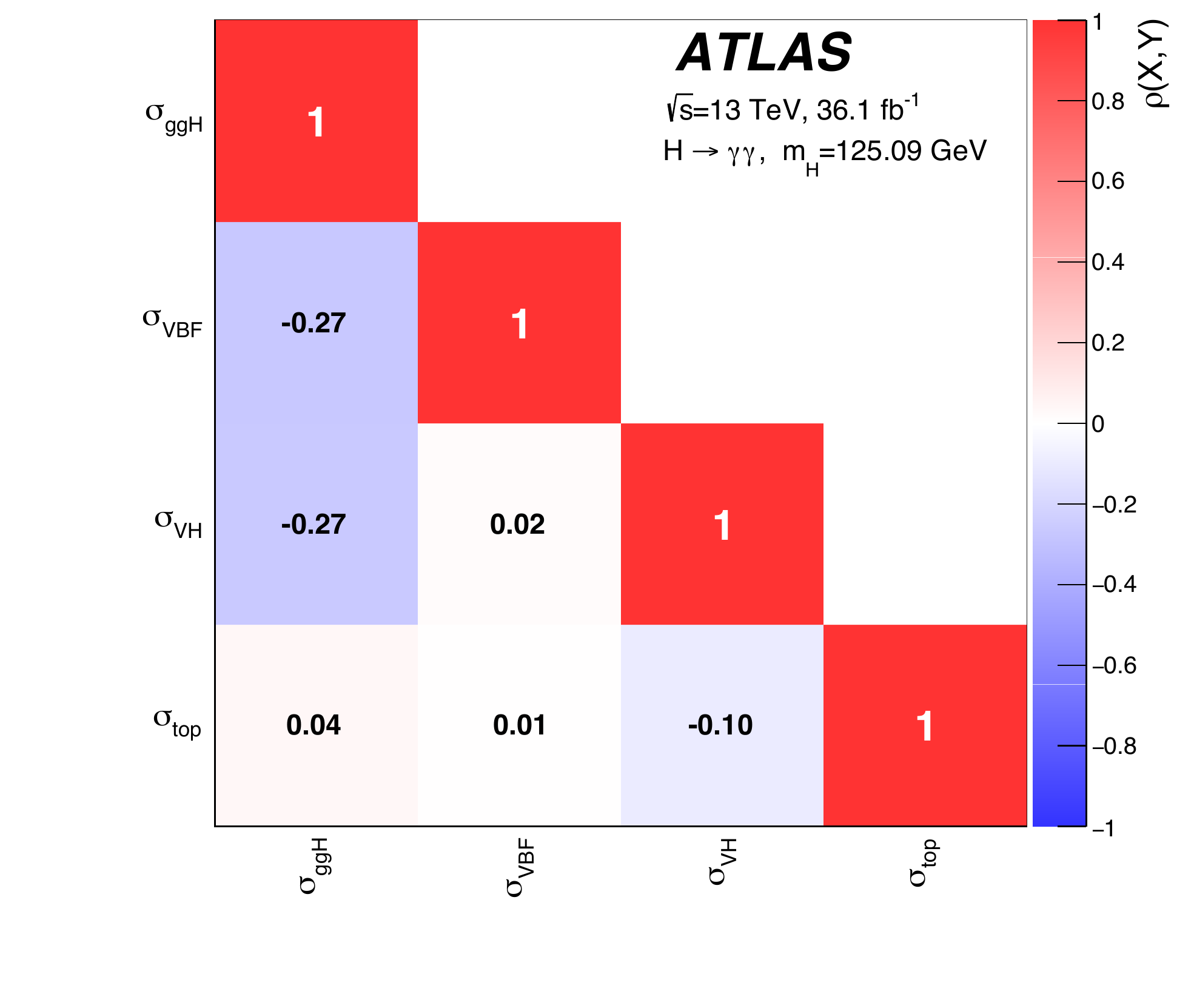}
\includegraphics[width=.48\textwidth]{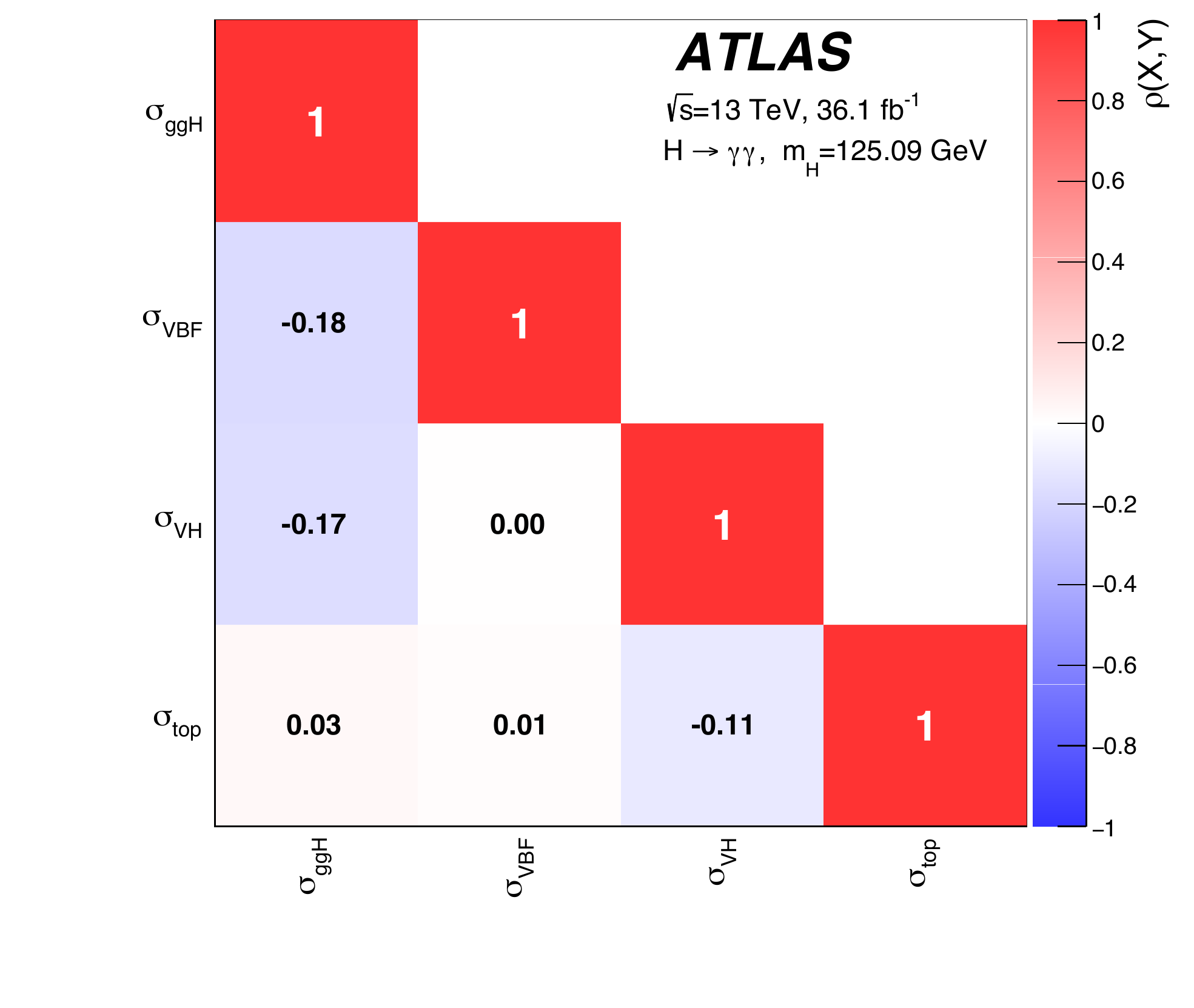}
\caption{Observed (left) and expected (right) correlations between the measured simplified template cross sections, including both the statistical and systematic uncertainties. The color indicates the size of the correlation. }
\label{fig:stxsCorr_prodXS}
\end{figure}

\begin{figure}[ht!]
\centering
\includegraphics[width=.48\textwidth]{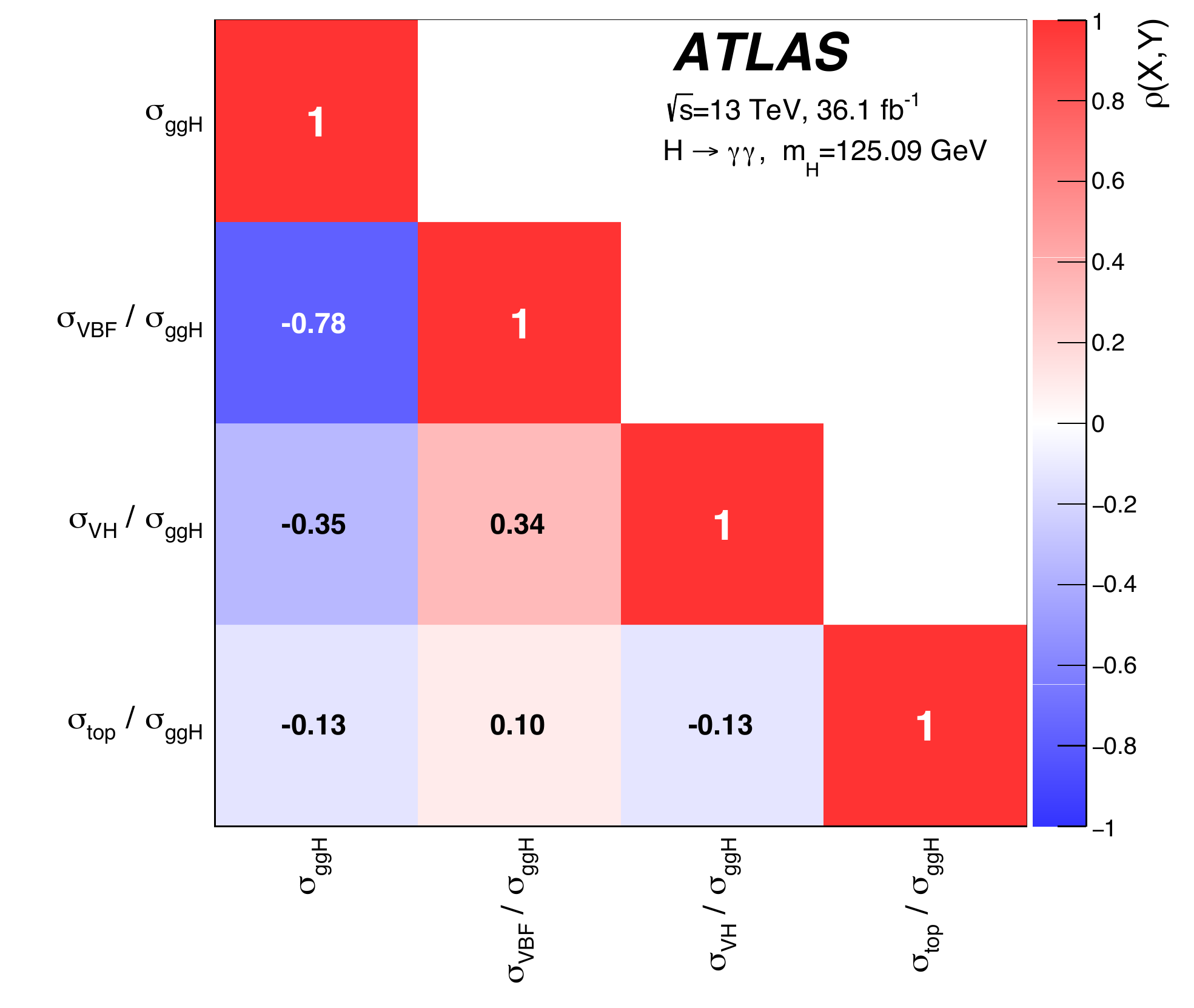}
\includegraphics[width=.48\textwidth]{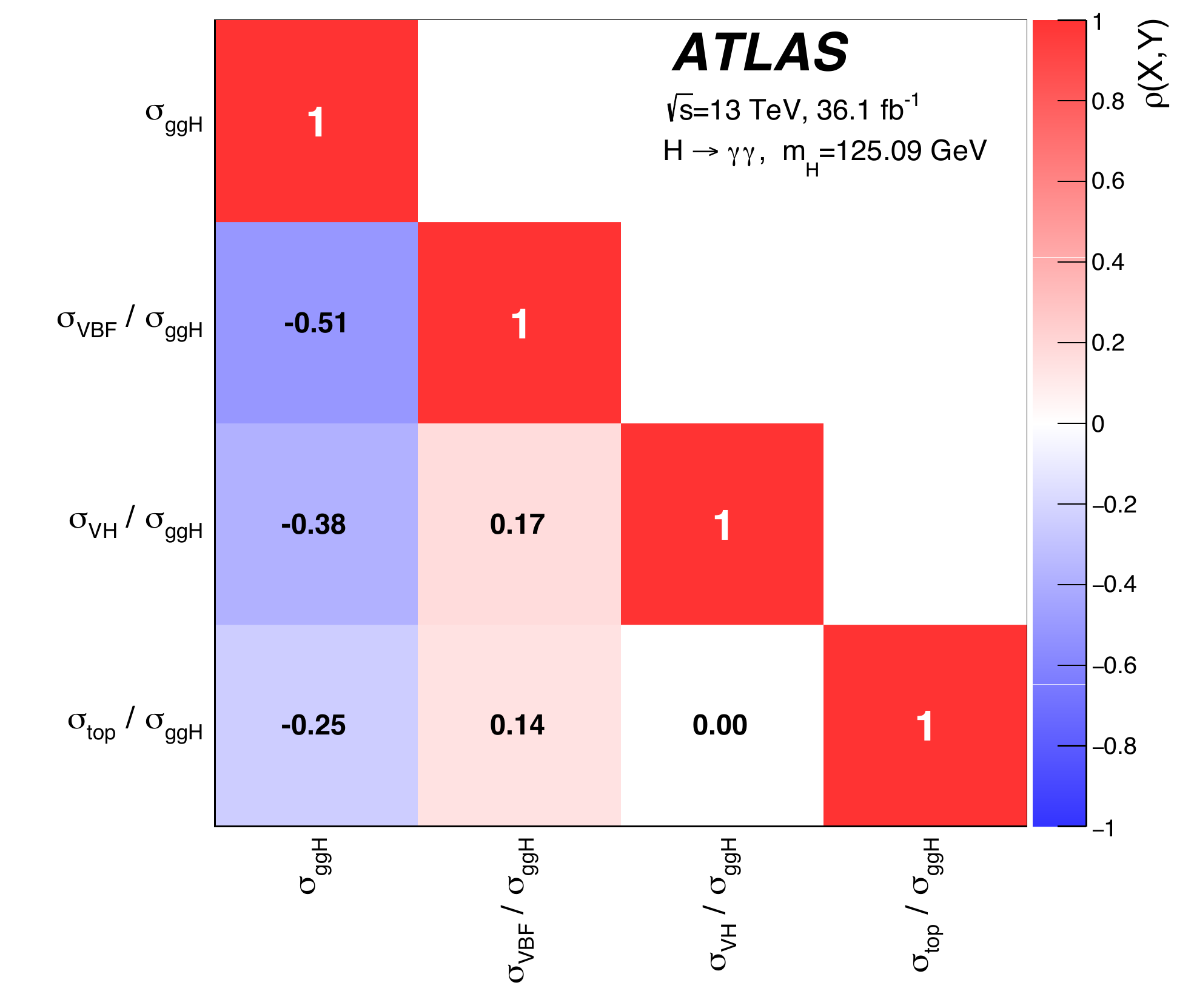}
\caption{Observed (left) and expected (right) correlations between the measured simplified template cross section ratios, including both the statistical and systematic uncertainties. The color indicates the size of the correlation. }
\label{fig:stxsCorr_XSratio}
\end{figure}

\begin{figure}[ht!]
\centering
\includegraphics[width=.75\textwidth]{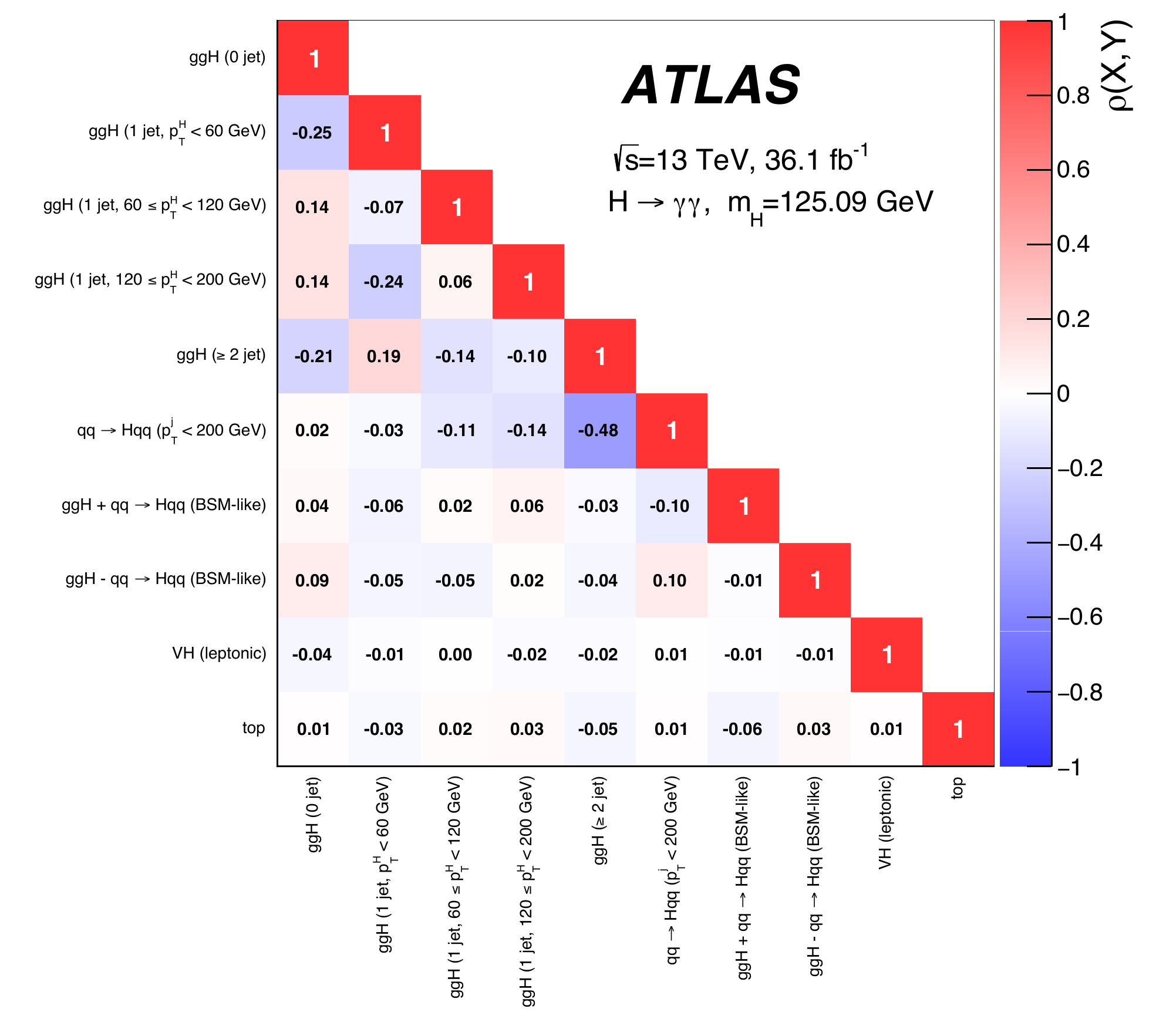}
\includegraphics[width=.75\textwidth]{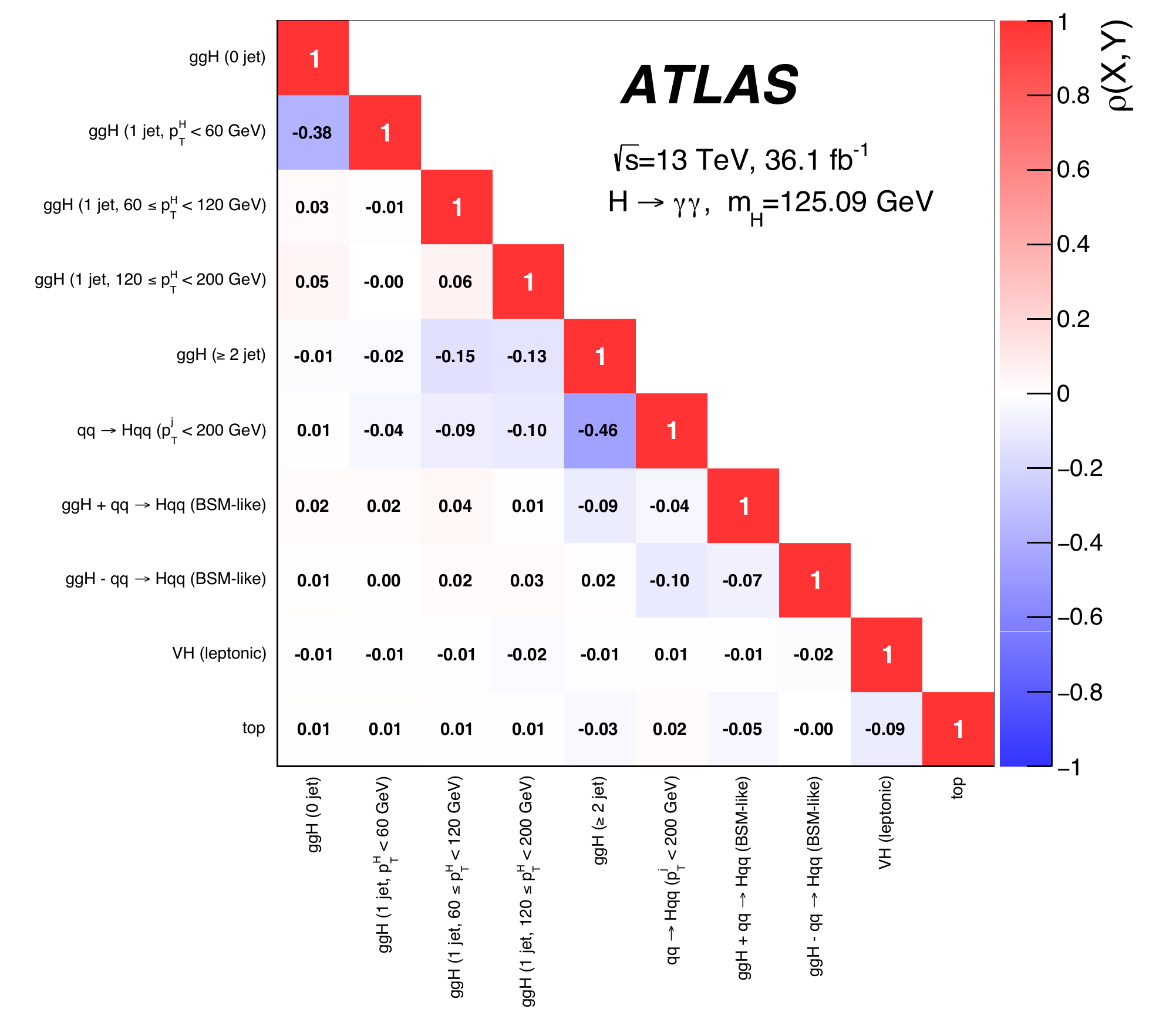}
\caption{Observed (top) and expected (bottom) correlations between the measured simplified template cross sections, including both the statistical and systematic uncertainties. The color indicates the size of the correlation. }
\label{fig:stxsCorr_STRONG}
\end{figure}

\begin{figure}[ht!]
\centering
\includegraphics[width=.75\textwidth]{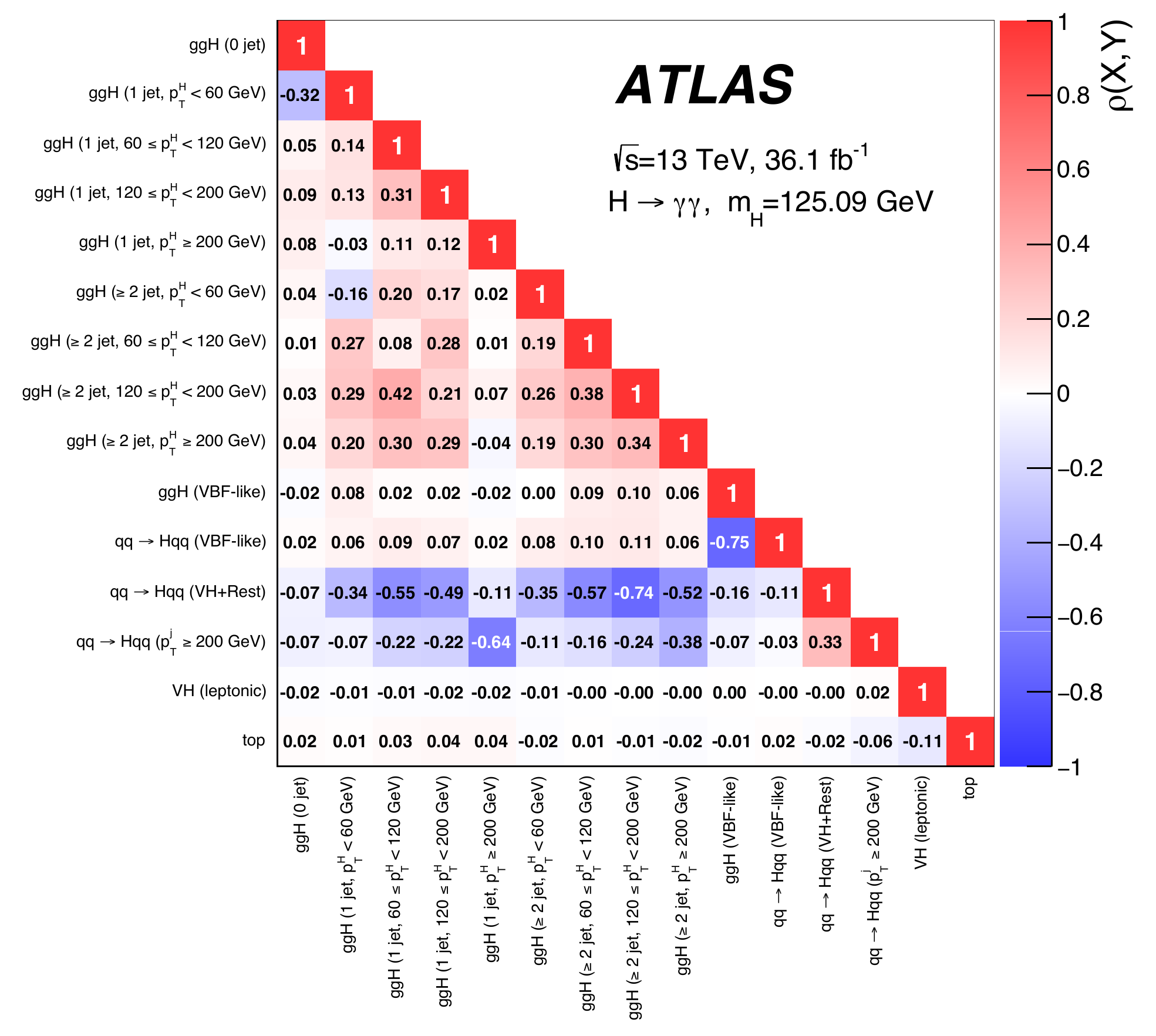}
\includegraphics[width=.75\textwidth]{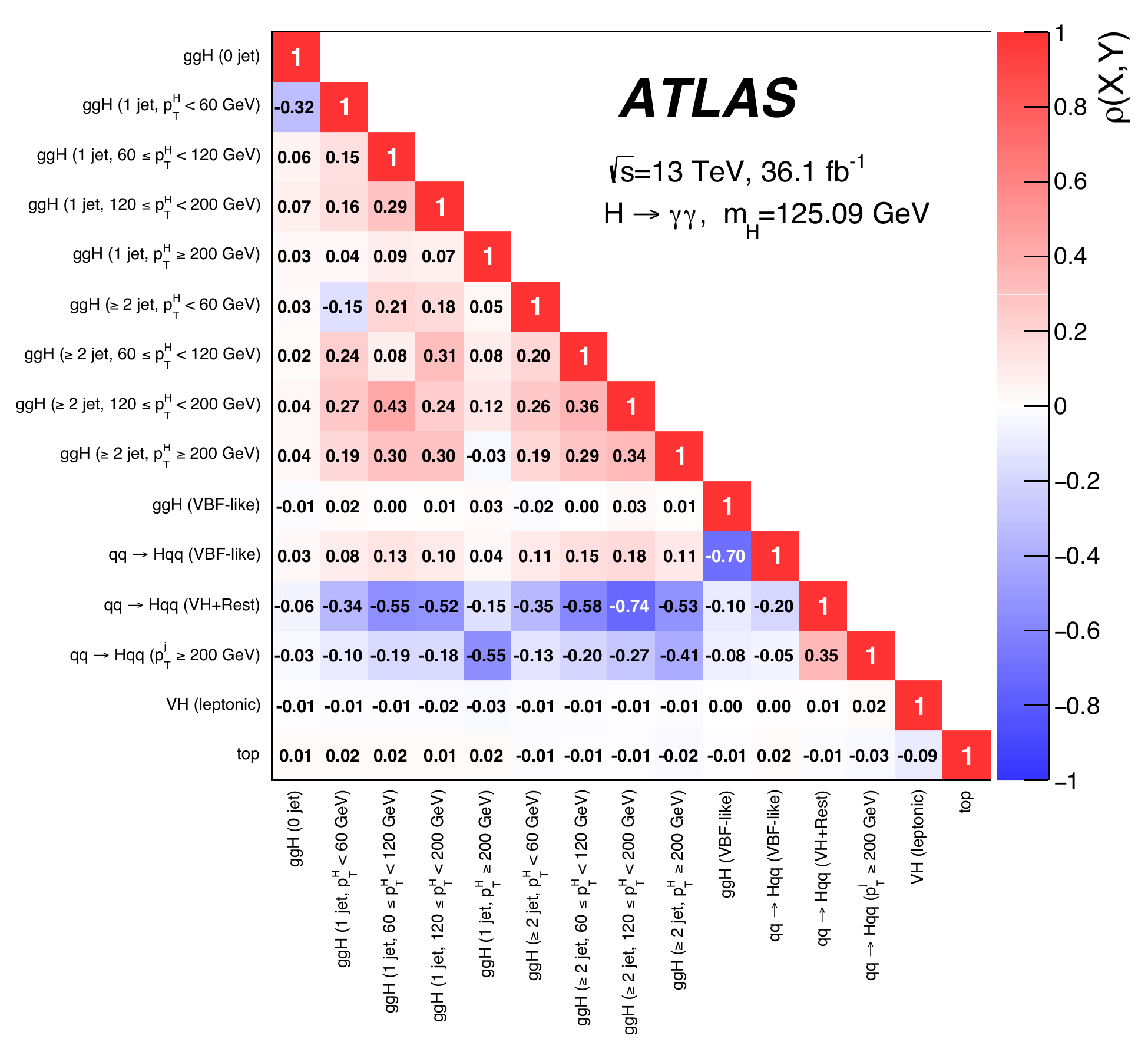}
\caption{Observed (top) and expected (bottom) correlations between the measured simplified template cross sections, including both the statistical and systematic uncertainties. The color indicates the size of the correlation. }
\label{fig:stxsCorr_WEAK}
\end{figure}

This appendix summaries the observed and expected correlations between the parameters of interest of each of the 
measurements presented in Section~\ref{sec:results_coup} are given. The observed and expected
correlations for the production-mode cross sections and production mode cross-section ratios
are shown in Figures~\ref{fig:stxsCorr_prodXS} and ~\ref{fig:stxsCorr_XSratio}.
The observed and expected correlations for the simplified template cross sections and minimally merged
simplified template cross sections are shown in Figures~\ref{fig:stxsCorr_STRONG} and \ref{fig:stxsCorr_WEAK}.

\clearpage

\printbibliography

\clearpage

\begin{flushleft}
{\Large The ATLAS Collaboration}

\bigskip

M.~Aaboud$^\textrm{\scriptsize 137d}$,
G.~Aad$^\textrm{\scriptsize 88}$,
B.~Abbott$^\textrm{\scriptsize 115}$,
O.~Abdinov$^\textrm{\scriptsize 12}$$^{,*}$,
B.~Abeloos$^\textrm{\scriptsize 119}$,
S.H.~Abidi$^\textrm{\scriptsize 161}$,
O.S.~AbouZeid$^\textrm{\scriptsize 139}$,
N.L.~Abraham$^\textrm{\scriptsize 151}$,
H.~Abramowicz$^\textrm{\scriptsize 155}$,
H.~Abreu$^\textrm{\scriptsize 154}$
% ...
\bigskip
\\
% ...
$^{12}$ Institute of Physics, Azerbaijan Academy of Sciences, Baku, Azerbaijan\\
% ...
$^{88}$ CPPM, Aix-Marseille Universit{\'e} and CNRS/IN2P3, Marseille, France\\
% ...
$^{115}$ Homer L. Dodge Department of Physics and Astronomy, University of Oklahoma, Norman OK, United States of America\\
% ...
$^{119}$ LAL, Univ. Paris-Sud, CNRS/IN2P3, Universit{\'e} Paris-Saclay, Orsay, France\\
% ...
$^{137}$ $^{(a)}$ Facult{\'e} des Sciences Ain Chock, R{\'e}seau Universitaire de Physique des Hautes Energies - Universit{\'e} Hassan II, Casablanca; $^{(b)}$ Centre National de l'Energie des Sciences Techniques Nucleaires, Rabat; $^{(c)}$ Facult{\'e} des Sciences Semlalia, Universit{\'e} Cadi Ayyad, LPHEA-Marrakech; $^{(d)}$ Facult{\'e} des Sciences, Universit{\'e} Mohamed Premier and LPTPM, Oujda; $^{(e)}$ Facult{\'e} des sciences, Universit{\'e} Mohammed V, Rabat, Morocco\\
% ...
$^{139}$ Santa Cruz Institute for Particle Physics, University of California Santa Cruz, Santa Cruz CA, United States of America\\
% ...
$^{151}$ Department of Physics and Astronomy, University of Sussex, Brighton, United Kingdom\\
% ...
$^{154}$ Department of Physics, Technion: Israel Institute of Technology, Haifa, Israel\\
$^{155}$ Raymond and Beverly Sackler School of Physics and Astronomy, Tel Aviv University, Tel Aviv, Israel\\
% ...
$^{161}$ Department of Physics, University of Toronto, Toronto ON, Canada\\
% ...
$^{a}$ Also at Department of Physics, King's College London, London, United Kingdom\\
$^{b}$ Also at Institute of Physics, Azerbaijan Academy of Sciences, Baku, Azerbaijan\\
$^{c}$ Also at Novosibirsk State University, Novosibirsk, Russia\\
$^{d}$ Also at TRIUMF, Vancouver BC, Canada\\
$^{e}$ Also at Department of Physics {\&} Astronomy, University of Louisville, Louisville, KY, United States of America\\
% ...
$^{*}$ Deceased
\end{flushleft}


\end{document}